\documentstyle [12pt] {article}
\def\lsim{\lower.5ex\hbox{$\; \buildrel < \over \sim \;$}}
\def\gsim{\lower.5ex\hbox{$\; \buildrel > \over \sim \;$}}
\newcommand{\ce}{{\cal E}}

\newcommand{\cm}{{\dot{\cal M}}}
\newcommand{\cmp}{{\dot{\cal M}}_+}
\newcommand{\cmm}{{\dot{\cal M}}_-}
\begin{document}
\centerline{\bf Contents}
\bigskip
\noindent{\bf \ \quad  Abstract  \quad\dotfill\quad  4}\\
\bigskip
\noindent{\bf 1:\quad Introduction \quad\dotfill\quad  5}\\
\noindent{\qquad 1.1 Some General Issues Related to Black Hole Astrophysics 
\quad\dotfill\quad 7}\\
\noindent{\qquad\qquad 1.1.1 Formation of Black Holes \quad\dotfill\quad 7}\\
\noindent{\qquad\qquad 1.1.2 Fueling Active Galactic Nuclei \quad\dotfill\quad 9}\\
\noindent{\qquad\qquad 1.1.3 Evolution of Galactic Centers \quad\dotfill\quad 12}\\
\noindent{\qquad\qquad 1.1.4 Black Holes in Galactic Halos? 
\quad\dotfill\quad 13}\\
\noindent{\qquad\qquad 1.1.5 Some Signatures of Black Holes \quad\dotfill\quad 14}\\
\noindent{\qquad 1.2 Difference between Motions around a Newtonian Star and a 
Black Hole \quad\dotfill\quad 16}\\
\noindent{\qquad 1.3 Basics of Pseudo-Newtonian Geometries \quad\dotfill\quad 21}\\
\noindent{\qquad 1.4 Remarks About Units and Dimensions \quad\dotfill\quad 23}\\
\medskip
\noindent{\bf 2:\quad Spherical Accretion \quad\dotfill\quad 24}\\
\noindent{\qquad 2.1 Bondi Accretion on a Newtonian Star \quad\dotfill\quad 25}\\
\noindent{\qquad\qquad 2.1.1 Basic Equations \quad\dotfill\quad 25}\\
\noindent{\qquad\qquad 2.1.2 Phase Space Behaviour of the Bondi Flow
\quad\dotfill\quad 28}\\
\noindent{\qquad 2.2 Bondi Flow on a Black Hole \quad\dotfill\quad 30}\\
\noindent{\qquad\qquad 2.2.1 In Schwarzschild Geometry\quad\dotfill\quad 30}\\
\noindent{\qquad\qquad 2.2.2 In pseudo-Newtonian Geometry
\quad\dotfill\quad 31}\\
\noindent{\qquad 2.3 Bondi Flow with Simple Radiative Transfer 
\quad\dotfill\quad 32}\\
\noindent{\qquad 2.4 Bondi Flow with General Radiative Transfer 
\quad\dotfill\quad 32}\\
\noindent{\qquad\qquad 2.4.1 Single Temperature Solutions \quad\dotfill\quad 34}\\
\noindent{\qquad\qquad 2.4.2 Two Temperature Solutions \quad\dotfill\quad 35}\\
\noindent{\qquad\qquad 2.4.3 General Relativistic Optically  Thick Bondi Flow
\quad\dotfill\quad 40}\\
\noindent{\qquad\qquad 2.4.4 Inclusion of Pair Production and Preheating
\quad\dotfill\quad 44}\\
\noindent{\qquad 2.5 Shocks in Spherical Flows \quad\dotfill\quad 47}\\
\noindent{\qquad 2.6 Particle Acceleration at Shock Waves \quad\dotfill\quad 48}\\
\noindent{\qquad\qquad 2.6.1 Basic Physical Processes \quad\dotfill\quad 48}\\
\noindent{\qquad\qquad 2.6.2 Acceleration at Spherical Shocks and AGN 
Spectra \quad\dotfill\quad 50}\\
\medskip
\noindent{\bf 3:\quad  Thin Accretion Disk Models \quad\dotfill\quad 55}\\
\noindent{\qquad 3.1 Standard Disk Model and Its Spectra \quad\dotfill\quad 56}\\
\noindent{\qquad\qquad  3.1.1 Model Equations \quad\dotfill\quad 56}\\
\noindent{\qquad\qquad  3.1.2 Structure of a Standard Disk \quad\dotfill\quad 61 }\\
\noindent{\qquad\qquad  3.1.3 Emitted radiation from a Standard Disk 
\quad\dotfill\quad 61}\\
\noindent{\qquad \qquad 3.1.4 Stability of a Thin Disk \quad\dotfill\quad 63}\\ 
\noindent{\qquad \qquad 3.1.5 Viscosity in a Thin Disk \quad\dotfill\quad 64}\\
\noindent{\qquad  3.2 Two Temperature Disk Models  \quad\dotfill\quad 65}\\
\noindent{\qquad  3.3 Transonic Disk Models  \quad\dotfill\quad 65}\\
\noindent{\qquad  3.4 Magnetized Disk Models  \quad\dotfill\quad 66}\\
\noindent{\qquad 3.5 Thin Disks with Weber-Devis Magnetic Field Configuration
\quad\dotfill\quad 69}\\
\noindent{\qquad \qquad 3.5.1 Basic Equations and Magnetosonic Point Conditions
\quad\dotfill\quad 69}\\
\noindent{\qquad \qquad 3.5.2 Solution Topologies of Magnetized Accretion
\quad\dotfill\quad 72}\\
\noindent{qquad 3.6 Outflows by Extraction of Energy from Black Holes 
\quad\dotfill\quad 73}\\
\medskip
\noindent{\bf 4:\quad Thick Accretion Disks \quad\dotfill\quad 75}\\
\noindent{\qquad 4.1 Introduction \quad\dotfill\quad 75}\\
\noindent{\qquad 4.2 Pseudo-Newtonian Thick Disks \quad\dotfill\quad 77}\\
\noindent{\qquad 4.3 General Relativistic Thick Disks \quad\dotfill\quad 78}\\
\noindent{\qquad 4.4 Thermodynamic Conditions inside a Thick Accretion Disk
\quad\dotfill\quad 79}\\
\noindent{\qquad 4.5 Emitted Radiation from a Thick Accretion Disk 
\quad\dotfill\quad  80 }\\
\noindent{\qquad 4.6 Ion Pressure Dominated Thick Disks
\quad\dotfill\quad 81}\\
\noindent{\qquad 4.7 Magnetized Thick Disks
\quad\dotfill\quad 82}\\
\noindent{\qquad 4.8 Thick Accretion Disk -- To Be or Not to Be?
\quad\dotfill\quad 83}\\
\medskip
\noindent{\bf 5:\quad Advective Disks
\quad\dotfill\quad 85}\\
\noindent{\qquad 5.1 Model Equations and Multiplicity of Critical Points
\quad\dotfill\quad 87}\\
\noindent{\qquad 5.2 Adiabatic Disks With Discontinuities
\quad\dotfill\quad 90}\\
\noindent{\qquad \qquad 5.2.1 Steady State Solutions \quad\dotfill\quad 90}\\
\noindent{\qquad \qquad 5.2.2 Sonic Point Conditions \quad\dotfill\quad 91}\\
\noindent{\qquad \qquad 5.2.3 When Does a Flow Contain Shocks?
 \quad\dotfill\quad 92}\\
\noindent{\qquad \qquad 5.2.4 Shock Conditions \quad\dotfill\quad 93}\\
\noindent{\qquad \qquad 5.2.5 Examples of Complete Solutions with 
Discontinuities \quad\dotfill\quad 94}\\
\noindent{\qquad 5.3 Isothermal Advective Flow with Viscosity
 \quad\dotfill\quad 95}\\
\noindent{\qquad 5.4 Most General Advective Flow and Unification
of Accretion Disk Models \quad\dotfill\quad 96}\\
\noindent{\qquad 5.5 `Advection Dominated' Disk Models
 \quad\dotfill\quad 99}\\
\noindent{\qquad 5.6 Axisymmetric Shock Waves in a Few Other Systems
\quad\dotfill\quad 100}\\
\noindent{\qquad \qquad 5.6.1 In Kerr Geometry \quad\dotfill\quad 100}\\
\noindent{\qquad \qquad 5.6.2 In Magnetized Flows \quad\dotfill\quad 100}\\
\noindent{\qquad 5.7 Non-Axisymmetric Spiral Shock Waves
\quad\dotfill\quad 101}\\
\noindent{\qquad \qquad 5.7.1 Introduction \quad\dotfill\quad 101}\\
\noindent{\qquad \qquad 5.7.2 Basic Flow Equations \quad\dotfill\quad 102}\\
\noindent{\qquad \qquad 5.7.3 Examples of Spiral Shocks in Accretion Disks
\quad\dotfill\quad 105}\\
\medskip
\noindent{\bf 6:\quad Numerical Simulations of Accretion Disks
\quad\dotfill\quad 106}\\
\noindent{\qquad  6.1 Introduction  \quad\dotfill\quad 106}\\
\noindent{\qquad  6.2 Simulation of Bondi-Hoyle-Littleton Accretion
\quad\dotfill\quad 106}\\
\noindent{\qquad  6.3 Simulations of Inviscid Flows  \quad\dotfill\quad 108}\\
\noindent{\qquad  6.4 Simulations of Viscous Disks  \quad\dotfill\quad 110}\\
\noindent{\qquad  6.5 Simulations of Flows using Radiative Transfer
\quad\dotfill\quad 115}\\
\medskip
\noindent{\bf 7:\quad Observational Signatures of Black Hole Accretion 
\quad\dotfill\quad 117}\\
\noindent{\qquad  7.1 Introduction \quad\dotfill\quad 117}\\
\noindent{\qquad  7.2 Continuum Emission  \quad\dotfill\quad 118}\\
\noindent{\qquad  7.3 Variability of AGN Spectrum \quad\dotfill\quad 123}\\
\noindent{\qquad  7.4 Correlated Variabilities \quad\dotfill\quad 124}\\
\noindent{\qquad  7.5 Variability of Disk Line Emission \quad\dotfill\quad 125}\\
\noindent{\qquad \qquad 7.5.1 Temporal Variation \quad\dotfill\quad 126}\\
\noindent{\qquad \qquad 7.5.2 Spatial Variation \quad\dotfill\quad 127}\\
\noindent{\qquad  7.6 Quasi-Periodic Oscilaltions \quad\dotfill\quad 129}\\
\noindent{\qquad  7.7 Our Galactic Center \quad\dotfill\quad 129}\\
\medskip
\noindent{\bf \quad 8 Concluding Remarks \quad\dotfill\quad 134}\\
\noindent{\bf \quad 9 Note Added in Proof \quad\dotfill\quad 138}\\
\noindent{\qquad 10 References \quad\dotfill\quad 140}\\
\noindent{\qquad 11 Figure Captions \quad\dotfill\quad 157-167}\\
\newpage
\centerline{ \Large \bf  Accretion Processes On a Black Hole}

\centerline{\large Sandip K. Chakrabarti}

\centerline{Tata Institute Of Fundamental Research, Bombay, 400005 INDIA}

\vspace{2cm}

\centerline{Abstract}

We describe astrophysical processes around a black hole
keeping primarily the physics of accretion in mind.
In \S 1, we briefly discuss the formation, evolution and detection of
black holes. We also discuss the difference of flow properties
around a black hole and a Newtonian star. In \S 2, we present past and 
present developments in the study of spherically accreting flows. We study the
properties of Bondi flow with and without radiative transfer.
In the presence of significant angular momentum, which is especially
true in a binary system, matter will be accreted as a thin Keplerian disk.
In \S 3, we discuss a large number of models of these disks
including the more popular standard disk model. We present
magnetized disk models as well. Since the
angular momentum is high in these systems, rotational motion is the most
dominant component compared to the radial or the vertical velocity 
components. In \S 4, we study thick disk models
which are of low angular momentum but still have no significant radial
motion. The accretion rates could be very high causing the flow to become
radiation dominated and the disk to be geometrically thick. 
For low accretion rates, ion pressure supported disks are formed.
In \S 5, we extensively discuss the properties
of transonic flows which has with sub-Keplerian angular momentum. 
In the absence of shock discontinuities, these sub-Keplerian flows are
basically advecting, similar to Bondi flows, close to the
black holes, though far away they match Keplerian or sub-Keplerian
disks. In presence of shocks, the 
post-shock flow becomes rotation dominated similar to thick disks. In \S 6, 
we present results of important numerical simulations of accretion flows.
Significant results from the studies of evolution of viscous 
transonic flows are reported. In \S 7, we discuss some
observational evidences of the black hole accretion. We also present
a detailed model of a generalized accretion disk and present its
spectra and compare with observations. 
In \S 8, we summarize the review and make concluding remarks.

\newpage
\noindent {\large\bf 1. Introduction}

Among all the celestial objects, black holes are the simplest
to describe mathematically. The simplest among the black 
holes, known as a Schwarzschild black hole, is characterized by its
mass $M$ alone, and has occupied a very popular place in models of 
active galaxies, quasars and some compact X-ray binaries. 
Along with the constant of gravitation, $G$ and the velocity of 
light $c$, one can obtain a length-scale $l_{bh}=GM/c^2=1.5 
\times 10^{14} M_9$ cm, and a time scale $t_{bh}=GM/c^3=5000 M_9$s 
($M_9$ is the mass of the 
central black hole in units of $10^9 M_\odot$). Physical quantities close to a
black hole are expected to vary in these scales. The Schwarzschild radius 
$r_{bh}=r_g=2 l_{bh}$ of a black hole is identified as the radius
of a non-rotating black hole while $r_{bh}=l_{bh}$ is the radius of a 
rotating black hole with maximally allowed specific angular momentum
$J=GM/c=4.5 \times 10^{24} M_9$ cm$^2$ s$^{-1}$ (otherwise known as the 
extreme Kerr black hole). Information within $r<r_{bh}$ remains inaccessible 
to the observers located at $r>r_{bh}$. 

Since black holes are `black', there cannot be any direct observational 
evidence of them. However, there are many indirect ways to `see' if black 
holes are present at  galactic centers or in binary systems. Most of these
ways depend upon the peculiarities of the behaviour of matter close to a black
hole. In the present review, we shall study these behaviours.
There is a recent review on the identification of the black holes in binary
systems [1], and therefore, we shall mostly concentrate on the 
studies around super-massive black holes, although mathematical 
discussions will remain valid for black holes of any mass. An earlier
review on the black hole models for active galaxies [2] 
will be relevant for our present study and we shall refer it often. Because 
matter accreted on a galactic center as well as at least
some matter to the compact component of a binary system may 
have very little angular momentum, a great deal of our investigations 
will be made in studying low-angular momentum accretion flows, which
include spherical accretion, thick disks and the transonic accretion disks.

Nuclei of some galaxies undergo violent activity,
quasars being the most extreme instances of this phenomenon. Such
activity is probably short-lived compared to the galactic lifetime, and
was probably most prevalent when the universe was only about one-fifth of its
present age. A massive black hole is the inevitable end point of 
such activity [3].
Rees [2] depicted a few plausible scenarios showing that nearly
all the galaxies which have undergone some `violent' activities
in the past may have black holes at their centers. In Fig. 1.1
we reproduce a figure from Rees [2] to show the possible evolutionary routes.
A characteristic luminosity which may be observed from such
activity is the `Eddington limit', where the inward gravitational pull
on protons balances the outward force of radiation on electrons,
$$
L_{Edd}=\frac{4\pi G M m_p c}{\sigma_T} = 1.3 \times 10^{47} M_9 
{\rm erg\ s^{-1}} ,
$$
where, $m_p$ is the mass of the proton and $\sigma_T$ is the Thomson
scattering cross-section. The luminosity of the quasars and active galaxies
have indeed been observed to be sometimes as high as $10^{47}$ erg s$^{-1}$ and 
the usual explanation for such a high energy output is that the energy 
is mostly coming from the gravitational binding energy of matter accreting
onto a massive black hole with $M\sim 10^9 M_\odot$.
Accretion disks very close to the black hole (which may be extending from 
regions just outside the horizon up to a distance of a thousand of Schwarzschild
radii) have not been observed yet, but the molecular and ionized disks at 
distances of $\sim$ parsec have been observed. There are several models of 
accretion flows which reproduce observed spectra of active galactic nuclei (AGN)
and quasars very well. This may therefore indicate that accretions do take place
on black holes residing in galactic centers. For instance, the characteristic
black body temperature of the radiation with luminosity $L_{Edd}$ 
at $l_{bh}$ is,
$$
T_E \sim 2.8 \times 10^5 M_9^{-1/4} \ K.
$$
Indeed, radiation around this temperature is observed from accretion flows. 
Another way is to measure the rotational velocities around the center. 
The Keplerian velocity 
of a circular orbit of radius $r$ (in units of $l_{bh}$) around a mass $M$ 
is $v_{Kep}=3 \times 10^5 r^{-1/2}$ km s$^{-1}$.
Most recent photographs of M87 by Hubble Space Telescope indicate 
rotational velocities of several 
hundred kilometers at distances of only $60$ light years from the galactic
center. If the motion is truly around a black hole, the mass of it must be
a few times $\sim 10^9 M_\odot$ at the center of M87 [4-6].
There are also indications that in the center of our own galaxy, a black hole
of several thousand solar mass to several million solar mass may be present
[7]. Possible presence of a massive black hole in the nucleus 
of the Andromeda nebula (M31) is indicated by the
high-resolution measurements of the rapid rotation and large 
velocity dispersion of the stars [8-9]. It is estimated that a black
hole of mass $10^7M_\odot$ may reside at the heart of M31.
From the periodic flares observed over last few decades, it is suggested that
two super-massive black holes may be orbiting each other in OJ287 [10].
Such systems may be formed due to mergers of super-giant galaxies.
These binaries can be long lived against gravitational radiation
with life time well above the Hubble time.

The region around a massive black hole at the
galactic center, with a deep gravitational potential well,
is a cite of intense activity. Stars are created from the gas,
and some gas is lost from the stars directly or through collisions. 
The gas and the stars slowly spiral their
way into the center. In the following Sections, we shall present the dynamic 
and radiative properties of the gas flows. We shall show that
depending on the boundary condition, namely the amount of matter supply,
the average angular momentum of the flow, presence or absence of 
turbulence and viscosity in the disk, the nature of the disk as well
as the radiation coming from it may be very much different.
For instance, if the matter is distributed 
spherically symmetrically, the accretion process will be
spherical Bondi accretion. If the angular momentum is quite high, very little
matter may accrete unless turbulent transport of angular momentum
is very efficient allowing matter to be accreted in the form of a
Keplerian disk. If the angular momentum is intermediate, the
the matter would accrete quasi-spherically at a large distance and would
slowly take the form of a disk at a distance of tens or hundreds of
Schwarzschild radius as the angular momentum becomes comparable to the 
local Keplerian value. 
The present status of the knowledge of the accretion  process indicates 
that disks in a binary system consisting no compact stars, 
may be primarily similar to the standard Keplerian type
[11, 12].
However, the accretion process close to a black hole either in compact 
binaries or in active galaxies need not be Keplerian everywhere and the 
study of sub-Keplerian matter requires more attention. In some situations, 
discussed in \S 5, the viscosity and cooling parameters are such that the disk 
will become completely sub-Keplerian much before entering the black hole.

The physics of the accretion process on a black hole
consists largely of our understanding of the accretion disks.
Accordingly, we shall devote several Sections of this review on this aspect,
discussing some of the major works which have provided directions in this 
subject. However, `accretion' does not always mean the accretion of gas. It 
could be another star, or even another black hole! All these types 
of accretions are, in principle, verifiable observationally. Not only
that, all these processes should cause the black hole
and its surroundings to evolve. Thus, the question of a black hole accretion
encompasses varied issues such as 
formation and evolution of compact binaries, quasars,
evidence of the existence of black holes, the supply of fuels to the 
central black holes, structure and stability and radiative properties of
accreting flows, merger of galaxies, merger of black holes, 
gravitational waves generated from black hole accretion, 
gravitational lensing, nucleosynthesis around black holes,
formation of radio jets from the surroundings of a black hole,
continuum and line emissions from an accretion disk,  etc. etc.
We cannot possibly cover all these issues reasonably in a single
review. In the rest of this introductory Section, we shall briefly discuss 
the formation and evolution of central black holes in quasars, question 
of the supply of fuel and some other issues relevant to the accretion physics
which we do not intend to discuss in detail in subsequent Sections. 
We then devote a few paragraphs to discuss the salient differences 
between the particle and fluid dynamics around a black hole and around a
Newtonian star. Many of our current understanding of black hole physics
is from using simple minded models of the 
black hole geometries, namely, using the pseudo-Newtonian potentials. 
These have become very useful tools in studying astrophysics around black holes
in recent days and we shall be frequently using them in subsequent Sections.

\noindent {\Large 1.1 Some General Issues Related to Black Hole Astrophysics}

\noindent {\large 1.1.1 Formation of Black Holes}

As shown in Fig. 1.1, the formation of the black holes
usually involve catastrophic gravitational collapse.
Stellar mass black holes form in the same way as the neutron stars,
namely, by core collapse and supernovae explosion. The population
of the black holes in low mass X-ray binaries could be
at least as high as the population of neutron stars, or probably
more. Progenitor stars more massive than $ \sim 40-50 M_\odot$
probably form black holes [13, 14].
Several numerical simulations have also been carried out 
to study the process of formation of black holes.
Numerical simulations of head-on collisions of relativistic
clusters show that gas spheres with particles initially
at rest or boosted towards each other implode towards their
centers and may form black holes before colliding. 
When particles in a sphere are in circular orbits about their respective
centers, the collisions lead to either coalescence and virialization or 
collapse to a black hole [15]. 
These examples however ignore any quantum effects such as
particle creations in curves spacetime (which may have
significant astrophysical implications!).

Recent discovery of large population of millisecond pulsars, and hence
neutron stars, in globular clusters implies that several hundred stellar 
black holes (of about $10M_\odot$) should form within a typical cluster. 
In clusters of high central density, the rapid dynamical evolution
will cause them to segregate to the cluster cores, where they form binary and
multiple black-hole systems. During the subsequent evolution,
most of the black holes are ejected on a relatively short time scale. A 
typical cluster may  retain  up to four black holes in its core, and 
possibly a few black holes in its halo. The presence of binary, triple and 
quadruple black-hole systems in cluster cores can disrupt main-sequence 
and giant stellar binaries which is a probable cause for the observed 
anomalies in the distribution of binaries in globular clusters [16]. 

Active galaxies refer to the galaxies which show anomalous
energy output within the inner parsecs of the galactic nuclei.
Occasionally, the star burst galaxies may also show activities resulting from
high formation rates of stars in inner few kilo parsecs. 
When two gas-rich galaxies collide, huge amount of matter may fall 
towards the central black holes which trigger starburst activity and 
a quasar formation [17].
A super massive black hole may also form in multiple steps:
massive black holes ($\geq 10^5M_\odot$) 
can first form in a quasi-spherical collapse of gas clouds which came out of
non-linear density perturbations above the Jeans mass after the
cosmological re-combination ($200 \leq (1+z) \leq 1400$). 
Angular momentum and energy may be extracted efficiently from the 
electrons through Compton friction with the cosmic background
radiation. The black hole is produced, surrounded by a 
dark-matter halo. Later, additional mass can accrete
into the system, causing activity of star formation [18].
Alternatively, dynamical evolution of compact star systems may end 
up in a mixture of gas and surviving stars which unavoidably collapse into
a giant black hole. The subsequent evolution of the remnant star system with a 
massive black hole at the center can lead to the formation of a bright 
central source in the nuclei of active galaxies and quasars. It may also 
lead to the opposite case of a `dead' frozen black hole in the nucleus of a 
normal galaxy [19].

Assuming a standard black hole accretion model for quasars [2], it is 
possible to estimate the present total mass 
density $\rho_{BH}$ of quasar remnants using observations of quasar 
populations and of typical quasar spectra over a wide wavelength range.
It is found that $\rho_{BH} \geq (1.4-2.2) \times 10^5 M_\odot/ 
Mpc^{-3}$ for a quasar radiative efficiency $\eta$ ($=\frac {L}
{{\dot M} c^2}$, where, $L$ is the luminosity and
${\dot M}$ is the accretion rate) of $0.1$ [20].
A typical bright galaxy is expected to contain a central
black hole of mass $\geq 10^7 M_\odot$ and
$ \geq  50 \%$ of $\rho_{BH}$ may be contributed by objects of 
mass $\geq 10^8 M_\odot$ and $\geq 10 \%$ by those more massive
than about $6 \times 10^8 h^{-2}M_\odot$. 
During the lifetime of a galaxy, the accretion rate itself
may gradually evolve. The evolution of the accretion rate could be
determined by making a very simplistic assumption that the accretion rate
is either demand-limited (i.e., radiation pressure controls the accretion 
rate) when matter is available or supply-driven when matter is in meager
amount [21].
In this scenario, a newly formed galaxy quickly grows or acquires a black hole
and accretes at a rate close to the Eddington limit. The subsequent 
accretion takes place intermittently at an average rate that is a universal
function of black hole mass and time. With these assumptions the current mass 
distribution of relict black holes in the nuclei of nearby galaxies can be
derived [21]. Here, the accretion rate scales as $M^{-1.5}t^{-6.7}$.
The number of relict black holes per decade declines only as $M^{-0.4}$
for black hole masses between $3 \times 10^7 M_\odot$ and $3 \times 
10^9 M_\odot$. In order to verify such a distribution of central masses more
accurate estimate of the galactic masses are to be carried out.

An understanding of the stellar rotation and velocity dispersion profiles 
close to a nucleus can give a hint of the nature of the central concentration.
Figure 1.2 shows velocity dispersions and stellar rotations along
the major axis of the Andromeda galaxy M31 and along the major axis of its
bulge component [8]. The velocity dispersion is as high as $245$km s$^{-1}$
and the mass to light ratio of $23$ to $35$ in this
case is much higher than that in bulges and normal elliptical 
galaxies. Indeed, the rotational velocity fits with a Keplerian
model if the mass of the central black hole is a few $ \times 10^7M_\odot$.
Not all the luminous galaxies may have active past, however. 
For example, rotational velocities and velocity dispersions along the major and
minor axes of the spiral galaxies NGC 2613, NGC 4699, NGC 5746, and
NGC 7331 have been recently measured and the 
stellar kinematics were computed using maximum entropy models.
It was found that for all four galaxies, models without central black 
holes and with constant mass to light ratio as a function of radius are
able to fit the observational data. Dynamical models that include a
central black hole of about $10^9M_\odot$ are able to fit the observations of
NGC 2613, NGC 4699, and NGC 5746 but do not fit the observations of NGC 7331. 
Only one out of these three candidates, namely NGC 4699, 
is found to be a relatively strong case to have central black hole [22].

\noindent {\large 1.1.2 Fueling Black Holes}

In low mass X-ray binary systems, matter to the compact object can be
supplied from the companion in two different ways. When the
companion fills its Roche lobe, matter with Keplerian angular momentum
is accreted through the Lagrange point (\S 3). If the companion has a 
significant wind, and the compact object moves through it, than some
winds with a very little angular momentum is also accreted. Thus,
fueling low mass black holes in binary systems (galactic black holes)
is not a problem.

The fueling of massive black holes supposedly residing
at the center of the active galaxies is an well recognized
problem. This is because the most luminous of them have luminosities 
$L \gsim 10^{47}$ erg s$^{-1}$ requiring fuel of about a few 
$M_\odot $ yr$^{-1}$. The fuel could
come from  lost gases from dense star clusters or from interstellar medium
or from in-falling intergalactic matter from outside the galaxy. 
On an average, the Population II stars lose $ \sim 0.01 M_\odot$ yr$^{-1}$
per $10^9 M_\odot$ stellar mass which may be concentrated 
within a region of $r\sim 10$ pc [23]. This is clearly insufficient
for AGNs. When the star density is high enough, they collide and
gases are lost in the process. Star-star collision [24, 25]
and winds from clusters may provide very efficient fueling.
Stars can also be tidally squeezed as they approach a black hole [26-28].
Disruption of a single star may produce flares and may be indicative of the 
presence of a black hole [29, 30].
Detailed study of the formation of short lived flares 
due to the tidal disruption of stars by the massive black holes at the 
center of the galaxy shows that the flares may last for a few months with
the peak luminosity in the extreme ultraviolet region [31]. Such 
flares should be detectable and would constrain the
number of the remnant black holes in galactic centers.
Numerical simulations have attempted to study the nature of debris formed 
around black holes due to tidal disruptions.
The strength of the tidal encounter is related to the square root
of the ratio $\eta$ between surface gravity and tidal acceleration 
at pericenter (so that for $\eta <1$, the star is disrupted)
or the ratio between the duration
of the encounter and the hydrodynamic time scale [32]. 
Through such encounter
a flare of the order of Eddington luminosity can last for a period of
about an year ($1 M_\odot$ around $10^6M_\odot$ for $\eta=1$), after which
it fades away [33]. Low luminosity Seyferts may be explained in this way. 
An interesting study of the fate of the bound debris is made by Rees [29].
Investigations of the tidal
break-up of solar-type stars near massive black holes have shown that debris 
gas could be distributed throughout the galactic nucleus. Apparently, about
half of the stellar debris mass remains strongly bound to the black hole.
This is quickly swallowed and the rest is expelled [29]. This matter can fuel
black holes in galactic nuclei during periods of 
enhanced tidal disruptions induced by galaxy mergers. It is also
suggested that the broad line regions may be produced by debris clouds [34].
However, a detailed study of the
tidal capture of stars by a massive black hole [35] 
shows that the maximum ratio of the cross-section for tidal capture to that
for tidal disruption is $\sim  3$ for realistic systems. In other words,
tidal captures are probably more important than tidal disruptions. Inclusion
of general
relativistic non-linear effects on the tidal capture of stars by a massive 
black hole, however, seems to show that the non-linear effects significantly
re-inforce the absorption of the orbital energy by the star, and results in
tidal disruptions far beyond the Roche limit [36].
Once the mass of the central body is higher than 
$10^{8-9} M_\odot$, the tidal disruption is negligible since the 
Schwarzschild radius of the black hole is larger compared to the
tidal radius [24, 25], although the limit on the central black hole mass 
may be raised by a factor of $10$ when fully general 
relativistic calculation around a Kerr black hole is considered [37].
A typical simulation sequence using Smoothed Particle Hydrodynamics (SPH) 
is presented in Fig. 1.3 [37] which shows how a star ($1\ M_\odot$) is 
gradually deformed while passing close to the 
Roche radius of a black hole ($10^4 M_\odot$). 
The disk was not seen to form in these simulations, probably because
a realistic steller model was not used.
In a three dimensional general relativistic numerical study of
tidal disruption of a main-sequence star ($M_*=1 M_\odot$) by a super-massive
black hole ($M_h=10^6 M_\odot$), it is seen that during a close passage, as
a result of relativistic effects analogous to the perihelion shift,
the trajectories of the debris of the star fan out into a
crescent-like shape centered on the black hole [38]. 
In another extensive numerical simulation 
[39], over $8\times 10^5$ encounters of binary systems with black 
holes (having masses that range from $10$ to $10^4$ times that of the 
binary components) were considered using various collisional velocities
and impact parameters. This provided estimates of the probability of the
black hole tidal disruption in close encounters that produce
exchange collisions. It turns out that the exchange collisions
dominate over tidal dissociation if the sum of the radii of the
stars is less than the binary semi-major axis. Occasionally,
exchange collisions take place in encounters of black holes
with soft binaries leading to black hole capture of one binary
component and ejection of the other at a high speed. Observation of these
high speed stars are of extreme importance in understanding the nature of 
the galactic nuclei. In the case of multiple mergers of galaxies, some of the 
black holes may escape from the galaxies with a high scattering velocity [40].

In the case of a steady gas supply, the stellar orbits diffuse to the lower
angular orbits and enter a small loss cone of semi angle $\theta_{crit}
\approx (t_{dyn}/t_{rlx})^{1/2}$ where, $t_{dyn}$ and $t_{rlx}$ are the
dynamical and relaxation time scales respectively and $t_{rlx} \sim
t_{dyn} N_c / ln N_c$, where $N_c$ is the number of stars in the cluster [23].
For an almost isotropic velocity distribution of stars the tidally disrupted
matter supply rate is [41, 42],
$$
{\dot M}_{tid} \sim \frac{N_c M_*}{t_{rlx} ln (2/\theta_{crit})} \sim \frac
{M_*}{t_{dyn}}.
$$
Here, $M_*$ is the average mass of a star in the cluster. Whereas this
rate is not sufficient for fueling AGNs,
the matter lost from the stellar collision can be sufficient.
Indeed, the maximum rate of gas production by collision is given by [23],
$$
{\dot M}_{col} \sim \frac{N_c}{t_{col}}\sim \frac{M_*}{t_{dyn}} (\frac{v_c}
{v_*})^4 \sim {\dot M}_{tid} (\frac{v_c}{v_*})^4.
$$
Here, $v_c$ is the relative velocity of the colliding stars
and $v_*$ is the escape velocity from the stellar surface.
The velocity $v_c$ being comparable to the Keplerian velocity,
accretion is possible when some source of viscosity is present,
or when the supersonic winds formed passes through a bow shock
and its velocity randomized before falling onto a black hole.
In general, this seems to be more interesting possibility
of supplying fuel than the other mechanisms, provided a dense star cluster
is assumed a priori at the center of the galaxy.

One may imagine that matter is supplied from the extended
thin accretion disk up to distances of a few ($\sim 10$) parsec.
However, at this distance the angular momentum transport by the viscous 
process involves a time scale of $t\sim 10^9 \alpha^{-1} T_{100}^{-1} 
v_{\phi,100} r_{10}$ yr, where $\alpha$ is the Shakura-Sunyaev
viscosity parameter (usually $<< 1$). Here,
$r$ is measured in units of $10$pc, $T$ is in units of $100$K,
and $v_\phi$ is in units of $100$km s$^{-1}$. This means that the time scale is
large compared to the galactic age. Also, the thin disks
could be gravitationally unstable when the number density $n$ exceeds a critical
density [23], $n >n_{crit} \sim 10^6 v_{\phi, 100}^2 r_{10}^{-2}$ cm$^{-3}$.
This limits the accretion rate to only ${\dot M}_{crit} \sim 3\alpha c_s^3/G
\sim 5 \times 10^{-4} \alpha T_{100}^{3/2} M_\odot$ yr$^{-1}$.
The temperature is usually below $100$K, and radiation from the inner edge
cannot keep it warm enough to increase the accretion rate further.
Thus, the accretion rate computed from the thin disk assumption may remain
insignificant [23]. Since a self-gravitating disk would
be a place where efficient star formation might occur, this gives all
the more reason why fueling through the stellar winds may be taking place
in active galactic nuclei [23]. This gas may remain very hot at regions 
far away ($> 1$kpc) due to heating by supernovae, or 
the central X-ray source [43, 44]. 
Cloud-cloud collision may also be a viable mechanism, though
gas supplied by this process need not be very high [23].

If the gas is supplied from a number of randomly placed stars, then only those 
matter directed virtually towards the direction of the central black hole 
would be accreting on it, since otherwise it would have very high 
angular momentum. The escaping winds from a star may most 
certainly collide with winds from another star
placed nearby, canceling its velocity components normal to the direction
of the black hole.
Thus, the actual angular momentum of matter falling onto the black hole could
be very small. This situation could be very similar to the idea
of Hoyle-Lyttleton [45] accretion of matter by a moving star in a 
gas medium. Because of low angular momentum, the flow is
quasi-spherical far away from the black hole (primarily advecting, i.e., 
$v_r >> v_\phi$)  and becomes `disk'-like (primarily rotating, 
i.e., $v_\phi >>v_r$) only when the angular momentum becomes 
comparable to the local Keplerian angular momentum. Even in a stellar
system both types of flows are possible (Section 5).

\noindent{\large 1.1.3 Evolution of Galactic Centers }

Extensive numerical simulation results are present in the literature to study 
the dynamical and luminosity evolution of an active galaxy consisting of dense 
stellar system surrounding a massive black hole. A power law initial mass 
function, tidal disruption by central black hole, 
physical collisions, as well as the stellar evolutions are included
in these simulations [46]. The general result is that, for low initial central 
stellar mass densities, stellar evolution and tidal deformations
are dominant, but for high stellar densities, the physical collisions dominate. 
A similar study of the joint evolution of a stellar galactic nucleus and 
central massive black hole [47] shows that the final outcomes are
differentiated by the distribution and motion of stars
near the black hole which determine the rate of their penetrating into the
`loss-cone' and successive tidal disruption or swallowing by the black
hole. The stellar cluster may either contract or expand by these
evolution process, depending on the parameters of the total system. 
A massive black hole at the center of the host
galaxy is found to exist either in an excited state as a powerful central
source or in the ground state as a `dead' quasar. 

In our galactic nucleus, it is believed that there is a black hole
not exceeding $5 \times 10^6  M_\odot$ [7]. If the
mass were that high, the luminosity resulting from the disruption
and swallowing rate would have exceeded the upper limit of the 
observed luminosity of $10^{40}$ erg s$^{-1}$ [24, 48].
By immersing a black hole (of mass $5\times 10^6 M_\odot$) 
in an isothermal extended stellar system with dispersion
velocity $200$ km s$^{-1}$, Sanders \& van Oosterom [49] finds that the 
final black hole mass and luminosity are fairly independent of the
central density. In a second simulation, a mass of $2 \times 10^7M_\odot$
was used with a stellar system as that of the nucleus of M31.
In both the simulations, a swallowing rate of one star per $10^4$yrs 
is found and the time averaged luminosity was found to be $100$
times greater, i.e., about $10^{42}$ erg s$^{-1}$. Since the 
accretion flow time scale (of around a hundred years) is much 
shorter than the time scale of disruption of stars ($10^4$yrs),
one would expect that this energy be released in 
short luminous bursts on the order of $10^{44}$ erg s$^{-1}$. This is
consistent with the belief that at least some of the Seyfert
luminosities are generated due to the tidal disruption of stars.

Another important study has been to find out the end products of
the merger of two elliptical galaxies [50].
It is observed that the central black holes sink toward the
core of the merger and form a binary system. 
The orbital energy and the angular momentum of the binary black holes
are seen to be transferred to the field stars through the dynamical friction
and black hole merger can take place in a time shorter than $10^7$yr.
In the final phase, the orbit of the binary becomes
highly eccentric as dynamical friction is most effective at the
apocenter with the minimum orbital velocity. The periastron distance
decreases exponentially, and eventually the black holes merge when
the emission of the gravitational radiation becomes significant. 

\noindent {\large 1.1.4 Black Holes in Galactic Halos?}

While we are discussing the possibility of existence of black holes
in general, a question arises naturally: are there black holes in the galactic
halos and/or are there black holes scattered throughout the galactic disk?
The composition of the dark halos and the
heating of the stellar disks are still a debatable issue. There are
many controversies and no firm conclusions. If, for instance, our galactic 
halo is composed of massive black holes with masses
of the order of $10^6M_\odot$, they should heat the stellar 
population and the dispersion velocity should steadily go up. 
Estimates ignoring dynamical friction
suggest that the predicted velocity dispersion is in
good agreement with observations [51]. 
However, in this case the dynamical friction would 
cause approximately $100$ such objects to settle towards the Galactic
Centre. The most likely outcome would be the accumulation of a central 
mass $M >> 10^6 M_\odot$, which is much larger than is what is
predicted from the present observational constraints.   
This consideration argues against the
hypothesis that the halo is made of $10^6 M_\odot$ black holes [52].

Dynamical friction may help explaining the evolution of
the central massive black hole or even the formation of
a dense cluster out of small black holes close to the center. 
When the compact remnants of massive stars are themselves significantly 
more massive than the normal field stars in the galactic center, they may 
migrate inwards as a consequence of dynamical friction [53]. The
resulting mass segregation may lead to a concentration of
compact objects inside the central stellar core within a Hubble time.
The total mass of remnants concentrated into the inner few tenths of
a parsec, $0.4-5 \times 10^6 M_\odot$, is comparable to the
dynamically inferred mass in this region in our galaxy.

Could there be black holes forming all over the galaxy before the completion 
of the galaxy formation? Ipser \& Price [54] argue that if that was the 
case then any such object passing through the interstellar gas of a galaxy
should be highly luminous due to heating of accreted gas. It is found that 
both spherical and disk accretion should occur, with about $95 \% $ of the 
pre-galactic black holes in the Galaxy probably undergoing essentially 
spherical accretion. In this type of accretions,
typical luminosities of the order  $10^{37}$ erg s$^{-1}$ in the
far-IR are predicted which is yet to be observed making the 
pregalactic-black-hole hypothesis  untenable.

One can put other constraints on the amount of dark matter in the
form of black holes in halos.
COBE results rule out hot intergalactic model for the X-ray background (XRB)
and the possibility that XRB is due to discrete sources emitting at high
redshifts of about $10$. Indeed, there are suggestions 
that XRBs should be explained by the accreting massive black holes [55].
However, one may put a constraint that the X-ray emission by gas accretion onto
black holes should not exceed the observed soft X-ray background. 
Secondly, metals produced in stellar processes that lead to
black hole formation should not exceed the observed disk metal abundance.
Based on these constraints, it appears unlikely that the missing disk
mass could be contained in black holes [56]. 
Another independent constraint is that the 
nearest black hole in the interstellar medium should not be directly
observable in optical [57]. For this, the halo black holes
must be lighter than $10^3M_\odot$ and the dark matter in the galactic disk
cannot have black holes more massive than $10M_\odot$. These black holes
may be remnants of population III stars [58].

\noindent {\large 1.1.5 Some Signatures of Black Holes}

Below, we present a few paragraphs describing some of the
signatures of black holes about which we shall not make any further 
comments in the rest of this review. The observational signatures 
of black hole accretions will be discussed in detail in \S 7.

Since the ellipticals are thought to be end products of merger events,
merger of binary black holes should produce
intense bursts of gravitational waves detectable from earth.
The mean time interval between two bursts of gravitational wave is found to be
$2-5$yr [59]. These waves are detectable in proposed future space based 
interferometers. Numerical estimates of the frequency of mergers of binary
neutron stars and black holes show that the gravitational
wave detectors now under development, may be able to detect up to
approximately $100$ neutron star mergers per year at distances $\lsim 200$
Mpc [60]. More conservative estimate shows that 
if one uses a single Laser Interferometer Gravitational-wave Observatory 
(LIGO)-like detector and $1.4 M_\odot$ neutron star binaries, rate of the
coalescing binaries is $\sim 8\times 10^{-8}$ yr$^{-1}$ Mpc$^{-3}$. 
This implies that LIGO will observe approximately $50$ yr$^{-1}$ binary 
in-spiral events [61].
The rate of detection of black hole mergers are supposed to be comparable to 
that of neutron star mergers because of the strong dependence of the power of
outbursts on the masses of merging objects. Black holes
in low-mass X-ray binaries (LMXBs)
may be more common than neutron star binaries in LMXBs [13] 
and thus there may be more events with binary black hole mergers.

In case of coalescence of binary neutron stars, the duration of 
gravitational waves is very short and the detection of such short
bursts could be difficult. In an alternative system, where a lighter
black hole (say, $10M_\odot$) orbits around a massive black hole
(say,  $10^{7-8} M_\odot$), in the last Schwarzschild radius outside the
horizon, several thousands of orbits are traversed and the
detection of gravitational waves is easier. During this process
the companion is bound to interact with the accretion disk
(which could be sub-Keplerian or super-Keplerian in different
regions) and exchange angular momentum with it. 
A more interesting (but probably extreme) situation may prevail, when
the loss of angular momentum  of the lighter companion is exactly
compensated for by its gain from the disk at some radius (such gain is
possible close to the black hole since the disks can have more
angular momentum than Keplerian there; see \S 4, \S 6). The system becomes a 
steady emitter of gravitational waves [62]. Extensive numerical simulation
has recently verified that such systems are stable [63]. Observation
of such systems would verify the current black hole models of galactic nuclei.

Some of the cosmic radio jets are known to propagate
virtually in the same direction for hundreds of millions of years.
They are considered to be scaled version of the bi-polar outflows in 
protostars, only more powerful [64]. It is natural to assume that
the source of the stability and the power of these jets are super-massive
black holes at the center of active galaxies. However, except in very 
few cases, one cannot really obtain any constraint on the nature of the 
accretion disk and the black hole properties from the jet features.
One suggestion is that the wiggles in jets in extragalactic radio sources could
be partly due to orbital motions of two massive black holes.
Supposing the wiggles in the milliarcsec radio jet of the superluminal quasar
1928+738 (4C 73.18) are due to this, the period and amplitude of the wiggles 
can be easily explained as due to the orbital motion of a binary black hole 
with mass of order $10^8 M_\odot$, mass ratio larger than $0.1$, and
orbital radius approximately $10^{16}$ cm. 
The small orbital radius suggests that the binary has been losing a
significant amount of orbital energy during the last $10^7$ yr,
possibly by interaction with the matter which is flowing through the
active galactic nucleus [62-63, 65].

For massive black holes in galactic centers, the radiation pressure dominated
disks are not hot enough ($T\sim T_E$) to produce significant 
nucleosynthesis. But if the flow is gas pressure dominated,
the temperature of the gas would be very high  ($T \sim$ virial 
temperature of the protons, $T_p \sim 1.04 \times 10^{13} r^{-1}\ K$ 
where $r$ is in units of $l_{bh}$) and nucleosynthesis in the accreting 
gas could be important. If black holes form shortly after decoupling, this 
process may even influence the element abundances in the universe [66].

In Seyfert galaxies, such as NGC 6814, Iron K$\alpha$ line is
produced very close to the black hole as is evidenced by the
fact that the continuum X-ray variability and the intensity of line emission
are found to be well correlated by a time lag of less than $250$s.
Due to the variation of light travel time at various distances from the
center, the changes in the double-horned line pattern would take place,
first in the wings (which are formed by the innermost
disk region), and then the blue and red horns. Observation of such
changes may allow determination of the mass of the central black hole [67].
However, it has been recently suggested that these lines could come from 
outflows, rather than inflowing accretion disks [68].
Direct measurement of the disk emission lines also allows mass and inclination
determination as in M87 [4-6].

One could also imagine ways to
detect massive black holes ($M\sim 10^6 M_\odot$) in the halo.
It has been suggested that radio maps from the very large baseline array (VLBA)
of two milli-arcsecond jets of a gravitationally lensed quasar will show  the
signature of massive black holes. If there are no compact objects in this 
mass range along the line of sight, the two jets should be linear mappings 
of each other. If they are not, there must be compact objects in the 
halo of the galaxy that deform the images by gravitational deflection [69]. 

We now present a general background on some of the basics of the
geometries around a Newtonian star and a black hole.

\noindent {\Large 1.2 Difference Between Motions around a Newtonian Star 
and a Black Hole}

Around a Newtonian star, the spacetime can be treated as flat and the
four distance between any two points could be written as:
$$
ds^2=-dt^2+dr^2+r^2(d\theta^2 + sin^2 \theta d\phi^2).
\eqno{(1.1)}
$$
Here we use the spherical polar coordinate. Also we adopt
units where the gravitational constant $G$, the
central mass $M$ and the velocity of light $c$ are all unity ($G=M=c=1$).
In the study around a black hole, chosen here to be rotating (Kerr
solution) for generality, the distance between any two coordinate points
can be obtained from [70]:
$$
ds^2=-(1-\frac{2r}{\Sigma})dt^2 - \frac{4ar sin^2 \theta }{\Sigma} dt d\phi
+ \frac{\Sigma}{\Delta} dr^2 + \Sigma d\theta^2 + (r^2 + a^2 + \frac{2 r a^2 
sin^2 \theta}{\Sigma} ) sin^2 \theta d\phi^2 ,
\eqno{(1.2)}
$$
where, $a$ is the Kerr parameter of a black hole,
$$
a=\frac{J}{M} {\ \ \rm and\ \ } \ \ \Delta=r^2+2r+a^2, \ \ 
\Sigma=r^2+ a^2 cos^2 \theta .
\eqno{(1.3)}
$$
The deviation from Newtonian distance becomes significant as one
approaches the black hole $r \rightarrow 1$.

A second difference of importance arises out of the nature of the 
effective potential which the flow experiences. 
The gravitational potential that a test particle around
a Newtonian star feels is given by,
$$
\Phi_N = -\frac{1}{r}.
\eqno{(1.4)}
$$
The effective potential of a rotating gas with specific angular
momentum $\lambda$ is obtained by the summation of the gravitational
potential and the centrifugal potential.
$$
V_{N} (r) = -\frac{1}{r} + \frac{1}{2} \frac{\lambda^2}{r^2} .
\eqno{(1.5)}
$$
For small $r$, the rotational term causes the potential to diverge to 
positive infinity at the origin at $r=0$. Accretion onto a
star of finite size $r_*$, requires that the barrier due to 
the centrifugal force is weaker compared to the gravitational term, i.e.,
$$
F=\frac{d V_{N} (r)}{dr} = \frac {\lambda^2}{r^3}  - \frac {1}{ r^2} <0.
\eqno{(1.6)}
$$
In other words, around a Newtonian star, only those matter with 
angular momentum less than the Keplerian value on the surface of the star
$$
l_{Kep,N}= r_*^{1/2} .
\eqno{(1.7)}
$$
can accrete.

In the case of the black hole accretion, the situation is quite
different. Here the effective potential cannot be
obtained by addition of components.
For simplicity, let us first consider the Schwarzschild solution 
($a=0$ in eq. 1.2) of the metric around a non-rotating black hole. 
In this case, the effective potential is given by [70, 71],
$$
V_{S} (r)  = [(1-\frac{2}{r}) (1+\frac{{\tilde l}^2}{r^2})]^{1/2} .
\eqno{(1.8)}
$$
$V_{S}$  becomes zero on the horizon, {\it independent} of the 
angular momentum of the flow.
The conserved specific angular momentum for the particle dynamics is ${\tilde
l} =u_\phi$. The angular momentum $l=-\frac{u_\phi}{u_t}$ is conserved
for fluid dynamics as well as the particle dynamics. Here, $(u_t, u_r, 
u_\theta, u_\phi)$ are the components of four velocity, $u_t$ is the conserved
specific energy. The potential experienced by the fluid is:
$$
V_S (r)=[\frac{1-2/r}{1+\frac{(1-2/r)l^2}{r^2}}]^{1/2} .
\eqno{(1.9)}
$$
Figure 1.4 shows the effective potential felt by a particle
orbiting around a Schwarzschild black hole and around a Newtonian star
for various specific angular momenta. The solid curves 
are for marginally stable ($l_{ms}= 3.674$, lower curve) and for
marginally bound ($l_{mb}=4$, upper curve) angular momenta respectively.
The short-dashed curves are for, from bottom to top, for $l=3.51$, $3.837$ 
and $4.163$ respectively and the long dashed curve is for Newtonian case with
$l=l_{ms}$. If the specific angular momentum
of the flow is less than that of the marginally stable value, the flow can 
fall onto the black hole, without any barrier whatsoever! If the angular 
momentum is higher than the marginally bound value, the potential barrier
is higher than $1$ -- the rest mass of the particle. This means that for
accretion to take place, matter must have significant energy to begin with.
Thus, matter must have significant radial velocity or thermal energy at a 
large distance. In any case, since the potential turns around and passes 
through zero, matter can always be made to accrete when `pushed' sufficiently
hard. Matter which has angular momentum between the marginally bound and
marginally stable values would form accretion disks.

As mentioned above, unlike in Newtonian geometry, the effective potential in 
general relativity is not obtained by addition of its `constituents'
(the quote mark is to indicate that it may not be meaningful
in general relativity to separate the constituents, such as the
potential energy, rotational energy... etc.). This is because various
energies in general relativity couple together to form new terms. 
For example, the above potential could be written as
$$
V_S(r) = [1-\frac{2}{r} + \frac{{\tilde l}^2}{r^2} -
\frac {2{\tilde l}^2}{r^3}]^{1/2} .
\eqno{(1.10)}
$$
The presence of the $1/r^3$ term, which can be thought of as the
coupling between the gravitational energy and the rotational energy 
or simply due to attraction of rotational `mass' by gravity of the central
object. These terms become important close to a black hole. For example,
$1/r^3$ term becomes very strong near the horizon and causes the reversal of the
effective potential to zero [Fig. 1.4] as one goes closer to the black hole. 
Thus, as the angular momentum is increased, its rotational mass and therefore 
the coupling term are also increased.
As a result, the net radial force is always attractive at the horizon [72]. 

In the case of a Kerr black hole, the effective potential energy of a particle
in terms of conserved particle angular momentum ${\tilde l}=u_\phi$ on the
equatorial plane is given by,
$$
V_K (r) = \omega {\tilde l} + (\alpha^2 + {\tilde l}^2 \alpha^4 /\Delta )^{1/2}.
\eqno{(1.11)}
$$
where, 
$$
\omega (r)= \frac {2a}{r^3+a^2r^2+2a^2r} ,
\eqno{(1.12)}
$$
is the rotational velocity of the zero angular momentum observer, and
$$
\alpha = \frac {r^2 + a^2 - 2r}{r^2 a^2 + 2a^2/r} 
\eqno{(1.13)}
$$
is the red-shift factor in the Kerr geometry and
$$
\Delta= r^2+a^2-2r .
\eqno{(1.14)}
$$
This potential also goes to zero on the horizon as seen from the
zero angular momentum observers. As the matter crosses the horizon,
its $\omega$ becomes the same as the rotational velocity of the black 
hole on the horizon.

In terms of the angular momentum which is conserved both for the
particles as well as for the fluids, $l$, this potential takes the form,
$$
V_K= \frac{\alpha}{ (1-\omega l)^2 + \frac{\alpha^4 l^2}{\Delta}} .
\eqno{(1.15)}
$$
The coupling term between the angular momentum of the black hole and 
the flow generates new force components [72, 73]
which are responsible for various effects, such as
the formation of shock waves at the outer distances in contra-rotating
flows or the reversal of the trend in ellipticity of collapsing spheroids.
Figure 1.5 shows the potential felt by a particle orbiting around
a Kerr black hole ($a=0.95$) for a range of specific angular momenta of the
particle. The solid curves are for the marginally stable ($l_{ms}=2.331$,
lower curve) and marginally bound ($l_{mb}=2.447$, upper curve) angular 
momenta respectively and the dashed curves are, from bottom to top, for 
$l=2.273$, $l=2.389$ and $l=2.505$ respectively. 

The amount of energy which the in-falling matter can dissipate during 
the accretion process depends upon the minimum of the effective potential.
In the case of a Newtonian star, this minimum may occur close to the axis well
within the star. On the surface of a star, the angular velocity coincides with 
the angular velocity of the star and the specific energy becomes,
$$
{\cal E}_s=\frac{1}{2}v_{\phi,*}^2  - \frac{1}{r_*}.
\eqno{(1.16)}
$$
If the accretion disk were Keplerian, without any radial velocity, the 
energy of the particle just before hitting the surface would be,
$$
{\cal E}_*=\frac{1}{2}v_{Kep,*}^2-\frac{1}{r_*} .
\eqno{(1.17)}
$$
But, 
$$
v_{Kep,*}=\frac{1}{{r_*}^{1/2}} .
\eqno{(1.18)}
$$
Thus,
$$
{\cal E}_*=-\frac{1}{2 r_*} .
\eqno{(1.19)}
$$
In other words, matter accreting through a Keplerian disk lose half of its
potential energy (calculated on the star surface) in the accretion disk 
itself. For example, for a star of radius $r_* \sim 10^{12}$cm, and a mass 
as that of our sun, energy loss would be $7.5 \times 10^{-6}\%$ 
which is very small indeed.
Within the boundary layer which is formed on the surface of the star
the amount of energy dissipated would be,
$$
{\cal E}_s - {\cal E}_* = \frac{v_{Kep,*}^2 - v_{\phi,*}^2}{2}  .
\eqno{(1.20)}
$$

On the contrary, the energy loss in accretion disks around black holes
could be much higher. The lowest of the minima occurs  at the
marginally stable orbits. These correspond to the point of inflection of the
effective potentials drawn for marginally bound angular momenta 
$V_{S}$ and $V_K$ (See, Figs. 1.4 and 1.5). In Schwarzschild 
geometry, this occurs at $r_{ms}=6GM/c^2$ and $V_{s, ms}\sim 0.94$.
This means that about $6$ per cent of the rest mass can be lost 
from the surface of the disk. 
In the case of a rotating black hole, the energy release could be 
higher. For example, for the rotation parameter $a=0.95$ (see Fig. 1.5),
the marginally stable orbit is at $r_{ms}=1.9372$, and correspondingly,
$V_{K,ms}=0.81$. Thus, about $19$ percent of the energy can be released in the
accretion disk! For $a=1$, $V_K(r_{ms}) \sim 0.6$, i.e., about $40$ percent 
energy could be released. Of course, the exact figure depends on the 
inner edge of the disk which may be located at positions other than
$r=r_{ms}$ due to pressure effects (\S 4).

In Newtonian dynamics, various components of the energy of matter can
be simply added up to obtain the total energy. Thus, for example, the 
conserved specific energy of a gram of in-falling rotating fluid 
would be given by,
$$
{\cal E} = \frac{1}{2} v_r^2 + \frac {1}{2} v_\phi^2 
-\frac{1}{r} +e+\frac{P}{\rho}
\eqno{(1.21)}
$$
where, $P$ and $\rho$ are the isotropic (total) pressure and density
respectively, $v_r$ and $v_\phi$ are the radial and azimuthal velocity
components, $e=\frac{P}{(\gamma-1) \rho}$ is the specific internal energy. 
This expression is obtained by integrating the radial momentum equation
in the absence of any dissipative processes. 

The conserved specific energy of the rotating flow
in a Kerr geometry, on the other hand, is given by ([70, 71])
$$
{\cal E} =h u_t
$$
where, $h$ is the specific enthalpy $\frac{P+\epsilon}{\rho}$ with
$\epsilon$ as the internal energy density. Due to non-linearity
in the expression of $u_t$ in curved space-time, this cannot be expressed
as a sum of different constituent energies.

Another important difference between the fluid dynamics around a black hole
and that around a Newtonian star is that,
as the flow accretes on a Newtonian star, it can directly hit the
surface of the star subsonically (when the radial velocity of matter
is $less$ than the adiabatic sound speed) or supersonically 
(when the radial velocity of matter is {\it greater} than the
adiabatic sound speed) depending on the location of the star surface.
But in the case of black hole accretion, the flow
is supersonic on the horizon since the velocity of sound
(even for the steepest equation of state) is less than the
flow velocity on the horizon, which is the velocity of light. What this 
implies is that a black hole accretion is necessarily transonic. This 
inner boundary condition has a very important implication
on the properties of the accretion flows. For instance, the relativistic
converging inflow passing through the inner sonic point will have sufficient
radial momentum to deposite on cooler photons in an optically thick flow
[74] and produce a characteristic hard spectra which are observed
from galactic black hole spectra in soft states [68]. This will not be 
possible for a neutron star accretion since in the latter case,
although the flow would be converging, the momentum deposition
would be negligible and would not be seen against the background noise.
We will discuss this difference in greater detail in later Sections,
particularly in \S 7.

\noindent{\Large 1.3 Basics of Pseudo-Newtonian Geometries}

In the case of most of the astrophysical systems involving
a rotating compact star or a black hole, it is not essential that one solves 
the problem using full general relativity. Fortunately, a few tools
are now available which allow one to use the Newtonian concepts
(such as Equations in flat geometry, additivity of the energy components ..
etc.) at the same time retaining the salient features of a black hole
geometry. As long as one is not interested in astrophysical
processes `extremely' close (within $1-2r_g$)
to a black hole horizon, one may safely
use these tools and obtain satisfactory results. The idea here is to
modify the gravitational potential in a manner that the derived force
`roughly' matches the  actual force in curved spacetime in a given
coordinate system. This method is significantly better than the
back-of-the-envelop calculations and easily accessible by
astrophysicists (who have no formal background in general relativity)
without sacrificing very much of physics.

Paczy\'nski and Wiita [75]
first suggested that for many practical purposes one need not use the
Schwarzschild solution and one could use a pseudo-Newtonian potential,
$$
\Phi_{PN}= 1-\frac{1}{r-2} 
\eqno{(1.22)}
$$
instead of the usual $-1/r$ potential to capture most of the physical 
properties of the black hole. (First term $1$ denotes the rest mass 
of the particle, and can be dropped for Newtonian description.)
Barring purists, the astrophysical community in general seems to 
appreciate this potential very much and a large number of
recent astrophysical literature, particularly dealing with accretion
disks use this potential [76-77].
The energy per unit gram of the flow can be simply expressed as:
$$
{\cal E} = \frac{1}{2} v_r^2 + \frac {1}{2} v_\phi^2 -\frac{1}{r-2}
+ \frac{\gamma}{\gamma-1}\frac{P}{\rho} ,
\eqno{(1.23)}
$$
which is same as eq. (1.21) but the gravitational
potential energy is taken from eq. (1.22).
The Keplerian distribution of angular momentum obtained by
equating the centrifugal force with the gravitational force,
$$
l_{Kep,PN}=\frac{r^{1/2}}{(1-2/r)} 
\eqno{(1.24)}
$$
is exactly same as in Schwarzschild geometry, provided one 
equates $l_{Kep, PN}$ with conserved angular momentum for the
flow $l_{Kep}=-u_\phi / u_t $ for the circular time-like
geodesics. Note however that this potential {\it does not} satisfy the
boundary condition on the horizon, i.e., the potential does not become
{\it zero} at $r=2$, but only at $r=3$. This becomes the effective
horizon, as if. However, accretion flows are supposed to be
so supersonic near the horizon (\S 5-6), and the infall time so short that they do 
not significantly radiate very close to the black hole, or even if they do,
the radiation is mostly red-shifted away and contribute very little
to the radiation that escapes the disk. Thus, these errors
close to the horizon should have very little effect on the emitted spectrum
from the disk, still one should be cautious against taking results
close to the horizon ($r\sim 2GM/c^2$) very seriously.

It is generally believed that the black holes in galactic centers may have 
some rotation. Even at the test particle level, 
solutions using the Kerr metric become difficult and cumbersome,
particularly when the particle is not confined on the equatorial plane.
Recently, an attempt has been made to simplify studies around a
rotating black hole by using a pseudo-Kerr potential:
$$
\Phi_k= 1 - \frac{1}{r-r_0} ,
\eqno{(1.25)}
$$
where, $r_0$ is in general a function of $a$, the Kerr parameter,
to be chosen in a manner that $r_{ms}$, $r_{mb}$ roughly agree
with actual Kerr results.
When the test particle has a specific angular momentum ${\tilde l}$,
the effective potential to describe it takes the form [73],
$$
\Phi=
1 - \frac{1}{r-r_0}+\frac{r_1a {\tilde l}}{r^3}+\frac{(1-2/r){\tilde l}^2}
{2r^2 sin^2 \theta},
\eqno{(1.26)}
$$
where, $r_1$ is a suitable `spin-orbit' coupling constant, which is in general
a function of $a$, the rotational parameter.
In eq. (1.26), the first term is clearly the rest mass of the particle. 
The second term describes the gravitational potential.
This term is singular at $r=r_0$. 
When the particle has angular momentum, the energy due to
coupling between the angular momentum $a$ of the black hole and 
that of the particle ${\tilde l}$ would be proportional 
to $a{\tilde l}/r^3$ as in the spin-orbit coupling term in a hydrogen atom.
This is the third term. Similarly, the rotational energy
of the particle can be described by $\frac{(1-2/r){\tilde l}^2}{2r^2}$. 
(A better result for smaller Kerr parameter is obtained by choosing
$\alpha^2$ in stead of the factor (1-2/r); see Chakrabarti \& Khanna [73].)
This appears as the fourth term in eq. (1.26). This potential mimics the 
geometry around a rotating black hole tolerably well (provided $r_0$ is 
carefully chosen for each a so as to reproduce the marginally stable and
marginally bound orbits well.).

Since there already exists an exact form of the potential in Kerr geometry,
($V_K$, eq. 1.11)
one may imagine it superfluous to `construct' another potential
which mimics the actual potential! However, if one wants to describe
the flow using Newtonian concept, one needs a potential whose
derivative is directly equated to the force (unlike in general relativity
where the derivation of force computed from the potential is tricky). 
$V_K$ does not represent such a potential. Hence a pseudo-Kerr potential 
is helpful.

An important usage of these potentials is that one can study more complex 
problems such as an ensemble of black holes, or numerical hydrodynamics,
or magnetohydrodynamics around a black hole practically as simply as in 
Newtonian spacetime. For instance, a numerical code written for a Newtonian 
flat geometry, could be upgraded to study flows in a black hole
geometry by just changing the potential and the force in relevant
equations. One must use these potentials with some caution, however. Since
they are not directly derivable from Einstein equations, they are not exact.
They could be used to obtain more accurate `correction' terms over 
and above the Newtonian results to study physical processes much 
closer to the black hole than what a usual post-Newtonian
approximation would provide. Any radically new discovery using these 
potentials must be checked with exact theory. One of the problems, for example,
which {\it cannot} be studied using the above potential is the evolution of 
the angular momentum of a rotating black hole in a galactic center.
This is because, for a fixed function $r_0$, the variation of the salient
features with $a$, do not follow exactly as in Kerr solution [73].
So, problems which involve evolution of $a$ should not be studied 
with these potentials. The choice of the potential is not unique, and
different choices may yield different results depending on the
application. The best way is, of course, to use the exact Kerr geometry
if at all possible.

\noindent {\Large 1.4 Remarks About Units and Dimensions}

In a broad review such as the present one, 
results of different workers in the field are compiled
and often the figures (with various symbols) are borrowed. It is 
therefore difficult to keep the
choice of units same in all the Sections. We have made efforts
to choose units as uniform as possible. In any case, in
each Section, the units used are spelled out so as to avoid confusion.

In general, the gravitational units $G=c=M =1$ would be chosen, where
$G$ is the gravitational constant, $c$ is the velocity of light
and $M$ is the mass of the black hole. Thus the unit of 
velocity would be $c$, the unit of distance would be $GM/c^2$,
the unit of time would be $GM/c^3$ and the unit of angular momentum
would be $GM/c$. However, sometimes, we may choose $2GM/c^2$, $2GM/c^3$
and $2GM/c$ as the unit of length, time and specific angular momentum.
Sometimes we may find it instructive to bring back  $G, \ M, \ $ 
and $c$ in the equations so as to see the explicit dependence on 
these quantities. So, we will mix these approaches when 
there is no confusion. Readers can easily
check the units used as well. For instance, if the
pseudo-Newtonian potential is written as $-1/2\, (r-1)^{-1}$ then the length
unit is $2GM/c^2$, but when it is written as $-(r-2)^{-1}$ then the
length unit is $GM/c^2$.

Yet another cause for confusion is that we shall use $M$ to represent the
Mach number of the flow, particularly in \S\S 2,5 \& 6. However, context
will be made sufficiently clear.

\newpage
\noindent{\large\bf 2 Spherical Accretion }

The process of capture of matter by a gravitating object is called
accretion. The rate at which matter is accreted is known as accretion
rate (usually denoted by ${\dot M}$ with units gm s$^{-1}$). 
The luminosity $L$ emitted is given by,
$$
L=\eta {\dot M} c^2 \ \ \ {\rm erg \ s}^{-1} \ \ 
\eqno{(2.1)}
$$
where, $\eta$ is the efficiency of conversion of matter into energy and $c$ is
the velocity of light. For a spherically symmetric Newtonian star of 
$M=1 M_\odot$, the quantity
$$
L_{Edd}=\frac{4 \pi G M m_p c}{\sigma_T}= 1.3 \times 10^{38} \ \ \
{\rm erg\ s}^{-1}
\eqno{(2.2)}
$$
known as the Eddington luminosity, is a very useful benchmark for luminosity.
It is obtained by equating the outward force exerted on electrons
by radiation due to deposition of momentum through Thomson scattering 
with the inward gravitational force on the protons, which are coupled to these
electrons via coulomb force. In the above equation, $\sigma_T$ is the
Thomson scattering cross-section, $m_p$ is the mass of the proton, $G$
and $M$ are the gravitational constant and the mass of the central
star respectively. Though it is a Newtonian concept and the
computation is done using spherical geometry, the definition
is generally used unchanged in measuring luminosity of accreting
matter around a black hole.

In the present Section, we shall be interested in the steady behaviour
of accreting matter which has no angular momentum. The flow is spherically
symmetric. The importance of the study of accretion of moving gas
by a compact object was realized nearly six decades ago.
Imagine a compact object of mass $M_*$ moving in a medium which has 
density $\rho_\infty$ and velocity $v_\infty$ at a large distance, 
the rate at which matter is intercepted by the star would be [45]
$$
{\dot M}=\pi R_A^2 \rho_\infty v_\infty
\eqno{(2.3)}
$$
where, $R_A$ is so-called accretion radius,
$$
R_A = \frac{2GM_*}{v_\infty^2}
$$
where the velocity $v_\infty$ is roughly the escape velocity of the star.
This result assumes that the gas is cold and collisionless, i.e., pressureless
and therefore valid for flows which are highly supersonic $M\sim \infty$.
The angular momentum of matter which could in principle be accreted
onto the compact object is $\sim R_A v_\infty \sim GM_*/v_\infty$.
In a medium which is homogeneous far away, most of the transverse 
component of the angular momentum would be canceled when gases from both
sides collide in the downstream side of the star and therefore
the actual angular momentum transfer could be very low [45]. 
In an inhomogeneous medium (with a finite density gradient), the rate
of angular momentum transfer to the star is finite [78-79],
$$
J\sim {\dot M} R_A^2 v_\infty/H
\eqno{(2.4)}
$$
where, $H$ is the density scale height. Clearly, $J \rightarrow 0$ 
as $H \rightarrow \infty$. 

The effect of finite pressure (finite sound velocity $a_\infty$ at a large 
distance) was studied in more detail by Bondi [80]. As we show in the next 
sub-Section, the accretion rate in this case is given by,
$$
{\dot M}= 4\pi\lambda R_B^2 \rho_\infty a_\infty
\eqno{(2.5)}
$$
where, 
$$
R_B=GM_*/a_\infty^2
$$
is the Bondi radius and $\lambda$ is a function of the polytropic index 
of the gas (see \S 2.1). In this case, the flow must have $M \rightarrow 0$
at a large distance. For an intermediate Mach number a `bridging'
formula  is suggested [80-81],
$$
{\dot M}= 4\pi\lambda a_\infty \frac {(GM_*)^2}{(a_\infty^2+v_\infty^2)}
\eqno{(2.6)}
$$
These results have become the central pieces of both analytical and
numerical investigations of spherical accretion processes onto compact objects.
There are many developments in the literature concerning 
the mathematical and the emission properties of the flow.
We shall now study them in increasing order of complexity.

\noindent{\Large 2.1 Bondi Accretion on a Newtonian Star}

To understand the problem of the spherical accretion on a black hole
better, it may be worth while to spend some time on the properties of
accreting flows on a Newtonian star which is well studied and well 
understood. These are the so called `Bondi solutions' [80].
We shall use $G=M=c=1$
so that our use of $M$ as the Mach number of the flow may not confuse the
readers.

\noindent {\large 2.1.1 Basic Equations}

Let us consider a spherical star of radius $R_*$ surrounded by 
a spherically symmetric infinite cloud of gas which is rest
at infinity. The equation of motion of this infalling matter is given by [80],
$$
\frac{\partial u}{\partial t} + u \frac {\partial u}{\partial r}+\frac{1}
{\rho}\frac{\partial P}{\partial r} + \frac{1}{r^2}=0,
\eqno{(2.7)}
$$
where $u$ is the radial velocity, $\rho$ is the density of the gas, $P$ is 
the isotropic pressure, $r$ and $t$ are the radial and the time coordinate
respectively. The equation is in dimensionless unit where mass, length and
time are measured by $M$, $GM/c^2$ and $GM/c^3$ respectively. The first 
term is the Eulerian time derivative of the velocity at a given $r$, the
second term is the advective term, the third term is the momentum
deposition due to pressure gradient and the final term is due to the
gravitational acceleration.

The continuity equation where the divergence of the mass flux
is equated with the temporal variation of density at a given $r$, is given by,
$$
\frac{\partial \rho}{\partial t}+\frac{1}{r^2}\frac{\partial}{\partial r}
(\rho u r^2) = 0.
\eqno{(2.8)}
$$
In the steady state solution, the terms
containing $\frac{\partial}{\partial t}$ may be deleted
and ${\frac{\partial}{\partial r}}$ may be replaced by $\frac{d}{dr}$.
Assume that there is no loss of energy during the accretion and the 
flow is adiabatic. The flow has the equation of state:
$P \!=\! K \rho^\gamma$, where $K$ is a constant
measuring the entropy of the flow, and $\gamma$ is the adiabatic index
which is the ratio of the specific heats and is assumed to be a constant
throughout the flow. Using this equation of state eq. (2.7) could be
integrated to obtain the conserved specific energy of the flow:
$$
{\cal E}= \frac{1}{2} u^2 + n a^2 - \frac{1}{r} = n a^2_\infty .
\eqno{(2.9)}
$$
Here, $n$ is the polytropic index, $n\!=\!{\displaystyle\frac{1}{\gamma-1}}$.
Integrating eq. (2.8) the mass flux is obtained as
$$
{\dot M}= \rho u r^2 .
\eqno{(2.10)}
$$
apart from a geometric constant $4\pi$.
Here $a\!=\!\sqrt{\displaystyle{\frac {\gamma P}{\rho}}}$ is the adiabatic 
sound speed and
$a_\infty$ is its value at infinity, i.e., 
$a_\infty\!=\!\sqrt{\displaystyle{\frac {\gamma P_\infty}{\rho_\infty}}}$.
Using $\rho\!=\!( \displaystyle{\frac{a^2}{\gamma K}} )^n$ the eq. (2.10)
can be re-written as,
$$
{\dot{\cal M}}=a^{2n} u r^2 .
\eqno{(2.11)}
$$
Hereafter, the quantity $\cm \!=\! {\dot M} \gamma^n K^n$, which is also 
conserved in the flow, will be called the `accretion rate'.
Equations (2.9) and (2.11) are the governing equations for two unknowns:
$u(r)$ and $a(r)$. Differentiating these equations
with respect to $r$ and eliminating $\displaystyle{\frac{da}{dr}}$
the equation for the velocity variation is obtained as,
$$
\frac{du}{dr}=\frac{\displaystyle{\frac{1}{r^2}}-
\displaystyle{\frac{2a^2}{r}}}{\displaystyle{\frac{a^2}{u}}-u}=\frac{N}{D} .
\eqno{}
$$
Since the flow is assumed to be smooth everywhere, if the denominator vanishes
at any radial distance $r$, the numerator must also vanish there.
This radius is called the critical radius or the sonic radius of the flow.
One therefore arrives at the so-called {\it sonic point conditions}:
$$
u_c=a_c ,\  \   \   \ \ {\rm i.e.,} \ \ \ M_c=1,
\eqno{(2.12a)}
$$
and
$$
r_c=\frac{1}{2a_c^2}.
\eqno{(2.12b)}
$$
The subscript $c$ is to denote quantities at the critical radius.
The spherical surface of radius $r=r_c$ is
also known as the {\it sound horizon} because 
for $r\!<\!r_c$, $u\!>\!a$ and any acoustic disturbances
created in this region are advected towards the star. Thus, no disturbance
created within this radius can cross the sound horizon
and escape to larger distance. This is analogous to the event horizon of a 
black hole where no electromagnetic disturbance can escape outside.
In terms of $a_\infty$, the sound speed (or, the
radial velocity) at the critical radius is obtained as,
$$
a_c=\left ( \frac {n}{n-3/2} \right )^{1/2} a_\infty .
\eqno{(2.13)}
$$
This shows that a transonic adiabatic flow is possible only if $n > 3/2$,
i.e., $\gamma < 5/3$. It is to be noted that a transonic flow has two extra
conditions, namely, eqs. (2.12a) and (2.12b) to satisfy, whereas only one extra 
unknown, namely, $r_c$ is introduced. Therefore, two conserved quantities, 
i.e., $\ce$ and $\cm$ cannot be independent, i.e., $\cm\!=\! \cm ( \ce )$. 
The accretion rate $\cm$ for the transonic flow is easily calculated from 
$a_\infty$ as,
$$
{\dot{\cal M}}_c= \frac{1}{4}\left ( \frac{n a_\infty^2}{n-3/2} 
\right )^{n-3/2}.  
\eqno{}
$$
In terms of the density of the flow at infinity $\rho_\infty$, the Bondi mass
flux ${\dot M}$ is written as,
$$
{\dot M}_c = \frac{1}{4} \frac{\rho_\infty}{a_\infty^3} (\frac{n}
{n-\frac{3}{2}})^{n-\frac{3}{2}}.
\eqno{}
$$
which is the same as eq. (2.5).
Figure 2.1a shows the integral curves of the flow
where the Mach number is plotted against the logarithmic radial
distance [77]. The curves $ABC$ and $A'BC'$ are
the physically significant transonic flows as they connect infinity with the
surface of the star. The contours are of constant accretion rate $\cm$. The 
radial velocity $u$ along $ABC$ is subsonic at infinity 
and supersonic on the surface of the star. This is the Bondi
accretion flow. The radial velocity $u$ along $A'BC'$
is subsonic on the surface of the star and supersonic at infinity. This 
is the corresponding wind flow. Other solutions such as $DD'$ or $EE'$ are 
not transonic: the former is subsonic everywhere and the latter is supersonic
everywhere. In fact, in the realistic flows, $EE'$ can also be excluded.
Later, we shall show that the properties of the integral curves
strongly depend upon the polytropic index $n$. Figure 2.1a is drawn for 
$n\!=\!3$ and Fig. 2.1b is drawn for $n\!=\!3/2$. Note that
in Fig. 2.1b, the critical curve
passes through $r\!=\!0$. Therefore, a neither accretion
nor wind type transonic solution is possible.

The accretion rate ${\dot{\cal M}}_c$ associated with a transonic flow is the
highest accretion rate that the star can accrete from the surrounding cloud
[70, 77]. To see this, we use the independent variable $M$, the Mach number,
instead of the velocity $u$. We eliminate the sound speed $a$ from eqs. 
(2.9) and (2.11) to obtain,
$$
f(M){\cm}^{\frac{2}{2n+1}}=g(r),
\eqno{(2.14)}
$$
where,
$$
f(M)=[\frac{1}{2}M^{\frac{4n}{2n+1}}+\frac{n}{M^{\frac{2}{2n+1}}}],
\eqno{}
$$
and
$$
g(r)=n {a_\infty^2 r^{\frac{4}{2n+1}}} +r^{\frac{3-2n}{1+2n}}.
\eqno{}
$$
$f(M)$ and $g(r)$ each has a minimum. $f(M)$ is minimum at $M\!=\!1$,
and $f_{min}\!=\!f(1)\!=\!\displaystyle{\frac{2n+1}{2}}$.
$g(r)$ is minimum at $r\!=\!r_c$ and
$$
g_{min}=g(r_c)=\frac{2n+1}{4}\, a_\infty^{\frac{4n-6}{2n+1}}
(\frac{2n-3}{4n})^{\frac{3-2n}{2n+1}} .
$$
At $r_c$, we have
$$
f(M)={\cm}^{-\frac{2}{2n+1}} \, g_{min}.
\eqno{(2.15)}
$$
Since $g_{min}$ is already minimum, the right hand side could be further
minimized by increasing $\cm$. But it
cannot be less than $f_{min}$, the minimum of the left hand side.
This fixes the maximum value of $\cm\!=\!{\dot{\cal M}}_{max}\!=\!
(\displaystyle{\frac {g_{min}}{f_{min}}})^{\frac{2n+1}{2}}$ 
which turns out to be the same as ${\dot{\cal M}}_c$ derived above.

\noindent{\large 2.1.2 Phase Space Behaviour of the Bondi Flow}

In this review we shall use the phrase `phase space'
rather loosely. In the strict sense, a phase space describes the
variation of velocity  with spatial coordinate. But we shall
use this phrase, even when we describe Mach number variation 
in stead of velocity. For isothermal flows, where the sound speed
is constant, these definitions are identical.
In order to understand the phase space behaviour of more 
complex flow described in \S 5, we spend some time
here to understand the nature of figs. 2.1ab and their variations
since the Bondi flow is one of the simplest possible astrophysical flows. 

Consider first the behaviour of a simple harmonic oscillator (SHO).
In the case of an undamped SHO
the conserved specific energy $E$ is the sum of kinetic and potential terms:
$$
E=\frac{1}{2}{\dot x}^2 + \frac{1}{2} C x^2 ,
\eqno{(2.16)}
$$
where $C$ is the stiffness constant, and ${\dot x}$ denotes the
derivative with respect to time.  In the phase space ($x$ versus 
${\dot x}$ diagrams), the trajectories of constant energy are 
{\it elliptical}. This is because the 
expression for energy (written for $x\! \sim\! 0$), is
the {\it sum} of two squared terms. In the neighbourhood of $x\!=\!0$, the
trajectories are circular in topology and $x\!=\!0$ is called a `circle' type,
`center' type or `O'-type point. In the case of Bondi flows, the potential
term is negative i.e., the `stiffness' is negative, as if. The trajectories 
are hyperbolic and the critical point is `saddle' type, or `X'-type. In 
the presence of damping, the trajectories of the 
SHO {\it spiral in} towards $x\!=\!0$. That is, the behaviour at $x\!=\!0$
changes from circle type or `O'-type to `spiral' type. As we shall discuss
in \S 5.3, the presence of viscosity in the flow causes the center type point
to change into a spiral type point. Actually, the behaviour
of the SHO at the origin could be one of the numerous kinds depending upon 
the relative magnitudes of the two parameters: the
damping factor $b$, and the stiffness constant $c$.
Figure 2.2 schematically shows various types of critical points [77].
It is interesting to study the behaviour of the critical point in
the Bondi flow which has only one parameter, namely, the
polytropic index $n$. How does the behaviour of a critical point depend on $n$?
To answer this, consider once again the case of a damped harmonic oscillator and
classify its critical points. We start with the equation of motion:
$$
{\ddot x} + b {\dot x} +  c x = 0.
\eqno{(2.17)}
$$
Assuming a solution of the form, $x\!=\!Ae^{\lambda t}$ with $A$ and 
$\lambda$ as constants, one obtains the equation for the characteristics:
$$
\lambda^2+b\lambda+c=0.
\eqno{(2.18)}
$$
Two roots of this equation,
$$
\lambda_\pm=\frac{-b\pm (b^2-4c)^{1/2}}{2} ,
\eqno{(2.19)}
$$
determine the nature of the solution. Let $D\!=\!
b^2-4c$ denote the discriminant. The classification 
is done in the following manner:

\indent  $D\!<\!0$:  `spiral' type if $b\!\ne\!0$, 
`O' (center)-type if $b\!=\!0$; \\
\indent  $D \! = \! 0$:  `Inflected nodal' type; \\
\indent  $D\! >\!0$:  `nodal' type if $c\! >\!0$, `straight line' if $c\!=\!0$,
`X'-type $c\!<\! 0$.

In the presence of damping in the SHO (or, viscosity in the accretion
flow), the nodal type point may form. In that case both the slopes
have the same sign, i.e., there are two possible solutions of 
accretion (or, winds). The center type point
is unphysical in accretion as the flow cannot pass through this point.
Figure 2.3 (adapted from [77] and [82] ) shows the nature of the 
critical points in various regions of the parameter space. We have marked
`Gravity' and `Viscosity' to indicate the similarity with negative
stiffness and damping respectively. Viscosity would be
important when angular momentum is present. In the case of accretion,
viscosity should transport angular momentum outwards, and in the case
of winds, the transport must be inwards. Thus, for wind solution
one may think of a `negative' viscosity as compared to the accretion.
In these respects, the entire comparison with simple harmonic oscillators 
including the concept of negative stiffness and damping remains valid.

To determine the behaviour of Bondi solutions near the critical point,
one needs to study $(\displaystyle{\frac{dv}{dr}})_c$ (note that in the
SHO problem we were studying the
behaviour of $\lambda\!=\! \displaystyle{\frac{d{\dot x}}{dx}}$ near $x\!=\!0$).
Since $\displaystyle{\frac{dv}{dr}}\!=\!N/D\!=\!0/0$
at a critical point, one must apply l'Hospital's rule. Thus,
$$
(\frac{dv}{dr})_c=\frac{\frac{dN}{dr}}{\frac{dD}{dr}}.
\eqno{(2.20)}
$$
\noindent After expanding this equation and using the critical
point conditions 2.12(a-b) one obtains,
$$
(\frac{dv}{dr})_c=\frac{4a_c^3}{2n+1}[1\pm\sqrt{n(n-3/2)}] .
\eqno{(2.21)}
$$
In the present problem, the discriminant $D\!=\!n(n-3/2)$ determines the 
nature of the critical points. For $D\!>\!0$, namely, for a 
physical transonic solution, one must have $n\!>\!3/2$. For $0\!<\!n\!<\!3/2,
\  D\!<\!0$, 
and the critical point is of `spiral' type. For a `saddle' type or `X'-type
point two derivatives $(\displaystyle{\frac{dv}{dr}})_c$ must be
of opposite signs, hence one must have $n\!  > \!2$. 
For $2 \! > \! n \! >3/2$, the critical point is `nodal'.
For $n\!=\!3/2$, the critical point is `inflected nodal' 
type and for $n\!=\!2$, the critical point is `straight line' type. 

From the solutions just presented, one can now study the asymptotic
properties of the Bondi flow. Concentrating on the supersonic branch $BC$,
it is easy to show that the Mach number varies as,
$$
M \sim r^{-\frac{2n-3}{4n}}.
\eqno{(2.22a)}
$$
Thus, as $r \rightarrow 0$, $M \rightarrow \infty$ only if $n > 3/2$.
Similarly, for the branch $BC^{\prime}$, we have,
$$
M \sim r^{n-3/2},
\eqno{(2.22b)}
$$
so that $M \rightarrow 0$ as $r \rightarrow 0$ only if $n > 3/2$. 

At a large distance, the solution on the branch $B A^{\prime}$ behaves as
$$
M \sim r^{1/n}.
\eqno{(2.22c)}
$$
Thus, $M \rightarrow \infty$ as $r \rightarrow \infty$ for any value of $n$.

\noindent{\Large 2.2 Bondi Flow on a Black Hole}

\noindent{\large 2.2.1 In Schwarzschild Geometry}

In the case of a Bondi accretion on a Schwarzschild
black hole, the outer boundary condition remains the same
(namely, matter is at rest at infinity), but the inner boundary 
condition is different. Here,
matter must cross the event horizon with the velocity of light $c$.
In flows with realistic equations of state, the maximum value of $a$ 
is $\frac{c}{\sqrt 3}$. Thus a black hole accretion is transonic [77].

The stationarity condition $\displaystyle{\frac{\partial}{\partial t}}\!=
\!0$ [83] implies a conserved energy $\ce$ given by,
$$
\ce=hu_t=\frac{p+\epsilon}{\rho} \left ( \frac {1-2/r}{1-u^2} \right )^{1/2} .
\eqno{(2.23)}
$$
It is to be remembered that $\ce$ includes the rest mass of matter.
Thus, the specific energy in terms of the quantities at infinity is
$\ce\!=\!1+n a_\infty^2$. $h(=\frac{P+\epsilon}{\rho} 
\!=\!\displaystyle{\frac{1}{1-na^2}})$ and $u_t$ are the specific enthalpy
and the specific binding energy
respectively, $\epsilon$ is the internal energy density, $u$
is the radial velocity in the local Lorentz frame. Equation (2.23) is the
relativistic generalization of Bernoulli's equation. The conservation of 
mass and entropy along the flow line implies that
$$
\cm=(\frac{a^{2}}{1-na^2})^n u u_t r^2 .
\eqno {(2.24)}
$$
(Here $\cm$ has the same interpretation as in the Newtonian case.)
By differentiating the above two equations with respect to $r$ and eliminating
$\displaystyle{\frac{da}{dr}}$ one obtains,
$$
\frac{du}{dr}=\frac{u(1-u^2)[1-a^2-2a^2(r-2)]}{r(r-2)(a^2-u^2)} .
\eqno {(2.25)}
$$
The sonic point conditions are
$$
u_c=a_c , \  \; {\rm i.e.,} \ \; M_c=1 ,
\eqno {(2.26a)}
$$
and
$$
r_c=\frac{3}{2} +\frac{1}{2a_c^2} .
\eqno {(2.26b)}
$$
When $a_c$ is as large as the velocity of light (such as when 
$p=\epsilon$ equation of state is chosen) the sonic point lies just on the
horizon, independent of any other properties of the flow. 

Although it is easy to study the adiabatic Bondi accretion with full
general relativity, a rigorous study with complete radiative
transfer cannot be carried out. However, for most of the astrophysical purposes
this is not essential either. As a prelude to some of the study of accretion 
disks in the next few Sections, we shall briefly
mention here the study of Bondi flow using the pseudo-Newtonian potential.

\noindent{\large 2.2.2 In pseudo-Newtonian Geometry}

In the pseudo-Newtonian geometry, the conserved energy is expressed as
$$
{\cal E}= \frac{1}{2} u^2 + n a^2 - \frac{1}{r-2} = n a^2_\infty 
\eqno{(2.27)}
$$
and the mass flux as,
$$
{\dot M}= \rho u r^2.
\eqno{(2.28)}
$$
From these, the velocity gradient at any point of the flow becomes,
$$
\frac{du}{dr}=\frac{\frac{1}{(r-2)^2}-{\frac{2a^2}{r}}}
{\frac{a^2}{u}-u}= \frac{N}{D}.
\eqno{(2.29)}
$$
The sonic point conditions are derived as, 
$$
u_c=a_c, \ \ \; {\rm i.e.,} \ \; M_c=1 ,
\eqno{(2.30a)}
$$
and
$$
r_c=2+\frac{1}{4a_c^2} + \sqrt {\frac{1}{a_c^2} + \frac{1}{16a_c^4}} .
\eqno{(2.30b)}
$$
Note that this being a Newtonian treatment, $a_c$ is allowed to become
infinity, so that the sonic point stays at the `horizon', namely, at
$r\!=\!2$ in the same way as in the Schwarzschild geometry. Other properties
remain similar to the original Bondi solution in the Newtonian geometry. 

\noindent{\Large 2.3 Bondi Flow with Simple Radiative Transfer}

As matter accretes onto a hole, radiations emitted would have to
escape through the matter itself, and therefore will interact with 
the incoming matter.
To see the `zeroth' order effect of the radiation on matter, 
consider a spherical accretion in which the luminosity
at any radial distance $r$ is assumed to be a constant fraction of the
Eddington luminosity, i.e., $L_\gamma (r)\!=\! C L_{Edd}$.
Since the radiation force is opposite to gravity, the net force on 
the accreting matter due to the gravity and the radiation will be given by,
$$
F_{net}= -\frac{1}{r^2} + \frac {C}{r^2} = -\frac{1-C}{r^2}.
\eqno{(2.31)}
$$
Thus the effective potential is given by,
$$
\Phi_{net}= -\frac{1-C}{r}.
\eqno{(2.32)}
$$
Using this potential instead of $-\displaystyle{\frac{1}{r}}$ in equation 
(2.5) one can repeat the analysis. The location of the critical point is 
found to lie closer to the black hole, as, $r_c\!=\!\displaystyle
{\frac{1-C}{2a_c^2}}$ in this case. However, the sound speed at the critical
point ($a_c$) does not change. Clearly, a stationary transonic solution is
not possible when $C > 1$, i.e., when the luminosity is super-Eddington [77].
The properties of the solution remain otherwise similar to what was discussed
in the previous Section.

\noindent{\Large 2.4 Bondi Flow with General Radiative Transfer}

Observations of extremely large energy output
from active galactic nuclei and quasars
suggests that the radiation efficiency $\eta$ has to be of the order of
a few percent of the rest mass. Much of the research in accretion physics 
is focussed to find out this efficiency 
of the conversion of the rest mass to the radiated energy (eq. 2.1). The 
efficiency is a function of the accretion rate (opacity) and the heating and
cooling mechanisms in the flow. Over the past two decades, several efforts
have been made to estimate this quantity as accurately as possible.
First, we present a brief overview of the major developments in this topic. In
the Bondi accretion solution presented above, the properties of a spherical
adiabatic accretion is studied. In principle, energy is conserved in this
case and no radiation is emitted from this flow, i.e., entropy
is also advected. Several works have been reported in the literature 
which study spherical accretion self-consistently adding more 
physical processes, such as various heating and cooling mechanisms, 
pair productions, pre-heating of incoming gas, etc. 
Since the luminosity and efficiency of the emission of the gas 
accreted from an interstellar medium cannot be high,
Shvartsman [84] suggested that the presence of magnetic field could
increase the efficiency of radiation. 
Subsequently, magnetic field dissipation was introduced in the
computation and efficiency was increased [85-86]. 
Earlier attempts to solve optically thick flows in 
general relativistic framework using 
radiative transfer showed divergent behaviour [87-88], till
it was realized that the luminosity  of the flow as 
seen from infinity is also an eigenvalue of the problem [89, 90]. 
High accretion rate solution of the general relativistic models do not seem 
to have significantly higher efficiency [91]. Soffel [92] found most of the
the low luminosity (low temperature, effectively optically thick)
solutions connecting some of the earlier solutions. In these works, 
Comptonization was not taken into account and the solutions were steady.
Ostriker et al. [93] pointed out that the pre-heating, i.e., heating of the
far away gas through Compton scattering of the radiation emitted close to 
the black hole could stop the flow and possibly make it unstable. 

Comptonization of the flow was studied in greater detail by several authors
including Maraschi, Roasio \& Treves [94], Colpi, Maraschi \& Treves [95]
and  Wandel, Yahil \& Milgrom [96] (hereafter, referred to as
MRT, CMT and WYM respectively). These results show that higher luminosity
and higher efficiency solutions are indeed possible when accretion rate
is not necessarily very high. MRT considered magnetic dissipation 
and therefore obtained very high luminosity. CMT studied two temperature
models without the magnetic dissipation and WMY self-consistently studied
the emitted radiation (by balancing Compton heating with
bremsstrahlung cooling) in a non-relativistic freely falling gas. They ignored
the pre-heating of the gas. WMY also found a new high luminosity solution 
branch in the optically thick flows.

Park [97] and Nobili, Turolla \& Zampieri [98] (hereafter NTZ)
studied the self-consistent solutions with pre-heating included and 
within a single self-consistent framework verified the presence
of both the high and low luminosity branches.
Park \& Ostriker [99] include the effects of pair production
and pre-heating and were able to locate yet another branch of solutions
when the pair density is very high.

In general, the efficiencies of the
spherical solutions were not found to be sufficiently high
to explain the observed luminosity from active galaxies.  Secondly,
they cannot reproduce the observed  bump in the ultraviolet in the
continuum. However, it is a kind of `paradigm' problem and should be studied 
in greater detail so that these may be subsequently generalized to
more realistic flows, namely, flows with some angular momentum.
Below, we discuss some of the spherically symmetric models in greater detail.

\noindent {\large 2.4.1 Single Temperature Solutions}

Earliest solutions of this kind which include cooling due to bremsstrahlung
are due to Shapiro [85-86]. Apart from the continuity equation and the radial
momentum equation (as we discussed earlier), 
in the presence of heating and cooling one must
include the energy equation which is obtained from the thermodynamic
identity,
$$
v(\frac{d e}{dx} - \frac {e+P}{n}[\frac{dn}{dx}])= \Lambda - \Gamma ,
\eqno{(2.33)}
$$
where, $x$ is the non-dimensional radial coordinate, $v$, $n$, $e$, $P$
are the velocity, number density of protons, proper internal energy 
density and isotropic pressure respectively, and $\Lambda$ and $\Gamma$ 
are the combined cooling and the heating effects respectively. Here,
the decrease in entropy of the in-flowing gas is equated with the
energy lost through the cooling effects ($\Lambda$) minus the 
energy generated ($\Gamma$). The above equation replaces the
adiabatic equation of state used earlier.

As a first approximation, the electron and proton
temperatures are assumed to be the same, although in reality,
when cooling of electrons are efficient,
the temperature of electrons would be a few orders of magnitude 
lower ($T_p$ close to its virial value $m_p c^2 /k r$,
and $T_e \sim m_e c^2/k r$, see below).
It is possible that the collective plasma processes, which can be present 
when magnetic field is significant, could couple electrons and protons 
more strongly than Coulomb coupling [100]. In this case, the electron and proton
temperatures can become the same. 

In the simplest of solutions one can assume no heating other than that due to
compression ($\Gamma=0$) and the cooling is only due to bremsstrahlung. 
Since bremsstrahlung cross-section is low, the efficiency of radiation turned
out to be very small: $\eta \sim 10^{-10}$ for a $1 M_\odot$ black hole 
accreting from H I region with gas number density $n_\infty=1$cm$^{-3}$ [85]. 
This situation is appropriate for a black hole immersed inside
an interstellar medium. Overall efficiency $\eta$ varies with number
density and the mass of the black hole as: $\eta \propto n_\infty M$. 
The luminosity varies as $L \propto n_\infty^2 M^3$.
Hence, one could imagine that raising the accretion rate,
luminosity and efficiency could be enhanced. However, higher density increases
the optical depth and therefore the accurate inclusion of the
interaction of radiation with matter becomes necessary.
Furthermore, at high optical depth radiation is trapped at some distance,
reducing the efficiency. This happens at the so called `trapping radius' 
$r_{trap}$ when the accretion flow advects photons inwards before they 
could diffuse outwards [2], i.e., $\tau_T > 1/v_{r}$. The radial velocity 
and the in-fall time scale varies as $v_r = r^{-1/2}$
and $t_{infall} = r/v_r = r^{3/2}$. From the mass conservation,
the density varies as $\rho \sim \frac {\dot M} r^{-3/2}$. The optical
depth due to Thomson scattering at radius $r$ is $\tau_T \sim
{\dot M} r^{-1/2}$ giving rise to $r_{trap} \sim {\dot M}$.
Thus higher radiation obtained by increasing accretion rate
would be advected towards the black hole.

In the presence of a radial magnetic field, cooling due to synchrotron
radiation from the thermal electrons go up. Assuming that
well ordered radial field should always be close to the equipartition value,
($P_{field} \sim P_{gas} \propto r^{-5/2}$)
at regions close to a black hole one can obtain the variation of the 
magnetic field component as [84]:
$$
B= {\dot M}^{1/2} G M^{1/4} r^{-5/4} {\rm Gauss} .
\eqno{(2.34)}
$$
Inclusion of synchrotron cooling using this field
variation increases the efficiency very much. Furthermore, the 
frequency at which the luminosity was highest in case of bremsstrahlung 
alone, is shifted to the thermal synchrotron frequency [86],
$$
\nu_s = \frac {3}{4} \frac{e B}{\pi m_e c} (\frac {k T}{m_e c^2 })^2 ,
\eqno{(2.35)}
$$
where, $m_e$ is the mass of the electron and $k$ is the Boltzman 
constant. 

The magnetic field considered so far is ordered, which is less likely. 
In order to keep the equipartition with the gas pressure,
the fields may develop chaotic components and would render the flow
turbulent in small scales. The turbulent dissipation would convert parts of 
the gravitational energy of the accreted matter to heat the flow at a rate
[101],
$$
\Gamma = \frac{1}{8 \pi} \frac{G M {\dot M}}{r^4} .
\eqno {(2.36)}
$$
This variation is obtained by the consideration that the turbulent
energy between $r$ and $r/2$ be dissipated in one free fall time.
Inclusion [94, 102, 103] of this heating increases the efficiency  of the
emission to about $\eta \sim 0.1$ 
when the accretion rate is high (${\dot M} > 10^{-4} {\dot M}_{Edd}$). 
These latter models include Comptonization of the photons. Below,
for the sake of completeness, we devote a few sentences to describe
this important process.

Comptonization is the problem of energy exchange in the scattering of 
photons off the electrons. As the accretion rate is increased and the 
Thomson opacity,
$$
\tau_T = \int_{r_i}^{\infty} \sigma_T n_e d r 
\eqno{(2.37)}
$$
approaches unity, the escaping photons gain energy due to repeated 
scattering with the hot electrons.
For a thermal non-relativistic electron distribution with 
a temperature $T_e$, the average energy exchange per scattering is given 
by [104], ($h\nu, kT_e\ll m_ec^2$),
$$ 
{{<\Delta\nu>}\over{\nu}}~=~{{4kT_e-h\nu}\over{m_ec^2}}.
\eqno{(2.38)}
$$
When $h\nu \ll kT_e$, photons gain energy due to the Doppler effect
and when $h\nu\gg kT_e$, photons lose energy because of the 
recoil effect. As the radiation passes through the medium, 
the probability of repeated scattering by the same photon decreases 
exponentially although the corresponding gain in energy is exponentially 
higher. A balance of these two factors yields a power law distribution 
of the energy density: 
$$
F_\nu \propto \nu^{-\alpha}
\eqno{(2.39)}
$$
In the limit of the high energies ($h\nu \gg kT_e$ ) the exponential 
hard tail is formed as result of the recoil effect (the photons are unable
to gain more energy than what the electrons actually have: $h\nu\propto kT_e$).
The limit can also be set by $\nu^+$, the cyclotron self-absorption frequency
[105] (the frequency at which the plasma becomes transparent to cyclo-synchrotron
emission). 
The index $\alpha$, which varies between $1$ to $0.5$ for an wide range of
accretion rates ${\dot M}$=($1$ to $10^{-2} M_{Edd}$) is obtained analytically 
in the non-relativistic limit [106] as,
$$
\alpha= [\frac{9}{4} + \frac{\pi^2}{3(\tau_T + \frac{2}{3})^2 
\Theta }]^{1/2} -\frac{3}{2}
\eqno{(2.40a)}
$$
and numerically [107] in the relativistic limit as,
$$
\alpha= - ln A / ln [1+4 \Theta+16 \Theta^2 ]
\eqno{(2.40b)}
$$
where,
$$
A=\frac{3}{4} \tau_T  \ \ \   {\rm for} \ \ \  \tau_T <1
$$
and
$$
A=(1-\frac{3}{4\tau_T} )  \ \ \  {\rm for} \ \ \  \tau_T > 1 .
$$
and $\Theta=\frac{k T_e}{m_e c^2}$ with $T_e$ as the electron
temperature. The efficiency of radiation increases to about 
$\eta \sim 10^{-2}$ in this case.

It has been recently shown that the power laws are the exact solutions
of the radiative transfer kinetic equations taking into account the Doppler
effect only [108]. Accurate spectral indices
are derived for an wide range of the optical depths and electron
temperatures of the plasma cloud as follows:
$$
\alpha=\frac{\beta}{\ln [1+ (\alpha+3)\Theta/(1+\Theta)
+4d_0^{1/\alpha} \Theta^2]}
\eqno{(2.40c)}
$$
where, 
$$
d_0 (\alpha) =\frac{3[(\alpha + 3)\alpha + 4 ] \Gamma (2\alpha + 2)}{(\alpha
 + 3) (\alpha + 2)^2}
$$
and $\Theta=k T_e/m_e c^2$ which represents the appropriately
weighted average of the electron temperature in the plasma cloud. Here,
$\beta$ is obtained from the eigenvalue $\zeta\!=\!exp(-\beta)$
of the corresponding eigenfunction for the appropriate radiative
transfer problem. Assuming plasma geometry to be disk-like,
for optical depths $\tau_0>0.1$,
$$
\beta= \frac{\pi^2}{12(\tau_0 + 2/3)^2}(1-e^{-1.35 \tau_0})+
0.45 e^{-3.7\tau_0} \ln\frac{10}{3 \tau_0}
\eqno{}
$$
and for very low optical depth ($\tau_0 <0.1 $),
$$
\beta= \ln\left[\frac{1}{\tau_0 \ln(1/2\tau_0)}\right] .
\eqno{}
$$
If on the other hand, the plasma has a spherical geometry,
then for  optical depths ($\tau_0>0.1$),
$$
\beta= \frac{\pi^2}{3(\tau_0 + 2/3)^2}(1-e^{-0.7\tau_0})+e^{-1.4\tau_0} 
\ln\frac{4}{3\tau_0}
\eqno{}
$$
and for very low optical depth ($\tau_0 <0.1$)
$$
\beta=\ln\frac{4}{3\tau_0}.
\eqno{}
$$ 

\noindent {\large 2.4.2 Two Temperature Solutions}

In the case when
ion-electron collision time is larger than the free-fall time,
protons and electrons can be thermally decoupled and therefore
one has to consider the equations for ions and electrons separately. 
It is important to determine the electron temperature very
accurately because the emitted radiation depends on it.
Protons will behave roughly adiabatically and
the temperature reached by the protons (in adiabatic accretion) could 
reach the threshold for $\pi^0$ production so that the black hole could become
a $\gamma$-ray source due to pion decay. The pion production rate and therefore
the gamma ray luminosity depends strongly upon the proton temperature [95].

Since the proton temperature is higher than the electron temperature,
energy can be lost by protons and gained by electrons due to the collisional
energy exchange term. Electrons can lose energy due to
thermal bremsstrahlung, cyclo-synchrotron radiation and Comptonization.
We present here a typical analysis which includes these processes.
Following CMT, we assume the length scale in the following discussion
to be in units of $r_g=2GM/c^2$. The zero energy gas, cooler at a large 
distance, yields a velocity (without relativistic corrections), 
$$
v(x) \sim  x^{-1/2}.
\eqno{(2.41)}
$$
The mass conservation law gives the number density of hydrogen as,
$$
n(x) \sim x^{-3/2} {\dot M} M^{-2} \ \ {\rm cm}^{-3} .
\eqno{(2.42)}
$$
The optical depth for Thomson scattering, 
$$
\tau_T (x) \sim 1.44 x^{-1/2} {\dot M}_{18} M_{1}^{-1} .
\eqno{(2.43)}
$$
The magnetic flux is assumed to be radial (monopole like). The
flux conservation requires that the field varies as $B \propto x^{-2}$.
With the assumption of the equipartition between
the magnetic energy density and the gravitational energy density
gives the field variation as:
$$
B(x)=(\frac{8 \pi m_p n G M}{3 r_g})^{1/2} x^{-2} \ \ {\rm Gauss}= 4.75 \times
10^7 {\dot M}_{18}^{1/2} M_{1}^{-1} x^{-2}\ \  {\rm Gauss}.
\eqno{(2.44)}
$$
Here, $M_{1}$ is the mass of the central black hole in units of $10M_\odot$,
and ${\dot M}_{18}$ is the mass accretion rate in units of $10^{18}$gm s$^{-1}$.
Maxwellian distribution is established through Coulomb collisions
between electrons and protons in time [109]:
$$
t_{ee}= \frac {C}{n ln \Lambda_0} m_e^2 (\frac  {T_e}{m_e})^{3/2} \ \ {\rm s},
\eqno{(2.45)}
$$
$$
t_{pp}= \frac {C}{n ln \Lambda_0} m_p^2 (\frac  {T_p}{m_p})^{3/2} \ \ {\rm s},
\eqno{(2.46)}
$$
where $\Lambda_0$ is usual Coulomb logarithm, $T_e$ and $T_p$ are the
electron and proton temperatures and $C$ is a constant,
$$
C=\frac{3k}{4 (2 \pi )^{1/2} e^4} .
\eqno{(2.47)}
$$
Protons and electrons will reach the thermal equilibrium in
$$
t_{ep} = \frac{C}{2 n ln \Lambda_0} m_e m_p (\frac{T_e}{m_e} 
+ \frac {T_p}{m_p})^{3/2} \ \ {\rm s}.
\eqno{(2.48)}
$$
Close to a black hole, protons reach the virial
temperature $T\sim \frac {GMm_p}{ r k}$ of $10^{12}$K and the electrons
reach a temperature of $10^9$K  ($\frac{T_e}{m_e} \approx
\frac{T_p}{m_p}$). With these parameters, it is interesting to compare 
different time scales [95],
$$
t_{ee} = 2.4 \times 10^{-5} {\dot M}_{18} M_{10}^{-2} x^{3/2} (T_e/10^9)^{3/2}
(ln \Lambda_0)^{-1} \ \ {\rm s} ,
\eqno{(2.49)}
$$
$$
t_{ep}=\frac{m_p}{m_e} t_{ee} ,
\eqno{(2.50)}
$$
$$
t_{pp}=\frac{m_p}{m_e}^2 t_{ee} ,
\eqno{(2.51)}
$$
with the free fall time scale
$$
t_{ff} = 10^{-4} M_{10} x^{3/2} {\rm s}.
\eqno{(2.52)}
$$
For optical depth near unity, a Maxwellian description is justified
for the electrons, but $t_{ep}$ and $t_{pp}$ are both larger than $t_{ff}$
so that for the protons, the Maxwellian distribution may break down.

The equation for energy (eq. 2.33) is written down separately
for protons and electrons. Assuming $v(x)$ and $n(x)$ as above 
appropriate for a Bondi flow (thus ignoring the effect of 
radiation on the distribution of velocity and density) and 
from the definition of the internal energy $e=P/(\gamma-1)$
it is easy to write down the equations as
[84, 86, 95, 101]:
$$
\frac{d T_p}{dx} + \frac{T_p}{x}+\frac{8 m_p r_g^3 \pi x^2 (\Gamma_p - 
\Lambda_p )}{3 k {\dot M}} = 0
\eqno{(2.53)}
$$
and 
$$
\frac{d T_e}{dx} + \frac{3}{2} (\gamma -1 ) \frac{T_e}{x}+\frac{4 m_p r_g^3 
\pi (\gamma -1 ) x^2 (\Gamma_e - \Lambda_e )}{ k {\dot M}} = 0
\eqno{(2.54)}
$$
Here, $\gamma=5/3$ is the adiabatic 
index for non-relativistic temperatures ($T_e \leq \frac {m_e c^2}{k}$) 
and $\gamma=4/3$ for relativistic regime. The quantities $\Gamma_{p,e}$ and 
$\Lambda_{p,e}$ in above equations represent  energy gains and energy losses 
due to various processes. If one neglects heating due to dissipation for 
protons, $\Gamma_p=0$. The proton equation consists of two
loss terms: 
$$
\Lambda_p = \Gamma_{ep} + \Lambda_{ib}
\eqno{(2.55)}
$$
where, 
$$
\Gamma_{ep}=\frac{3}{2} \frac{n k (T_p - T_e )}{ t_{ep}}
\ \ {\rm erg \ \ cm}^{-3}\ {\rm s}^{-1}
\eqno{(2.56)}
$$
is the energy transfer rate from the protons to the electrons, and 
$$
\Lambda_{ib}=1.4  \times 10^{-27} n^2 (\frac{m_e}{m_p} T_p )^{1/2}
\eqno{(2.57)}
$$
is the radiative energy loss for protons due to inverse bremsstrahlung.

In the electron equations $\Gamma_e= \Gamma_{ep}$ can be taken as the energy
gain, and the cooling term
$$
\Lambda_e = \Lambda_b + \Lambda_{cs} + \Lambda_{mc}
\eqno{(2.58)}
$$
contains contribution from the
the bremsstrahlung, cyclo-synchrotron and multiple-Compton
scattering of the cyclo-synchrotron photons by the thermal electrons
which are in units of erg cm$^{-3}$ s$^{-1}$:
$$
\Lambda_b= 1.4 \times 10^{-27} n^2 T^{1/2} (1+4.4\times 10^{-10} T) ,
\eqno{(2.59)}
$$
$$
\Lambda_{cs}=\frac{\pi k T_e}{GM} \frac{\nu^3}{x} ,
\eqno{(2.60)}
$$
and
$$
\Lambda_{mc}=\Lambda_{cs}  [ \frac{\alpha}{\alpha -1} 
\frac {(\nu_{max}/\nu^+)^{1-\alpha} - 1 }
{(\nu_{max}/\nu^+)^{-\alpha} - 1 } -1 ] ,
\eqno{(2.61)}
$$
where, the expression for $\alpha$ is as given above 2.40(a-c) and $\nu^+$ is 
the synchrotron self-absorption frequency [105].

Figures 2.4(a-c) show variations of temperature of electrons and protons
as functions of logarithmic radial distance from the black hole.
This cases are computed for (a) ${\dot M}=10^{16}$ gm s$^{-1}$, 
(b-c) ${\dot M}=10^{18}$ gm s$^{-1}$ 
and the black hole mass is $10M_\odot$. The difference between (b) and (c)
is that in (c), the electron-proton coupling term is artificially 
multiplied by a factor $20$ to show the effect of strong cooling.
The outer boundary is chosen at $x=10^6$, with the  electron and proton
temperatures same as in adiabatic ($\gamma=5/3$) Bondi solution:
$$
T_e|_{out}=T_p|_{out}=T_{ad} = \frac{3}{20} \frac{m_p c^2}{k} \frac{1}{x}.
\eqno{(2.62)}
$$
In Fig. 2.4(a), temperatures behave as in an adiabatic solution initially, 
but close to the black hole, radiative losses of the electrons become so
important that energy gained through the compression of the gas as well 
as that obtained from protons through collisions are radiated away. 
Even closer to the black hole the loss is further increased
due to Comptonization. In Fig. 2.4(b) the optical depth is higher
and the bremsstrahlung loss reduces the overall temperature. In Fig. 2.4(c),
due to the enhanced exchange of energy, protons lose more energy to electrons
which lowers their temperature substantially reaching a maximum 
$ T << 10^{11}$ K. 

In Fig. 2.5(a-d) the spectrum of radiation emitted from electrons is shown.
The spectrum has a maximum at around $\nu^+$, the synchrotron self-absorption
frequency. At higher frequencies, 
the spectrum is dominated by Comptonization and is a sum of power law
in frequency range from $\nu^+$ to $3\kappa T/ h$.
In Figs. 2.5(a-b) the accretion rates are chosen to be $10^{16}$ gm s$^{-1}$
and $10^{18}$ gm s$^{-1}$ respectively on a black hole of mass $10\ M_\odot$. 
In Figs. 2.5(c-d) the accretion rates are chosen to be $10^{23}$ gm s$^{-1}$
and $10^{25}$ gm s$^{-1}$ respectively on a black hole of mass $10^8\ M_\odot$. 

It is observed that at a fixed optical depth, the results depend very
weakly on the mass of the black hole. The efficiency of the
conversion of the rest mass energy to the outgoing luminosity radiated
by electrons seems to be very low: $ \eta \sim 0.005 - 0.01 $.
This is higher than one-temperature solutions [85-86]
which neglected cooling due to Comptonization. 
Inclusion of dissipative heating increases the efficiency 
by a factor of $\sim 10$ [94, 101, 103], although the
dissipation of fields considered is possibly an upper estimate.

The luminosity of $\gamma$ rays
which are produced from the pion decay is computed from [95]:
$$
p + p \rightarrow p + p + \pi_0 (\rightarrow 2 \gamma ) .
\eqno{(2.63a)}
$$
This pion production rate per unit volume
$$
R=\frac{1}{2} n^2 <  \sigma v_{th} > 
\eqno{(2.63b)}
$$
(where $\sigma$ is the p-p cross section for $\pi_0$ production and
$v_{th}$ is the proton thermal velocity and $<>$ denotes average taken
over the Maxwellian distribution) strongly depends on temperature [110]
and therefore strongly depends on the optical depth. The inset of Figure 2.5a
provides the luminosity due to gamma rays. As the accretion is
increased, till the optical depth $\tau_T$ (Thomson optical depth
at $r_g=1$) reaches a value of $0.3$, the luminosity due to $\gamma$
rays goes as $L_\gamma \approx {\dot M}^2$ and with the
increase of $\tau_T$, $L_\gamma$ decreases again. Figure 2.6 gives the  
the variation of $L_\gamma$ with the accretion rate, which shows the
maximum at around $\tau_T \simeq 0.3$.

\noindent{\large 2.4.3 General Relativistic Optically Thick Bondi Flow}

In the adiabatic Bondi flow  model (\S\S 2.1-2.2) no outflow of energy 
or entropy was allowed. The inclusion of photon diffusion effects
changes the flow behaviour qualitatively. The solutions corresponding to this
situation is described below [89-90].

Basic equations governing the optically thick Bondi flow are the 
generalization of the equations in a black hole geometry presented 
in \S 2.2. However, because of the
loss of radiation, the specific energy and entropy do not remain conserved
within the flow and one has to keep the momentum and the entropy equations
in the differential form. The continuity equation, the radial momentum 
conservation equation, the Bernoulli's equation, and the diffusion equation 
of photons respectively take the forms:
$$
r^2 \rho u u_t= {\dot M}={\rm constant},
\eqno{(2.64a)}
$$
$$
\frac{1}{u_t}\frac{du_t}{dr}=-\frac{1}{p+\epsilon}\frac{dp}{dr},
\eqno{(2.64b)}
$$
$$
hu_t{\dot M}- {\hat L}={\dot E}={\rm constant} ,
\eqno{(2.64c)}
$$
$$
{\hat L}=- r^2 u_t^3(1+u^2) \frac{4 p_r}{\rho \kappa_1}[
\frac{1}{T}\frac{dT}{dr}+\frac{1}{u_t}\frac{du_t}{dr} ] .
\eqno{(2.64d)}
$$
Here $\kappa_1$ is the first-moment opacity [83, 89, 90]. 
In the case when Thomson scattering opacity is dominant, $\kappa_1\!=\!
\displaystyle{\frac {\sigma_T}{m_p \mu_e}}$; $\sigma_T$, $m_p$ and $\mu_e$ being
the Thomson scattering cross-section, the mass of a proton and the
mean molecular weight per electron respectively. Also, ${\hat L}$ is related
to the diffusive luminosity $L$ by,
$$
{\hat L}=(1+u^2)u_t^2 L .
\eqno{(2.65)}
$$
The quantity $\dot E$ represents the net rate of the mass-energy accretion,
allowing for the effects of the `PdV' work and the heat transfer. 
Because of the presence of the `extra' unknown over and above what was discussed
in \S 2.2, one has to supply two unknowns, say, ${\dot E}$
and ${\dot M}$ to describe the flow completely.

Unlike the results for adiabatic solutions in \S 2.2, one now has the basic 
equations in differential form and the complete solution is impossible 
analytically. However, the trend of the solution is easily obtained by 
expansion of the equations around the critical point. After differentiating
above equations, one can cast them in the form of eq. (2.25),
and therefore similar logic holds in finding critical point
conditions. Since we need to supply only two unknowns, only two of the 
differential equations are truly independent. Flammang chooses the velocity
and the luminosity equations. The solution is a trajectory
in the three-dimensional space spanned by $(r, u, {\hat L})$ instead of just
$(r, M)$ plane as in the adiabatic case (Fig. 2.1ab). The critical points are
found to lie on the critical curve formed by the intersection of the 
three surfaces:
$$
u=a_t,
\eqno{(2.66a)}
$$
$$
{\cal L}=\frac{1-a_t^2}{1-u^2} \frac{1}{ru_t^2}-2a_t^2 ,
\eqno{(2.66b)}
$$
and
$$
{\cal L}=\frac{a_s^2-a_t^2}{a_s^2-u^2}C .
\eqno{(2.66c)}
$$
where,
$$
C = \frac{1}{ru_t^2}-2 u^2 ,
\eqno{(2.67a)}
$$
$$
{\cal L} \equiv \frac{b}{1-b} \frac {\rho}{4 p_r} \frac {\kappa_1}{r}
\frac{{\hat L}}{u_t^3 (1+u^2)}=-\frac{b}{1-b}\frac{r}{Tu_t}\frac{d(Tu_t)}{dr} ,
\eqno{(2.67b)}
$$
$$
a_t^2 = (\frac{a}{1-b} ) ,
\eqno{(2.67c)}
$$
is the isothermal sound speed squared, and
$$
a_s^2=qb+a ,
\eqno{(2.67d)}
$$
is the adiabatic sound speed squared, with
$$
a \equiv \frac{\rho}{p+\epsilon} (\frac {\partial p}{\partial \rho} )_T ,
\eqno{(2.67e)}
$$
$$
b \equiv \frac{T}{p+\epsilon} (\frac {\partial p}{\partial T})_\rho ,
\eqno{(2.67f)}
$$
$$
q \equiv (\frac{\partial p}{\partial \epsilon})_\rho ,
\eqno{(2.67g)}
$$
On the critical curve ${\cal L}\!=\!C$, there are two branches of the solution
and one is good for the accretion and the other for the wind. Note that
unlike the adiabatic case, the critical point occurs when the in-fall velocity 
equals the isothermal sound speed $a_t$, rather than the adiabatic sound speed
$a_s$. The {\it sonic} point, which Flammang calls the {\it sub-critical} point
lies where $u\!=\!a_s$ [77, 89, 90] and the physically significant
solution must pass through this point. A simple observation from eq.
(2.66c) is as follows: since in a black hole accretion,
velocity runs from zero at infinity to unity at the
horizon, at some point of the flow, one must have $u\!=\!a_s$.
Clearly, from eq. (2.66c), at the sonic point, $C$ must vanish to avoid any 
catastrophic enhancement of luminosity at that point (${\hat L}\!=\! 
{\hat L}_s$). It is easy to show that $C\!=\!0$ at $u\!=\!a_s$ always.

To understand the trend of the solution, for example, the variation of the
luminosity function as a function of the radial distance, one notices from
the combination of eqs. 2.64(c-d) that the luminosity ${\hat L}$ may have 
an exponential behaviour with the rate:
$$
k=-\frac{u^2-a_s^2}{u^2-a_t^2}\frac{hu_t{\dot M}}{qr} \frac{{\cal L}}
{{\hat L}}.
\eqno{(2.68)}
$$
In other words,
$$
\frac{d{\hat L}}{dr}=k ({\hat L} - {\hat L}_s ) .
\eqno {(2.69)}
$$
One can choose a new variable $\tau$ along the integral curve such that,
$$
d \tau=k\, dr .
\eqno {(2.70)}
$$
Hence eq. (2.69) becomes
$$
{\hat L}(\tau)-\frac{d}{d\tau}{\hat L}(\tau)={\hat L}_s (\tau) ,
\eqno{(2.71)}
$$
which is integrated along the integral curve to yield:
$$
{\hat L}(\tau) = {\hat L}(\tau_0)\, exp(\tau-\tau_0)
+\int_{\tau}^{\tau_0}{\hat L}_s ({\tau}')\,exp(\tau-{\tau}')\,d {\tau}' .
\eqno {(2.72)}
$$
Since one is interested only in the integral curve which has finite
luminosity as $\tau_0 \rightarrow \infty$, one must have for such solution,
$$
{\hat L}(\tau) = \int_{\tau}^{\infty}
{\hat L}_s ({\tau}')exp(\tau-{\tau}')\,d {\tau}' .
\eqno {(2.73)}
$$
Expanding ${\hat L}_s ({\tau}')$ around $\tau$ one obtains
$$
{\hat L}(\tau)=\Sigma_{n=0}^{\infty}(\frac{d}{d\tau})^n {\hat L}_s(\tau) ,
\eqno {(2.74)}
$$
or,
$$
{\hat L}(r)=\Sigma_{n=0}^{\infty}(\frac{1}{r}\frac{d}{d r})^n {\hat L}_s(r) .
\eqno {(2.75)}
$$
$k$ itself could be written in a more useful form by using eqs. 
(2.64a), (2.67b) and (2.68):
$$
k=-\frac{u}{\xi}\frac
{(u^2-a_s^2)}{(u^2-a_t^2)}\frac{a_t^2}{a_s^2}\frac{1}{u_t(1+u^2)} .
\eqno{(2.76)}
$$
where $\xi$ is the thermal diffusivity given by,
$$
\xi=\lambda_\gamma\frac{4p_r}{\rho}\frac{A}{b}\frac{q}{h}\frac{a_t^2}{v_s^2} ,
\eqno{(2.77)}
$$
where
$$
\lambda_\gamma=\frac{1}{\rho \kappa_1} ,
\eqno{(2.78a)}
$$
is the photon diffusion length, and
$$
A \equiv \frac{\rho}{p+\epsilon} (\frac {\partial \epsilon}{\partial \rho} )_T .
\eqno{(2.78b)}
$$
The important point to note is that $k$ changes sign
at the sonic point $u\!=\!a_s$ by passing through a zero and at the critical
point $u=u_c$ by passing through a pole. The resultant solution is shown in
Figure 2.7 (from [77]). The solution of ${\hat L}_s$ obtained from eq. 
(2.75) blows off at $u\!=\!a_s$, as $k\!=\!0$ at that point. 
A typical 
solution which satisfies the condition that it should pass through the point
$u\!=\!a_s$ when $C\!=\!0$, should look like the dashed curve which 
asymptotically merges with the ${\hat L}_s$ curves. Figure 2.8 shows 
a complete solution [89].
Cross marks are the critical points and open circle marks are the sonic points. 
Parameters chosen in CGS units are $M\!=\! 3 M_\odot$, ${\dot M}\!=\!2.954
\times 10^{-5} M_\odot yr^{-1}$, ${\hat L}_\infty \!=\! 9.795 \times 10^4
L_\odot$, ${\hat L}_{s,\infty}\!=\!9.795 \times 10^4
L_\odot$, $\rho_\infty \!=\! 1.878 \times 10^{-9} g \ cm^{-3}$, $T_\infty\!=\!
4.313 \times 10^5 K$, $a_{s,\infty} \!=\! 8.303 \times 10^{-3} \times c$,
$a_{c,\infty}\!=\! 2.814 \times 10^{-4} \times c$, $[p_{gas}/p_{rad}]_\infty
\!=\!1.532 \times 10^{-3}$.

\noindent {\large 2.4.4 Inclusion of Pair Production and Pre-heating}

Observational results of $\gamma$-ray emission from a wide range of
objects indicate that the flow reaches relativistic and semi-relativistic
temperature regimes. At these temperatures, pair creation takes
place via photon-photon, photon-particle and particle-particle interactions.
The pairs in turn participate in other photon and pair producing processes
or just annihilate into photons. Recent theoretical and observational 
works suggest that AGNs and quasar spectra (close to $ \sim 1 MeV$) may be
better explained with pair
production taken into account [111]. A large number of works are present 
in the literature where pair production effects are considered 
in a flow whose dynamical and thermodynamical properties 
are provided a priori and are not self-consistently determined [112-114].
Detailed  procedure to solve for the equilibrium pair density
in thermal plasma is beyond the scope of this review and the readers 
are recommended to consult works referred above.

Some self-consistent models have been recently 
constructed where the generated radiation 
reacts back on the flow and controls its dynamic and thermodynamic properties.
We describe below one such model, which claims to be self-consistent
[99]. These solutions in spherical geometry are obtained in the range
$3 \le {\dot m} \le 40{\dot M}_{Edd}$. Pair production by all the three
major processes, namely, photon-photon, photon-particle, 
and particle-particle interactions as well as the pair annihilation are 
considered. As a specific example, a flow with $90\%$ hydrogen
and $10\%$ helium by number is considered and is assumed to be
fully ionized. Electron temperatures are
assumed to be different from ion temperatures ($T_i \ne T_e$). 

Hydrodynamic equations which the ions and electrons satisfy are [99]:
$$
\frac{d e_i}{dx} - \frac{e_i + P_i}{n_i} \frac{d n_i}{dx} =
\frac{3}{2} \frac{n_i k (T_i - T_e)}{t_{ep} u }
\eqno{(2.79)}
$$
and
$$
\frac{d e_l}{dx} - \frac{e_l + P_l}{n_l} \frac{d n_l}{dx} =
\frac{\Lambda - \Gamma}{u}+ \frac{3}{2} \frac{n_i k (T_e - T_i)}
{t_{ep} u } + \frac{2}{u}  \frac {e_l + P_l}{n_l} ({\dot n}_C -
{\dot n}_A ) ,
\eqno{(2.80)}
$$
where, $n_i$ is the proper ion number density, $n_+$ is the  positron density,
$n_e$ (=$1.1n_i+ n_+$) is the electron density, $n_l=n_e+n_+$, $P=P_i + P_l$,
$P_i=n_i k T$, $P_l=n_l k T_e$, (the subscript $l$ is to denote
leptons), and $t_{ep}$ is the electron-proton Coulomb time (eq. 2.50),
$u\equiv cU_r$ is the radial velocity, $e_i =\frac{3}{2}
n_i k T_i +1.3 n_i m_p c^2$, $e_l=\frac{3}{2} n_l x^* k T_e + n_l 
m_e c^2 $, $x^*$ is the coefficient to incorporate relativistic effects
($x^*$ equals to $2$ for ultra-relativistic case and $1$ for non-relativistic
case). Also, ${\dot n}_C$ and ${\dot n}_A$ denote the rate of creation and
the annihilation of pairs.
These equations are generalizations of eq. (2.33) described above. The heating
and cooling processes are the usual bremsstrahlung and thermal Comptonization
[113, 115].  

The radiation field is calculated from the radiative transfer equations 
[99, 116]:
$$
\frac{1}{x^2}\frac{d}{dx} (x^2 F) = \frac{u}{c} \chi_{co} 
[F+ u (E+P)] - (\Gamma -\Lambda),
\eqno{(2.81)}
$$
$$
\frac {dP}{dx}+\frac{3P-E}{x} = - \chi_{co} 
\frac{F}{c} - \frac{u}{c} \chi_{co}
(E+P) +  \frac{u}{c}  + \frac{u}{c} \frac {\Gamma - \Lambda }{c},
\eqno{(2.82)}
$$
and
$$
f_E\equiv \frac {P_{co}}{E_{co}}=1-\frac{2}{3} e^{-1/x^2},
\eqno{(2.83)}
$$
where, $\chi\equiv n_l \sigma_{KN}$ is the Klein-Nishina electron scattering
opacity, $\tau \equiv \int_x^\infty \chi_{co} dx$, and $F, \ E \ P $ are
radiation energy density, flux and pressure in fixed frames respectively
(subscript $co$ refers to quantities in co-moving frame). These relativistic
radiation moment equations are correct up to an order of $(u/c)^1$.

The density of pairs is governed by the continuity equations which
include specific pair creation rate and pair annihilation rate [99],
$$
-\frac{1}{x^2}\frac{d}{dx} (x^2 n_+ u)={\dot n}_C - {\dot n}_A .
\eqno{(2.84)}
$$
Photons are assumed to follow Wien distribution $n_\gamma ({\nu} )
\propto \nu^2 e^{-\frac{h\nu}{k T_\gamma}}$. 
The radiation temperature $T_\gamma$ is defined as,
$$
T_\gamma= \frac{1}{4k}\frac{\int_0^\infty e_\gamma^2 n_\gamma d e_\gamma}
{\int_0^\infty e_\gamma n_\gamma d e_\gamma } ,
\eqno{(2.85)}
$$
where $e_\gamma$ is the photon energy. The cross section ($\sigma_{Th}$) for
photon-photon pair creation is chosen to be constant as $T_\gamma$ is close or
below the threshold temperature $m_e c^2 / k $  and most of the pairs
would be produced by photons which are just above this threshold.
The pair creation rates from photon-photon process  [111],
$$
{\dot n}_{C, \gamma - \gamma }= \frac{3}{8} \ ln 4\  \sigma_{Th} c 
n_{\gamma, 1}^2 \ , \ \  n_{\gamma , 1} \equiv \int \int_{h \nu_1 \ h \nu_2
> (m_e c^2 )^2} n(\nu_1) n(\nu_2) d \nu_1 d \nu_2 
\eqno{(2.86)}
$$
and lepton-photon process
$$
{\dot n}_{C, pl - \gamma }= \frac{1}{3} \sigma_{Th} \ c \ \alpha_f 
n_l \  n_{\gamma, 2} \ , \ \  n_{\gamma , 2} \equiv \int_{h \nu > (m_e c^2 )^2/
{k T_e} } n(\nu) d \nu ,
\eqno{(2.87)}
$$
and the pair annihilation rate [112] ,
$$
{\dot n}_A = \frac{3}{8} \sigma_{Th} [ 1 + 2 \theta_e^2 /ln 
(1.12 \theta_e + 1.3)]^{-1} c n_+ n_e
\eqno{(2.88)}
$$
where, $\theta_e\equiv \kappa T_e / m_e c^2 $.
All the other processes such as particle-particle interactions 
and proton-photon interactions are neglected.
The radiation temperature $T_\gamma (x) $ depends upon the
degree of Comptonization. Hence its change depends upon how far
it is away from electron temperature at each $x$ [99],
$$
\frac{d \ ln (T_\gamma )}{d \ ln x} = - \frac {T_e -T_\gamma}{T_e} Y_c,
\eqno{(2.89)}
$$
where $Y_c\equiv 4\theta_e (\tau_{es} + \tau_{es}^2)$ is the Compton
parameter. 

Above equations are solved  iteratively with an outer boundary
condition close to free-fall flow.  For a given ${\dot m}$,
two self-consistent pair densities were found which give rise to
two different models [99]. The higher pair density solution produces
higher luminosity ($l=L/L_{Edd}\sim 10^{-2}$) and higher efficiency
($\eta \sim 10^{-4} - 10^{-2}$). This luminosity is higher
by a factor as much as {\it 10-100} than the lower pair density models.

Figure 2.9 compares the results of a collection of 
`self-consistent' models computed during last two decades.
The figure shows the accretion rate ${\dot m}$ as a function of the
resulting luminosity. These numbers are in units of Eddington rate 
and the Eddington luminosity respectively. The symbols are the following: 
crosses and `circross'es (`cross' with a `circle' at the center) represent
respectively the low and high luminosity branches of the solution of Nobili, 
Turolla \& Zampieri [98].  
Open diamonds are the solutions by Blondin [91]  
while filled diamonds are the results of Wandel, Yahil \& Milgrom [96].
Filled circles and open circles are the high and low pair density solutions of
Park \& Ostriker [99].  
Filled squares and open squares are the 
one temperature optically thin and optically thick solutions of Park [97].
Shapiro [85] solution is shown in solid lines at the lower left.
Dash-dotted line represents low temperature solution of Park [97]. 
Long dashed curves are the solutions of Soffel [92]. 
Time dependent solutions of Ostriker, McCray, Weaver 
\& Yahil [93] and Cowie, Ostriker \& Stark [117] 
fall within the regions I, II, and III bounded by short dashed lines. The 
dotted region in between two solid lines represents the parameter space
with low or no pairs. Solutions by MRT [94], which include the magnetic
dissipation, are of high luminosity and are shown with `cirplus' (`plus'
with a `circle' at the center) and the solutions by CMT [95] (two-temperature
models including Comptonization) are shown with `plus' symbols.

A comparison  of these results indicate that the self-consistent
model with pair production may be relevant to the observed sub-Eddington
luminosity close to the interesting accretion rate of ${\dot m}\sim 1$.
Solutions without pair (e.g., NTZ high) produces luminosity $\sim 10^{-4}$.
Although WYM produces much higher luminosity, it requires very high
accretion rates to do so. The results of CMT shows very high luminosity
($l\sim 10^{-1}$), but they require efficient magnetic dissipation
in the infall time scale. Rest of the steady solutions are likely to be 
unimportant because of low efficiency. The high luminosity time-dependent
models could become important in explaining some long time scale
AGN variability and require further investigations.

\noindent {\Large 2.5 Shocks in Spherical Flows}

It is clear from our discussion in \S\S 2.1-2.2 that in a spherical 
accretion of adiabatic matter onto a black hole, a
flow may have exactly one sonic point. The sonic surface is
spherically symmetric. Such a flow has been proven to be stable by
several workers. Using the most general form of the perturbation,
in a complete general relativistic framework, Moncrief [118] has
shown that: (a) a suitable energy norm of the perturbation outside the
sonic surface remains bounded by its initial value,
(b) no perturbation introduced inside the sonic surface extends to regions
outside the sonic surface, (c) there is no unstable mode
corresponding to a standing shock at the sonic
surface, thus no unstable mode exists inside the sonic surface.
The perturbations within the sonic surface are advected within the black 
hole through the horizon. No part of it is reflected
back, hence it is never amplified. Therefore, a flow, subsonic at a large
distance and becoming supersonic at the sonic point, cannot have a shock.
If there were a shock, the post-shock flow would be subsonic
and it would not have any other sonic point which would allow it
to become supersonic again before disappearing
into the black hole, thus violating the inner boundary condition at the horizon. 
However, when the flow is not adiabatic (or, when the
spherical symmetry is broken, \S5-\S6),
and locally there are sources of heating and cooling, the flow may
pass through more than one sonic point and have shocks in between.
Chang \& Ostriker [119] have investigated this possibility.
In this case, the polytropic index is not assumed to be strictly
constant but a function of the radial distance, i.e.,
$$
n=n(a).
\eqno{(2.90)}
$$
Following the same procedure used in deriving (2.13), one finds in this case
that,
$$
a_c^2[n(a)-\frac{3}{2}(1+\frac{a_c}{2n_c}\frac{dn}{da}|_c)]=n_\infty 
a_\infty^2 .  
\eqno{(2.91)}
$$

Chang \& Ostriker [119] concludes that if $n$ is a decreasing function
of $a$ then there could be more than one sonic point.
Particularly, if $n$ oscillates around $3/2$, there could be more than
one sonic points in the flow and thus shock formation would be possible.
Since for a perfect monatomic gas $n\!>\!3/2$
corresponds to cooling of the flow and $n\!<\!3/2$ corresponds to heating
of the flow, a combination of bremsstrahlung cooling and Compton
heating allowed them to construct quite a few self-consistent
cases where spherical shocks may form.

In their quest to obtain the highest possible luminosity for a given 
accretion rate, Chang \& Ostriker [119]
concentrated their search in the region of parameter space
where the previous attempts [117]
did not find any stationary solutions. The reason for not obtaining stationary
solution without a shock is that shortly after the flow passes through the
sonic point, it is heated so rapidly that the adiabatic sound speed
becomes higher compared to the infall velocity and the flow becomes
subsonic. It remained so in the post-shock region and the inner boundary
condition on the black hole horizon could not be satisfied. A choice of a 
different cooling curve as used by Stellingwerf \& Buff [120] 
allowed them to obtain stationary solutions which do not have standing 
shock waves.

Figure 2.10 shows the region of the parameter space where Chang
\& Ostriker [119] 
find solutions of stationary spherical accretion which contained shock waves 
(shown by open circles). In these solutions, shortly after the flow passes
through the isothermal sonic point, the gas is heated and becomes
subsonic. Subsequently, as the temperature approaches $T_\gamma$, the radiation
temperature (chosen here to be $10^8$K), the Compton cooling becomes more
important and the flow passes through another sonic point. Also presented 
in the figure are
the self-consistent shock solutions of Babul, Ostriker and M\'esz\'aros [121] 
where the effects of pair processes are included as well. Initial 
conditions are chosen from the spherically symmetric solutions of Park [97].
(Kazanas and Ellison [122] also studied
shocks which are supported by ion pressures.)
It is observed that the
luminosities of the order $10^{-1}$ times the Eddington luminosity could 
be obtained. The results seem to show very weak dependence on the exact 
form of the velocity profile in the shocked region of the disk.
The region where high-pair density with shocks 
is achieved is shown in horizontal dashed lines and the region where
very low or no pair density is achieved is shown in slant dashed lines.
Figures 2.11(a-d), show variations of (a) velocity, (b) temperature,
(c) density and 
(d) Mach Numbers in a specific case, clearly showing the formation of a 
standing shock at few $\times 10^6 r_g$. The computation is done for a
$1 M_\odot$ black hole.  The luminosity (in units
of Eddington luminosity) is $0.01$ and the efficiency $\eta = 0.003$.
The dashed curve in 2.11(a) denotes the free-fall velocity and the
dashed curve in 2.11(b) denotes the virial temperature.

Results of spherical flows which include shock waves
indicate that for efficient conversion of the energy, the
disk accretion may not be absolutely essential.  In reality,
of course, the matter would always have some angular momentum,
albeit very low. The angular momentum supported shocks
are located much closer to the black hole (typically, at tens of
Schwarzschild radius). It would therefore
be useful to carry out these analysis for the
more realistic cases of low-angular momentum solutions. Some of these
works will be presented in \S\S 5-7.

\noindent {\Large 2.6 Particle Acceleration at Shock Waves}

\noindent {\large 2.6.1 Basic Physical Processes}

Highly relativistic particles are observed in cosmic rays.
It is generally believed that the shock waves could be
responsible to accelerate particles and to produce non-thermal
radiations. To understand the basic physical mechanisms involved
in the acceleration we first assume plane parallel shock waves [123, 124].
We also assume that the flow has  a weak magnetic field parallel to the
flow direction. We consider only those energetic particles whose gyroradii 
are larger compared to the thickness of the shock waves so that they can 
freely pass from the pre-shock region of the flow to the post-shock region. At
the shock front, the bulk velocity is randomized and the flow becomes
more turbulent and heated up. Turbulent dissipation takes place
in the post-shock region. In the pre-shock region, there are turbulences in the
form of Alfv\'en waves excited by highly energized particles which 
diffuse through the shock and try to escape through the
pre-shock region. These waves scatter the energetic particles
and bring them back in the post-shock region. Thus the same 
particles are scattered again and again by the turbulent wakes
(from post-shock region to pre-shock region) and by the
Alfv\'en waves (from pre-shock region to post-shock region)
and their energies are enhanced every time such crossing takes
place [123, 124].

A cartoon picture of the shock wave is drawn in Fig. 2.12. Mean velocities
of the scattering centers in the pre-shock and post-shock regions
of the shock are $u_-$ and $u_+$ respectively. All the energized particles
in the pre-shock region return to the post-shock region due to scattering,
but only some of these particles diffuse back to the pre-shock
region. Let this probability be $(1-\eta)$ with $\eta <1$. $\eta$ could be
calculated from the diffusion equation:
$$
\frac{\partial n}{\partial t} + u_+ \frac{\partial n}{\partial x} =
\frac{\partial}{\partial x} [D(x) \frac{\partial n}{\partial x}]
\eqno{(2.92)}
$$
where, $n (x,t)$ is the particle density and $D(x)$ is the 
diffusion co-efficient. Assuming that the mass flux
is constant at the shock front as well as in the region far away
down stream, the steady state solution  takes the form:
$$
n(x,t)= A+ B \  exp [ \int_0^x \frac{u_+}{D(x')} dx' ] .
\eqno{(2.93)}
$$
If the energetic particles have velocities $v >> u_+$, then the
region of scattering in the post-shock region extends to infinity. 

The flow of particles in the post shock region away from the shock is 
$u_+ n (x,t) - D(x) \frac {\partial n} {\partial x} $ which is equal to  
the mass flux of the flow, $u_+ n(0,t)$, $x=0$ being the location of the 
shock. Assuming isotropy in the velocity distribution, the rate at which 
particles cross the  shock is $\frac {1}{4} v n(0,t)$ out of which $u_+ 
n(0,t)$ escape into the post-shock region. In other words, $\frac{1}{4} v \eta 
n(0,t) = n(0,t) u_+$ and
$$
\eta \sim \frac {4 u_+}{v} .
\eqno{(2.94)}
$$
During scattering, particle energy, as seen from the
rest frame of the scattering centers remains intact. However, the 
velocities of the centers are different in the two regions. Every time
a particle completes the cycle of scattering from pre- to post-shock
regions and back, the energy is increased by a factor:
$$
\frac{E_{k+1}}{E_k}= \frac {1+v_{k1} (u_- - u_+) cos {\theta_{k1}}/c^2}
{1+v_{k2} (u_- - u_+) cos {\theta_{k2}}/c^2} .
\eqno{(2.95)}
$$
where, $k$ denotes the number of times a particle has 
completed the cycle of crossing and re-crossing the
shock front. Here, $v_{k1}$ and $v_{k2}$ denote the velocities
of the scattered particles from pre- to post-shock region and vice versa
and $\theta_{k1}$ and $\theta_{k2}$ denote the angles made by the
particle path with the shock normal. Assuming $v_{k1} = v_{k2}$,
and averaging over all the incident angles with the
assumption of isotropic velocity distribution of particles,
one obtains the differential energy distribution as,
$$
N(E)dE= \frac {\mu -1}{E_0} (\frac {E}{E_0})^{-\mu} dE ,
\eqno{(2.96)}
$$
where,
$$
\mu = \frac {2 u_+ + u_-}{u_- - u_+} .
\eqno{(2.97)}
$$

For non-relativistic particles of $\gamma=5/3$, the compression
ratio $\rho_+/\rho_-=u_-/u_+\sim 4$, so that $\mu=2$. A power
law slope with the exponent $\mu \sim 2$ is close to what is observed
and therefore confirms the picture of particle acceleration at the shock 
front. In the case of accretion onto a black hole, the velocities
in the pre- and post-shock regions  as functions of the
shock strength (\S 5) are easily obtained from the solution of transonic 
flows, and accordingly, the exponent $\mu$ could be self-consistently
determined.

In the above derivation the relativistic particles are assumed
with $v >> u$. When protons and electrons are both injected,
the protons may be non-relativistic, while the electrons may be relativistic.
When the particles are non-relativistic, 
the probability $\eta$ of the return of the particles depends upon
the energy, which makes computations of $\mu$ more complex.
The energy spectrum for
non-relativistic particles is less steep compared with the energy spectrum
for non-relativistic particles [124]. Thus, for example,
for an equal number of protons and electrons injected at the shock front,
the proton number density would be higher at energies in the range of
$m_e c^2$ to $m_p c^2$. The density of accelerated protons may exceed
that of the electrons by the factor of a few hundred. 

A similar conclusion is drawn from the spectrum of
accelerated particles at relativistic shock fronts by assuming that the 
particles undergo only the pitch angle scattering
on each side of the shock so that only the first order Fermi process
may take place [125, 126]. The spectral index depends
sensitively on the shock velocity in the case of relativistic shocks.
Non-relativistic shocks produce a spectrum with index 
$ N(p) \propto p^{-\gamma}$, with $\gamma = 2.0$  but the relativistic treatment
produces $\gamma$ slightly larger than $2$, in agreement with
observed index of $2.4$. Detailed discussion on the shock acceleration
processes is beyond the scope of this review and 
interested readers may consult Blandford \& Eichler [127].

\noindent {\large 2.6.2 Acceleration at Spherical Shocks and AGN Spectra}

In AGNs and quasars, emitted energy from radio to $\gamma$-ray band
is found to be roughly equal in each decade of frequency range.
This signifies a non-thermal origin of most of the radiation.
A plausible suggestion is
that the production of the relativistic particles which
emit these radiations in AGNs and quasars may be due to
acceleration of protons at the accretion shocks [128].
Some self-consistent study of shock are already in the
literature [129, 130] where the linear, test particle
theory of first order Fermi acceleration is extended to include
the energy losses by the energetic particles, the back pressure
of the accelerated particles on plasma, and a self-consistent
determination of the effective adiabatic index $\gamma_{eff}$. 
These models can be used to determine the luminosity of
radiation emitted from shock accelerated particles [122].

Consider a standing shock at some radius $R_1$ in a spherically
symmetric accretion. In-flowing 
thermal particles of velocity $u_-$ may be shock accelerated via
first order Fermi acceleration and relativistic protons may be
produced provides the pressure to support the shock. The
protons collide in the post-shock region and produce pions
through $p + p \rightarrow p + p + \pi^\pm + \pi_0$. These
pions decay to produce $\gamma$ rays and  relativistic electrons
which subsequently produce observed radiation through 
synchrotron and inverse Compton emissions. Luminosity of the
radiation emitted largely depends upon the ability to pressure
support a shock by the relativistic protons and in the most
plausible cases, it may be $\lsim 10\%$ of the Eddington luminosity.

In the case of spherical flow geometry, the free fall velocity is
given by,
$$
\beta = \frac{u}{c} = (\frac {2GM}{c^2 R})^{1/2} = x^{-1/2} .
\eqno{(2.98)}
$$
and the number density is given by,
$$
n \sim 1.15 \times 10^9 \frac {\dot m}{x^{3/2} M_9^2}  \ \ {\rm cm}^{-3}.
\eqno{(2.99)}
$$
where, ${\dot m}$ is the  accretion rate in solar masses per year
and $M_9 = M/(10^9 M_\odot)$. For reference, the Eddington rate
is ${\dot M}_{Edd}$ is $\sim 2 M_\odot M_9$ yr$^{-1}$ and
the radius of the shock can be expressed as,
$$
R \sim 3 \times 10^{14} x M_9 \ \ {\rm cm} .
\eqno{(2.100)}
$$

As the flow passes through a shock,
the density is increased to $<n_i> \sim 4 n_-$, where $n_-$ is the
number density in the pre-shock flow. The characteristic time scale for the
$p-p$ interaction is given by,
$$
\tau_{pp} \sim \frac{1}{<n_i> \sigma c}\sim 2.4 \times 10^5 \frac{x_1^{3/2}
M_9^2}{\dot m} \ {\rm s} ,
\eqno{(2.101)}
$$
where, $\sigma \sim 3 \times 10^{-26}$cm$^{-2}$ is the $p-p$ cross section
and $x_1$ is the location of the shock in units of the Schwarzschild radius.

Let $Q$ denote the fraction of the energy flux falling onto the
shock which is converted into radiation via $p-p$ collision and pion
decay. The luminosity of the radiation will then be,
$$
L_{tot} = 2 \pi R_1^2 n_- m_p u_-^3 Q .
\eqno{(2.102)}
$$
Here, $L_{tot}$ includes the loss due to neutrinos, $m_p$ is the
mass of the protons and $R_1$ is the shock radius.

This is same as the volume integral of the emissivity $\epsilon=
\frac{<E_{rel}>}{\tau_{pp}}$, i.e.,
$$
L_{tot}\sim \frac {4 \pi R_1^3 <E_{rel}>}{3 \tau_{pp}} ,
\eqno{(2.103)}
$$
where, $<E_{rel}>$ is the average energy of the relativistic particles
($= 3 P_{rel}$, $P_{rel}$ being the average pressure for 
the relativistic particles). Equating these two expressions for
the luminosities we get the radius of the shock as,
$$
R_1= \frac{Q \tau_{pp}}{P_{rel}} \frac{1}{2} n_- m_p u_-^3.
\eqno{(2.104)}
$$
Elimination of $R_1$ and $\tau_{pp}$ using above equations yield
$$
{\dot m} \sim 11 Q M_9 /\eta ,
\eqno{(2.105)}
$$
where, $\eta={P_{rel}}/ n_- m_p u_-^3 $ is the ratio of the post-shock
particle pressure to the incoming ram pressure. 

The quantities $Q, \ \eta$, and the compression ratio $r=\beta_-/ \beta_+$  
for high Mach number ${M}_1$ of the pre-shock flow are [122], 
$$
r \sim 1.4 {M}_1^{3/4}, \ \ \ \ {M}_1 \geq 4,
\eqno{(2.106a)}
$$
$$
Q\sim 1 -3.2 {M}_1^{-0.62}, \ \ \ \ {M}_1 \geq 10,
\eqno{(2.106b)}
$$
and
$$
\eta \sim 1 -2.4 {M}_1^{-0.68}, \ \ \ \ {M}_1 \geq 10 .
\eqno{(2.106c)}
$$
Figure 2.13 shows the variation of the compression ratio $r$ as a function
of the Mach number. The solid curve is drawn when losses are included.
The dotted and dashed curves are drawn when adiabatic flows with
$\gamma=4/3$ and $\gamma=5/3$ are considered.
Figure 2.14 presents the efficiencies $Q$ and $\eta$ as functions of
the Mach number. Since at high Mach numbers, $Q$ and 
$\eta$ approach $1$, one obtains in this limit,
$$
{\dot m} \sim 11 M_9 ,
\eqno{(2.107)}
$$
above which no steady state solution should be possible under the
physical situation described here.

Using the derived quantities above, 
and using definitions of $L_{tot}$, $R$, $n$, and the ratio ${\dot m}/M_9$,
the observed non-thermal luminosity becomes,
$$
L \sim 1.5 \times 10^{47} \frac{Q^2 M_9}{\eta x_1} .
\eqno{(2.108)}
$$
Or, in units of Eddington luminosity, $L_E=
1.3 \times 10^{47} M_9$ erg s$^{-1}$,
$$
l=\frac{L}{L_{E}}\sim 1.2 \frac{Q^2}{\eta x_1} .
\eqno{(2.109)}
$$
Thus the luminosity is inversely proportional to the location of the shock.

The Mach number of the flow is obtained from the ratio of the gas velocity
and the sound velocity,
$$
{M}_1^2 =\frac {u_-^2}{c_{s-}^2}\sim 6.5 \times 10^4 \frac{\beta_1^2}{T_8},
\eqno{(2.110)}
$$
where,
$c_{s-}= [\gamma (T_p+T_e)/m_p]^{1/2}$ and $\gamma=5/3$ is assumed.
Or, substituting $\beta$,
$$
x_1 = 6.5 \times 10^4 \frac{1}{T_8 {M}_1^2}.
\eqno{(2.111)}
$$
The pre-shock temperature is obtained from the balance of the
Compton heating and cooling and is roughly constant $T_8\sim 1$ [131].
Thus the location of the shock mostly depends on the Mach number of the flow 
in the pre-shock region. Non-thermal luminosity drops off rapidly as the 
location of the shock is increased and conversely, increases with the decrease
of $x_1$. For $x_1 \sim 5, \ \  l\sim 0.1$. In the case of extreme
Kerr black hole, the minimum shock location could be much lower: $x_1 \sim 1$,
raising the luminosity to a much higher value.

The total non-thermal luminosity as derived above is directly proportional
to the mass of the black hole and the luminosity is expected to be
much below the Eddington rate. Figure 2.15
shows the comparison between the luminosity as predicted by this model and the 
correlation between mass and luminosity found by Wandel \& Yahil [132].
Wandel \& Yahil [132] assume that the widths of $H_\beta$ lines
are due to the dynamical motions around central black holes, 
and show that the luminosity of a large number ($\sim 70$) 
of quasars and Seyferts actually correlate with the black hole masses.
Open circles and crosses are the computation results of Wandel \&
Yahil [132] for quasars and Seyferts respectively
and the slant solid lines are from the shock acceleration model [122].

If the pairs remain coupled to the accreted matter,
their contribution to the opacity is likely to reduce the effective Eddington
luminosity [133]. As a result, the emitted radiation could contribute strongly
to the pressure support of the shock. This effect is clearly demonstrated
by using a simple model of a shock that 
is mediated by Fermi-accelerated relativistic protons and radiation 
and in which $e^+$-$e^-$ pair are produced through photon-photon collisions.
When pair dominates, the maximum mass
accretion rate for the steady state solutions are found to be significantly 
lower than the value predicted from the model of Kazanas \& Ellision [122].

It is possible that these models are probably irrelevant for mass ranges other
than $2 <\frac{{\dot m}}{M_9} <10$, since the steady spherical 
shock solutions ceases to exist in this scheme. However, in more realistic
accretion flows, other effects such as the presence of angular momentum 
(\S\S 5-7) and magnetic field are important. Here, 
the region of parameter space where shocks can exist is increased, 
and therefore a similar mechanism for the production of non-thermal 
radiation may work for a wider range of accretion rates as well.

\newpage
\noindent {\large\bf 3 Thin Accretion Disk Models}

In the last Section, we discussed the properties of spherically symmetric
accretion flows. The infall velocity was very high and hence for a 
given accretion rate, the density was very low. Therefore,
these flows were found to be inefficient in 
converting the rest mass energy of inflowing matter into radiation.  When 
a flow has some angular momentum, the inflow velocity becomes 
much smaller and the density much higher. The infall time being higher,
viscosity has time to dissipate angular momentum (except in
regions very close to the black hole) and energy. 
As matter loses angular momentum, it sinks deeper 
into the potential well and radiates more efficiently.
In the case of an accretion disk around a Schwarzschild black hole,
the disk can radiate up to six percent of the rest mass of the infalling 
matter and the case of an accretion on a Kerr black hole, the 
efficiency could be up to forty percent depending upon the
rotation parameter (\S 1). However, the actual efficiency
depends on quantities such as viscosity parameter and the cooling
processes inside the disk. This energy is released in the
entire electromagnetic spectrum and the success of a disk model depends on
its ability to describe the way this energy is distributed in various 
frequency bands. The temperature and density distributions as well as 
the geometrical shape of the disk dictates the nature of the
emerging radiation spectrum. These, in turn, depend on the outer boundary 
condition, namely, the rate of matter supply, the specific angular momentum,
as well as the energy content of the matter.

In the case of a binary system, where
one of the components is compact, i.e., a white dwarf, neutron star
or a black hole, the companion is stripped of its matter due to the tidal
effects. The matter, with a specific angular momentum
equal to that of the companion  gradually falls towards the central body
as the angular momentum is removed by viscosity.
The flow encircles the compact primary and forms a quasi-stationary
structure around it, preferably in the orbital plane. This quasi-stationary
structure, commonly known as an accretion disk, may contain very 
little matter compared to the binary components, but it is the most 
important ingredient of an accreting binary system
from the observational point of view. 

To understand the circumstance in which an accretion disk may form
around a black hole, we consider a binary system with component
masses $M_1$ and $M_2$ having angular velocity $\Omega$. The effective
potential of the corresponding Newtonian system is given by,
$$
\Phi_{eff} = -\frac{M_1}{| {\vec r}- {\vec r}_1 |} - \frac{M_2}{| 
{\vec r } - { \vec  r}_2 |} - \frac{1}{2} ({\vec \Omega} \times {\vec r}).
\eqno{(3.1)}
$$
Fig. 3.1 shows the equipotential surfaces (surfaces of constant $\Phi_{eff}$)
of a typical binary system with mass ratio $M_1/M_2=1/3$ drawn with
the center of mass as the origin. The distance is measured
in units of $GM_1/c^2$ and the separation of the stars is chosen
to be $6GM_1/c^2$ for illustration. Five points of four distinct types marked
as $L_1$, $L_2$, $L_3$ and $L_4$ are the so-called Lagrange points where 
$\Phi_{eff}$ is locally or globally an extremum. The value of $\Phi_{eff}$ 
passing through these points are $-1.290$, $-1.187$, $-1.082$ and $-0.9376$ 
respectively. The innermost self-intersecting contour marks the 
Roche lobes of two stars. The lobes meet at  $L_1$, the inner
Lagrange point. Matter from the normal star $M_2$ overflows its
Roche lobe and enters within the Roche lobe of the compact star $M_1$ through
$L_1$ while remaining in the same plane as that of the binary orbit
and eventually forming an accretion disk. Since the flow has a considerable
angular momentum to begin with, it is reasonable to assume that the
disk will form and the viscosity would transport angular momentum 
from the inner part to the outer part to allow matter to accrete onto
the primary. The standard accretion disks, which are Keplerian
in nature, describes the behaviour of the disks of this type.
Of course, some matter from the winds of the companion will also be
accreted by the primary, and the Keplerian flow could 
become sub- or super-Keplerian close to the black hole due to
the terms (such as advection, pressure and cooling efficiency)
neglected in standard disks. Therefore, flows close to the
compact object will be an admixture of Keplerian and sub-Keplerian
matter. Behaviour of such disks will be discussed in \SS 5-7.

In the case of active galaxies and quasars, the situation could be somewhat
different. Matter may be supplied to the central black hole very 
intermittently, and the angular momentum of the supplied matter
at the outer edge is {\it not necessarily} as 
high as Keplerian. This is because the matter may originate from winds 
of star clusters and the supersonic winds may pass through shock waves which
randomize some of their velocity components. They may also lose much of 
the angular momenta through collision  with other
winds. Thus the flow is expected to be of low angular momentum,
quasi-spherical and mostly
advecting ($v_r >> v_\phi$). Matter tend to be almost freely falling
till it `hits' the centrifugal barrier, which brakes the flow
and causes a strong shock formation in the disk. The post-shock flow
has initially very little infall velocity dominated by circular motion,
and can be described by the so-called `thick' accretion disk models. These 
disks have very little viscosity and therefore radiation efficiency
could be somewhat intermediate  between a spherical flow and a
thin Keplerian disk. As the viscosity of these
disks are increased, the shock wave becomes weaker, the
efficiency of radiation becomes higher, and the disks become quasi-Keplerian. 
These disks will be discussed in subsequent Sections. 
In this Section, we shall deal with the standard Keplerian
disk models and its variations.

\noindent {\Large 3.1 Standard Thin Disk Model And Its Spectra}

The standard disk model, which was originally conceived to describe 
Roche lobe accretion in a binary system is discussed extensively 
in the literature [11-12, 70, 134], and so we shall discuss 
this model only briefly. 

\noindent {\large 3.1.1 Model Equations}

In this model, the disk is assumed to be thin, i.e, $H(r) <<r$. 
The heat generated by the 
viscous stress is radiated away efficiently in the vertical direction
and the disk becomes cool $k T << GMm_p/r$, contrary to
what we have seen in the spherical accretion where the temperature
is virial $k T \sim GMm_p/r$. Thus, the thin disks are highly
non-adiabatic. In the thin disk limit, vertical
velocity could be neglected compared to the radial velocity
or azimuthal velocity. The vertical equation is solved independent
of the radial equations. Equations in two directions are therefore decoupled.
The accretion rate is assumed to be much lower compared to the Eddington 
rate and the pressure is negligible so that the radial force balance
equation dictates the specific angular momentum distribution to become
Keplerian. A Keplerian circular orbit of radius $r$ around a 
Newtonian star has an angular momentum, 
$$
{\tilde l}= (GMr)^{1/2}.
\eqno{(3.2)}
$$
Let $2H$ be the vertical height of the disk and $\Sigma$ be the surface
density of the disk at radius $r$. Therefore,
$$
\Sigma \equiv \int_{-H}^H \rho dz .
\eqno{(3.3)}
$$
$\rho$ is computed on the mid plane of the disk. Replacing the integral of
products by the product of the  averages, the above integral becomes,
$$
\Sigma \approx 2 H \rho .
\eqno{(3.4)}
$$
For a Keplerian disk, the stress tensor is 
$$
t_{r\phi}= \eta r \frac{d\Omega}{dr}=-\frac{3}{2}\eta \Omega ,
\eqno{(3.5)}
$$
where, 
$$
\Omega^2=GM/r^3
\eqno{(3.6)}
$$
is the Keplerian angular velocity. Let $f_\phi$ denote the viscous 
stress exerted in the
$\phi$ direction by the fluid element at $r$ on neighbouring element at $r+dr$
(Fig. 3.2).
The viscous stress is related to the stress tensor according to $f_\phi
=-t_{r \phi}$ and therefore,
$$
f_\phi=-t_{r \phi}=\frac{3}{2}\eta \Omega=\frac{3}{2} \eta (GM/r^3)^{1/2} .
\eqno {(3.7a)}
$$
Here $\eta$ is the dynamic viscosity coefficient.

To obtain a steady state disk structure, one has to solve four equations
simultaneously which describe 
the conservations of the rest mass, the specific angular momentum, the 
specific energy and the vertical momentum balance condition. In addition, a 
viscosity law must be specified
which should transport angular momentum outwards allowing matter to fall in.

\noindent (a) Equation describing the rest mass conservation:

As the flow approaches the compact star, it is compressed
due to geometry and the density is increased. Assuming baryons are conserved
the variation of density is obtained from the integral of the continuity 
equation which defines the accretion rate ${\dot M}$ as,
$$
{\dot M}=2 \pi r \Sigma v_r = {\rm constant} .
\eqno{(3.7b)}
$$

\noindent (b) Equation describing angular momentum conservation: 

As the flow approaches the compact star, its angular momentum must
be transported outwards, since the angular momentum distribution
is supposed to be maintained at the Keplerian value. This requires the 
presence of significant viscosity in the disk. The necessary viscous stress
is obtained by forcing the flow to have Keplerian angular momentum everywhere.
The torque applied by the stress is given by (see, Fig. 3.2),
$$
{\cal G}=f_\phi (2  \pi r . 2 H) r = {\dot M} (GM)^{1/2} (r^{1/2}-r_I^{1/2}) ,
\eqno{(3.7c)}
$$
where, $r_I$ is the inner edge of the disk which is usually assumed at the
marginally stable orbit at $6GM/c^2$ in the Schwarzschild geometry.

\noindent (c) Equation governing the energy conservation:

Energy generated by viscous dissipation at each radial distance
must be radiated away. The heat is generated by the viscosity at the rate,
$$
{Q_+} \def  \rho T {\dot s} \sim (t_{r\phi})^2 / \eta = - \frac {f_\phi
t_{r \phi}} {\eta} .
\eqno{(3.7d)}
$$
Using eqs. (3.7a) and (3.7c), the flux (half $\times$ vertical 
height $\times$ heat per unit volume per unit time) is obtained as,
$$
F(r)=H{Q_+}=\frac{3 {\dot M} GM}{4 \pi r^3} [ 1 - (\frac{r_I}{r})^{1/2}].
\eqno{(3.7e)}
$$
Emitted energy does not depend upon the exact viscosity law
because whatever be the law, viscosity has to work in a manner that
the disk remains Keplerian. The luminosity of the disk is obtained by
integrating over the radial extent of the disk,
$$
L =\int_{r_I}^{\infty} 2 F \times 2 \pi r dr = \frac{1}{2}
\frac{GM{\dot M}}{r_I}.
\eqno{(3.7f)}
$$
Note that this luminosity is exactly half the potential energy of the matter
at the inner edge of the disk. This magical factor is due to
the fact the Keplerian  distribution is assumed.
If there were no loss of energy, the rotational velocity at the inner edge
should have been obtained from,
$$
\frac{1}{2} v_\phi^2=\frac {GM}{r}.
$$ 
But instead it is, 
$$
\frac{1}{2} v_{\phi K}^2= \frac{GM}{2r} ,
$$
because of the choice of angular momentum distribution at the inner edge. 
So, half of the energy must come out of the disk
whatever be the physical viscosity, in order to maintain a Keplerian disk.

\noindent (d) Equation governing the vertical momentum balance:

Since the disk is thin, the vertical velocity component must be
much smaller compared to other velocity components. In this case, one can
ignore the advection term (containing
$v_z$) in the vertical ($z$) component of the Euler equation:
$$
\frac{1}{\rho}\frac{dP}{dz} = -\frac{GM}{r^2}\frac{z}{r} .
\eqno{(3.7g)}
$$
Setting $\Delta  P \sim P$ and $\Delta z \sim H$, we obtain,
$$
H\sim \frac{a_s}{\Omega} .
\eqno{(3.7h)}
$$
Thus,
$$
\frac{H}{r} \sim \frac {a_s} {v_\phi} .
$$
Hence, the thin disk condition $H(r)<<r$ boils down to assuming that the flow is
subsonic with respect to the azimuthal velocity.

It is generally agreed that the exact nature of the physical viscous 
mechanisms which may be working in an accretion disk is very poorly
understood. Molecular and radiative viscosities are usually too low
for a cool Keplerian disk.
One of the possibilities is to consider small scale turbulent 
dissipation. In this case, the coefficient of viscosity is given by:
$$
\eta\sim\rho \ v_{turb} \ l_{turb},
\eqno{(3.7i)}
$$
where, $v_{turb}$ is the velocity of turbulent cells relative to the mean
gas motion and $l_{turb}$ is the size of the largest turbulent cells. 
Shocks dissipate energy into heat for supersonic turbulence
to enforce $v_{turb} \leq c_s$. The largest cell sizes are bounded by the 
disk thickness, so $ l_{turb} \leq H$, thus the viscous stress is bounded by,
$$
f_\phi=-t_{r\phi} \leq \rho v_s H \Omega \sim \rho c_s^2 \sim P .
\eqno{(3.7j)}
$$
Therefore, in general,
$$
f_\phi = \alpha P ,
\eqno{(3.7k)}
$$ 
with $\alpha <1$. This is the $\alpha$ disk prescription 
of Shakura and Sunyaev [11]. Some attempts have been made
to obtain $\alpha$ using actual physical processes involved in the disk.
We shall discuss this towards the end of this Section. 

This generally used prescription is inadequate in describing flows which
include discontinuity such as shock waves. We shall discuss this in \S 5.

\noindent (e) Opacity of the disk:

In the standard accretion disk model, first presented
in the context of binary systems, the mass of the black hole may 
be small -- a few times $M_\odot$. The temperature of the flow at the
inner part of the disk is hot enough to have the whole disk fully ionized 
and the opacity may be assumed to be dominated by the
Thomson electron scattering with the coefficient,
$$
\kappa_{es}=0.40 \ {\rm cm}^2 \ {\rm gm}^{-1}.
\eqno{(3.7l)}
$$
In the cooler part of the disk (far from the black hole), absorption dominates
over scattering and in the inner part of the disk, scattering dominates over
absorption. The frequency averaged Rossland mean absorption opacity is:
$$
\kappa_{abs}\sim \kappa_{ff}\sim 0.64 \times 10^{23} \rho T^{-7/2} .
$$
The mean opacity which can be used throughout the disk is then,
$$
\frac{1}{\kappa_{mean}}=\frac{1}{\kappa_{scat}}+\frac{1}{\kappa_{abs}} .
\eqno{(3.7m)}
$$

\noindent (f) Total pressure inside the disk:

Total pressure of the disk is the sum of the gas pressure and the
radiation pressure. For pure ionized hydrogen, 
$$
P(\rho ,T )=\frac{2 \rho k T} {m_p} +P_{rad}.
\eqno{(3.7n)}
$$
If the radiation is in local thermodynamic equilibrium, its pressure
is obtained using Planck distribution: $P_{rad}=\frac{1}{3} a T^4$.

\noindent (g) Radiative Transport Equation:

The heat flux goes out of the disk, mostly by radiation, rather than convection
or conduction. In the optically thick regime, $\tau >> 1$,
and the radiation is transported via diffusion,
$$
F=-\frac{c}{3}\frac{d (aT^4)}{d \tau} .
$$
Here, $\tau = \int_0^H \kappa_{mean} \rho dz \sim K \Sigma$ is the optical
depth computed from the Rossland-mean opacity ($K$ is a suitable 
vertical average of opacity). Replacing differentials by finite differences,
we obtain,
$$
F \approx \frac{acT^4}{\tau} = \frac{acT^4}{K\Sigma} .
\eqno{(3.7o)}
$$
This regime applies to the outer part of the disk, where the thickness 
is high, and the innermost part where the temperature is high.

\noindent {\large 3.1.2 Structure of a Standard Disk}

The eqs.  3.7(a-c), 3.7(e), 3.7(h), 3.7(k), 3.7(m-o) 
can be solved for nine unknowns $\rho(r)$, $H(r)$, $\Sigma (r)$,
$v_r(r)$, $P(r)$, $T(r)$, $f_\phi (r)$, $\kappa_{mean}(r)$, $F(r)$ as functions
of $r$, ${\dot M}$, $M$. The solutions are obtained by 
Shakura \& Sunyaev [11]  and Novikov \& Thorne  [12].
The general result is that for a given $M$ and ${\dot M}$, 
there are three regimes depending upon $r$ [11-12, 70].
The outer region (large $r$) is dominated by the gas pressure and
free-free absorption is the major contribution to opacity.
The middle region (at an intermediate $r$) gas pressure may still
dominate the radiation pressure but the opacity is due to electron
scattering. The transition radius between these two regions
is obtained by the condition $\kappa_{ff} \sim \kappa_{es}$.
At the innermost region (very small $r$) the radiation pressure dominates
and the scattering is the main cause of opacity. 
The transition from the middle region to the inner region
is obtained by the condition $P_{gas} \sim P_{rad}$ .

The expression for the thermodynamic quantities are [11, 70]:
$$
F (r) =5 \times 10^{26}  M^{-2} {\dot M}_{17} r^{-3}
[1-\sqrt{\frac{6}{r}}] \ \ {\rm \ erg\  cm}^{-2}s^{-1} ,
$$
$$
\Sigma (r) =7 \alpha^{-1} M {\dot M}_{17}^{-1}  r^{3/2} 
[1-\sqrt{\frac{6}{r}}]^{-1}
\ \ {\rm gm \ cm}^{-2} ,
$$
$$
H (r) =10^5 {\dot M_{17}} [1-\sqrt{\frac{6}{r}}]\ \  {\rm cm} ,
$$
$$
\rho (r)  = 3 \times 10^{-5}  \alpha^{-1} M {\dot M}_{17}^{-2}
r^{3/2} [1-\sqrt{\frac{6}{r}}]^{-2}\ \ \ \ {\rm gm \  cm^{-3}} ,
$$
$$
T (r) =5\times 10^7 (\alpha M)^{-1/4} r^{-3/4} \ {\rm K} ,
$$
$$
\tau_{es} (r)  =3  \alpha^{-1} M {\dot M}_{17}^{-1} r^{3/2} 
[1-\sqrt{\frac{6}{r}}]^{-1} ,
\eqno{(3.8)}
$$
In these expressions, $M$ is measured in units of the mass of the Sun, 
${\dot M}_{17}$ is in units of $10^{17}$ gm s$^{-1}$ 
and the $r$ is measured in units of $GMM_\odot/c^2$.

\noindent {\large 3.1.3 Emitted Radiation from a Standard Disk}

In case the disk is optically thick and the opacity due to
free-free absorption is more important than the opacity due to scattering,
then the  emission is black body with the surface temperature computed
from the local effective temperature, i.e. [11, 70],
$$
T_s(r) =[\frac{F(r)}{\sigma}]^{1/4} \approx
5 \times 10^7 (\frac{M}{M_\odot})^{-1/2} {\dot M_{17}}^{1/4} r^{-3/4}
(1-\sqrt\frac{6}{r})^{1/4} \ \ {\rm K}.
\eqno{(3.9)}
$$
where $\sigma=ac/4$, and $a$ is the radiation density constant.
The intensity of the emerging radiation  is
$$
I_\nu=B_\nu (T_s)=\frac{2h\nu^3}{c^2} \frac{1}{e^{\frac{h\nu}{k T_s}} -1}.
\eqno{(3.10)}
$$
\noindent The flux of radiation crossing outwards through the surface is 
then related to the intensity $I_\nu$ by,
$$
F_\nu \sim \int_0^{\pi/2} I_\nu cos \theta d \Omega \sim 2 \pi B_\nu (T_s) .
\eqno{(3.11)}
$$
To obtain the intensity distribution, one computes $T_s$ at each radius 
using eq. (3.9). At each frequency, the intensity due to black body emission
at each radius is then calculated from eq. (3.10). It is then
multiplied by the surface area of each annulus at radius $r$.
The contributions are then summed over. However, the above assumption  
of black body radiation is satisfied only in the outer parts of the disk.
In the inner region, where the scattering opacity dominates over free-free
absorption, the temperature
is somewhat higher than the black body temperature and the intensity
of emerging radiation is expressible as,
$$
I_\nu \sim B_\nu (T_s) (\kappa_\nu^{ff}/\kappa_{es})^{1/2} ,
\eqno{(3.12)}
$$
where, $\kappa_\nu^{ff}$ is the opacity coefficients for free-free 
absorption which for ionized hydrogen is given by,
$$
\kappa_\nu^{ff} \sim 1.5 \times 10^{25} \rho T_s^{-3.5} 
\frac{1-e^{-x}}{x^3} \ \ {\rm cm}^2 \ {\rm gm}^{-1} .
\eqno{(3.13)}
$$
Here, $x=h\nu/k T_s$, $k $ being the Boltzman constant. 
In this regime one obtains,
$$
F_\nu \propto \frac{x^{3/2} exp (-x/2)}{{(e^x-1)}^{1/2}} .
\eqno{(3.14)}
$$
This is the modified black body spectrum. Integrating over all range 
of $\nu$, 
$$
F\sim 6.2 \times 10^{19} \rho^{1/2} T_s^{9/4} \ \ {\rm erg\ cm}^{-2} 
\ {\rm s}^{-1} .
\eqno{(3.15)}
$$
Using this expression for flux along with eqs. 3.7(e) and 3.8, 
the temperature is obtained as,
$$
T_s = 2 \times 10^9 \alpha^{2/9} (M/M_\odot)^{-10/9} {\dot M}_{17}^{8/9}
r^{-17/9} [1-\sqrt{\frac{6}{r}}]^{8/9} \ \ {\rm K}.
\eqno{(3.16)}
$$
This is much hotter than the corresponding black body temperature
at the same radial distance $r$. This is because black body emission
is the most efficient mechanism of radiation. Any other 
mechanism must be much hotter in order to emit the same amount of flux.

The integral radiation spectrum of the disk computed for different
accretion rate ${\dot M}$ and viscosity parameter $\alpha$ is shown in 
Fig. 3.3. The upper curves are drawn for accretion rate comparable to the
critical rate ${\dot M}_{crit}=3 \times 10^{-8} \frac{0.06}{\eta} M_\odot$
yr$^{-1}$ with $\eta=0.06$ for Schwarzschild black holes. The spectrum
depends on $\alpha$: higher $\alpha$ causes the spectrum to be flatter.
The lower set of curves are drawn for accretion rate a hundred times smaller
where the outcome is basically insensitive to $\alpha$ parameter as well.
The region `BB' is contributed by the outer region of the disk which
emits black body spectrum: $L_\nu \sim \nu^{1/3}$.
The region marked `Modified BB' is contributed by the inner and the
central regions of the disk where electron scattering opacity is 
dominant over the free-free absorption and the resultant spectrum is flat: 
$L_\nu \sim \nu^0$. The exponential tail $\sim exp(-h\nu/k T)$ is from the 
very hot innermost part of the disk.

\noindent {\large 3.1.4 Stability of a Thin Disk}

Inner region of the thin accretion disks is found to be
notoriously unstable [134] due to thermal and viscous instabilities: 

\noindent (a) {\it Thermal instability}: Consider first the radiation pressure
and Thomson scattering-dominated inner disk. In the time scale we are
interested, the mass transfer is not significant, so that the surface
density $\Sigma$ could be treated as constant. The internal pressure of the gas 
$P$ varies as $P \sim \Sigma \Omega^2 H \propto H$. Hence, the rate of viscous
heating per unit area of disk is given by,
$$
Q^+ = \frac{3{\cal G} \Omega}{4 \pi r^2}= 3 \alpha P \Omega H \propto 
\alpha H^2.
\eqno{(3.17)}
$$
Here, ${\cal G} \sim {\dot M} (GMr^{1/2})$ is the viscous torque (eq. 3.7c)
assumed to be $ \alpha P$. Now, the radiative cooling rate is
$$
Q^{-} = P/\tau_T \propto  H .
\eqno{(3.18)}
$$
When a disk is in thermal equilibrium, $Q^+=Q^-$. As the thickness is slightly
increased, the rate of energy production will be faster than the 
rate of radiative cooling, thereby heating the disk and increasing the
disk height even further. When the disk thickness is slightly reduced, the 
heating would be reduced much faster than the rate at which cooling would
take place. This will cause the disk to become thinner still. This behaviour
shows that the disk is thermally unstable. The instability depends strongly
upon the assumption  of viscosity prescription with {\it constant } $\alpha$.
In a realistic accretion disk, which will advect away energy and
entropy, such instability may not exist.

The general criterion for thermal instability is [11, 134]:
$$
[\frac{\partial (Q^+ - Q^-)}{\partial H}]_\Sigma >0.
\eqno{(3.19)}
$$
Such an instability produces corrugated disk surface. The amplitude of the 
corrugation may be limited by non-linear effects. The growth time scale 
is given by,
$$
t_{th} \sim \frac{2\pi}{\Omega \alpha}.
\eqno{(3.20)}
$$

\noindent  (b){\it Viscous Instability}: This instability develops in a 
much shorter, accretion time scale. In this case, as the accretion rate of 
the disk is suddenly changed, the density should also change in a manner that,
$$
\frac{d( \nu \Sigma)}{d \Sigma} >0 ,
\eqno{(3.21)}
$$
in order to remain stable. If the density cannot increase, for example, with 
the increase of the accretion rate, then the disk will have viscous 
instability. A time dependent behaviour of a thin disk resulting
out of this which has been found 
to be of particular importance in explaining the dwarf novae outbursts
is called a limit cycle behaviour. In this case, the mass accretion rate 
varies with the variation of density (at a given 
radius of the disk) as in Figure 3.4. As discussed
above, the region of the curve $BD$ having negative slope would be viscous
unstable and the disk density would oscillate between the two branches $AB$ 
and $CD$ which have positive slopes [135]. Such behaviour is found to be present
even when one considers accretion disks around super-massive black holes
[136]. In this case, the oscillatory behaviour is seen with period
$10^{4-7}$yr operating at distances of $10^{15-16}$cm. Of course, 
self-gravity of the disk might play an important role in active galaxies
so that the simple dwarf-novae type computation need not hold.

\noindent {\large 3.1.5 Viscosity in a Thin Disk}

There is a big uncertainly in the nature of the viscosity 
inside a thin accretion disk. One of the major
problems is to explain the origin of sufficiently large
viscosity that seems to be present inside accretion disks
in the binary systems. The generally accepted
view is that the viscosity may be due to magnetic transport
of angular momentum [137-139] or due to turbulent transport [140] which can
operate inside the disk. If magnetic field is coherent
and threads the disk as in proto-stellar disks, the transport
through the wind is efficient [137]. If the accretion flow contains
stochastic flux tubes, they would be repeatedly sheared and re-connected.
Some blobs with larger angular momentum would diffuse radially outwards
and some other would be advected inwards. The net effect causes
the transport with an effective $\alpha\sim 0.01$.
Other methods of magnetic transport include the production of violently
turbulent disk due to rapid growth of weak magnetic
fields by an instability in presence of differentially
rotating Keplerian disks [139]. This so-called Balbus-Hawley
instability works in the following way: Imagine two blobs of matter
attached to a vertical  magnetic flux tube (at $R=R_0$)
inside the differentially rotating disk which rotates faster at 
smaller radius (Fig. 3.5). Let the blob
positions be displaced radially, in a way that one blob is inside the
other (radially). Magnetic flux tube will force them to rotate
with the angular velocity $\Omega (R_0)$ 
of the average position $R_0$ of the fields. Thus, the blob inside 
(at $R_0-dR$) will be forced to rotate slower than its ambience.
This will reduce its centrifugal force and it will sink further.
The outer blob 
(at $R_0+dR$) will rotate faster than its ambience. This blob will have
more centrifugal force than the average value and will move  further
outwards. This instability causes net transport of angular momentum
outwards and would be a source of viscosity. The calculated
$\alpha$ is about $\sim 0.01$. In presence of toroidal
magnetic field, such argument may be applicable provided magnetic
tension is not important (Fig. 3.5). Close to a black hole,
where the tension may be the dominent force (if the temperature
is high enough: $T \gsim 10^{10}$K) [141] the whole flux tube would 
be advected towards the black hole in a much shorter time than
the growth time scale of this perturbation.
Another detailed model which attempts to compute
$\alpha$ parameter assumes that the angular momentum is transported 
by internal waves inside an unmagnetized
disk. Vishniac \& Diamond [140]  
use perturbation analysis of a thin, inviscid accretion disk
with adiabatic vertical structure, including the effects of 
shearing, but ignoring radial structure to calculate the so called 
$\alpha$ parameter in a quantitative manner. They classify the 
waves into two classes: sound waves and internal waves. 
The non-axisymmetric mode $m=\pm 1$ inward traveling internal waves deposit
negative angular momentum in the disk. These are excited at the outer rim
of the disk at the location of the accretion stream hitting the disk.
An estimate of the effective viscosity is $\nu=(H/r)^2 (H/c_s)$ where $H$
is the disk scale height and $c_s$ is the velocity of sound. 
These internal waves are similar to the spiral shock waves
generated in simulations of Sawada, Matsuda and Hachisu [142]
in the context of a two dimensional thin disk. The internal waves are 
incoherent as opposed to the spiral shock 
simulations [142] where a coherent shock pattern forms.

\noindent {\Large 3.2 Two Temperature Disk Models}

Thermal, geometrically thin and optically thick disks
which have been discussed so far, can be
unstable at the inner edge and are too cool to produce X-rays observed
in binaries and active galaxies and quasars. 
There are some models where the disks are much hotter. These are
optically thin, two temperature Comptonized soft photon disks 
and pure bremsstrahlung disks [143].
As before, 
there are two parameters in these models, viscosity parameter $\alpha$ and
accretion rate ${\dot M}$. For physically reasonable values of these
parameters, the temperature could become very high since the potential
energy is virialized, particularly when the accretion rate is
low (\S 4.6). Soft photons from the
outer disk is intercepted by the inner region and are Comptonized.
An interesting variation of these models is discussed by
White \& Lightman [144] who consider two temperature Comptonized bremsstrahlung
disks. In these disks, there is a critical accretion rate ${\dot M}
(\alpha )_{crit}$ above which there is no steady state solution 
for $r \leq 10GM/c^2$ (runaway pair production). Depending on
$\alpha$, ${\dot M}_{crit}$ corresponds to a rate ranging from $0.03$
to $0.1$. For ${\dot M}< {\dot M}_{crit}$, there are two 
solutions, one with a high density and the other with a low density of pairs.
Another important result from these studies is that for Comptonized soft 
photon disks, steady solutions are possible for all ${\dot M}$ and $\alpha$.
Reasonable amount of X-rays could be produced by this mechanism
though the details of the observations required further refinement
which we discuss on \S 7.2. These results hold for two temperature disks.
If electron
and protons couple more strongly than the Coulomb coupling, the temperatures
become equal. In this case, there is also a critical accretion rate, 
and depending upon $\alpha$, the luminosity varies from $6 \times 10^{-5}$ to 
$0.03 L_{edd}$. Single temperature disks are very hot and might be ruled 
out by the constraints from $\gamma$-ray background and therefore no efficient
coupling between protons and electrons may be allowed in these disks. 

\noindent {\Large 3.3 Transonic Disk Models}

The standard Shakura-Sunyaev thin accretion disk models [11-12]
do not accurately describe the flow close to the
black hole as the solution is terminated at the marginally
stable orbit $r_{ms}$. Since the black hole accretion is
necessarily transonic, it may be important to know how a Keplerian
flow behaves when it passes between the marginally stable orbit and the
horizon of the black hole. A number of efforts have been 
made to study the so-called `transonic' solution [145-146].
A general feature of these solutions is that, unlike Shakura-Sunyaev disks, 
these disks pass through the inner sonic point close to the horizon. 
They are otherwise Keplerian at larger distances. Though the first goal, 
i.e, to have a solution which passes through the inner sonic point
is achieved with success, smoothly matching of Keplerian solutions
has been a major problem and this usually introduces some
arbitrary parameters, such as the matching radius, matching radial
and angular velocities etc. Figure 3.6, taken from Abramowicz et al. [146]
shows  a few self-consistent angular momentum distribution inside disks 
which are Keplerian at large distances. The distribution seems to depend
strongly upon the radial distance where Keplerian distribution is matched. 
The important feature is that the angular
momentum remains almost constant close the black hole as the flow
becomes supersonic. Another important aspect of this work is that it
attempts to `stabilize' thin accretion disks against viscous and thermal
instability [134] by advecting the perturbations away as quickly
as they develop. Such features are present also in globally complete
advective disks discussed in \S 5.  This supersonic inner region is probably 
less relevant for the observational point of view as the infall time in this
region is much smaller compared to the viscous time scale, and therefore
this region does not radiate efficiently. The spectrum and other properties of 
these smooth transonic solutions are expected to be similar to the standard 
thin accretion disks and therefore we shall not discuss them any further.
Recently, this model is generalized by removing the assumption of vertical 
integration and solving the equations self-similarly. We shall discuss
them in \S 5.5. 

\noindent{\Large 3.4 Magnetized Disk Models}

It is usually believed that the origin of the bipolar outflows and jets
are related to the properties of accretion disks. Outflows and jets
can extract angular momentum from disks very efficiently (especially
when there is no binary companion), and thereby aid accretion process.
Since various properties (such as velocity, kinematic luminosity, etc.)
of an outflow must be directly related to the properties of the underlying
disk, theoretical disk models as well as numerical simulations
must produce both the disk and the jet {\it simultaneously}. There are a few
theoretical models which achieve this in the context of bi-polar
outflows [137] 
and in the context of radio jets [147-150]. 
Pudritz \& Norman [137] show that angular momentum transport in a bipolar 
outflow is consistent with its loss from the proto-stellar disk. 
In the latter case [147-150], magnetohydrodynamic equations are solved
inside as well as outside of the disk and the solutions are
matched on the disk boundary. So far,
no satisfactory numerical work is present which successfully produce
collimated and accelerated outflows/jets {\it  which are originated}
from magnetized disks, even though separate simulations of magnetized disks
and jets are present in the literature. Some of the simulations have obtained
new insights into old problems -- examples being the Balbus \& Hawley
instability [139] and the detailed instabilities at the
working surface of a radio jet. 
Below, we present a comparative study of the  typical self-consistent solutions
of the magnetized accretion disk which also produces collimated radio jets.

Blandford \& Payne [147] consider centrifugally driven self-similar MHD winds 
originating from infinitesimally thin, Keplerian disks. 
In this model, matter is ejected
using a `sling-shot' mechanism if the field line is sufficiently
bent outwards, much like the outward motion of a `bead' on an
inclined and rigidly rotating wire. The effective potential
for the bead on a wire attached at $r=r_0$ on a Keplerian disk is
[147],
$$
\Phi= - \frac{GM}{r_0}[\frac{1}{2} (\frac{r}{r_0})^2 + \frac {r_0}
{(r^2+z^2)^{1/2}}]
\eqno{(3.22)}
$$
The `bead' will be unbound if the wire makes
an angle $\theta >60^{o}$ with the vertical axis. In a magnetized
disk, the wire is likened to be the magnetic field lines which may be thought
to rotate rigidly till the Alfv\'en point and the bead is likened to be
lumps of matter ejected from the disk. This work shows that collimated
bipolar jets may be formed from a Keplerian disk. In reality disks
need not be Keplerian and in presence of a black hole geometry the potential
would be modified, particularly close to the black hole. Close to a black hole,
one would require  a much higher inclination (than $60^{o}$) in order that
matter may come out of the strong gravitational field. 

Contopoulos \& Lovelace [151] assume a power law distribution of the
magnetic field, $B(r,z) \propto r^q B(z)$ and a self-similar
flow out of an infinitesimally thin Keplerian disk. They find that 
asymptotically collimated jets are obtained when $q <-1$. K{\" o}nigl [149]  
numerically determines the self-similar magnetic field  structure
inside a cold, partially ionized Keplerian disk. He also shows that there exists
disk parameters which can, in principle, produce Blandford-Payne [147]
type winds that attain super-Alfv{\'e}nic velocity at a finite distance from
the disk surface.  

A more general solution of the field where the field inside
the disk, as well as outside the disk are solved simultaneously, are 
presented in Chakrabarti and Bhaskaran [150].
Similar to Blandford \& Payne [147] and K\"onigl [149] solutions, this work 
also assumes that the field is self-similar in radial direction, but the 
functional relationship inside the disk is different from that of the
outside. The solutions are obtained analytically by Taylor series expansion
of the field variables inside and outside the disk and matching the
solutions at the disk boundary, the location of which is obtained
self-consistently. The important result which emerges from this
exercise is that it proves analytically the existence of
a complete outflow solutions  from a magnetized
disk which pass through the Alfv\'en surface at a finite distance.
Solutions are obtained using a non-Keplerian angular momentum distribution
$l =C_1 r^n$, where $n$ need not be $0.5$ and $C_1$ need not be $\sqrt (GM)$.
Indeed, more promising solutions are obtained only for sub-Keplerian disks.

A typical solution of magnetized accretion disk  is shown in
Fig. 3.7 which contains all the generic features found in these
solutions. The solution is drawn in $R-Z$ plane.
The solid curves are the resultant field lines. 
The long-dashed line indicates the disk boundary where two solutions
are matched and the inclined short-dashed line indicates the Alfv{\'e}n 
surface. One notices that inside the disk
the magnetic field lines are bent as if advected by the inflow. Outside
of the disk, the field lines
seem to diverge due to centrifugal acceleration. 
But this is soon overcome by the tension of the azimuthal 
field [which ($\sim 1/r$)
dominates over poloidal component ($\sim 1/r^2$) as the flow propagates 
outwards], because of which the field lines bend back towards the axis. 
It is expected that after the initial collimation, the field lines 
should become parallel to the $Z$-axis, the
axis of the outflow, as the magnetic pressure is likely to rise near the axis.
The inclusion of higher order expansion in Taylor series would be needed
to show such effects. According to 
the criterion of Contopoulos \& Lovelace [151],
$q=-0.5$ in these solutions (flux is assumed constant with radial
distance [150]) and therefore these represent well behaved collimated
asymptotic solution.

The jet solution presented above has all the
necessary ingredients to be a `good' solution. They are bent along the
flow inside the disk, diverge outside the disk due to centrifugal force and 
then collimate back towards the axis to form the jet. A large number of other
recent works have been able to reproduce some of these aspects [152-154].
Particularly interesting is the time evolution of the 
advected magnetic field which vertically threads an accretion disk
[154]. This work shows that field lines, initially uniform in the vertical
direction are carried with the flow towards the center inside the disk and
form an hour-glass pattern.
Some other studies also [148, 155] show that azimuthal component of 
the magnetic field does contribute to collimate jets. 

Another class of magnetized disk solutions are present in the
literature in which the background flow is assumed to be the same
as the standard Keplerian disks and the magnetic field is assumed
to be sheared and advected by the flow [156-157]. These works show that
magnetic dynamo is not essential in disk models if sufficient
magnetic flux is steadily fed into the disk. The field configurations
close to the disk surface become Blandford-Payne [147] type and 
can, in principle, launch cosmic jets. In Figures 3.8(a-b), the field topologies 
(with even and odd symmetries) inside a disk are shown which 
can probably be matched with jet solutions.
Apart from these disk models showing generic effects, some completely
new physical results have been discovered in Keplerian disks around a Kerr
black hole. The gravitomagnetic potential of a rotating black hole
in the presence of a differentially rotating disk is seen
to drive a self-excited dynamo and amplify weak (even axisymmetric)
magnetic field to a higher field strength [158]. The source term for the 
dynamo effect is present in the Maxwell's equation in Kerr metric [159] 
and therefore does not violate Cowling's theorem. Figure 3.9
shows the poloidal flux $\Psi$ at $t=4.6$ times the diffusion time scale
$$
t_{diff} = \frac{r_g^2}{\eta_0}\sim 4 \times 10^5M_9 (0.1/\alpha_T)
(r_g / 2 H_{in})^2 (c / 2r_g \Omega_H) \ \ \ {\rm s},
\eqno{(3.23)}
$$
where, $\eta_0$ is the turbulent diffusivity at the horizon for extremely 
rotating black hole, $M_9$ is the 
mass of the black hole in units of $10^9M_\odot$, $\alpha_T$ is the 
$\alpha$ parameter due to the turbulent viscosity, $r_g$ is the Schwarzschild
radius, $H_{in}$ is the height of the disk at the inner edge and $\Omega_H$
is the rotational velocity of the black hole on the horizon. The Figure
is drawn for the Kerr parameter $a=0.98$.

The jet solutions presented above have been constructed in a manner 
that the angular momentum transport by the jet is such that
the disk is maintained Keplerian or quasi-Keplerian. In reality,
disks in active galaxies need not have so much angular momentum 
in the first place and therefore, so-called `angular momentum' problem
does not apply. In \S 4.6 we shall discuss magnetic thick disks
where the toroidal fields are ejected from the disks due to combined
effects of magnetic tension and buoyancy and the jets may be produced
without significant poloidal component.

\noindent{\Large 3.5 Thin Disks with Weber-Devis Magnetic Field Configuration}

One of the simplest magnetized disk models could be a
conical flow which has ordered radial and azimuthal magnetic field components.
Such a flow passes through the magnetosonic points and can also 
have shock waves in them. In the context of solar winds, such field
configurations were first studied by Weber \& Devis [160]. Although these
studies were originally made for wind type flows, recent 
studies of accretion and wind flows with these
field configurations show very rich topologies in the phase space
and they have now been extensively studied [77, 161]. We present below
the general procedure of studying these solutions using a pseudo-Newtonian
potential. These solutions (which are not self-similar)
have completely new topologies with fast-magnetosonic points
forming at finite distance. They have since been generalized in the external
Kerr black hole geometry and are seen to have the same features [161]. 

Let us start with a conical wedge shaped flow which is assumed to have
an adiabatic equation of state $P\!=\!K\rho^\gamma$, 
where $P$ and $\rho$ denote the isotropic gas plus
radiation pressure and the matter density respectively, $K$ is the adiabatic
constant which measures the entropy of the flow and $\gamma$ is the
adiabatic index given by the ratio of specific heats. We use
$\gamma\!=\!4/3$ for the relativistic flows. $K$ remains constant
throughout the flow if it is shock-free, otherwise it changes at the
shock. The solution with shock waves will be discussed in \S 5.

\noindent {\large 3.5.1 Basic Equations and Magnetosonic Point Conditions}

The basic equations are [cf. \S 2]: 

\noindent (a) the energy conservation,
$$
E = \frac{1}{2} u^2 +\frac{1}{2}\vartheta_\phi^2 + \frac{\gamma}
{\gamma-1}\frac{p}{\rho} +\Phi(r)-
\frac{B_\phi B_r \Omega r}{4 \pi \rho u} ,
\eqno{(3.24a)}
$$

\noindent (b) the angular momentum conservation,
$$
L = r \vartheta_\phi - \frac {B_r B_\phi r}{4 \pi \rho u} ,
\eqno {(3.24b)}
$$

\noindent (c) the mass flux conservation,
$$
{\dot M}= \rho u r^2 ,
\eqno {(3.24c)}
$$

\noindent (d) the radial magnetic flux conservation,
$$
C_1=B_r r^2 ,
\eqno{ (3.24d)}
$$
and finally,

\noindent (e) the Maxwell's equation for a perfectly conducting fluid, 
${\vec e}\!=\! - {\vec v} \times {\vec B} \!=\! 0$,
$$
r (u B_\phi - \vartheta_\phi B_r ) = -\Omega C_1 .
\eqno{ (3.24e)}
$$
Here $u$ is the radial component of the velocity,
$\vartheta_\phi$ is the azimuthal component of velocity,
$B_r$ is the radial component of magnetic field,
$B_\phi$ is the azimuthal component of the magnetic field,
$\Phi(r)$ is the gravitational potential due to the central star, and
$\Omega$ is the constant angular velocity of the surface of the star.
In the energy conservation eq. (3.24a), the final term
represents energy transported out by magnetic fields and is equal to
the Poynting energy flux. The
specific angular momentum $L$ in (3.24b) consists of two terms:
the first term is the ordinary angular momentum and the second term
is due to the torque associated with magnetic stresses. 
For the present purpose we rewrite eqs. 3.24(a-c) as follows [162]:
$$
{\cal E}= E-L\Omega = \frac{1}{2} u^2 +
\frac{1}{2}\vartheta_\phi^2 + na^2 +\Phi(r)-
\vartheta_\phi \Omega r ,
\eqno{ (3.25a)}
$$
$$
L = r \vartheta_\phi - C_2  r B_\phi ,
\eqno{ (3.25b)}
$$
$$
{\dot{\cal M}}= a^{2n} u r^2 ,
\eqno{ (3.25c)}
$$
where $a\!=\!({\gamma p}/\rho)^{1/2}$, the adiabatic sound speed and
$$
C_2=\frac{B_r}{4\pi\rho u}=\frac{C_1}{4\pi{\dot M}}.
\eqno{(3.26)}
$$
$\Phi(r)$ is given by $\Phi_N\!=\!-GM/r$, or $\Phi_{S}\!=\! -GM/(r-2GM/c^2)$ 
depending upon whether the flow is around a Newtonian star or a compact,
general relativistic non-rotating star or a black hole. $n\!=\!1/(\gamma-1)$
is the polytropic index. The specific energy $E$ has been
re-defined after incorporating the constant rotational contribution.
Thus, ${\cal E}$ is allowed to be negative even for an unbound flow.
${\dot{\cal M}}\sim K^n {\dot M}$ measures the entropy of the flow
and usually changes at the shock waves. As in \S 2, we shall use the phrases
`accretion rate' for ${\dot{\cal M}}$ and `mass flux' for ${\dot M}$.

First, we eliminate $B_\phi$ from eqs. (3.24e) and (3.25b) to obtain
$$
\vartheta_\phi=\frac{uL-\Omega C_1C_2}{ru-\displaystyle{\frac{C_1C_2}{r}}}.
\eqno{(3.27)}
$$
Since $\vartheta_\phi$ is expected to be smooth everywhere one obtains the
condition that wherever denominator vanishes, numerator must
also vanish there. This condition is satisfied at the Alfv\'en radius,
$r\!=\!r_a$, where $u\!=\!u_a\!=\!C_1C_2/r_a^2\!=\!\Omega C_1C_2/L$. 
The total angular momentum $L$ and the angular velocity of the star $\Omega$ 
are related: $L\!=\!\Omega r_a^2$. Because the flow has to pass through
the Alfv\'en radius $r_a$ with Alfv\'en velocity $u_a$, 
we use below these as the units of distance and velocity.
Time is measured in units of ${\frac{r_a}{u_a}}$.
Thus, in these units $L\!=\!\Omega$, $B_{ra}\!=\!C_1$ and
the product of the constants $C_1C_2\!=\!1$. Equation (3.49) reduces to
$$
\vartheta_\phi=\frac{(u-1)\Omega r}{ur^2-1}.
\eqno{(3.28)}
$$
We continue to use the same notations for the dimensionless
quantities since there arises no confusion. To simplify the expressions,
we specify the field $B_r$ at the Alfv\'en point to be unity prior
to any calculations so that $C_1\!=\!C_2\!=\!1$, $4\pi{\dot M}\!=\!1$. 
Also, $g_0\!=\!GM/(u_a^2r_a)$ is a constant which measures the mass of 
the central body and ${\bar c}\!=\!c/u_a$ is the
velocity of light in units of the Alfv\'en speed. Differentiating
eqs. (3.24a) and (3.24b) with respect to $r$, and eliminating 
${da}/{dr}$, we get
$$
\frac{du}{dr}=\frac{N}{D} ,
\eqno{(3.29)}
$$
where
$$
N=ur[(ur^2-1)^2(u-1)\Omega^2+(\frac{2a^2}{r^2}-\frac{\Phi ' }{r})
(ur^2-1)^3+u\Omega^2(1-r^2)(u-1)(ur^2+1)] ,
\eqno{(3.30a)}
$$
and
$$
D=(u^2-a^2)(ur^2-1)^3-u^2r^2\Omega^2(r^2-1)^2 .
\eqno{(3.30b)}
$$
Here
$$
{\Phi '}={ \Phi ' }_N = g_0/r^2 ,
$$
for a Newtonian star, or,
$$
{\Phi '}=\Phi_S=g_0/(r-2g_0/{\bar c}^2)^2 ,
$$
\noindent for a Schwarzschild black hole or a compact nonrotating star. Prime
indicates derivative with respect to $r$. It is clear that
both numerator and denominator vanish identically at the
Alfv\'en point. However, there are other points in the phase space
[spanned by the pair $(r, u)$] where they vanish also. These
points ($r_c, u_c$) are the slow and the fast magnetosonic points.
These locations are obtained by eliminating sound speed
$a$ from the numerator and the denominator for a given ${\cal E}$ and $L$.
In Fig. 3.10, the contours of constant ${\cal E}$
and $L$ in the $(r_c, u_c)$ plane are shown. Solid contours are drawn for
$L\!=\! 1.2, \  1.8,\ 2.2345  ,\ 3.0 ,$ and the
dashed contours are drawn for ${\cal E}\!=\!1.2  ,\ 1.8 , \  2.355,
\ 2.8$ respectively. A non-dissipative flow preserves energy and angular
momentum. Therefore, the intersections of given ${\cal E}$ and $L$ curves 
give the locations of all critical points corresponding to
these conserved quantities. The points where
$\displaystyle{\frac{du_c}{dr_c}}|_{\cal E}< 0$
are the `X'-type critical points and the points where
$\displaystyle{\frac{du_c}{dr_c}}|_{\cal E} >0$ are the `O'-type critical 
points. For example, five critical points formed at the intersections of
${\cal E}\!=\! 1.8$ and $L\!=\!1.8$ curves are marked by `X' and `O'.
The points where $u_c r_c^2  < 1$ are the slow magnetosonic,
and the points where $u_c r_c^2 > 1$ are the fast magnetosonic.

\noindent {\large 3.5.2 Solution Topologies of Magnetized Accretion}

For a given set of parameters a flow can have
five critical points other than regular Alfv\'en point at $u\!=\!r\!=\!1$.
Figures 3.11(a-f) show distinctly different topologies of the solution.
Here the radial velocity $u$ is plotted against $r$. Contours are
of different `accretion rates' ${\dot{\cal M}}$. Each diagram is drawn for
a different energy. In
(a) $L\!=\!10^{-4}$, ${\cal E}\!=\!2.0$,  Bondi-like slow point is present. In
(b) $L\!=\!1.0$, ${\cal E}\!=\!-0.5$, the rotational slow and
the fast `O'-type points are present, in (c) $L\!=\!1.0$, ${\cal E}\!
=\!1.8871$, all the five sonic points are present. 
Notice that the lower left branch of Bondi-like point
is not connected to infinity. This branch opens up
as $L$ is increased as in (d-f). In
(d) $L\!=\!1.7$, ${\cal E}\!=\!-2.0$, the Bondi-like slow and
the fast `O'-type points are present. 
In (e) $L\!=\!1.7$, ${\cal E}\!=\!0.0$, the Bondi-like
and the rotational slow points, and slow as well as the fast `O'-type
points are present. In
(f), $L\!=\!1.7$, ${\cal E}\!=\!4.0$, all the fast points are absent.
The critical wind solutions are marked by $u_{\alpha 1}$, $u_{\alpha 2}$, 
$u_{\beta 1}$, and $u_{\beta 2}$. 

In the case of accretion flows, any of the topologies in Fig. 3.11(a),
3.11(c-f) could be used as the flow passes through the magnetosonic
point close to the black hole. The choice depends upon the
accretion rate. The topology 3.11(c) is important for
shock transitions, which we shall discuss in \S 5.

\noindent {\Large 3.6 Outflows by Extraction of Energy from Black Holes}

So far, we discussed the formation of winds
using thermal, kinetic, and/or magnetic energies close to the
origin of the outflows on the accretion disks. It is well known that
within the ergosphere [defined by $r=M+(M^2-a^2 cos^2 \theta^2)^{1/2}$]
where any flow is forced to co-rotate with the black hole, energy of 
test bodies could be negative. This property can be exploited to extract
energy directly from the black holes, as was conceived by Penrose [163].
In this case, matter with positive energy could accrete from infinity
and `somehow' split into two components with one component having
negative specific energy and the other component having 
specific energy more than that at the initial flow. If this latter
component could be expelled from the ergosphere, then an energized
outflow could be created. The energy extraction can continue until
the black hole attains the so called irreducible mass [164] $M_{ir}$ defined
by 
$$
M^2= M_{ir}^2+a^2M^2/4M_{ir}^2.
$$
Thus, $M_{ir}= (r_+M/2)^{1/2}$ ($r_+$ being the horizon).
It is clear from the definition that only the rotational energy could be
extracted. For an extremely rotating Kerr black hole $a=1$ and 
the corresponding `reducible' energy is $\Delta M = M-M_{ir} \sim 29\%$. 

Several works are devoted to device models to extract this energy in 
astrophysically feasible way and in most of the cases the process involves
magnetic fields. Similar to Blandford-Payne process described 
in \S 3.4, where the disk is slowed down by the magnetic fields
passing through it, one could imagine slowing down the black hole itself
by threaded magnetic fields. Magnetic flux tubes advected by the
accretion disk may be collected close to the black hole which 
eventually threads the hole.
This is possible because a black hole has a finite resistance [165]
as the field decay time of a conducting sphere of radius $R$
cannot be shorter than the light crossing time $R^2/4\pi\sigma \sim
GM/c^3$. 

An important participant to the energy extraction process is
a new force close to a rotating black hole, known as the gravitomagnetic force.
This is derived from the so-called gravitomagnetic potential [166] $\beta^j$.
This is also called the `shift function' since this is the rate
at which the grid coordinates shift with respect to zero angular
momentum observers, i.e., observers moving with angular velocity,
$$
\omega \sim \frac{2aMr}{\Sigma^2}
$$
as seen from infinity, where $\Sigma=r^2+a^2cos^2\theta$.
The gravitomagnetic force interacts with the external magnetic field
${\vec B}$ to produce a magnetic-gravitomagnetic battery effect with EMF [166],
$$
EMF= \int_{\cal C} -{\vec \beta} \times {\vec B} . {\vec dl}
$$ 
the integral is taken along a closed curve suitably chosen to 
include regions along and across the threaded field lines.
In a special case, when the external field is uniform at infinity,
an exact solution of the electric and magnetic fields could be obtained [167]. 
The battery is very powerful, producing an electric field of about
$E \sim Ba/M \sim 3\times 10^6$ volts/cm $(a/M)$. Thus the voltage drop
between the horizon and a few Schwarzschild radii would be tremendous:
$V \sim E r_+ \sim Ba \sim 10^{20}$ volts for a $M=10^9M_\odot$
black hole [166]. Particularly interesting is the region where the
electric field is parallel to the magnetic field lines. Here,
similar to the pulsar model which produces a spark gap on the neutron star
surface, the charged particles from the accreting disk
could be accelerated along the field lines rapidly, causing
them to radiate high energy $\gamma$-ray photons tangential to the
field lines. Charges may also be formed by vacuum breakdown
due to high voltage and hard $\gamma$ rays may also be 
supplied by hard bremsstrahlung radiation from accretion
disks with a very small accretion rate (which keeps the disks
hotter). As the field lines bend, these photons 
scatter off these fields producing electron-positron pairs
which  are again accelerated and the same process continues
creating a magnetosphere around a rotating black hole [168].
The field required for this process is around $10^4$G which is 
roughly the equipartition field of the plasma. The Poynting flux
transports power from the black hole and the accretion disk
along the flux tubes rotating roughly at an angular velocity
half of that of the horizon [169]. This power could be sufficient to
drive the outflows observed in active galaxies.

An insight into the energy extraction process through Blandford-
Znajek mechanism could be obtained in the following way [168]. Consider
a thermally conducting disk rotating with an angular velocity $\Omega_H >0$
is surrounded by a thermally insulated annual ring which 
rotates with an angular velocity $\omega$. Suppose these two systems are
coupled by friction which is proportional to the difference
between these two angular velocities: $K(\Omega_H - \omega)$.
This frictional couple will increase the rotational energy of the
ring at a rate $K\omega(\Omega_H-\omega)$ and decrease the
rotational energy of the disk at a rate $K\Omega_H(\Omega_H-\omega)$.
The rest will be spent in heat generation $K(\Omega_H-\omega)^2$.
When $\Omega_H > \omega$, the efficiency of mechanical energy transfer 
to the ring is $\omega/\Omega_H$.

In the present situation, similar transfer takes place between the
rotating hole (with angular velocity $\Omega_H$) and the
surroundings magnetosphere. The frictional coupling is
thought to reside inside the event horizon where the 
electrical circuit is presumably complete. In Fig. 3.12, we show
schematically these processes. The dashed circle is the event horizon 
$r=r_+$. The solid curves are the surfaces of constant poloidal field 
lines $A_\phi=$const. Some of them pass along the polar region ($A_p$) 
and some other pass through the equatorial plane of the disk ($A_e$). 
A current flows down the magnetosphere into the hole and comes out into the
disk on the equatorial plane. In the magnetosphere, SG is the
spark gap where positrons and electrons are created. Positrons flow
into the hole along the poloidal field and electrons fly away from 
the hole carrying energy and angular momentum. The short-dashed 
surfaces denoted by $L$ are the locations where the particles
slip out of the $A_\phi$ surfaces as otherwise they would be rotating 
faster than the speed of light.

\newpage
\noindent {\large\bf 4 Thick Accretion Disks}

\noindent{\Large 4.1 Introduction}

When the accretion rate is high (${\dot M} >>{\dot M}_{Edd}$), 
radiation emitted by the 
in-falling matter exerts a significant pressure on the gas. This pressure must
be incorporated to find the dynamical structure of the disk
and in determining the thermodynamical quantities
inside the disk. The radiation pressure inflates the disk, and makes it
geometrically `thick' [$H (r) \sim r$]. Incorporation of the
radiation pressure term in Euler equation modifies the angular momentum 
distribution from that of the Keplerian disk. To see this, let us consider
the radial component of the Euler equation,
$$
v_r \frac{d v_r}{dr}+ \frac{1}{\rho}\frac{dP}{dr} -\frac{l^2(r)}{r^3}+F(r)=0
$$
where $F(r)$ is the force due to gravity and $l(r)$ is the specific angular
momentum distribution. Ignoring the advection term for the moment, 
one finds the distribution of angular momentum as [170],
$$
l(r) \propto [ r^3F(r) + \frac{r^3}{\rho}\frac{dP}{dr}]^{1/2}.
\eqno{(4.1)}
$$
Clearly, wherever the pressure gradient term is 
positive(negative), the angular momentum is higher(lower) than Keplerian.
This term is significant when $GM/r \sim P/\rho \sim a_s^2$, where $a_s$ is the 
sound speed. This is the essence of thick accretion disk models. 
This is a general result and is valid even when a Newtonian treatment
is made. 

First quantitative study of the effect of radiation pressure on the
angular momentum distribution was carried out almost two decades ago [171].
Figures 4.1(a-b) show the variation of the angular velocity $\omega$
and radial velocity $v=v_r$ in units of the Keplerian angular velocity
[$\omega_K=(GM/r_0^3)^{1/2}$] the free-fall velocity $v_K=(GM/r_0)^{1/2}$ 
at the inner edge of the disk $r_0$ where the shear stress is assumed to vanish
($r_0=3r_g=6GM/c^2$ in the Shakura-Sunyaev disks [11]). In Fig. 4.1(a), the 
accretion rate is chosen to be the same as the critical rate, ${\dot M}_c = 
L_{Edd}/(c^2 \eta)$ with $\eta \approx GM/2r_0$, the efficiency factor. 
The dashed curve shows
the Keplerian distribution. The important conclusion drawn here is that close
to $r \sim (2-5)r_0$, the angular momentum is higher than Keplerian and at 
distances farther away, it becomes sub-Keplerian. The effect becomes stronger
when the accretion rate goes up. Figure 4.1(b) shows the  result for
accretion rate $25$ times higher than the critical rate and the
deviation from the Keplerian curve is also seen to be higher. The
velocity of the flow becomes smaller than Keplerian at $r\sim r_0$ as well.

In the case of critical or super-critical accretion around a black hole,
the effects of radiation are expected to be the same as that described above. 
One can study the subject in a reverse order. One can assume an ad-hoc angular
momentum distribution which deviates from the Keplerian distribution
as above, and find the radiation field required to produce such a 
distribution. In Fig. 4.2,  a typical angular momentum distribution 
inside a thick accretion disk (short dashed curve) around a black hole
is shown.
Typically, the angular momentum of the bulk of the disk is below Keplerian
(solid curve) value. Only within a region close to the marginally stable orbit
(minimum of the Keplerian distribution), the angular momentum is 
above Keplerian, or super-Keplerian. The pressure of the radiation pushes
the inner edge of the disk much closer to the black hole, i.e.,
the inner radius of the disk around a Schwarzschild black hole is situated
at $2r_g < r  <3r_g$. Since the potential (Fig. 1.4) is less deeper
at this range than at $r=3r_g$, these disks are less efficient than a 
Keplerian disk. Secondly,
matter is accreted by the strong pressure gradient term in the Euler equation,
hence viscosity is unimportant.

The thick disk angular momentum distribution intersects with the Keplerian 
distribution (solid curve)
at two locations, $r_i$ and $r_c$, the inner edge and the center
of the disk respectively. At $r_c$, the pressure not only 
does not vanish but it is maximum. However, still
the flow is Keplerian, because the pressure gradient term vanishes.
At $r<r_c$, the pressure gradient term is directed towards the black hole 
and the flow requires more angular momentum to fight against the
combined gravity and pressure gradient term. At $r> r_c$,
the pressure gradient term is directed away from the center and the flow
requires an angular momentum less compared to the Keplerian value since
a part of the pressure gradient term fights against gravity. 
One of the important characteristics of the equipotential surfaces
of a thick accretion disk around a black hole is the presence
of a cusp very similar to the Lagrange point in a Roche lobe overflow
(cf. Fig. 3.1). In this case, matter fills the closed potential and forms the
thick accretion disk, and the excess matter is accreted to the black hole
through the cusp. The cusp is formed at $r_i$ between the marginally bound
($2r_g$) and marginally stable ($3r_g$) orbits. For comparison,
the Keplerian distribution in a Newtonian geometry is also presented in the
Figure. The angular momentum in a thick disk is monotonic and intersects 
this distribution only once, and therefore, no cusp is expected to form in 
a thick disk in Newtonian geometry.

Before we go about discussing the model in detail, we draw a cartoon figure to
indicate forces acting on a blob of matter inside a thick accretion disk. 
Figure 4.3 shows schematically an equipotential surface. The gravitational
force acts radially inwards and the centrifugal force acts in a direction 
normal to the angular momentum vector. Addition of these two
vectors produces a net force along the effective gravity. In order to 
remain in hydrostatic equilibrium, a force of equal magnitude due to the
pressure gradient must act opposite to this direction. The tangents drawn 
normal to the pressure balance produce the equipotential surface.

Earlier attempts to mimic the super-Keplerian behaviour of the angular
momentum distribution were made by choosing the specific angular momentum
to be constant inside the disk. Lynden-Bell [172] 
showed that a flow with a constant angular momentum
would avoid the region inside two symmetric vortices (funnels or chimneys)
along the axis of the black hole, thought to be located at the center of
quasars from very early days [173]. These funnel-like regions, like two
giant vortices along the axis are devoid of matter and filled up with hot and 
anisotropic radiation field due to leakage of radiation from the
disk interior. These regions were considered to be suitable
places to accelerate matter to form radio jets. 
Intensive works followed [75, 174-180]
to study the properties of the thick accretion disks. 
Most of these fully general relativistic 
models study the properties of the barotropic fluid of constant 
angular momentum neglecting the self-gravity of the flow, though in more recent
works self-gravity has been included.
Below, we present a simple model of the thick accretion disk.

\noindent{\Large  4.2 Pseudo-Newtonian Thick Disks}

Paczy\'nski \& Wiita [75]  using a pseudo-Newtonian potential to mimic the 
Schwarzschild black hole geometry studies the thick accretion disk models
assuming that the angular momentum is distributed as a power law of the
axial distance. The force balance condition which determines the structure of a
thick accretion disk is given by, 
$$
\frac{\nabla P}{\rho} = -\nabla \phi +  \frac{l^2}{(r\sin\theta)^3} \nabla R ,
\eqno{(4.2)}
$$
where, $P$ and $\rho$ are isotropic pressure and density respectively,
$l$ is the specific angular momentum inside the disk,
$\phi$ is the gravitational potential, and $R=r \sin\theta$
is the axial distance. In the absence of a complete theory of viscous
flow around a black hole, the angular momentum distribution is chosen to be 
a power law $l(r, \theta)= l_0 (r \sin\theta )^n$ inside the disk 
($\theta$ is the angle measured from the $z$ axis). Here,
$l_0$ and $n$ are positive constants, with $0.5 \geq n \geq 0$. The positivity
of $n$ is required by the H\"oland stability criterion [181] under
axisymmetric perturbation: $dl/dr \geq 0$. For simplicity, 
assume that the equation of state is barotropic, i.e., $P=P(\rho)$. 
In this case, the left hand side of the equation becomes integrable. 
$W=W(P)=-\int{\frac{dP}{\rho}}$ surfaces are equipotential surfaces, 
which are also surfaces of constant pressure, density and temperature. 
These surfaces are obtained by integrating above equation as,
$$
W(P)-W(P_{out})=-\frac{1}{2(r-1)}-\frac{l_{out}^2}{2n-2}(r
\sin\theta)^{2n-2},
\eqno{(4.3)}
$$
where $W(P_{out})$ is the integration constant signifying the potential at the
outermost surface with pressure $P_{out}$. For a given pair ($l_0,\ n$) there
exists a point inside the disk where the potential is minimum which 
corresponds to the center of the disk. As in the case of stars, the 
potential gradually rises outwards till it reaches $W(P_{out})$.

Using above equation of state, one obtains the expression for 
temperature (in $^o$K) to be, 
$$
T_e(r) = T_e(R_{out}) - ({\mu\beta c^2\over{4R_G}}) \{{1\over
2}[{1\over{r_{out}-1}} - {1\over{r - 1}}] +
{l_0^2\over{2n-2}}[r_{out}^{2n-2} - (r \sin\theta)^{2n-2}]\},
\eqno{(4.4)}
$$
where $R_{out}=r_{out} r_g$ is the distance of the outer 
edge of the disk on the equatorial plane and $R_G$, the gas constant,
$\beta$ is the ratio of gas pressure to total pressure and $\mu$
is the mean electron number per baryon.
Temperature is maximum at the center of the disk and gradually falls 
off as one moves away from it. 

The equipotential surfaces are shown in Fig. 4.4a. I and C denote the 
inner edge (cusp) and the center of the disk respectively. Here, the
angular momentum is chosen to be higher than marginally stable value
and the cusp is present. When angular momentum
is farther reduced so that it is everywhere below marginally bound value,
the cusp is absent and the equipotentials are quasi-spherical, open 
and matter cannot stay in to form a disk. In Fig. 4.4b,
the equipotentials in the Newtonian geometry are shown to compare the
topological properties. As discussed before (cf. Fig. 4.2) no cusp is 
formed in this case, independent of angular momentum.

When the barotropic equation of state is chosen, surfaces of constant
pressure, density and temperature coincide. These quantities are
maximum at the center of the disk. 
Alternative disk models have also been  proposed where the 
accretion is in the equatorial plane [177]. These
models are found to be very sensitive to the external boundary condition 
and could be globally unstable with non-axisymmetric instabilities [182].
In some astrophysical models accretion disks are equipped with corona
similar to solar surface [141]. Here, magnetic flux tubes need to be anchored
deep inside the disk. This is possible in the
isothermal thick disks, or disks in which entropy rises outwards
are also considered. These magnetized disks will be studied in \S 
4.7.

\noindent {\Large 4.3 General Relativistic Thick Disks}

First few attempts to determine the structure of a thick accretion disk
were made using general relativistic equations [174-176]. The equations 
governing the rotational motion of the flow are:
$$
\frac{\nabla_i P} {P+\epsilon} = -\nabla_i ( ln u_t ) 
+ \frac {\omega \nabla_i l} {(1 -  \omega l)} ,
\eqno{(4.5)}
$$
where $P$ and $\epsilon$ denote the isotropic pressure and the
internal energy density respectively, $u_\mu$ (with $\mu = t,0,0,\phi $)
are the four velocity components and $\omega = \frac {u^\phi}{u_t}$
and $l=- \frac {u_\phi}{u_t}$ denote the angular velocity and specific
angular momentum respectively.

When the equation of state is chosen to be barotropic, $P=P(\epsilon )$. 
In this case, the left hand  can be written as the gradient of 
a potential. The first term on the right hand side is already
written in this manner. Hence the remaining term on the right
hand side also has to be the gradient of a scaler, i.e., $\omega=
\omega (l)$. This is so called von Zeipel theorem. 
The surfaces of constant $\omega/l$  are of great importance as they are the 
surfaces where angular momentum of the disk must remain constant [174-176, 
178-179, 183-184]. In the Schwarzschild geometry, $\lambda$ is obtained from:
$$
\lambda^2 = -\frac{g_{\phi \phi}}{g_{tt}},
\eqno{(4.6a)}
$$
and in Kerr geometry, it is obtained from,
$$
\lambda^2= -\frac{l g_{\phi\phi} + l^2 g_{t\phi}}{g_{t\phi} + l g_{tt}} .
\eqno{(4.6b)}
$$
Earlier attempts to study specific cases of these surfaces show cylindrical 
surfaces which are asymptotically flat cylinders [174], though in general 
they can be of remarkable complexity. A catalogue of these surfaces
[183] show that these surfaces could have toroidal topology as 
well. A knowledge of these surfaces finds a major application in studying
thick accretion disks [178, 179] as well as the study of structure 
and stability of relativistic rotating stars [184].
The Euler equations could be integrated if angular momentum
is chosen to be constant over these surfaces: $l=l(\lambda)$.
This prescription allows a generalized study of thick accretion disks
for any axisymmetric spacetime. The important point is to realize 
that a parameter $\lambda=\sqrt (l/\omega )$ could be defined which
is roughly the Newtonian equivalent of the axial distance. Similar to
the power-law distribution of angular momentum by Paczy\'nski-Wiita [75], 
one could assume the angular momentum distribution to be, 
$$
l( \lambda ) = c \lambda^n,
\eqno{(4.7)}
$$
with $c$ and $n$ as two positive constants. Vorticity free ($l=constant$) 
disks are obtained by putting $n=0$ and the rigid body ($ \omega= constant$)
disks are obtained by putting $n=2$. In general, in it
observed that for thick accretion disks, $0 \leq n<0.5$. 
The Rayleigh criteria of stability implies that the angular
momentum increases outwards. This is possible with $n \geq 0$ provided
$r> 3GM/c^2$, the circular photon orbit in the Schwarzschild geometry.
In the Kerr geometry also the function $\lambda$ behaves in a way that
it is monotonically rising outwards outside the photon orbit [179, 184].

\noindent{\Large 4.4 Thermodynamic Condition inside a Thick Accretion Disk}

The angular momentum of a thick accretion disk intersects 
with the Keplerian distribution at two different points (Fig. 4.2). The
innermost one of these two is at the inner edge ($r_i$) of the disk, where the
disk is pressure free. At the other intersection ($r_c$), the pressure gradient
is zero and the pressure of the disk is maximum. Density and temperature
are also maximum (for barotropic disks) at this point. 
In this respect the disk behaves as a toroidal star. Figure 4.5 gives the 
thermodynamic condition inside a thick accretion disk [2, 185-187].
The variation of central temperature and density in the disk are shown
as a function of the ratio of the gas pressure to the total pressure $\beta$.
Also shown are the optical depth $\tau$ (due to Thomson scattering) and 
entropy density $s$. Two regions are separated: in region $\tau <1$, the disk 
is optically thin and the radiation leaks out, so that thick disks do
not form. In the region marked as `self-gravity', the mass of the disk
becomes comparable or more than the central black hole and the self-gravity
of the disk cannot be ignored.  This happens when $\rho(r_c) \geq M/r_c^3$.

When the self-gravity becomes important, the Keplerian distribution of angular
momentum is qualitatively changed [188-190]. Assuming that most of the matter is
concentrated at the center of the disk where the density is maximum, 
and assuming that the rotational
energy of the disk can be ignored compared to the rest mass energy, it
is possible to compute the Keplerian distribution of angular momentum
fully general relativistically [188]. Figure 4.6 shows the distribution
when the mass of the disk is three (short dashed), five (medium dashed)
and ten (long dashed) percent as heavy as the
central compact object [188]. For comparison, the Keplerian distribution
in the absence of the disk is also shown by the solid curve.
The presence of the annular ring at $r_c=9.0$
increases the outward force on the particles for $2 <r <r_c$. Hence the
angular momentum required is lower. Similarly, in the presence of the ring,
the inward force on particles orbiting at $r_c <r < \infty$ is higher so that
the angular momentum needed to stay at a given orbit is higher. These are
reflected in this Figure. A thick accretion disk, if stable, is expected to be
much hotter when self gravity is important, since the deviation from Keplerian
is going to be higher. More realistic models of self-gravitating
thick accretion disks have recently been computed, with similar conclusions
[189-190]. 

It is clear that for low mass black holes, the temperature of a radiation
pressure dominated disks is very high for a given ${\dot m}
=\frac {\dot M}{{\dot M}_{Edd}}$. This gives rise to the possibility of 
significant nuclear reactions in the disk. In general, it is observed 
[185-187] that unless viscosity is negligible, 
nucleosynthesis is unimportant
in radiation pressure supported accretion disks around super-massive
black holes. When a stellar mass
black hole is considered, significant nuclear reaction may take place
inside the disk [187, 191]
and the composition of the gas and the outgoing jet may change
[187]. A number of such stellar mass holes in a galaxy
would be sufficient to produce the observed metalicity of the galaxy.

\noindent {\Large 4.5 Emitted Radiation from a Thick Accretion Disk}

One of the attractive features of a thick accretion disk is 
that it produces super-Eddington luminosity in the funnel [75].
On the surface of a funnel, 
the pressure gradient is balanced by the combined effects of the gravity 
and the centrifugal force. The effective gravity is vary strong
close to the axis, and therefore the luminosity exceeds
several times the Eddington luminosity. This is a useful feature
in the context of cosmic jet acceleration by radiation pressure in the funnel
itself [192]. However, excess directionality 
may also be a negative aspect of this model. The emergent radiation
may be highly anisotropic and is emitted mostly along the 
vertical z-direction. The torus viewed pole-on may show very large
soft X-ray/UV excess but the one viewed edge-on may be cooler. 

Knowledge of radiation pressure support and the electron scattering 
opacity allows one to compute the flux radiated from the surface
of the torus. The critical flux emitted from the disk surface 
(assuming it is in radiative and hydrostatic equilibrium), is given by [193]:
$$
F_r= \frac{c}{\kappa_T g_{eff}} = -\frac{c}{\kappa_T}
(-\nabla \phi + \omega^2 r ),
\eqno{(4.8)}
$$
where $g_{eff}$ is the effective gravity  which is perpendicular
to the disk surface (Fig. 4.3), $\phi=\phi_{s}$ the pseudo-Newtonian potential
for the Schwarzschild geometry and $\omega=l/r^2$ is the rotational velocity
of the flow. The radial dependence of the effective temperature is 
calculated by equating the flux above with  $\sigma T_{eff}^4$.

The luminosity of the disk, for the ratio $Q=r_{out}/r_{in} > 50 $ is 
found out to be,
$$
L/L_E \sim 3.8 \, log Q -2.43 .
\eqno{(4.9)}
$$
(Here the subscripts `out' and `in' refer to quantities at the outer
and inner edges respectively.)
Inside the torus, no local balance of energy is possible, as the radiation
transports energy in both the radial and the vertical direction.
The total energy liberated is given by,  
$$
L_g={\dot M} [ e_{in}-e_{out} - \omega_{out} (l_{out}-l_{in})] .
\eqno{(4.10)}
$$
As the inner edge goes to $r_{mb}$, the $e_{in} \rightarrow 0$, so an 
extremely large disk is needed to supply the same luminosity.

Characteristics of the spectrum resulting from a typical model [193]
is discussed below. Figures 4.7-4.10 
show the temperature variation on a thick disk surface, 
disk spectrum at various inclination angles, specific luminosity at 
various viewing angles and a comparison of a few model dependent spectra. 
In Fig. 4.7, temperature distributions in a thin disk 
(dotted curve) and in a thick disk (dashed curve) are
compared for the same central mass and the accretion rate.
In Fig. 4.8, the disk spectra at inclination angles (a) $i=90^o$, (b) $i=50^o$,
(c) $i=25^o$ and (d) $i=0^o$ are compared. Due to the presence of the funnel,
the spectrum depends very strongly on the inclination angle. Another
way of seeing this is to plot the specific luminosity at a few (observed) 
frequencies 
($\nu$) as a function of inclination angle. In Fig. 4.9, this is shown
for $log \, (\nu)$ equal to (a) $15.5$, (b) $16.0$, (c) $16.5$ and (d) $17$
respectively. In optical and ultraviolet regions, specific luminosity is almost
independent of the angle, but at higher frequencies the dependence is
very strong and may present some problem in modeling active galaxies.
In Fig. 4.10, a comparison of the thin disk spectrum (dashed curve) is made
with the spectra of a thick disk with various effects. The solid line
represents the spectrum without the reflection effects from the funnel
whereas the dotted line includes the reflection effects. The dash-dotted
curve is obtained by assuming a sum of black body spectrum from each
annulus of the disk. Compared to a thin disk, the thick disk
(with reflection effects) produces more emission at higher frequencies.

\noindent {\Large  4.6 Ion Pressure Dominated Thick Disks}

When the cooling time scale is longer compared with the
in-fall time the protons remain very hot --- close to the virial
temperature and the electrons cool through synchrotron or Compton
processes. The excess ion pressure can puff-up the disk and can produce a 
different kind of thick disk, the so called ion pressure
supported disk [194]. If Coulomb coupling is the only ion-electron
coupling, then the condition of longer cooling time scale is
satisfied if [2],
$$
{\dot M} \alpha^{-2} < 50 {\dot M_{Edd}}
\eqno{(4.11)}
$$
For any $\alpha\sim v_{infall}/v_{free fall}$ there is always some smaller 
${\dot M}$ for which above condition may be satisfied and ion pressure 
disk could be formed. This condition may, in particular be true,
if in some AGNs the matter supply is scanty. 

The above condition assumes the absence of pair-productions or 
collective plasma processes which may strengthen ion-electron coupling 
and cause rapid cooling of protons [195-196]. In the presence of
these collective processes (in presence of strong magnetic field,
for example) an ion torus may be rapidly cooled and deflated.
The restoring effects can bring the cool disk back to the hot torus
configuration. Such instability may also cause AGN variabilities [196].

In a gas pressure dominated thick disk  the gas may be completely 
photo-dissociated and the protons would
accrete onto a black hole (as they `feel' the magnetic viscosity),
while neutrons  would orbit till they decay. As a result, neutron
torus may form which combines with the incoming fresh matter
and produce neutron-rich elements in the galaxy [197].

\noindent {\Large 4.7 Magnetized Thick Disks}

We have already discussed a few models of magnetized thin accretion 
disks (\S 3.4). No similar detailed study has been made of magnetized
{\it thick} accretion disks and the associated jet production. A partial
reason is that the thick accretion disks (both the rotation
dominated radiation tori and the advection dominated ion tori)
may have very little angular momentum and matter may fall into
the hole without any need for a viscosity. Furthermore, the
thick accretion disks have funnels or chimneys where the anisotropic
radiation field could be sufficiently strong to accelerate radio
jets. Thus there was no immediate need to construct detailed
magnetized thick disk models. 
Nevertheless, magnetic fields are observed in jets and magnetic
activity in the funnel regions are thought to be responsible for the
rapid variation of polarization in BL-Lacs. Therefore, this 
requires some attention.

Recently, a thorough
analytical study has been made of the behaviour of the axisymmetric
flux tubes inside a thick accretion disk [141]. Magnetic field is
expected to be brought along with the in-falling matter and are expected
to be sheared to form mainly toroidal flux tubes in the disk. Due to 
the joint effects of drag, Coriolis force, magnetic tensions, buoyancy
force a significant fraction of these fields are found to be emerging
in the funnel whereas the remaining fraction would be expelled from the
outer parts of the disk. A few important observations of this study is
that (a) if the disk is sufficiently hot ($T_p > 4 \times 10^{10} K$),
the magnetic tension dominates all other effects and the tube
collapses
catastrophically on the axis (rapid pinch), possibly squeezing out 
plasma in the disk along the axis of the black hole in the process
to form radio jets. 
(b) Formation of coronae in an accretion disk is not {\it automatic}, namely,
presence of magnetic field inside a disk does not automatically
imply that a hot magnetic coronae would form. The formation 
requires the ability of the accretion disks to {\it anchor} flux tubes
inside the disk. This means that the disk should have an internal structure
akin to the solar interior and the entropy must increase outwards.
If the entropy condition is proper, the coronae would form, otherwise it
would come out of the disk as a whole without causing
any flare. In former case there would be
sporadic flaring events on the disk surface, whereas in the latter case,
the collapse of fields in the funnel would cause destruction of the
inner part of the disk and formation of blobby radio jets. Detailed
observation of GRS1915+105 shows these features [198]. Since the inner 
part of the accretion disk could literally disappear by this magnetic process,
radio flares should accompany reduction of X-ray  flux in this 
objects. Since the physical
process is generic, such processes could also be responsible for the 
formation of jets in active galaxies and similar anti-correlation
may be expected, though time delay effects are to be incorporated for 
a detailed modeling.

\noindent{\Large 4.8 Thick Accretion Disk -- To Be or Not To Be?}

Thick accretion disks have many `pleasant' features such as the
production of supercritical luminosity, possible ability to collimate 
jets, magnetic activity in the funnel as the explanation of the rapid 
variability in BL-Lacs, production of metalicity in the galaxy by
nucleosynthesis, etc. [180]. However,
as mentioned above, the strong anisotropic nature of the emission properties of
a thick accretion disk is a major disadvantage. Another problem is
that a non-accreting thick disk is found to be dynamically and globally 
unstable due to non-axisymmetric instability [199-201].
In this type of instability, disks with a large shear were found to be 
unstable due to non-axisymmetric perturbations. 
A necessary condition in this instability
is that the co-rotation radius (the radius where the mode rotates with the 
disk fluid) of the corresponding normal mode must be within the disk. 
Since usually the angular velocity of the flow decreases outwards, the 
mode rotates slower or faster compared to the flow depending on whether
it is inside or outside the corotation resonance. 
Angular momentum is transported from regions inside the 
resonance radius to the region outside which leads to amplification of the
perturbation and therefore the growth of instability. 
The instability grows faster if the angular momentum distribution 
is closer to a constant. It is shown in particular, that if the angular
momentum distribution were represented as a power law $l(r) \propto
r^n$, then, for $n <0.27$, the perturbation is found to grow. The
analytical work are carried out using non-accreting slender torus approximation
and the perturbation is found to grow fastest for $n \sim 0$.
Numerical simulations of slender tori verify the presence of this
instability [202-203] though  for very thick accretion disks
the instability grows much more slowly [204]. 
The presence of accretion processes is found to remove the instability
as well [205-206]. 

An ideal thick accretion disk  requires the flow to have a 
very low angular momentum and a very small
radial velocity. Typically,
the angular momentum required is a little above the marginally stable value
close to a black hole. Also, a radiation pressure supported
thick accretion disk ideally has negligible radial velocity. 
It is unclear how such a structure can be formed with matter supply
from stars and wind, given that either the incoming matter has
Keplerian angular momentum with a little
radial motion, or with low angular momentum but significant
radial motion. Thus matching this structure with outer boundary
cannot be smooth. What in reality
may be happening is that in the presence of small angular momentum
matter first `hits' the centrifugal barrier and forms
a strong shock (\S 5) where the flow thickens to form a torus
and subsequently
falls on a black hole after passing through the inner sonic point.
High accretion rate would cause radiation pressure dominated
tori, whereas low accretion rates would produce ion pressure supported tori.
The low angular momentum pre-shock flow is almost
freely falling. The flow in the immediate post-shock region
has a very little radial velocity and therefore it is rotation dominated.
This region resembles a thick accretion disk with an atmosphere and outgoing 
winds (\S 6). Thus, it is possible that in reality, thick disks, formed by 
strong shock waves or by sudden deviation of the disk from
Keplerian distribution, might be formed which becomes transonic at the
inner edge and therefore stable under non-axisymmetric instability.
It is also possible that since a well defined funnel with a geometrically
sharp boundary need not exist, the anisotropy of the emission might be
reduced, particularly because the disk is of much smaller in size ($r_{out}
\sim r_{shock} \sim 10-20r_g$). 
The collimating property of jets may suffer at the same time,
but this is not crucial since toroidal fields produced by the strong shear
in the post-shock region can collimate the jet as well.
The variation of the accretion rate changes the shock location and
the shape and size of the thick accretion disk. This might cause variabilities
observed in many active galaxies (\S 7).

\newpage
\noindent {\large\bf 5 Advective Disks}

We discussed in the previous Sections that in the case of accretion on
central black holes in binary systems and in active galaxies and
quasars, the angular momentum of the flow need not be Keplerian
everywhere. In a binary system, matter could be accreted both through
the winds (sub-Keplerian flow)
and through the Lagrange point (Keplerian flow). In the absence of
winds, specially in low mass X-ray binary (LMXB) systems where
winds may be absent, the sub-Keplerian matter could still
be produced from the Keplerian flow close to the black hole.
In an active galaxy, the same situation may prevail although 
it is likely that matter is accreted solely from the winds of stars 
very far away which has very little angular momentum.
The sub-Keplerian flow, whether it originates from Keplerian disk or
not, will have a significant velocity, since the
centrifugal force is not sufficient to overcome gravity. In either of the above
cases, the sub-Keplerian flow should behave in the following way:
at first, it accretes quasi-spherically with an infall time 
similar to the free-fall time $t_{infall} \sim r/v_{ff} \sim r^{3/2}$ until
the specific angular momentum  of the flow 
becomes comparable to the local Keplerian angular momentum $l^2(r) \sim
l_{Kep}^2 (r_s)$. At this point, $r_s\sim l^2(r)$ ($\sim 8r_g$ for 
marginally bound
angular momentum) the flow may be virtually stopped by the centrifugal barrier,
and a standing shock will form. Subsequently, the flow continues its journey
to the black hole and accretes on it supersonically. 
Due to pressure effects, the shock actually forms farther away and also
when the angular momentum is less that marginally stable value.
The shock can form anywhere between $10$ to $10^4r_g$
depending on the specific energy and angular momentum of the flow
and the heating and cooling mechanisms. 
The kinetic energy of the almost freely falling advecting
flow is thermalized rather abruptly at the shock which heats 
up the flow to about the virial temperature. The shock
heated region occupies a very large area of the disk since
it forms far away from the inner edge, and therefore it
influences the spectrum of the disk significantly. In fact,
this region can be compared with the boundary layer of a star where the  
soft radiations are hardened through Comptonization and other effects.
If the viscosity of the flow is small enough so that the shock is strong
and thin, the hot and the cold phases of the flow co-habit side by side.
The shock re-processes soft photons (either from synchrotron photons
from the preshock flow, or from soft photon from Keplerian disk on the
equatorial plane) and emit observed gamma rays.
The location and the strength of the shocks are very sensitive to the flow 
parameters, such as the energy and the angular momentum, and therefore, such
discontinuous flows may explain correlated variabilities in UV/X-ray and optical
frequency bands observed in many active galaxies and quasars. Shock 
acceleration of particles and photons can explain the origin of the high 
energy $\gamma $-rays and cosmic rays [122-130]. In recent years, 
observations of the continuum as well as line emission properties indicate
that many of the disks in active galaxies and quasars are non-Keplerian 
[207-212]. Black hole candidate spectra also show transition from soft 
to hard states which are described by varying degree of combination of
the power law spectra with soft `black-body' type spectra.
We shall briefly discuss the structure of these sub-Keplerian 
disks in the present Section and differ the discussion on the
observational properties till \S 7.

Although we have emphasized on shock formation by sub-Keplerian
flows, it is not the `only' choice. In a large range of parameter
space, spanned by angular momentum, entropy and the polytropic
index, the shocks are not formed at all. The flow passes either
through the outer sonic point, or through the sonic point. Sometimes,
for higher viscosity, the flow with have both the choices simultaneously,
though, in such cases, it chooses only the one passing through the inner
sonic point because of higher entropy at the inner sonic point of the flow.
An important reason why these shock solutions could
be relevant is that even when the steady shocks are not predicted by
theoretical analysis, they form but travel backwards to infinity while
making the disk subsonic and possibly Keplerian. Thus shocks are
important ingradients in undertanding a disk formation, specially
when matter supply is time dependent. When the flow only has the outer
sonic point, it forms ion supported advective flows as in [194].

We have already shown in \S 2
that a pure adiabatic spherical flow cannot have a shock as the flow
has only one sonic horizon. In the presence of heating and
cooling and other geometrical effects, shocks can form in a spherical
flow, but the efficiency of emission was not seen to be very high
as the shock forms very far away from the black hole: $r\sim 10^6 r_g$.
In the context of the study of the solar and stellar wind, where the 
variation of geometrical shape of the wind and deposition of radiation 
momentum can introduce more than one sonic point, shock waves are found 
to be common. Indeed, they are studied for quite some time [213].
In the context of the accretion flows, the presence of a significant angular 
momentum is found to increase the number of saddle type sonic points from one 
to two [214]. Early numerical simulations 
reveal the presence of traveling shock waves in low angular momentum 
accretion flows [215-216] (possibly due to the inherent viscosity, 
and the choice of input parameters). Subsequently, using a simple model of
adiabatic flows, it was pointed out 
that similar to the solar wind case, multiple sonic points may 
lead to the formation of shock waves in the disks as well [217]. 
Currently, shock waves in accretion disks (whether steady or 
non-steady) are considered to be indispensable parts of accretion 
of low angular momentum region of a disk [68, 77, 218]. 
In this Section, we present the theory of these discontinuous accretion
flows very briefly.

In the next subsection, we present model equations
and provide physical arguments as to why there are
multiple critical points in sub-Keplerian accretion flows around black holes.
In \S 5.2, we present solutions of adiabatic flows with discontinuities.
In \S 5.3, we present an example of the isothermal solution which includes
viscosity. In \S 5.4, we use most general heating and cooling processes
in shock study and discuss about the unification of disk models around
black holes. In \S 5.5, we discuss self-similar solutions of the model
equations in Newtonian geometry. In \S 5.6, we discuss the nature
of shock solutions in Kerr geometry and in magnetized disks. In \S 5.7, 
we deviate from axisymmetric situation considered so far, and discuss
self-similar non-axisymmetric spiral shock waves. In this case, a density
perturbation (induced by a passing companion, globular cluster, or tidal
capture of molecular clouds) can become steeper as it propagates towards the 
black hole and produce spiral shock waves in the disk. These disks
are sufficiently non-axisymmetric so as to influence the observational 
results. 

\noindent {\Large 5.1 Model Equations and Multiplicity of Critical Points}

The models discussed in \S 2 - \S 4 are somewhat limited in the sense
that either the angular momentum (\S 2), or the advective effects 
(\S 3 - \S 4) or the pressure effects (\S 3) are not fully taken 
into account.  A `grand unified' disk model, must not only 
include all these effects self-consistently, but also reproduce the
earlier disk models as special cases, and at the same time produce
new solutions (if possible) such as the solution with
discontinuities (shock waves). Above all, it should produce the
steady and time-dependent spectra compatible with observations. We shall 
present solutions of this model in this Section. To avoid complexity
of the most general equations, one usually solves vertically integrated
equations which reduces the dimension of the problem from three to
two and are therefore easily solvable.

The steady state model equations which one has to use are [68, 76-77, 219]:

\noindent (a) The radial momentum equation:

$$
\vartheta \frac{d\vartheta}{dx} +\frac{1}{\rho}\frac{dP}{d\rho} 
+\frac {\lambda_{Kep}^2-\lambda^2}{x^3}=0,
\eqno{(5.1a)}
$$

\noindent (b) The continuity equation:

$$
\frac{d}{dx} ( \rho x h \vartheta) =0 ,
\eqno{(5.1b)}
$$

\noindent (c) The azimuthal momentum equation:

$$
\vartheta\frac{d\lambda(x)}{dx} -\frac {1}{\rho x h}\frac{d}{dx} 
(\frac{\alpha P x^3 h} {\Omega_{Kep}} \frac{d\Omega}{dx}) =0
\eqno{(5.1c)}
$$

\noindent (d) The entropy equation:

$$
\Sigma \vartheta T \frac{ds}{dx} = Q^+ - Q^-
\eqno{(5.1d)}
$$

Here $\lambda_{Kep}$ and $\Omega_{Kep}$ are the Keplerian angular momentum and
Keplerian angular velocity respectively, $\Sigma$ is the  density $\rho$
vertically integrated, $h=h(x)$ is the half-thickness of the disk at radial
distance $x$, $\vartheta$ is the radial velocity, $s$ is the entropy density
of the flow, $Q^+$ and $Q^-$ are the heat gained and lost by the flow. 
$\lambda(x)$ is the angular momentum distribution.

First we argue in general terms that the flow satisfying the above set
of equations has multiple critical points.

In the case of a black hole accretion, matter which is at rest at infinity can
fall onto a black hole even when the specific angular momentum is as high 
as marginally stable value $\lambda <\lambda_{ms}$ (\S 1). 
When $\lambda_{ms}(x_{ms})
<\lambda<\lambda_{mb}(x_{mb})$, matter will form a disk which may accrete and 
lose binding energy $1-\phi(x_{in})$ where, $\phi(x_{in})$ is the
potential energy at the inner edge $x_{in}$ of the disk. When
$\lambda>\lambda_{mb} (x_{mb})$, matter at rest at infinity will be bounced
back by  the potential barrier (Fig. 1.4). Matter with energy higher than 
$\phi=\phi_{max}(\lambda)$ can overcome the barrier and fall into a black hole.

The above picture is valid for particle dynamics. When one considers a
fluid motion, the picture is somewhat different due to the presence of
flow pressure. Here, matter starts `feeling' the centrifugal barrier
even before actually reaching it, and starts piling up 
before falling onto a black hole. Here, the ram pressure of the advective,
low angular momentum, pre-shock flow roughly matches with the 
thermal pressure of the rotation dominated post-shock flow and 
a standing shock wave is formed.
In order to have a shock, however, the flow, subsonic at a large
distance, must first pass through a sonic point far away from the black hole
(the so-called outer sonic point similar to  what is seen in a Bondi flow) 
and becomes supersonic.
After this, the flow passes through a shock transition and becomes subsonic.
In order to satisfy the inner boundary condition, i.e., supersonicity
on the horizon, the flow has to pass through another sonic point
close to the horizon (inner sonic point).
Thus, to have a shock solution on an accretion flow, there should be
at least two saddle type (\S 1) critical points. If the gas has polytropic
index closer to $5/3$ (actually, $\gamma \gsim 1.5$ in the above model),
the outer sonic point does not exist, and shocks can form only if the
flow is already supersonic (such as from winds of another star).

To understand the origin of the multiple critical points in a rotating
accretion flow, we write the specific energy of an inviscid,
thin, isothermal flow (obtained by integrating eq. 5.1a 
with $P/\rho=$constant),
$$
{\cal E} = \frac{1}{2} \vartheta^2 + a_0^2 log (\Sigma ) + 
\frac{1}{2}\frac {\lambda^2}{x^2}-\frac{1}{2(x-1)}
\eqno{(5.2)}
$$
Here, the first term is the radial kinetic energy ($\vartheta$= radial
velocity), second term is the thermal energy ($a_0$=constant sound speed; 
$\Sigma$= surface density), third term is due to the rotational kinetic energy
($\lambda$= constant specific angular momentum)
and the final term is due to the gravitational potential energy,
as described by the Paczy\'nski-Wiita [75] pseudo-potential (\S 1). Ignoring
the slowly varying logarithmic term, we note that the constant
energy surfaces behave differently in three different regimes:

\noindent  (a) Close to the black hole ($x \sim 1$): 

When the flow is close to the horizon, the potential energy term is dominant 
compared to the rotational kinetic energy term and the energy behaves as,
$$
{\cal E} \sim \frac{1}{2} \vartheta^2 -\frac{1}{2(x-1)} .
\eqno{(5.3a)}
$$
The negative sign signifies that the contours of constant energy in the $v$ 
vs. $x$ space have hyperbolic character. The asymptotes cross to form a saddle
type (inner) sonic point [Fig. 2.2], very similar to a Bondi flow.

\noindent (b) Far away from the black hole ($x \sim \infty$):

At distances very far away from the horizon, the potential term is 
again dominant over the rotational energy term, which falls off as $1/x^2$.
As a result, the energy varies with $x$ exactly same way as in eq. 5.3(a):
$$
{\cal E} \sim \frac{1}{2} \vartheta^2 -\frac{1}{2(x-1)}.
\eqno{(5.3b)}
$$
\noindent In this case also, the contours of constant energy are of hyperbolic
type and the asymptotes cross to produce the outer saddle type sonic point
[Fig. 2.2]. 

\noindent (c) Intermediate distances ($x \sim \lambda$):

At intermediate distances, the rotational energy term dominates over the 
gravitational potential energy and energy equation becomes:
$$
{\cal E} \sim \frac{1}{2} \vartheta^2 +\frac{\lambda^2}{2 x^2}.
\eqno{(5.3c)}
$$
\noindent In this case, the positive sign indicates that the contours
of constant energy would be elliptical in shape, similar to the energy contours
in simple harmonic oscillators. An `O' type  or center type 
sonic point [Fig. 2.2] is formed at the center of the ellipse.

Thus the flow may have three sonic points,
two saddle type and one center type. Of course, whether or not all these 
three points would actually exist will depend on
the angular momentum of the flow and the heating and
cooling processes determining the polytropic index of the flow
(determining the thermal energy term, ignored in above
arguments). If the angular momentum is
significant (and $\gamma \leq 1.5$), 
all these three points would be present. In the 
absence of angular momentum, three critical points would merge
into a single one as in the Bondi flow (\S 2). 
Another point of importance is that the inner sonic point
exists only in the case of black hole accretion [77, 218]. A sufficiently
compact neutron star with a radius less than $\sim 2.5$, may also have 
the inner sonic point, tough such compactness may be unlikely. The
two outer sonic points may exist in the case of stellar accretion as well.
This can give rise to significant differences in observational results
between disks around a black hole and a neutron star, for instance [68, 77,
218]. The inner edge of a flow enters the black hole supersonically 
whereas it must be subsonic on a neutron star. Thus, the flow will Comptonize
soft photons in an optically thick, cool disk and produce
a spectrum characteristics of a convergent flow [74] with a spectral index
$\alpha\sim 2-1.5$ (\S 7). However, for a neutron star, the flow
is highly subsonic at the surface, and such a component should 
not be present. This is indeed what is observed.

Although the pseudo-Newtonian potential has been used in the above argument, 
it is easy to verify that a full general relativistic treatment [77]
also produces the same number of sonic points and display the same
behaviour. Similarly, arguments about the number of critical
points remain valid even when the more general flows are considered.
In the case of viscous flows, the `O' type sonic point is changed
to a spiral type (Fig. 2.2) as in the case of a damped simple harmonic
oscillator. In the case of magnetized flows,
the number of sonic points could be five (Fig. 3.11).

\noindent{\Large 5.2 Adiabatic Disks with Discontinuities}

\noindent {\large 5.2.1 Steady State Solutions}

We assume a thin, rotating, adiabatic accretion 
flow. In this model, the complete radial momentum equation and the accretion 
rate equation are written on the equatorial plane, while in the vertical 
direction only hydrostatic balance condition is used $h(x) \sim a x^{3/2}$.
The momentum balance equation which must be satisfied at the shock
is vertically integrated. This model is known as one-and-a-half dimensional
flow or hybrid flow [77, 218]. For simplicity, the pseudo-Newtonian potential
[75] (\S 1) is used in place of full general relativity. Flow is assumed to be
inviscid, and therefore the specific angular momentum is constant everywhere.
Matter is accreted only due to pressure gradient force. If a shock
wave exists, the entropy remains separately constant, before and after the
flow. The entropy generated at the shock as well as the entire energy
is completely advected towards the black hole. These flows are thus
raditively inefficient.

The dimensionless energy conservation equation can be written as (cf. eq.5.1a):
$$
{\cal E} = \frac{\vartheta_e^2}{2}+\frac{a_e^2}{\gamma-1}
 + \frac{\lambda^2}{2x^2}- \frac {1}{2(x-1)} .
\eqno{(5.4)}
$$
Here $\vartheta_e$ and $a_e$ are the non-dimensional radial and the sound
velocities on the {\it equatorial} plane measured in units of
velocity of light $c$, $x$ is the non-dimensional
radial distance measured in units of the Schwarzschild radius $r_g\!=\!
{2GM}/{c^2}$, and $\lambda$ is the constant specific
angular momentum in units of $2GM/c$. Subscripts $e$ refer to the
quantities measured on the equatorial plane. Apart from a
geometric factor, the mass flux conservation equation is given by
$$
{\dot M} = \vartheta_e \rho_e x h (x) ,
\eqno{(5.5)}
$$
If the flow is in hydrostatic equilibrium in the transverse
direction, the height of the disk $h (x) $ is given by,
$$
h(x)=a_e x^{1/2} (x-1) .
\eqno{(5.6)}
$$
We write the mass conservation equation in terms of 
$\vartheta_e$ and $a_e$ in the following way:
$$
\dot{\cal M} = \vartheta_e a_e^q x^{3/2} (x-1) ,
\eqno{(5.7)}
$$
where $q\!=\!2n+1$, $n$ being the polytropic index of the flow.
$q$ takes the value $2n$ if the flow boundary is fixed.
${\dot {\cal M}} \sim {\dot M} K^n$  would be called `accretion rate'
here, though it changes at the shock due the generation of entropy
(i.e., variation of $K$).
In a radiation dominated flow, the total pressure of written as the
sum of gas pressure and the radiation pressure. Defining $\beta$  to be the 
ratio of the gas pressure to total pressure we obtain [211]:
$$
\dot {\cal M} = 1.26 \times 10^{-21} \frac {M_\odot}{M} \frac{ 1-\beta}
{\beta^4} \frac{\dot M}{\dot M_{Edd}},
\eqno{(5.8)}
$$
in c.g.s. units.  Here, $n\!=\!3$ is used for the illustration purpose 
and pure hydrogen is used as the chemical composition of the gas. 
${\dot M}_{Edd}$ is the Eddington accretion rate given by,
${\dot M_{Edd}} = L_{Edd} / c^2 $. For $n\leq 2$, the shock solutions
do not exist in this model.

\noindent {\large 5.2.2 Sonic Point Conditions}

The sonic point conditions as obtained following the procedure in \S 2,
$$
{\vartheta_c}^2=\nu {a_c}^2 ,
\eqno{(5.9a)}
$$
and
$$
{a_c}^2 = \frac{2(x_c-1)}{\nu x_c^2}\frac{({{\lambda^2}_K}(x_c)-
\lambda ^2 )} {(5x_c-3)}.
\eqno{(5.9b)}
$$
where $\nu\!=\!\frac{2n}{2n+1}$. The energy of the flow ${\cal E}$ with angular
momentum $\lambda$ passing through a critical point at $x_c$ is given by,
$$
{\cal E} = \frac{\lambda^2}{2x_c^2}\left [ 1 -
\frac{(4n+4)(x_c-1)}{5x_c-3} \right ] + \frac{(n+1) x_c}
{(x_c-1)(5x_c-3)}-\frac{1}{2(x_c-1)} ,
\eqno{(5.10)}
$$
or, conversely,
$$
\lambda^2 =  \frac { [ {\cal E} (x_c-1)-2(n+1)x_c+1 ] (5x_c-3)x_c^2}
{[ (4n+1)-(4n-1)x_c ] (x_c-1)} .
\eqno{(5.11)}
$$

Although we started with three conserved quantities, namely,
$\ce$, $\cm$ and $\lambda$, not all of them can be specified independently
if the flow is transonic. This is because we have three unknowns
$\vartheta_e(x)$, $ a_e(x)$ and $x_c$, and we need three equations to solve
them uniquely. Together with the two transonic conditions 5.9(a-b),
we need only one more quantity: either the energy or the
accretion rate. Thus, exactly as in the Bondi flow,
$\cm\!=\!\cm ( \ce , \lambda )$. In other words, the parameters
for a stationary transonic solution lie on a hypersurface:
$$
{\cal F}( \ce, \cm , \lambda ) = 0 .
\eqno{(5.12)}
$$
Of course, if one is interested in the actual entropy density, or mass 
flux ${\dot M}$ (and not the combined form $\cm$), one has to supply
another quantity, such as $\rho=\rho_\infty$ at the outer boundary
as in the Bondi flow.

\noindent {\large 5.2.3 When does a Flow Contain Shocks?}

When there are two `X'-type critical points, $\cm$ (i.e., entropy)
calculated at these two points are not usually equal [77, 218].
That is, if we denote the outer and the inner critical
point locations as $x_o$ and $x_i$ respectively, then, $\cm$(${\cal E}$,
$\lambda$, $x_o$) $\ne$  $\cm$(${\cal E}$, $\lambda$, $x_i$). 
However, for a given $\lambda$, there exists one and 
only one energy ${\cal E}_m$ such that ${\dot{\cal M}}_m
$(${\cal E}_m$, $\lambda$, $x_o$) $\!=\!$  ${\dot{\cal M}}_m$
(${\cal E}_m$, $\lambda$, $x_i$). In Fig. 5.1, we plot
$\ce$ against $\cm$ for a given angular momentum $\lambda
\!=\!1.675$. This so-called ``swallow-tail" singularity
is produced by the projection of an otherwise
smooth curve, drawn for a constant angular momentum on the surface:
$$
x_c=x_c (\cm , \ce ) .
\eqno{(5.13)}
$$
A flow with parameters from the branch $AB$ passes through the
inner `X'-type critical point, and with parameters from the
branch $CD$ passes through the outer `X'-type critical point. The
parameters on the branch $BC$ are calculated at the `O'-type critical 
points and are unphysical since a flow cannot pass through these points.
The crossing point $M$ represents the case when $\cm$ at both the
critical points are identical. The parameter space division
in terms of the energy of the flow is also shown in this Figure.
For ${\cal E} \geq {\cal E}_i$ there is only one critical point. For
${\cal E}_m  \leq {\cal E} \leq {\cal E}_i$,
the accretion rate ${\dot{\cal M}_o} \geq {\dot{\cal M}}_i$. For
${\cal E}_s \leq {\cal E} \leq {\cal E}_m$, ${\dot{\cal M}_o}
\leq {\dot{\cal M}}_i$. For ${\cal E} < {\cal E}_s$, the flow passes through
only the outer critical point. Similar division can be  made in terms of 
the accretion rate. Referring to Figure 5.1, a shock
therefore connects flows with two sets of parameters
($\lambda,\  {\dot{\cal M}}, \ {\cal E}$), one from the branch $AB$ and the
other from the branch $CD$. The basic stationary equations above
contain no dissipative terms and therefore can be consistent only
with the Rankine-Hugoniot shock solutions which conserves energy. 
Schematic shock transitions in typical cases are shown in Figure 5.1
by horizontal lines $s_1s_2$ and $w_1w_2$. The former one is in accretion
and the latter one is in wind.  However, the shocks in thin, otherwise 
dissipation-free astrophysical flows may radiate away energy through the upper
and the lower boundaries simply because of the presence of the relatively 
strong temperature gradient. If the length scale in which such radiations
occur is small compared to the other (such as the diffusion and cooling) 
length scales of the flow, one can include such radiative effects by suitable
choices of the shock conditions.

\noindent{\large 5.2.4 Shock Conditions}

A shock is characterized by four $a \ priori$ unknown quantities: the shock
location $x_s$, and the possible jumps in the two independent velocities
$u$, $a$ and in the polytropic constant $K$ (entropy, or ${\cm}$) i.e.,
$$
x=x_s ,
\eqno{(5.14a)}
$$
$$
\Delta a = a_+ (x_s) - a_- (x_s) ,
\eqno{(5.14b)}
$$
$$
\Delta  u =  u_+ (x_s) -  u_- (x_s) ,
\eqno{(5.14c)}
$$
and finally,
$$
\Delta K = K_+ - K_- .
\eqno{(5.14d)}
$$
Subscripts minus ($-$) and plus ($+$) denote quantities before and 
after the shock. Independent of the nature of a shock, the conservation of
the fluxes of mass and momentum,
$$
\dot M_+ = \dot M_- ,
\eqno{(5.15a)}
$$
$$
W_+ + \Sigma_+ u ^2_+  = W_- + \Sigma_- u^2_- ,
\eqno{(5.15b)}
$$
provide two constraints on these four quantities. $W$ and $\Sigma$ denote the
pressure and density integrated over the vertical direction.
Since we are dealing with inviscid flows and axisymmetric shocks, 
$\lambda_+\!=\!\lambda_-$ is always fulfilled and gives no extra constraint. 
The pre-shock quantities are determined by the outside boundary conditions
in the case of accretion flows or by the inner boundary conditions in the
case of winds. The post-shock quantities, which are given by
($ \lambda_+,{\dot{\cal M}_+}, {\cal E}_+ $), must obey the sonic point
conditions 5.9(a-b) at the critical point after the shock. This provides the
third condition. The fourth and the final condition models the
nature of the dissipative processes (cf. eq. 5.1d) at the shock and uniquely
characterizes the shock as well as the post-shock flow properties.
A single parameter [219],
$$
f(x_s)=\frac{(Q_+-Q_-)_+}{(Q_+-Q_-)_-}
\eqno{(5.16)}
$$
determines the nature of the energy dissipation
and/or the entropy generation at the shock (eq. 5.1d).
Therefore, these shocks may differ qualitatively from the ``Rankine-Hugoniot"
shocks discussed in the standard text books 
in the way the fourth condition, connected with dissipation, is formulated.
This freedom is not surprising, as the astrophysical flow is 
in direct contact with the surroundings, as opposed to shock waves
in a standard shock tube problem where the flow is confined in a 
tube and preserves energy. Secondly, the entropy generated
at the shock need not be advected completely. A part of it could be
radiated away as well.

In a general case, the fourth shock condition can be formally written as
[77, 219], 
$$
({\cal E}_+ -{\cal E}_-) - A({\dot{\cal M}_+}-{\dot{\cal M}_-})=0.
\eqno{(5.17)}
$$
This can be considered as the most general shock condition. For a given set
of initial parameters, choosing the ratio $A$ from $0$ to $+\infty$,
one obtains all the possible shock solutions. This parameter basically
describes the behaviour of the entropy generated at the shock. 
Three special cases of eq. (5.17) are of significance. They are: (a) $A=0$. 
In this case the energy is conserved at the discontinuity and
Rankine-Hugoniot shocks (${\cal E}_+ \!=\! {\cal E}_-$) are formed.
The entropy generated at the shock is entirely advected by the flow 
inwards of the shock for the disks (and outwards of the shock for the winds)
and the raditive efficiency is completely negligible.
(b)$A=\infty$. Here the entropy generated at the shock is totally radiated away,
and the isentropic compression waves ($\cmp\!=\!\cmm$) are formed.
(c) At some intermediate  $A$ an isothermal shock wave ($a_+\!=\!a_-$) 
may form where the energy and entropy change in a manner that the
temperature of the flow at the shock remains constant [219, 77]. 

Assuming the Rankine-Hugoniot shock conditions,
the location of the shock is easily determined, by rewriting
shock conditions in terms of the Mach numbers before and after
the flow. This gives the following conditions [77, 218]:
$$
C= {{[M_+ (3\gamma - 1) + {2 \over M_+}]^2} \over {2+(\gamma-1)M_+^2}}
=
{{[M_- (3\gamma - 1) + {2 \over M_-}]^2} \over {2+(\gamma-1)M_-^2}} .
\eqno{(5.18)}
$$
The constant $C$ as defined above is invariant across the shock.
Mach numbers of the flow just before and after
the shock can be written down in terms of $C$ as,
$$
M^2_\mp=\frac{2(3\gamma-1)-C \pm \sqrt{C^2-8C\gamma}}{(\gamma-1)C
-(3\gamma-1)^2}.
\eqno{(5.19a)}
$$
The product of the Mach numbers is given by,
$$
M_+M_-= {2 \over {[(3\gamma-1)^2-(\gamma-1)C]^{1/2}}}.
\eqno{(5.19b)}
$$
The weakest shock which may form in a transonic flow in vertical equilibrium
has the pre-shock Mach number $M_-\!=\!\sqrt{2 /{(\gamma+1)}}$.  The 
corresponding location of the shock coincides with the sonic point. It
is easy to verify that the post-shock Mach number is also the same. 
Thus, the minimum shock strength is unity, 
which is expected from a self-consistent model of this type.

\noindent {\large 5.2.5 Examples of Complete Solutions with Discontinuities}

Returning to Fig. 5.1, we note that for ${\cal E} \geq {\cal E}_i$, there 
is only one critical point and no shock is possible.
For ${\cal E}_m  \leq {\cal E} \leq {\cal E}_i$,
${\dot{\cal M}_o} \geq {\dot{\cal M}}_i$ and a shock in an accreting
flow is not possible either, since the entropy will $decrease$ in the process.
However, shocks can form in winds with the change
of entropy corresponding to ${\dot{\cal M}}(w_1)-{\dot{\cal M}}(w_2)$.
For ${\cal E}_s \leq {\cal E} \leq {\cal E}_m$, shocks form in accretion
(but not in winds) with the change of entropy calculated from the two 
intersection points $s_1$ and $s_2$.
For ${\cal E} < {\cal E}_s$, a flow passes only through the 
outer critical point and the shocks are not possible.
This figure was drawn for a given angular momentum of the flow. 
For other angular momentum the region of interest will be different
[77, 218]. A general procedure to obtain the shock locations
must involve verifying if the Mach number relation (eq. 5.18a) is satisfied
at any point in the flow. This is done by plotting the contours of constant
$\dot{\cal M_-}$ and constant $\dot{\cal M_+}$ (i.e., contours of
pre-shock and post-shock entropies) for a given set of energy
$\cal E$ and angular momentum $\lambda$ and see whether/where they intersect.
An example of a self-consistent shock solution
(in the limit of zero viscosity) is shown in Fig. 5.2 for the flow parameters
${\cal E}\!=\!0.689\times 10^{-2}, \ 
{\dot{\cal M}_-}\!=\! 0.831 \times 10^{-5}, \lambda\!=\!1.65$.
The four possible shock locations denoted
by $X_{s1}, \ X_{s2}, \ X_{s3},$ and $X_{s4}$ are shown
where all the shock conditions are simultaneously satisfied.
In the case of black hole accretion flows,
which are subsonic at a large distance and supersonic on the horizon, 
$X_{s2}$ and $X_{s3}$ are the only two possibilities for shock locations.
However, for accretion onto the neutron stars, $X_{s1}$
is also a formal possibility provided the inner sonic point actually
forms. In reality, in a neutron star accretion, the inner sonic
point may not exist, and the shock transition jumps to a branch
which will remain subsonic till it hits the surface. The contours are of 
constant `accretion rate' ${\dot{\cal M}}$.
Only the solution with arrows is the allowed stationary solution, the others
are just the integral curves. 

In Fig. 5.2 one notices that there are four locations, $X_{s1}$, $X_{s2}$, 
$X_{s3}$ and $X_{s3}$, where all the shock conditions 
are satisfied and both the shocks are of same strength with the same amount of
entropy generated. However, since the flow cannot pass through all the
shocks, only one of them can be stable at the most. Which one of these
two shocks is stable? A local stability analysis
does not seem to be able to answer this question uniquely [77, 218].
However the riddle is easily resolved by using a simple consideration of varying
the (total) pressure at the post-shock flow. If a shock
is perturbed on either way from its equilibrium location,
(where pressure balance is maintained), and the excess pressure is
directed in such a way that the shock returns to its stationary location,
then that shock must be stable [220]. Indeed through extensive numerical 
simulations (\S 6), as well as analytical works [220-223]
it has been established beyond doubt that in accretion,
the shock located at $X_{s3}$ is stable, while solutions at other locations
are unstable. The stable shock generates exactly the right amount of entropy 
which is advected with the flow to enable it to pass through the inner 
sonic point. Since the flow is chosen to be adiabatic, entropy is separately
constant before and after the shock. Another important feature
of these solutions is that, the flow is radiatively inefficient
(in fact, energy and entropy are completely advected
with the flow towards the black hole). 
In the case of the winds, the situation is completely opposite. Here,
the entropy at the outer sonic point is higher than the entropy
at the inner sonic point. The shock located at $X_{s2}$ is 
found to be stable [220]. Details are already in the literature [77, 218]
and we do not discuss them here.

\noindent {\Large 5.3 Isothermal Advective Flow with Viscosity}

Important properties of viscous transonic flows emerge
only when global solutions of eqs. 5.1(a-d) are obtained.
Most complete global solution was first found in the 
context of viscous, isothermal flows [77, 224-225].
The inviscid isothermal flow has two possible locations where shocks
may form, exactly as in the adiabatic case. However, when the
viscosity is added, the most important change that occurs is the
change of the `O' type sonic points to the spiral type sonic point (Fig. 2.2).
At the same time, the stable shock at $X_{s3}$ (away from the black hole)
becomes weaker, breaking the ambiguity. At a critical viscosity
parameter (in this case, $\alpha_c \sim 0.015$, which depends 
upon the disk temperature, sonic point location $x_{in}$ and 
$\lambda (x_{in})$ ), the shock at $X_{s3}$ disappears altogether.
At the same time, the topology of the solution changes so as to allow
two smooth solutions, one with a higher dissipation
passes through the inner sonic point and the other with a lower
dissipation passes through the outer sonic point. Figures 5.3(a-d)
show this transition of topologies. In Figs. 5.3a and 5.3c, low
viscosity parameter ($\alpha=0.01$) is used. The shock at $X_{s3}$ 
(which is stable) is weaker but it is still present. The shock-free solution
can pass only through the outer sonic point, exactly as in the inviscid case,
but in reality it does not since it has a choice to pass through
the shock. In Fig. 5.3(b) and Fig. 5.3(d) drawn for $\alpha=0.02$, 
the flow becomes shock-free, and passes through the
inner sonic point. In both the cases, the flow becomes
Keplerian at a large distance. The flow also has a solution to pass 
through the outer sonic point, but the energy required is higher. 
The critical $\alpha$ parameter above which the shock disappears 
and the flow becomes Keplerian depends upon other assumptions,
such as the viscosity prescriptions and the accretion rate and angular
momentum at the inner sonic point as we discuss below. 
The numerical simulation  results of this case 
will be discussed in the next Section \S 6 [225]. 

\noindent {\Large 5.4 Most General Advective Flow 
and Unification of Accretion Disk Models}

Fig. 5.4(a-b) show the first attempt to unify different sub-Keplerian
disk models in the limit of very small viscosity [77, 218]. In 5.4a, the 
`accretion rate' (which is a combination of true accretion rate and entropy, 
see Eq. 5.7) is plotted against angular momentum. The zone marked by I,
represents the region where the flow has only the inner sonic
point. The zone marked by W, has two sonic points but,
$\cm$(${\cal E}$, $\lambda$, $x_o$) $\geq$  $\cm$(${\cal E}$, $\lambda$, $x_i$),
so that shocks are possible in winds but accretion can also occur
through the inner sonic point. The zone markes by A has two sonic points but,
$\cm$(${\cal E}$, $\lambda$, $x_o$) $\leq$  $\cm$(${\cal E}$, $\lambda$, $x_i$)
so that shocks are possible in accretion if shock conditions are satisfied.
The zone marked by O has only outer sonic point respectively.
Fig 5.4b is further divided to show whether or not
shocks actually form. If viscosity is higher than $\alpha_{c}$
shocks disappear altogether since the topology changes, and this particular
sub-division is modified on a case by case basis.

Referring to Eqn. 5.1(a-d), one notes that typically, for optically
thick flow, one has, in the limit of diffusion approximation,
$Q_-=4acT^4/3k\Sigma$, and for optically thin flows,
$Q_-$ is given by bremmstrahlung, Comptonization and other processes
(such as synchrotron processes if magnetic field is present).
The heat generation with $\alpha$ viscosity is simply, $Q_+\propto
\alpha Px \frac{d\Omega}{dx}$. The result is however same as above.
The critical viscosity $0<\alpha_{c}<1.0$ becomes a function of
the relative importance of $Q_-$ vis-a-vis $Q_+$, the viscosity
parameter, location of inner sonic point and the angular momentum
at the inner edge of the disk. For $\alpha<\alpha_{c}$,
the flow may have a shock but for $\alpha>\alpha_{c}$ the 
flow simply passes through the inner sonic point similar to what was discussed
\S 5.3. Thus, the most general transonic flow shows no new topology
other than what was seen in the context of isothermal flows.

To understand the shock formation properties, one needs to know
how entropy and energies behave at the sonic points ([226], Fig. 5.1).
Fig. 5.5 shows entropy $s$ {\it vs} energy $E(x)=0.5 \vartheta^2+
(\gamma-1)^{-1} a^2 + 0.5\  \lambda^2/x^2 - 0.5\ (x-1)^{-1}$ plots at the
sonic points for $\gamma=4/3$ (left) and $\gamma=5/3$ (right)
for $\alpha=0, \ 0.4 \ 0.8$ and $f=(Q_+-Q_-)/Q_+=0$ (cooling
dominated) and $f=1$ (heating dominated) respectively
(as marked on the curves). Angular momentum at the inner edge
$\lambda_{in}=1.6$ is chosen. Solid, dashed and dotted
regions of the curves are the saddle, nodal and spiral
(circle type for $\alpha=0$) type sonic points respectively
(\S 2). For $\gamma=4/3$, the branches $AMB$ and $CMD$
(or the like in other curves in this figure) are the
loci for inner ($x_{in}$) and outer ($x_{out}$) sonic points 
(77, 218) similar to what was discussed about Fig. 5.1.
For a shock in accretion to be possible, pre-shock flow
parameter must lie somewhere on the branch $MD$ and the post-shock
flow parameter must lie somewhere on $MB$ as long as the energy and
entropy conditions: $E(x_{in}) \leq E(x_{out})$ and $s(x_{in}) \geq
s(x_{out})$ are satisfied. A typical shock transition ${\vec ss}$ is
shown (cf. Fig. 5.1). For $\gamma=5/3$,
the outer sonic point does not exist in this model as discussed
above (but can exist in a thin disk with a constant height, for example).
Thus shocks can form only if the flow is already supersonic
(such as, when originates from winds of a companion).
For both $\gamma$s, one notices that as the viscosity is
increased, the outer sonic point no longer remains saddle type
and only the inner sonic point exists. Thus, shock transitions
are no longer possible. Flow parameters (e.g., the inner sonic point)
originally on the branch of type $AM$, enters in the branch of type
$MB$ (i.e., below $M$) as $\alpha$
is increased from $0$ to $\alpha_{c1}$ and shocks become possible.
However, it escapes the shock region for
$\alpha>\alpha_{c}$. Thus, for $\alpha >\alpha_{c}$,
only allowed solutions are those which pass through the inner sonic
point and therefore no shock solution in accretion is possible.

Fig. 5.6 shows an example of a most general solution with a shock
wave, where $Q_-/Q_+$=0.5 is chosen everywhere and $\alpha=0.05$
(here, $\alpha_c\sim 0.1$).
The stable shock transition is shown by the vertical dashed curve.
For $\alpha>\alpha_c$, the faster rate of angular momentum transport
in the post-shock region eventually drives the
shock wave to a large distance making the flow subsonic and
Keplerian. Secondly, if $\alpha>\alpha_c$ in the equatorial plane
and $\alpha<\alpha_c$ above, then the pre-shock and post-shock
solutions are described by optically thick Keplerian disk
and optically thin transonic disk respectively.
This property is used in \S 7 where the most general disk model 
and its observational properties will be discussed. 

One important aspect of viscous, transonic flow  which was not adequately
stressed so far, is that an originally Keplerian disk
could  become sub-Keplerian very close to the black hole. This could be
seen easily from the angular momentum distribution of the
viscous flow (from eq. 5.1c) ([224-226]):
$$
\lambda (x) - \lambda (x_{in})= \frac{\alpha a^2 x}{v}
$$
Fig. 5.7 shows examples of the ratio of the disk angular momentum
distribution to the Keplerian distribution (for $x<3r_g$,
$\lambda_{Kep}(x) = \lambda_{3r_g}$ is kept for stability reasons)
as a function of distance from the black hole [68].
The ratios show deviation from Keplerian due to advection, pressure and
viscous effects. Three diverse cases have been chosen to illustrate
how non-Keplerian the flow can get.
In Case A (marked `A'), $\lambda (x_{in})=1.88$, $x_c=2.2$,
$\alpha=0.005$, $\gamma=4/3$ (radiation pressure dominated)
and $Q_-=Q_+$. The flow deviated from Keplerian disk at $x_{Kep}=7.5r_g$
and even becomes super-Keplerian (ratio $>1$) close to the black hole.
In Case B (marked `B'), $x_{c}=2.3$, $\lambda (x_{in})=1.7$,
$\gamma=5/3$ (gas pressure dominated), $\alpha=0.02$ and $Q_-=Q_+$.
Here the flow deviates from Keplerian at $90r_g$ and always remained
sub-Keplerian. In both the cases above, there is no shock formation.
In Case C (marked `C'), the flow passes through a shock at $x=13.9r_g$
(angular momentum distribution remained continuous) and remains
completely sub-Keplerian after deviating from the Keplerian disk at 
$x=480r_g$. The parameters chosen are $\lambda(x_{in})=1.6$, $x_{c}=2.87$, 
$\gamma=4/3$, $\alpha=0.05$ and $Q_-=0.5 Q_+$.
This important result enables one to construct a general accretion disk
model as discussed in \S 7.

To be concrete about heating and cooling processes, 
we present a simpler example here with a power-law cooling $ \Lambda \propto 
\rho^2 (T/T_0)^{\alpha_b}$ in the flow and the polytropic
index $\gamma=5/3$. For $\alpha_{b}=0.5$, this represents 
bremmstrahlung cooling. Assume that $\alpha=0$,  so that $\lambda$=const.
The governing equations could be written as:\\
\noindent (a) The entropy equation:
$$
\frac{\partial (e+\frac{P}{\rho})}{\partial x}
- \zeta_\alpha \frac{a^{2 \alpha_b} \rho}{\vartheta}=0
\eqno{(5.20a)}
$$
\noindent (b) The radial momentum equation:
$$
v \frac{\partial \vartheta}{\partial x}+
\frac{1}{\rho} \frac{\partial P}{\partial  x} + {\psi^{\prime}}(x)=0 ,
\eqno{(5.20b)}
$$

\noindent (c) The continuity equation:
$$
\frac{1}{x} \frac{\partial}{\partial x} (\rho \vartheta x)=0 .
\eqno{(5.20c)}
$$
Here, $x$ is the radial coordinate, $\psi(x)$ is the effective potential 
energy $\psi(x)=g(x)+\lambda^2/2x^2$ with $g(x)$ as the radial force 
potential, which in the pseudo-Newtonian
model takes the form: $g(x) = - \frac{1}{2} (x-1)^{-1} $, $e$ is the internal
energy $e=\frac{P}{(\gamma-1) \rho}$, $\vartheta$ is the radial component of
velocity and $a$ is the adiabatic sound velocity: $a^2=\frac{\gamma P}{\rho}$ .
Prime denotes derivative with respect to $x$.
$\zeta_\alpha$ is the non-dimensional bremsstrahlung loss coefficient,
$$
\zeta_\alpha = \frac{j \rho_{ref} x_{ref} \mu^\alpha }{(m_p)^{2-\alpha} 
\gamma^\alpha \kappa^\alpha v_{ref}^{2}}
\eqno{(5.21)}
$$
where, $\mu=0.5$ and $j=1.4 \times 10^{-27}$ c.g.s. unit for ionized
hydrogen,
$m_p$ is the mass of the proton, and $\kappa$ is the Boltzman constant. 
The cooling law to be used is $\Lambda= \zeta_{1/2} \rho^2
(T/T_0)^\alpha $. In this case, the coefficient is constant and only the 
cooling exponent $\alpha$ varies. $\rho_{ref}$ is the reference
density at the outer edge, $v_{ref}=c$, $x_{ref}=2GM/c^2$, and
$T_0$ is the reference temperature.

Figure 5.8 shows the example of the shock solution
with $\alpha_b=0.5$ (solid) and  $\alpha_b=0.6$ (long-dashed). The short-dashed
curve is without cooling for comparison. Due to cooling, the shock becomes 
weaker and comes closer to the black hole. A thorough understanding requires 
a fully time dependent solution [227], because the flow is expected to
show interesting behaviour.
As the shock is perturbed and propagates outwards, the post-shock region
becomes hotter since the relative velocity
between the shock and the incoming flow is higher. 
The cooling rate goes up eventually stopping
the outward motion of the shock when the flow has a sufficient 
time to cool. As the shock returns towards the black hole, lower temperature
and therefore lower pressure in the post-shock region 
is  unable to balance the pre-shock ram pressure and the shock
collapse continues till the higher temperature closer to the black hole
is reached. The shock then overshoots this region and
bounces from from higher pressure region and the cycle continues.
Thus the shock may show oscillations whenever the cooling
time scale roughly agrees with the infall time scale: $t_{cool} \approx
t_{infall}$. In the next section (\S 6) we show some results of 
time dependent simulations which show these oscillations [227].
Similar oscillatory behaviour has also been seen
in shocks in accretion columns of white dwarfs [228]. 

\noindent {\Large 5.5 `Advection Dominated' Disk Models}

Recently, it has been stressed that a particular aspect of the 
above solutions, namely, the advection of energy and entropy
of the transonic equations may be more important than realized
before. A sub-class of the general solution is thus termed
`advection dominated disks'. As we stressed before, because
of advection, the energy release is very inefficient as it is advected 
towards the black hole. However, the assumption of the vertical 
integration of the above model was removed by Narayan 
and his collaborators [229]. This latter study
shows that the disks may or may not be advection 
dominated depending on the cooling processes operating inside the disk. 
The importance of these works lies in their ability to provide
a global solution of quasi-spherical viscous flows
(albeit using very restricted assumptions, such as
self-similarity) and to establish that the
quasi-spherical solutions obtained by these models are similar 
in every respect to the vertically averaged viscous transonic 
models. This confirms earlier numerical simulation works [259]
as discussed in the next Section.

A global solution is obtained where a number of physical processes
are included. (a) The assumption of self-similarity is made
with a constant ratio of heating and cooling [229] $Q_+/Q_-$, 
(b) wherever this assumption is dropped,
solution is obtained at a given radial distance $R$ of a Keplerian disk [230]
and was shown that advection will stabilize the flow (an attempt to
show this earlier was made in [146]),
(c) bremsstrahlung, is added self-consistently, (d)the `bridging formula' 
of Wandel and Liang [231] is used to take into account $\tau\sim 1$ 
regions. The basic conclusions of these models are the followings: 
(1) that the advection of entropy is very important in establishing a stable 
solution of accretion disk,
(2) the winds may be formed if the flows have positive specific energies,
(3) the angular velocity $\Omega$ is far less than the 
Keplerian velocity $\Omega_K$,
(4) the efficiency of radiation is found to be very small, since
most of the energy is advected away.

These newly discovered results completely confirms fully consistent 
transonic advective flows solutions discussed earlier (\S 5.2-5.4)
and therefore do not represent any fundamentally new solution.
Importance, however, lies in their ability to separate out
the effects of advection. An artifact of the self-similar transonic
flow model is that it has constant Mach number throughout. This sharply
contrasts with the exact nature of Mach number variation shown in Figs.
5.3(a-b) and 5.6, where the flow Mach number varies considerably while 
passing through the sonic points and shock waves. In an exact 
solution [226], even around $r \sim 10r_g$, generally $v_r <<v_\phi$
and so advection is not dominated.

\noindent {\Large 5.6 Axisymmetric Shock Waves in a Few Other Systems}

\noindent {\large 5.6.1 In Kerr Geometry}

Solution including shock waves using fully general 
relativistic equations in Kerr geometry produce very similar results 
as in Schwarzschild geometry. However, the sonic points and the location of the
shock waves are found to be very sensitive to the Kerr 
parameter $a$. In particular, in the contra-rotating flows, 
the shocks were seen to be formed farther away from the black hole [77, 232].
Numerical simulations by smoothed particle hydrodynamics
show that these shocks are stable [233] as well. 

\noindent {\large 5.6.2 In Magnetized Flows}

In the case of a magnetized accretion, other than regular Alfv\'en point at 
$u\!=\!r\!=\!1$, five critical points may be present in the flow (\S 3).
Accordingly, a large number of new topologies emerge (Fig. 3.11a-f).
These new topologies allow the formation of shock waves in accretion 
as well as in winds. In a complete solution of accretion disk
which includes a standing shock wave, the flow must satisfy a set of 
conditions on either side of the discontinuity. These are [77, 161]:\\
\noindent (a) the total energy flux is conserved:
$$
\frac{1}{2} \vartheta_{r+}^2 + \frac{1}{2}\vartheta_{\phi+}^2 + na_+^2 -
\vartheta_{\phi_+} \Omega r_+
=\frac{1}{2} \vartheta_{r-}^2 + \frac{1}{2}\vartheta_{\phi-}^2 + na_-^2 -
\vartheta_{\phi_-} \Omega r_- ,
\eqno {(5.22a)}
$$
\noindent (b) the total mass flux is conserved:
$$
\rho_+\vartheta_{r+}=\rho_-\vartheta_{r-} ,
\eqno {(5.22b)}
$$
\noindent (c) the radial momentum is balanced:
$$
p_+ + \rho_+\vartheta_{r+}^2 + \frac{B_{\phi+}^2}{8\pi}
=
p_- + \rho_-\vartheta_{r-}^2 + \frac{B_{\phi-}^2}{8\pi} ,
\eqno {(5.22c)}
$$
\noindent (d) the transverse momentum is balanced:
$$
\rho_+ \vartheta_{r+} \vartheta_{\phi+} - \frac {B_{r+}B_{\phi+}}{4\pi}
=
\rho_- \vartheta_{r-} \vartheta_{\phi-} - \frac {B_{r-}B_{\phi-}}{4\pi} ,
\eqno {(5.22d)}
$$

\noindent (e) the radial magnetic flux is conserved:
$$
B_{r+}=B_{r-} ,
\eqno {(5.22e)}
$$
and finally,

\noindent (f) the field equation is independently satisfied on either side
of the shock:
$$
\vartheta_{\phi+}B_{r+}-\vartheta_{r+}B_{\phi+}=
\vartheta_{\phi-}B_{r-}-\vartheta_{r-}B_{\phi-} .
\eqno {(5.22f)}
$$
\noindent Six conditions 5.22(a-f), together with the transmagnetosonicity
of the post-shock and the pre-shock flows are enough to calculate all the 
variables uniquely including the location of the shock [77, 161].

In Figure 5.9, an example of the shock formation in an accretion flow is 
presented [77, 161]. Here, two critical solutions passing through rotational
and Bondi-like slow magnetosonic points are connected by a shock transition.
Paczy\'nski-Wiita potential [75] is chosen for computational
purpose. The parameters are:
$L\!=\!\Omega\!=\!1.45$ and $g_0\!=\!1.334$. 
The solution is shown in arrows and the shock transition is shown by a vertical 
dashed line. Other parameters are: ${\cal E}\!=\!0.2$, 
${\dot{\cal M}}_-\!=\!0.4857$, ${\dot{\cal M}}_+\!=\!
0.5074$, $B_{\phi +}\!=\!0.556$, $ B_{\phi-}\!=\! 1.106$, 
$\vartheta_{\phi+}\!=\! 1.875$, $\vartheta_{\phi-}\!=\!2.424$, 
$u_+\!=\!0.417$, $u_-\!=\!0.621$, $r_s\!=\!1.0995$,
$r_s$ being the location of the shock. 
Dashed curves are the contours of constant
slow magnetosonic wave velocities. The pre-shock
flow is super(slow)magnetosonic and the post-shock flow is
sub(slow)magnetosonic and both are sub-Alfv\'enic, thus satisfying the
condition of the formation of the slow shocks [77].
Shock location is found to be unique. These works are carried
out with pseudo-Newtonian potentials. However, no new 
solution topologies appear when the exact Kerr geometry is used [162]
and therefore, possibly no new types of shock solutions are expected.

\noindent {\Large 5.7 Non-Axisymmetric Spiral Shock Waves}

\noindent {\large 5.7.1 Introduction}

So far, we have concentrated on specific model flows which are axisymmetric.
However, in a binary system with one component a compact star, the disk moves
in a non-axisymmetric potential because the secondary normal star exerts 
non-axisymmetric tidal force on the accretion disk around the 
compact primary. Numerical simulations have shown formation of spiral 
density waves eventually steepening to produce spiral shocks [142, 234-236]
In the case of active galaxies, spiral shocks 
could be induced due to the passage of a massive object,
such as a globular cluster, molecular cloud, dwarf galaxy or another
black hole [6, 237]. These shocks may also be induced
by non-axisymmetric instabilities inside the disk or due to tidal capture
of a molecular cloud.

Study of spiral shocks are important because they allow accretion to take 
place in the absence of any explicit viscosity in the flow. Furthermore, 
recent observations of the temporal [239-247] and spatial
variability [4-5] of line emissions from ionized
disks around black holes indicate that the disks around super-massive
black holes may have spiral shock waves [6, 212].
We shall discuss it here very briefly for the sake of completion
of the review on accretion disks which include shock waves. 

\noindent {\large 5.7.2 Basic Flow Equations }

Consider some matter which is accreting  on a Newtonian star
non-axisymmetrically. Assume that the flow equations are two-and-a-half
(2.5) dimensionally correct in the sense that whereas radial and
azimuthal components of the momentum equations are kept exactly,
momentum of the flow in the vertical direction (direction off the
equatorial plane of the disk) is neglected. Assume that the flow is
in hydrostatic equilibrium at each point in this direction, 
i.e., vertical pressure balance is maintained throughout.
Assume that the shocks are self-similar [248]. 

In this sub-section, the radial distance $r$ is measured in units of
$\frac{GM}{c^2}$, the velocity components are measured in units of $c$,
time is measured in units of $\frac{GM} {c^3}$.
The equations of motion in cylindrical polar ($r,\ \phi ,\ z$) coordinate 
are first written in the spiral coordinate
($x=r$ and $\psi=\phi+\beta(r)$) and the solutions
are assumed of the form [77, 248]: 

$$
u=x^{-n_u}q_1(\psi),
\eqno{(5.23a)}
$$
$$
v=x^{-n_v}q_2(\psi),
\eqno{(5.23b)}
$$
$$
\rho=x^{-n_\rho}q_\rho (\psi),
\eqno{(5.23c)}
$$
$$
p=x^{-n_p}q_p (\psi),
\eqno{(5.23d)}
$$
$$
\frac{d \beta}{d x}=x^{-n_\beta} B .
\eqno{(5.23e)}
$$
Here $u$ and $v$ are the radial and the azimuthal velocity components,
$\rho$ and $p$ are the density and the pressure of the flow, $2h$ is the
local vertical thickness of the flow. 
${n_u}, \ n_v, \ n_\rho, \ n_P$, $\ n_\beta $ and $B$ are constants.
Gravitational potential is that of a point
Newtonian body. Pseudo-Newtonian potential used in the
previous Sections cannot be used here, because that would introduce
a preferred length scale in the system, breaking self-similarity. Let
the equation of state be: $P\!=\!K\rho^\gamma$, where $K$ is the 
adiabatic constant, a measure of entropy of the flow and $\gamma$ is the 
adiabatic index. In terms of the polytropic index $n$, $\gamma\!=\!1+{1}/{n}$.
The quantity $K$ remains constant in the azimuthal direction in
between the two shocks. It can change only at a shock. In the
absence of the radiative loss, the generated entropy is advected by the flow. 

Using above form of the solutions in the respective 
hydrodynamic equations one obtains [77,  248],
$$
n_u \!=\! n_v = \frac{1}{2}, n_\beta=1, n_\rho-n_p=-1.
\eqno{(5.24)}
$$
The adiabatic sound speed $a\!=\!({\partial p}/{\partial \rho})^{1/2}$ varies
as
$$
a\!=\!(\gamma p/\rho)^{1/2}=x^{-1/2}q_3^{1/2}({\psi}) .
\eqno{(5.25)}
$$
Accordingly, from the consideration of the hydrostatic
balance, the thickness of the disk is obtained as
$$
h=ax^{3/2}=x{q_3^{1/2}(\psi)}  ,
\eqno{(5.26)}
$$
apart from a geometrical constant. Using such variations, the continuity
equation takes the form:
$$
q_1q_\rho q_3^{1/2}(3/2-n_\rho)+\frac{\partial}{\partial \psi}[
q_\rho(q_2+Bq_1){q_3}^{1/2}]=0.
\eqno{(5.27)}
$$
Integrating the above equation, one obtains
$$
(3/2-n_\rho)\int_0^{2\pi} q_1q_\rho {q_3}^{1/2} \, d\psi +
[q_\rho(q_2+Bq_1){q_3}^{1/2}]_0^{2\pi}=0.
\eqno{(5.28)}
$$
The bracketed term must be equal at both the limits
since it is the mass flux in the azimuthal direction.
The first term, i.e.,
$$
{\dot M}_\parallel=\int_0^{2\pi} q_1q_\rho {q_3}^{1/2} \, d\psi ,
\eqno{(5.29)}
$$
represents the net mass flux in the radial direction, which is non-zero only if
$n_\rho\!=\!3/2$. We choose this value of $n_\rho$. However,
$n_\rho \ne 3/2$ (where the net flux is zero) case may also be important in 
certain circumstances where both the wind and the accretion take place 
simultaneously. If instead of the assumption of vertical equilibrium, the
flow is chosen to be conical or of constant thickness [236], the value of
$n_\rho$ would be $3/2$ and $1/2$ respectively and $h$ would be independent
of $\psi$.

A simple solution is obtained by choosing  the `spirality'
$B$ to be given by $B\!=\!tan\, \theta$, where $\theta$
is the constant winding angle (angle between the radial direction and the 
outward tangent of $\psi\!=\!$ constant curve). With this choice,
flow variables take the forms:
$$
u=x^{-1/2}q_1(\psi) ,
\eqno{(5.30a)}
$$
$$
v=x^{-1/2}q_2(\psi) ,
\eqno{(5.30b)}
$$
$$
a=x^{-1/2}{q_3}^{1/2}(\psi) ,
\eqno{(5.30c)}
$$
$$
\rho=x^{-3/2}q_{\rho}(\psi) ,
\eqno{(5.30d)}
$$
$$
p=x^{-5/2}q_p(\psi) ,
\eqno{(5.30e)}
$$
and
$$
{\frac{d \beta}{d x}}=x^{-1}B .
\eqno{(5.30f)}
$$
The adiabatic constant $K$, which is a measure of entropy, remain constant on
a flow line in between two shocks, but changes at the shock itself. 
The variation of $K$ with the radial distance
can be calculated using the definition of the sound speed and is given by,
$$
K(x)=x^{3\gamma/2-5/2}K_o (\psi),
\eqno{(5.30g)}
$$
where $K_o=q_p/q_\rho^\gamma$ is constant on a given spiral coordinate $\psi$.
Since entropy must increase inwards in accretion and outwards in winds, one
must have
$$
\gamma \leq 5/3 , \ \ \ \ \ {\rm or \ equivalently },\  n > 3/2,
\eqno{(5.31a)}
$$
for accretion, and,
$$
\gamma \geq 5/3 , \ \ \ \ \ {\rm or \  equivalently },\  n < 3/2,
\eqno{(5.31b)}
$$
for winds. 

The location of the sonic surface could be studied following procedures
as described in earlier Sections. The  nature of sonic points
depends strongly upon the conserved quantities
of the flow. The usual practice is to supply two parameters,
$B$-- measuring the spirality of the shock surface and $q_{2c}$-- 
the azimuthal velocity coefficient on the sonic surface. Other
quantities, such as $q_1$, $q_3$, $\gamma$ are determined self-consistently.
One also assumes that when shocks are present they are equidistant in 
the angular scale $\psi$. Sonic surfaces which lie in between the shocks 
are also equidistant.
Thus, if there are $n_s$ number of shocks, the angular distance in between
two successive shocks as well as sonic surfaces is given by $\delta\psi\!=\!
{2\pi}/{n_s}$. Figure 5.10 shows schematically an accretion disk which 
includes two shocks at $\psi\!=\!\psi_{s1}$ and $\psi\!=\!\psi_{s2}$ 
and two sonic surfaces at $\psi\!=\! \psi_{c1}$ and $\psi\!=\!\psi_{c2}$. 
One can write down the location of a shock formed just prior to a sonic 
surface as $\psi_{s1}\!=\! \psi_{c1}-\epsilon \delta\psi$,
and the location of a shock formed just after the sonic surface as
$\psi_{s2}\!=\!\psi_{c1}+(1-\epsilon)\delta \psi$, where 
$0 \leq \epsilon \leq 1$ is
to be determined for the given conserved quantities of the flow. 

Some of the important properties of the self-similar shock solutions are:
whereas the velocity components individually depend upon the radial distance,
Mach numbers of pre-shock and post-shock flows as well as
the strength of the shock do not. Since the disk is self-similar,
the height of the disk is directly proportional to the distance 
on the surfaces of constant $\psi$. The height varies in the azimuthal
direction as $\psi$ changes. Angular momentum per unit mass $\lambda$ 
dissipated at the shock is given by,
$$
\frac{\lambda_+}{\lambda_-}=\frac{q_{2+}}{q_{2-}} .
\eqno{(5.32)}
$$

Recently, this self-similar procedure to solve vertically averaged
flow has been adopted to solve axisymmetric flows as well [229].
This self-similar vertically averaged flow is found to be very
useful to study quasi-spherical flows [229].

\noindent {\large 5.7.3 Examples of Spiral Shocks in Accretion Disks}

To obtain a solution which includes spiral shock waves, one supplies
the number of spiral shocks $n_s$, the spiral angle 
parameter $B=tan\, \theta$, and the azimuthal velocity coefficient
$q_{2c}$. The polytropic index $n$, the shock location $\epsilon$
and the entire velocity field, namely, $q_i( \psi )$ becomes known.

Figures 5.11(a-b) show an example of solution with 
a two-armed spiral shock.
In Fig. 5.11(a), a 2-shock  solution in accretion is shown with
dashed vertical lines ($\psi_{s1}$ and $\psi_{s2}$) [77]. 
Crosses represent locations of the sonic surfaces 
($\psi_{c1}$ and $\psi_{c2}$). Figure 5.11(b) shows the variation
of velocity components in between two shocks. The
input parameters are: $B\!=\!-1.0, \ q_{2c}\!=\! 0.02 $, and the
results are $n\!=\!6.1842,\ \epsilon\!=\!0.7274, \ M_-\!=\!1.0394, \ 
M_+\!=\!0.8909$. The strength of the shock ${M_-}/{M_+}$ is $1.17$.
The shock is therefore reasonably weak.

One intriguing conclusion in the study of non-axisymmetric shocks
is that there are two distinctly separate regions in the parameter space,
one marked by a large rotational velocity and a small radial velocity
($v_r<<v_\phi$) and the other by a large radial velocity and a
small rotational velocity ($v_r>>v_\phi$), for which shock solutions
exist in accretions and winds [77, 248]. These are the `high'
and the `low' states respectively. Furthermore, it is observed
that in the `high' state, shocks are substantially weak for small $B$,
and very strong for large $B$. The existence of these bifurcations in states
may be significant in the astrophysical context. The accretion
rate is found to be lower in the `high' state and higher in the `low' state.
Many of the variabilities observed in the disk spectra could be due to
such switching. In numerical simulations of accretion disks,
(\S 6), spiral shocks are found  to be unstable under various
circumstances and it is possible that these are the manifestations
of the instabilities discussed in this Section.

\newpage
\noindent {\large\bf 6  Numerical Simulation of Accretion Processes}

\noindent{\Large 6.1 Introduction}

In the last few sections we have concentrated on the theoretical models of
the accretion flows. Though analytical models are always welcome, they
basically describe the stationary properties of an accretion disk.
There are increasing evidence that most of the astrophysical systems which
are believed to harbour accretion disks and black holes
are also variable in some time scales or the other. Occasionally,
the same system shows variabilities in completely different time scales
(an example being the X-Ray flickering and optical micro-variabilities 
in Blazers, see \S 7). Sometimes, analytical stability analysis helps to
sort out which of the models might be applicable, but such a procedure
is often very difficult and scopes are limited.
Theoretical models investigate various types of different
detailed {\it physical} processes (e.g., radiative transfer) in an
accretion disk (which is why there are so many models
in the literature), but fail to study detailed dynamical behaviour.
The numerical simulations, on the other hand,
have been very successful in obtaining the {\it dynamical} behaviour
(e.g., presence of various instabilities, shock waves, etc.), though
many of the important physical ingredients are missing (which is why
there are only a few disk simulation results available).
In fact, the goal of these two directions so far have been complementary
and it is possible that neither of them can ever be totally independent.
Though it might seem obvious that the results from numerical models should
correspond more closely to the observations and therefore they must play a
major role, so far, the aim and scope of this approach
have been found to be limited as well. 
In future, one has to include more physical
processes in the code so that both the physics and the dynamics
emerge from the same simulation.
In certain situations, the exact nature of physical processes
(such as, viscous process which is essential for transporting 
angular momentum) may be illusive and the theoretical input
to numerical works has been very little. This has resulted in
the fact that many of the disk simulations
have no `physical' dissipation. They are  energy and angular momentum
conserving, and show violent behaviour such as strong shock waves,
contrary to some of the more popular theoretical disk
models which do not have these discontinuities.  

Some of the theoretical models of accretion disks
which we discussed in the past have been put to rigorous tests of
numerical simulation. Below, we present some important developments 
in this direction in recent years.

\noindent {\Large 6.2 Simulation of Bondi-Hoyle-Lyttleton Accretion}

In \S 2, we discussed steady state
accretion on a compact star immersed inside an infinite medium.
In a variety of astrophysical situations, such as (a) 
in massive X-ray binaries containing neutron stars and black holes
which have wind producing OB companion, (b) in symbiotic binaries where
white dwarfs accrete from the winds of cool giants, etc. 
such accretion processes have been found to take place. Since
these are typically time dependent situations, 
a number of recent works carry out numerical simulations 
of these systems. From these results we have learned
the dependence of mass and angular momentum
accretion rates on various physical
parameters [249-257]. We present here a summary of our present
understanding of the numerical results.

In case of the flow with a finite density and velocity gradients,
Matsuda et al. [250] and Taam \& Fryxall [251] show that a steady state
is not achieved. Blondin et al. [252] performed
a two dimensional simulation of gas flow in the orbital
plane of a massive X-ray binary system. The mass accretion
is primarily fueled by the radiation driven wind from an
early-type companion. In these simulations a new
phenomenon was discovered. It was noticed that instead of a disk-like 
accretion, the flow makes an unstable channel to accrete onto the compact
object. The channel itself switches sides or `flip-flops'
episodically. In between two states, a transient disk structure is formed
which is found to change its sense of rotation. These simulations were 
further repeated with higher resolutions with adiabatic or nearly
isothermal [254-255] and 
isothermal gas [255] and the same unstable behaviour was verified. In these
mostly two dimensional simulations, matter accretion rate is found to be
similar to the Bondi rate (eq. 2.6), but the angular momentum transfer
rate is only about thirty percent of the angular momentum deposited
on the accretion cylinder $\sim 0.3 GM/v_\infty$.

Fig. 6.1 presents a typical simulation result of the two dimensional
wind accretion. Here the medium at infinity is chosen to be homogeneous.
The density contours at times (a) 28.35, (b) 30.00, (c) 35.10 and 
(d) 37.50 are shown. Here the distances and times are measured
in units of $R_{A}$ and $R_A/a_\infty$, respectively. The sound crossing
time through the whole grid is $8$ in this unit. The flow is 
chosen nearly isothermal: $\gamma=1.005$. The flip-flop occurs
in a time scale of the sound crossing time of the flow close to the
compact object, and possibly directly proportional to the 
sound crossing time of the central object itself, as a large
object is found to develop instability only very slowly.

The growth of this instability in two-dimensional accretion
always follow this sequence [254]: radial modes are first excited which 
generates `pumping' type accretion where sometimes
more matter is dumped than the compact object can accrete. This
leads to the formation of a dome-shaped shock on the upstream
side of the object. The entire shock cone moves entirely on one
side or the other while forming small accretion disks which 
switches its sense of rotation. Eventually, however, the shock cone
settles down to the symmetric position and the accretion is less violent.

In the more realistic case of simulations in a three-dimensional flow,
the flip-flop instability was not found to be very strong, or 
perhaps negligible [254, 257]. The angular momentum transfer is also
found to be smaller than what is observed in the two dimensional simulations.
This behaviour has an important bearing on the study of the quasi-spherical
accretion flows with some angular momentum as well as to construct a 
generalized accretion disk model around black hole as we shall see below.

\noindent {\Large 6.3 Simulation of Inviscid Flows}

The first significant attempt to study the behaviour of 
matter around black holes was made by Wilson [258]. 
An Eulerian, fully general relativistic, first-order backward space difference
technique was used. The spatial resolution was low and the system was 
evolved till $\sim 100 GM/c^3$. 
It was shown that the large angular momentum accretion is 
accompanied by shock waves which travel outwards. This code was later improved
upon, with the number of grid points as well as the evolution time higher by
several orders of magnitude. A series of very important simulations were made 
with this code to show that thick accretion disks can indeed form in inviscid 
flows [215-216]. These simulations also confirm the results of Wilson [258]
that non-steady shock waves are formed which travel outwards.
From the post-shock flow, a very strong wind is generated which is hollow in
nature and `hugs' the funnel wall. Due to the inviscid nature of the simulation,
centrifugal force kept the flow away from the axis of symmetry. In Fig. 6.2
we reproduce a typical simulation result of Hawley, Smarr \& Wilson [216].
The flow is inviscid and the angular momentum of the flow at the 
outer boundary is chosen to be $l=3.77GM/c$. The contours of constant 
density and the velocity vectors are shown. The figure is drawn at 
$t=300GM/c^3$. Several features of the 
thick accretion disk, such as the accretion through the cusp, the formation 
of the funnel and the outflow are very clearly recognizable.
A large number of simulations of the disk were subsequently
carried out [202-205] to check if the thick accretion disks were 
unstable due to the so-called `Papaloizou-Pringle' instability 
[199-201] discussed earlier (\S 4.8).

These simulation results are understood clearly
in the light of what is discussed in \S 5 about the
shock formation in accretion flows.
In above simulations, no standing shock waves were found because
the parameter space (spanned by energy and angular momentum)
for which standing shocks might form were not used at the
boundary.
Numerical study of standing shock waves are obtained in a series of very recent
simulations [220, 225, 259]. These authors use Smoothed 
Particle Hydrodynamics (SPH) code written in axisymmetric co-ordinate system
using Paczy\'nski-Wiita [75] potential. The shocks are found to form {\it 
exactly} where they are predicted theoretically, particularly in a
one-dimensional, axisymmetric, thin disk.
At the beginning of the simulation, a transient behaviour is experienced
by the flow, after which the flow 
assumes only the supersonic branch of the analytical solution. After a 
small perturbation is introduced at the outer edge, a shock appeared
close to the inner edge of the disk. It then traveled outwards
until it reached the analytical solution where it 
settled down. Thus, it was clear that in the presence of
perturbations (which are normal for a realistic flow), the flow
chooses the solution which include shock waves (higher entropy
solution). In the case where the
outer boundary condition does not allow shock solution at all, the perturbation
is damped out and only the shock-free transonic solution is chosen.

Figure 6.3 shows an example of the simulation of a thin accretion disk
which includes standing shock waves [220].
Mach number of the flow is plotted against the radial distance (in units
of the Schwarzschild radius of the central black hole).
Solid curves are simulation results and the dashed curves are the
supersonic and subsonic branches respectively. Two vertical dashed
lines indicate locations of the analytically predicted shock
transitions, the outer one being stable. After a transient phase, a shock
forms close to the inner edge which then travels outward till it reaches
the outer stable shock. The specific energy and the specific 
angular momentum were chosen to be ${\cal E}=0.011c^2$ and $\lambda
=3.8 GM/c$ respectively. 

In the thick 2-D inflow simulations, shocks are found to form  without
explicit introduction of the perturbation, probably because in 2-D
systems turbulence generated act as a perturbation [259]. As the
simulation starts, the low angular momentum advective flow
is almost freely falling
till it hits the centrifugal barrier. Here,  a part of the flow bounces back
and the shock wave formed travels outward and settles down at a location 
slightly outside the location predicted by a one-and-a-half dimensional model 
(\S 5.2). This deviation is probably due to the turbulent pressure in the flow
just behind the shock and/or inaccuracy of the 1.5D model which ignores 
the vertical component of
velocity altogether. These simulations also find strong wind formation
(as in the finite element methods discussed above), and the wind is found to 
become supersonic within a finite distance. These simulations verify all the 
assertions made in the theoretical work, such as the stability of the shock 
wave at $X_{s3}$ and the fact that the flow does not choose a shock free
solution if it has another solution which includes a shock.  
It also verifies that the flow with a positive energy and higher entropy
would produce strong winds which may contribute  to jet formation a
conclusion which was also verified by subsequent self-similar models [229].
The flow passed through both the sonic points and does not
switch back-and-forth from one sonic point to the other contrary to
what was suggested earlier [260].

Figures 6.4-6.6 show results of a 2-Dimensional simulation after the
steady state is reached. The time is  $T=700GM/c^3$ [259]. 
Here, $60,000$ particles are used in each quadrant. Figure 6.4 shows the 
particle distribution in which the standing shock at $X\sim 16$ is clearly 
visible. Note the presence of the oblique shock also. The specific energy and 
angular momentum used were $0.006 c^2$ and $3.3GM/c$ respectively.
Figure 6.5 shows the contours of constant Mach number. Note that the flow
which is subsonic after the shock, becomes supersonic very close to the 
black hole. Also,  winds which are originated subsonically on the disk 
surface, eventually become supersonic. Figure 6.6 shows the contours of
constant temperature (labeled in geometric units). At the post-shock region,
the temperature of the flow becomes high and the velocity becomes
very low satisfying all the assumptions of a thick accretion disk (Fig. 
4.4a-b). The contours in the immediate vicinity of the post-shock region 
indeed resemble that of a thick accretion disk. The shock heating causes
a strong pressure gradient term which pushes matter out of the disk to form a 
wind. These solutions, particularly 1-Dimensional work which is understood
more completely, can be used as `test problems' in future for any new 
numerical code 
which are meant to study shock waves in astrophysical conditions. These
tests are expected to be more rigorous than 1-Dimensional shock-tube
problems in Cartesian grid. 

An important set of simulations, which pertains to the formation
of the non-axisymmetric shock waves, are carried out
mainly by Kyoto group and others [142, 234-237, 261-262].
In this case, the spiral shock waves are induced by the tidal force of 
the companion. Non-axisymmetric modes amplify and steepen to produce spiral
shocks inside the disk. It was shown that spiral shocks may remove 
angular momentum quite efficiently and it is possible to achieve an 
effective $\alpha \sim 10^{-2}$. Figures 6.7(a-b) show examples of the spiral 
shock waves. Density contours are plotted. In both the figures
the ratio ($Q$) of the compact object to the companion is chosen to be unity.
In Fig. 6.7a, $\gamma=1.2$ and in Fig. 6.7b, $\gamma=5/3$ were used.
Matter is injected through the inner Lagrange point on the left (cusp
of the smooth curve) with sound speeds $c_o=0.15$ (in 6.7a) and
$c_o=0.05$ (in 6.7b) respectively. (Velocities are measured in units
of the Keplerian velocity of the companion and the distances are
measured in units of the binary separation.) The tidal perturbation
of the companion produce density waves which 
steepen into spiral shocks. In presence of a companion as in a binary 
system, usually a steady shock pattern forms. In Fig. 6.7a,
a two armed spiral is formed whereas in Fig. 6.7b, a three-armed spiral
is formed. However, depending upon boundary conditions (the dependence
is as yet poorly understood), the number of arms may change in the same 
simulation. Figures 6.8(a-c) show 3-D surfaces of density plots for $Q=1$ and
$\gamma=1.2$ case at times $T=2.3$, $3.2$ and $3.7$ respectively.
This case is similar to above but the injected gas is somewhat
cooler ($c_o=0.1$). Also matter is supplied all over the outer boundary
(at $r=0.65$) with the angular momentum gradually increased
from zero to $0.5$ ($0.8$ would correspond to Keplerian in this unit).
The number of arms changed from $2$ (6.8a) to $3$ (6.8b) to $2$ (6.8c).
The simulations are done with a code obtained from Kyoto group.
When the companion is present only for a short time (e.g., 
just passing by), multi-armed fragmented shocks may form which may
cause variabilities in active galaxies [237]. 
When the companion is more massive than the primary compact object, 
as is believed to be the case in SS433 system, the perturbation is more 
violent and causes variability in very short time scales [262]. Similarly,
the results with $\gamma \rightarrow 5/3$ tends to be unstable 
and consists of fragmented and multi-armed as well. This is
probably because the entropy function $K(x)$ becomes more flat (eq. 5.30g).
Theoretical results on self-similar spiral patterns have been discussed 
in \S 5.7. 

\noindent {\Large 6.4 Simulation of Viscous Disks}

It has been realized rather recently, both from the theoretical and numerical
simulation results that the process of shock formation in accretion disks 
depends strongly upon the viscosity [224] 
as well as the viscosity prescriptions [225].
As discussed in \S 5.3 (see, Fig. 5.3a-d),
when the viscosity is increased, shock waves slowly recede from the 
black hole and the strength of the shock gradually diminishes. 
Finally, for a large viscosity, the shock disappears and the disk
becomes standard Keplerian type. Below, we present the results of the
numerical simulation of viscous isothermal disks as their
properties are understood reasonably well. The results with
general radiative transfer are expected to be similar as suggested
by theoretical considerations (\S 5).

A serious problem with the Shakura-Sunyaev viscosity prescription 
is that the efficiency of transport of angular momentum becomes
very high in the post-shock region of the disk compared with the
pre-shock region. To see this, consider
a thin, isothermal, axisymmetric, accretion flow onto a black hole.\\
\noindent The radial momentum equation is (cf. eq. 5.1a):
$$
\frac{\partial v_r}{\partial t} + v_r \frac{\partial v_r}{\partial r}
+ \frac{1}{\Sigma}\frac{\partial W}{\partial r}-\frac{\lambda^2}{r^3} +
\frac{\partial \Phi}{\partial r}=0 .
\eqno{(6.1a)}
$$
\noindent The continuity equation is given by (cf. eq. 5.1b),
$$
\frac{\partial\Sigma}{\partial t}
+\frac{1}{r}\frac{\partial}{\partial r} (\Sigma r v_r)=0 .
\eqno{(6.1b)}
$$
\noindent The azimuthal momentum equation is given by (cf. eq. 5.1c),
$$
\frac{\partial \lambda}{\partial t} + v_r \frac {\partial \lambda}{ 
\partial r } = \frac{1}{r \Sigma} \frac{\partial}{\partial r} (r^2 W_{r \phi})
\eqno{(6.1c)}
$$
\noindent Here, $v_r$ and $\lambda$ are the radial velocity and the azimuthal
angular momentum respectively, and $W$, $\Sigma$ and $W_{r \phi}$
are the pressure, density and the $r\phi$ component of the viscous stress 
tensor respectively. $\Phi(r,\theta)$ is the gravitational potential of the
central object. This may be any of the pseudo-Newtonian potentials
discussed in \S 1.3.

If one considers the solution to be steady, the time derivatives of the
above equations vanish and we get the following conservation equations,

\noindent (a) Conservation of energy:
$$
{\cal E}= \frac{1}{2} v_r^2 + a_0^2 log(\Sigma) + \frac{1}{2} \frac
{\lambda^2}{r^2} + \Phi ,
\eqno{(6.2a)}
$$
\noindent (b) Conservation of baryons,
$$
{\dot M}= \Sigma v_r r
\eqno{(6.2b)}
$$
and,\\
\noindent (c) Conservation of the angular momentum
$$
{\dot M} (\lambda-\lambda_{e}) = -r^2 W_{r \phi}
\eqno{(6.2c)}
$$
Here, we have used equation of state $W= const. \ \Sigma$
appropriate for isothermal flows and $a_0=(W /\Sigma)^{1/2} $ is the 
isothermal sound speed, $\lambda_e$ is the angular momentum at the
inner edge of the disk. Note that this distribution suggests sub-Keplerian
disks will become Keplerian at larger distance.

In an inviscid, axisymmetric flow,
$$
W_{r \phi}= 0 ,
\eqno{(6.3)}
$$
and the location and strength of the shock is determined by the shock 
conditions:

\noindent (a) Temperature of the flow is constant across the shock,
$$
a_{0-}=a_{0+} ,
\eqno{(6.4a)}
$$
\noindent (b) Baryon flux is conserved,
$$
{\dot M}_- = {\dot M}_+ ,
\eqno{(6.4b)}
$$
\noindent (c) The total pressure is balanced,
$$
W_-+\Sigma_- v_{r-}^2 = W_++\Sigma_+ v_{r+}^2 ,
\eqno{(6.4c)}
$$
and,\\
\noindent (d) Angular momentum  flux is conserved,
$$
\lambda_- = \lambda_+ ,
\eqno{(6.4d)}
$$
Here, $-$ and $+$ signs represent quantities in the pre-shock and
post-shock flows respectively.

In the presence of viscosity, the angular momentum is transported according to 
eq. 6.2c. In the following subsections, we discuss how the transport process
depends on the viscosity prescriptions.

Let us first consider the effects of Shakura-Sunyaev prescription of
$\alpha$ (denoted hereafter as $\alpha_s$) viscosity. In this case, we have 
$W_{r\phi} = \alpha_{s} W \frac {r}{\Omega_K}\frac{d\Omega}{dr}$,
and the angular momentum distribution is obtained from [77, 224],
$$
\lambda - \lambda_e = \frac{\alpha_{s} r^3 W }{\Omega_K \dot M}\frac{d\Omega}
{dr} = \frac {\alpha_{s} r a_0}{M\Omega_K}\frac{d\Omega}{dr} ,
\eqno{(6.5)}
$$
where, $M=v_r/a_0$ is the Mach number of the flow, and $\lambda_e$ is the
angular momentum at the inner edge of the disk and $\Omega_K$ is the local 
Keplerian angular velocity. At the shock, since 
the Mach number $M$ varies from supersonic 
($M_- >1$) to subsonic ($M_+ <1$), the rate of transport of angular
momentum would be quite different on both sides of the shock.
In particular, in the post-shock region, the angular momentum
would be transported faster and would `pile' up as the 
pre-shock flow is unable to transport it away to outer radius very
efficiently. Hence, the angular momentum must be
discontinuous and $\lambda_+ $ must be higher compared to $\lambda_-$. 
Thus the shock formed would be a mixture of compressional type as well
as shear type (`mixed' shock).

When the viscosity is very small, transport rates of angular
momentum on both sides of the shock can match and the shock can remain steady.
In this case the shock is weaker and forms farther away from the
black hole. When the viscosity parameter
is increased, the shock wave is driven outwards and the disk
in the post-shock flow becomes Keplerian [225]. For very high viscosity,
the shock disappears and the disk becomes almost Keplerian except very
close to the inner edge of the disk.
In Figs. 6.9(a-b), we compare Mach number and angular momentum variations
in viscous (solid) and inviscid (dashed) flows. The viscosity parameter is
chosen to be $\alpha_s=0.01$ everywhere in the simulation.
Note that the shock in the viscous disk forms farther out. It is weaker
and more spreaded out. Figure 6.9b shows that a mixed shock is formed
with a jump in angular momentum at the shock ($\lambda_+ > \lambda_-$).
The angular momentum is super-Keplerian in the post-shock flow
and almost constant in the pre-shock flow.
As the $\alpha_s$ parameter is increased, the flow behaviour changes
dramatically. Figures 6.10(a-b) show the evolution of 
Mach number and angular momentum when $\alpha_s=0.1$ is chosen.
Dashed curves marked $\alpha=0$ represent solutions for inviscid flows.
Successive solid curves
are drawn at intervals of $\Delta t=500$. The in-fall time-scale is $t_{i}=75$
in the same unit. What clearly happens in this case is that 
due to higher efficiency of the transport of angular momentum in the subsonic
post-shock region, angular momentum is piled up in this region.
This drives the shock outwards continuously and no steady shock solution 
at a finite distance is possible. 

To understand this behaviour, we refer to the Fig. 5.3(a-d) of \S 5.3
where the topological properties of the phase space in viscous flows
have been discussed. It was shown that
when the viscosity is high enough, a flow at a large distance
has two choices. In one (higher energy) solution, the flow passes through the
outer sonic point and in the other (lower energy) solution, 
the flow passes through the inner sonic point. The latter behaviour 
is clearly chosen by the system. 

Results presented above indicate the following behaviour which are the 
general characteristics of any accretion flow (\S 5.4), not 
necessarily isothermal: (a) if the flow is provided with two choices, one
shock free (low entropy or high energy) and the other with shocks (high 
entropy and low energy), the flow chooses the solution which includes shocks.
(b) If the flow is highly viscous, it is again provided with two choices.
Both are shock free, but one (with high energy and low entropy) passing 
through the outer sonic point and the other (with low energy and high entropy)
passing through the inner sonic point, the flow chooses the latter solution.
Thus generally speaking, (a) A black hole solution must
be transonic, and any  solution which 
does not satisfy this criterion must be rejected, 
(b) if the flow has a choice to pass through the
inner sonic point without violating any physical principle, whether or not
the flow has any shock wave, it would like to choose that solution.

It is possible that when the transport of angular momentum is not
limited by the sound velocity  so that the transport
becomes efficient in the supersonic region of the flow, standing
shock waves may be allowed in disks with larger $\alpha$ parameter as well. 
This can be achieved through the transport of angular momentum via
Alfv\'en waves (in presence of magnetic fields)
or through radiative transport (which is specially
possible in the hot, radiation dominated post-shock region).
In the presence of efficient transport processes on either sides
of the shock, the angular momentum would be continuous. Continuity of
$\lambda$ implies continuity of the viscous stress (cf. eq. 6.2c) and one
notes that at the shock,
$$
W_{r \phi - } = W_{r \phi +} .
\eqno{(6.6)}
$$
In the usual form:
$$
W_{r \phi} = \nu \Sigma r \frac{\partial \Omega}{\partial r} .
\eqno{(6.7)}
$$
Here, $\nu$ is the kinematic viscosity coefficient, and $\Omega =\lambda/r^2$ 
is the local angular velocity. Since $\lambda$ is assumed to be
smooth and continuous, $\Omega$ is also a smooth and continuous 
function. One way to achieve the continuity of $\lambda$ across the shock is
to define the kinematic coefficient as,
$$
\nu_p = \frac{\alpha_p (a_0^2 + v_r^2)}{\Omega_K}
\eqno{(6.8a)}
$$
so that the viscous stress is,
$$
W_{r \phi} =\frac {\alpha_p (a_0^2 + v_r^2) \Sigma r }{\Omega_k}
\frac{\partial \Omega}{\partial r}.
\eqno{(6.8b)}
$$
In this case, the pressure balance condition (eq. 6.4c)
ensures the continuity of $W_{r\phi}$. A subscript $p$ of $\alpha$
is used to denote the pressure balance.
One could also use the balance of the mass flux in defining the 
kinematic viscosity,
$$
\nu_m= \frac{\alpha_m v_r}{\Omega_K}
\eqno{(6.8c)}
$$
so that the viscous stress becomes,
$$
W_{r \phi} =\frac{\alpha_m |v_r| \Sigma  r }{\Omega_K}
\frac{\partial \Omega}{\partial r}
\eqno{(6.8d)}
$$
which also ensures continuity of $W_{r\phi}$. A subscript $m$ of
$\alpha$ is used to denote the mass flux balance.

Simulations using this prescription shows continuity in
angular momentum even across a shock [Fig. 5.7, 68, 225]. This indicates
that in the presence of efficient transport of angular momentum in the
supersonic branch, shocks would exist when viscosity is relatively high.
This is important since the observations of X-rays and $\gamma$-rays 
from black hole candidates require hot regions close to the black hole
which is self-consistently provided by the post-shock flow.

In the case when the angular momentum supplied at the outer edge is
Keplerian, and the viscosity is reasonably high, the flow becomes
Keplerian throughout, except close to the black hole. At around the
marginally stable orbit, the flow can become supersonic also.
Below marginally stable radius the angular momentum transport
becomes relatively inefficient and the
flow accretes onto the black hole with almost constant angular momentum
[76, 225].
Figure 6.11 shows the formation of a Keplerian disk [225] in this way.
In this simulation, matter is injected at the outer edge of the disk
at $r_{o}=100$ with angular momentum $7.00$ which is close to 
Keplerian value at $r_o$. The (constant) sound speed
is chosen to be $a_0=0.005$ and the injection velocity was $v_0=0.003$.
$\alpha_s=0.25$ was used in this simulation. 

\noindent{\Large 6.5 Simulations of Flows using Radiative Transfer}

An important piece of work which includes $\alpha_s$ viscosity as well as 
the radiative transfer in a sub-Eddington {\it Keplerian} disk shows a 
completely different behaviour [263] 
than what is shown above. Evolution of the inner part of a Keplerian disk 
around a $3M_\odot$ black hole shows that the disk
collapses into a thin sheet on the equatorial plane. This is expected since
radiation pressure supported $\alpha_s$ disks are known to be unstable in
viscous and thermal time scales (\S 3.1.4). The collapsing behaviour
propagates till the outer boundary. Figure 6.12 shows the density contour
of the simulation at different times ($t$ marked in each panel). Here the 
length and the time units are $GM/c^2$ and $GM/c^3$ respectively. Initially,
the disk is radiation pressure supported but eventually it becomes dense,
cool and fluid pressure supported. The accretion rate is reduced as well.
An interesting scenario can emerge in this type of disks: suppose that 
initially viscosity is low. Material which is injected at the outer edge
cannot work their way to the black hole efficiently, and therefore
accumulate. The accumulated material is not able to radiate the internal energy
and becomes hot. This hotter disk is of higher pressure and of higher 
viscosity which causes the accumulated material to
accrete catastrophically and emitting very high radiation. 
The disk can therefore resemble a rapid burster.

In another important simulation [264], super-Eddington flow is studied 
which produces 
a thick accretion disk. Figure 6.13 shows the density contours and unit
velocity vectors in a simulation with ${\dot M}= 4 {\dot M}_{Edd}$ at
time $t\sim 6000$. Contours are labeled with densities in
CGS unit. Large convective cells form on either side of the 
equatorial plane which drives the flow along the axis, as well as along the 
equatorial plane. The physical origin of the latter 
type of outflows is not clear. Four regions are distinguished and marked 
on the Figure: (A) a convective core, (B) an accretion zone, 
(C) a photocone and (D) a jet.

A number of simulations which include bremsstrahlung cooling have recently
been carried out [227] using Smoothed Particle Hydrodynamics (SPH).
The location of the shock wave is
found to oscillate in a time period comparable to the 
in-fall time scale from the shock wave as discussed in \S 5.4. 
Fig. 6.14(a-b) show results of three-dimensional (axisymmetric) simulations
with bremsstrahlung cooling process. The two figures show two different
phases of oscillation of the shock and the post-shock corona. The upper halves
provide the velocity vectors and the contours of constant Mach numbers
($M=1$ are thick curves) while the lower half provides the contours of
constant density. The `breathing' of the corona appears to take place when the 
cooling time-scales roughly agree with either of the dynamical/in-fall
time scales and are possibly responsible for quasi-periodic behaviours
seen in galactic and extra-galactic black hole candidate spectra [227].
The post-shock halo may be responsible for the production of
X-rays and $\gamma$-rays seen in spectra of galactic and extragalactic 
candidates.

The agreement of above results with analytical works provide strong
support for the SPH simulations. One
possible uncertainty could be due to the deviation of the geometry as described
by a pseudo-Newtonian potential from a full general relativistic description. 
The theoretical problem in hand using a full general relativistic framework 
is not solved yet. A few preliminary tests (of spherical accretion, for
instance) have been carried out recently using a relativistic SPH code [265].
For a concrete understanding of the above simulations, it is essential that
the above mentioned studies be carried out using a fully 
general relativistic code.

\newpage
\noindent{\Large\bf 7 Observational Signatures of Black Hole Accretion }

\noindent {\Large 7.1 Introduction}

In most of the models of active galaxies, black holes are
considered to be essential. A black hole is believed to 
be an indirect (but most relevant) participant and the catalyst
of a large number of activities which take place close to the
galactic nuclei. In cases where galaxies are no longer active,
such as our own galaxy, the observers still look for signatures
for activities which may suggest the presence of black holes.
Though the interpretations are made assuming the presence of black holes,
in general, they do not constitute full proofs. In an alternate model, most
of the activities are supposed to be taking place due to star bursts at the
galactic center [266]. This model can also explain many 
of the observations. We shall comment on this as we proceed.

Assuming the efficiency $\eta \sim 0.06$ to $0.40$ for the conversion
of the accreted mass into radiated energy, the critical mass accretion 
rate could be several times the Eddington rate
${\dot M}_{crit}= \frac{ {\dot M}_{Edd}}{\eta} \sim 
2 M_\odot M_9 /\eta$ yr$^{-1}$,
where, $M_9$ is the mass of the black hole in units of $M=10^9M_\odot$.
Black hole being dense and compact, the length scale of the 
observable quantities,
such as the intensity of emission at different wave bands should be small and
typically of the order of the Schwarzschild radius $2GM/c^2$. Similarly, 
the intensity at a given band could vary in time scales of
light crossing time of the horizon $\sim 2GM/c^3$. The stability of the 
direction and the superluminal behaviour of the cosmic radio jets [64, 267]
are also signatures of a very powerful and stable compact object 
`sitting' at the center of the jet-producing galaxies. Also, in cases
where direct determination of radial velocities are possible,
large velocity variation within a small region would certainly be a
strong signature of a black hole nearby [4-5]. 

In the case of X-ray binaries, 
such velocity measurements are common and provide the mass function
of the binary system. A successful candidate of a black hole
must have $M> 3M_\odot$  [1, 268]. Hence a dynamical mass determination
usually is the strongest criterion to fix a stellar black hole candidate [269].
The strong candidates include Cyg X-1, LMC X-3 and A0620-00 and the
candidates with relatively weaker confirmation
include CAL 87 and LMC X-1. A recent inclusion in this list is
V404 Cygni [270] 
whose mass function $f(M)=6.26 M_\odot$ is very large and gives
the candidate a mass of about $6.5 M_\odot$ or more [271]. 

In last few Sections, we have discussed the theoretical nature of the
accretion processes on a black hole. If a black hole is indeed present,
and matter is available, then the accretion should 
take place, and the emission from the hot, radiating matter
should be observable. The radiation could be in continuum and/or in lines.
It could be stationary as well as non-stationary. If the source of matter is
a single companion such as that in a stellar binary system, the angular
momentum of the matter is expected to be high, but still the disk will comprise
of the Keplerian disk from the accretion through Lagrange point and 
a sub-Keplerian halo from the wind accretion, particularly in the case of
high mass X-ray binaries. In the case of low mass X-ray binaries sub-Keplerian
matter may be produced from Keplerian disk itself (Fig. 5.7).
In the case of accretion onto a black hole 
at galactic centers, the source of matter
could be winds from star clusters at distances of parsecs. These winds
are expected to collide and their angular momentum  may be mostly `canceled' 
in the process, before they can proceed to the galactic center. This way one
expects the angular momentum to be much smaller, which may further be
removed by means of small viscosity or spiral density waves. The
supplied matter could also be an admixture of Keplerian flows
along with some low angular momentum matter.
Since the present report is not a review on active galaxies (on which
there are several reports such as Begelman, Blandford \& Rees [64])
we shall not describe various aspects of these objects. 
Below, we shall review some models which require a compact black hole at the 
nucleus to explain the continuum spectra, line emission and the variabilities. 

\noindent {\Large 7.2 Continuum Emission}

A dominant feature of the continuum spectrum is a hump in the spectrum
near blue region known as the `big blue bump'. This emission feature is
seen in radio--weak sources as well as in many radio--loud quasars. 
In galactic black hole candidates, the spectra in the soft-state also 
has the bump is soft X-rays (see below).
The `big blue bump' in the spectrum of many QSOs as well
as the soft X-ray bump in galactic black hole candidates are usually attributed
to the quasi-thermal emission from accretion disks [1, 272-273],
and good fits to this part of the spectrum can be obtained from standard,
optically thick but geometrically thin `$\alpha$' disk models [274-275], 
thick accretion disks [193], 
and disks incorporating shock waves in the disk [211] among others.
Sun \& Malkan [274] fits 60 quasars and AGNs from IR to UV region of the
continuum by using standard accretion disk models [11-12] around black holes.
The relativistic
effects of disk inclination, including Doppler boosting, gravitational focusing
and gravitational red-shift on the observed spectra for both Kerr and
Schwarzschild black holes are considered. The apparent disk spectrum
becomes harder when the disk is viewed at high inclination angles due to
the Doppler boosting and the focusing effects. The relativistic corrections
are significantly higher when a rotating black hole is considered. Low 
red-shift Seyfert galaxies were found to require relatively low accretion 
rates, only a few percent of the Eddington rate, while the most luminous
quasars are found to be accreting close to the Eddington rate.
This model, which includes a power law emissivity in the infra-red region is
an improvement over the earlier works [276] which required super-Eddington
accretion rates.
The model of Sun \& Malkan [274] produces reasonably good fit of the spectrum. 
Figure 7.1, shows examples of fits in optical/UV  [277] of nine typical quasars
using emissions from a thin Keplerian disk
and contributions from the dust (modeled as a power law in infra-red region),
from starlight, from re-combination lines and Fe II lines [277]. 

In many cases, the standard model seems to underestimate
the production of UV intensity. In the presence of shock waves [211], 
UV flux produced in the post-shock region is higher compared to that produced
in a standard disk. Excess UV could also be produced
by cooling electrons at the shock by the soft photons from 
underlying disk. In above mentioned models simple black body spectrum
were used to compute the spectrum. 
If more sophisticated radiation transfer processes are used one develops a 
better understanding of the emergent spectra from the entire disk. We shall
discuss this below.
There are several models which do not include well defined disks.
For instance, it has been suggested that the blue bumps 
might be due to optically thin thermal emission of predominantly
free--free origin [278]. 
Both the thermal and non-thermal models are discussed in detail by
Odell, Scott \& Stein [279]. 
The thermal disk models give a flat visual-UV region up to a maximum
temperature of the disk. In the non-thermal models,
the synchrotron emission from electrons coming out of the 
proton-proton collisions are also included. This model predicts several 
observed properties of the non-blazer quasars and AGNs such as a relatively 
flat spectrum up to a high frequency break, synchrotron losses at least 
comparable to the Compton losses, weak polarization and no Lyman discontinuity.

One way to verify that the active galaxies accrete through
accretion disks is to measure as many line ratios as possible from the 
emission region [280]. These ratios should depend upon the nature of 
the incident spectrum and may be able to distinguish the thermal from 
non-thermal disk continuum emission.
Different subcritically accreting disk models could also be distinguished 
by the polarization properties of the emitted light. In a Thomson
scattering dominated region of the disk, the polarization is significantly 
higher than the classical limit of $11.7\%$ [281] 
when scattering half-thickness of the disk is less than unity
corresponding to subcritical accretion. The degree of polarization 
varies inversely with the optical depth in this regime and the
radiation pattern is anisotropic as the optical depth decreases [282].  

Some of the active galactic nuclei, especially those which are radio-quiet,
show strong evidence of X-ray emission. Based upon the analysis of the
Ginga data at least four (model dependent) components 
have been tentatively assumed to be present [283]:
(i) an incident non-thermal power law component with spectral
index $\alpha_X \simeq 0.9$; (ii) a broad emission excess at energies 
$\geq 10$ keV as a Compton reflected component from cold matter; 
(iii) fluorescent Fe-K emission line at $6.4$ keV 
and (iv) the absorption (Fe-K) edge at $7 - 8$ keV. ROSAT in its all sky
survey has identified numerous active galactic nuclei in the soft X-ray range
A consistent  model should satisfy the following features [284]: 
(a) the $2-10$ keV X-ray spectral 
index $\alpha$ is distributed in the range $0.4 \leq \alpha_X \leq 1.0$;
(b) the conversion efficiency of the mass to luminous energy is
about $10$ percent if the X-ray luminosities of typical quasars and
Seyferts are maintained over cosmological time; 
(c) the extension of hard X-rays of AGN into $\gamma$ regime must be
truncated at dimensionless photon energy $x= h \nu / m_e c^2 < 5-10$ in
order not to conflict with the observations of the $\gamma$ ray background.
A popular model of the origin of  X-rays is due to Zdziarski [285, 286].
In this model, electrons of the plasma cloud of energy $\Gamma m_ec^2$ 
scatter with soft photons and lose their energy
to these photons whose energy, in turn, is  boosted by a factor of
$\Gamma^2$ per scattering ($\Gamma$ being the Lorentz factor). The high
energy photons produce pairs depending upon the compactness of the 
interaction region. When the compactness is small, the optical
depth to pair production is small enough and X-rays become observable.
Thus, the observation of X-rays may testify to the existence of compact
hot source of energized electrons as well as soft photon source from the
underlying pre-shock cool accretion disks. The energetic electrons could also
be produced from magnetic corona of the accretion disks [141]. 
However, interpretation of the observational results need not be
unique, particularly the existence of the Compton reflection component
is found to be not required by more recent analysis of Ginga data [287]
In this new interpretation, the X-ray is considered to be of
thermal origin. Single component thermal models [287] 
(as opposed to multi-component Compton reflection models [285, 288-289])
have been successfully able to 
explain a variety of hard X-ray sources such as the galactic
black hole candidate Cyg X-1 and the Seyfert galaxy NGC 4151. Fig. 7.2a
show fits of the EXOSAT, GRANAT and OSSE observations of Cyg X-1 [290] with
various models [285, 287-289].  The basic conclusion is that the
electron temperature of the emitting region must be of the order
of a hundred keV ($\sim 10^9$K) and the optical depth is 
on the order of $\sim 1$. In Fig. 7.2b, the  OSSE [291] and
Ginga data [292] are fitted with the single component model [287].
Here, the electron temperature is about $44$keV ($\sim 4 \times 10^8$K).
These results indicate the presence of a very compact source for most
of the hard emissions.

In the spectra of the galactic black hole candidates, such as LMC X-3, LMC X-1,
GS1124-38 etc, the so-called soft component can be 
attributed to the thermal emission from the standard disk [1].
However, GS1124-68 [293] 
Cyg X-1 and GX339-4  [293-295] 
are occasionally observed to change states: they are sometimes in the hard
state (low spectral index $\alpha \sim 0.5-0.8$: $F_\nu \sim \nu^{-\alpha}$)
and sometimes in the soft state ($\alpha >1$). 
Judging from the success of the thermal models [287], it may be possible to
imagine that the one does not require non-thermal processes after all.
Since the hard component cannot be produced using a standard disk,
the only way to produce (without invoking ad hoc components, such as
corona, plasma clouds etc.) would be to produce a hot re-processing site,
such as a post-shock region, close to the black hole, self-consistently. 
To obtain a self-consistent
disk model one has to take the following points into considerations:

\noindent (a) Some objects are sometimes in the hard state and some 
times in the soft state (e.g., GS1124-68, Cyg X-1, GX339-4) [293-295]. \\ 
\noindent (b) The hard and soft components may vary almost independently (GS1124-68)
 [293]. 
\noindent (c) Sometimes the soft luminosity may vary by several orders of magnitude
while leaving the spectral index almost unchanged (GS2023+338, GX339-4, GS1124-68)
[293-295].
\noindent (d) Sometimes the hard flux is seen to rise with the spectral index 
(NGC 4151) [292]. 
\noindent (e) In general, in most of the black hole candidates, the soft
state has a very weak power low hard component of spectral slope 
$\alpha \sim 1.5$ apart from the usual black body bump.

Perhaps it is possible to understand these observational features by a single
model. Fig. 7.3 shows a plausible accretion disk model [68, 170, 296, 297]
constructed taking into account the  following considerations.
In a black hole binary system, the matter supply at the outer
edge of the accretion disk could be a mixture of Keplerian component
(with specific angular momentum distribution $l_{Kep}(r)$)
from the companion and a sub-Keplerian component (with
a specific angular momentum  $ l \lsim l_{Kep}(r_{out})$) from the
winds [170, 249-257, 296] of the companion or from the
Keplerian disk itself (Fig. 5.4). In active galaxies, sub-Keplerian matter 
supply could come from the constantly colliding stellar winds (which is
mostly devoid of angular momentum) at a distance of a few parsec away.
Higher viscosity near the equatorial plane creates a
Keplerian disk out of this (\S 6) flow but lower viscosity
away from the plane leaves the flow sub-Keplerian.
(Contrarywise, higher viscosity cooling dominated
flow deviates from Keplerian disk close to the black hole, but
low viscosity, radiatively inefficient flow deviates farther away.)
At a distance $r_s$ from the black hole, where $l\sim l_{Kep}(r_s)$, 
the effectively freely falling sub-Keplerian halo component
forms a standing shock due to the centrifugal barrier.
The post-shock halo, which acts as the boundary layer 
of the black hole, becomes hot as the radial kinetic energy of the in-fall is 
thermalized: $\rho v_r^2(r_s) \sim P \sim n_H \kappa T_p$.
The solution with a shock wave is preferred to the alternative
shock free solution as the entropy at the inner sonic point close to the
black hole is higher compared to the entropy of the shock free flow [77].
For matter with marginally stable angular momentum,
the shock forms at $r_s \sim 10r_g$ ($r_g={2GM/c^2}$,
the radius of the black hole) in Schwarzschild geometry.
The flow remains dissipative up to the marginally stable radius $r_{ms}=3r_g$
beyond which the flow rapidly passes through the inner sonic point [77].
For a flow around a Kerr black hole of angular momentum
parameter $a=0.99$, these length scales are roughly half as large.
For higher angular momenta shocks form much
farther away [77]. A flow in vertical equilibrium with a
polytropic index $\gamma=5/3$ has a post-shock temperature of
$T_p (r_s) \sim {1.6\times 10^{12}}/ {x_s}$, ($x=r/r_g$).
The post-shock flow becomes geometrically thick with a height
$h(x_s)\sim a_+(x_s) x_s^{3/2} \sim x_s/2 $ ($a_+ (x)$ being the
post-shock sound speed) which intercepts a few ($\lsim 5$) percent of the 
soft photons from the optically thick, Keplerian
component of the pre-shock accretion disk. The post-shock, optically
slim region with a Thomson optical depth of $\tau_h \sim
\int_{x_{ms}}^{x_s} 0.4 \rho dx $ ($\sim 1-3 $ for 
${\dot M} \sim {\dot M}_{Edd}$) rates
heats up these soft photons through inverse Compton scattering and 
the photons are re-emitted as hard radiations [68, 106, 108, 287, 297].

Let $L_{I}=f_{ds} L_{SS}$ denote the fraction of
the Kepleian (i.e., Shakura-Sunyaev [11] type) 
disk luminosity $L_{SS}$ of the soft radiation 
intercepted by the bulge of the shock [68, 297]
$F_e=L_{H}/L_I$ denote the enhancement factor of this flux due to cooling
of the electrons through inverse Comptonization [106, 108]
($L_{H}$ is the hard component luminosity),
and $f_{sd}$ ($\sim 0.25$ for a small spherical bulge) denote the
fraction of $L_{H}$ intercepted back by the disk.
The soft component observed from a disk around the black hole candidate is 
therefore contributed by the original disk radiation plus the absorbed
intercepted radiation: $L_{S}=L_{SS} + L_{SS} f_{ds} F_e f_{sd} {\cal B}$,
and $L_{H} \sim L_{SS} f_{ds} F_e$,
where ${\cal B}=1-{\cal A}$, ${\cal A}$ being the albedo of the Keplerian disk.
The enhancement factor $F_e \sim 3(T_{e}/3T_{d})^{1-\alpha} \sim 10-30$
because, typically,
the electron temperature $T_{e} \sim 50$ keV and the disk temperature
$T_d \sim 5$eV for parameters of active galaxies and $T_{e} \sim 150$keV
and $T_d \sim 100$eV for stellar black hole candidates. Hence, one easily 
achieves a convergence, $f_{ds} F_e f_{sd} {\cal B} < 1$  
since $f_{sd}\sim 0.25$, $f_{ds} \sim 0.05$, and ${\cal B} \sim 0.5$.
Here, $\alpha$ is the energy index ($F_{\nu} \sim \nu^{-\alpha}$)
in the the up-scattering dominated region of the hard spectra.
Soft photons from the disk component in the post-shock hot region do not 
participate in  cooling the electrons since the
Keplerian and sub-Keplerian components behave identically in this region
for three reasons: (1) the angular momentum of the
sub-Keplerian and Keplerian components are the same at $r \lsim r_s$;
(2) $v_r \approx 0$ in the immediately post-shock region of the halo
whereas $dv_r/dr \approx 0$ in the disk component --- in either case
the advection term $v_r dv_r/dr$ (cf. eq. 5.1a) is negligible and 
both the flows accrete downstream as a single component,
(3) since the post-shock halo is much hotter than the Keplerian component,
it is likely to evaporate the disk underneath and together they
behave as a single component, though the components need not find
time to mix effectively.

Figure 7.4 shows a comparison of four spectra around
a black hole of mass $M=5 M_\odot$ and the halo
accretion rate ${\dot m}_h=\frac{{\dot M}_h}{{\dot M}_{Edd}}
=1.0$. The disk accretion rates ${\dot m}_d=\frac{{\dot M}_d}{{\dot M}_{Edd}}$
are $1.0$ (dotted), $0.1$ (short dashed), $0.01$ (long-dashed) and
$0.001$ (solid) respectively [68]. The dotted curve is overlapped
with another curve (dot-dashed) including the effects of the convergent
inflow in the soft-state.
With the increase of the disk accretion rate, the number of soft photons
intercepted by the post-shock bulge is increased, cooling this region
efficiently. As a result, the temperature $T_{e}$ of the electron
is reduced and the energy index $\alpha$
is increased. The luminosity and the peak frequency of the soft component
go up monotonically with ${\dot m}_d$.
The intensity of the hard component rises with ${\dot m}_d$ below $\nu \sim
10^{18-19}$Hz (corresponding energy $\sim 5-50$keV) but it falls at higher
energies since the electrons become cooler.
The ratio $L_H/L_S$ decreases with the disk accretion rate.
Increasing the halo rate ${\dot m}_h$ increases the optical depth of
the post-shock region. The spectra are hardened due to
the saturation of Comptonization. Here $\alpha$ decreases and the
hard luminosity increases. Results in the case of massive black holes
are similar, since the electron temperature weaky depends on the
black hole mass ($ \sim M^{-0.1}$).

The transition from the hard state to soft state could be
seen by plotting the spectral index as functions of the
disk accretion rate (${\dot m}_d$) (Fig. 7.5). 
Halo accretion rates are marked on curves.
In the hard state, $\alpha \sim $constant, 
even when the disk accretion rate is increased by a couple of orders of 
magnitude. On the right hand side, the plots of spectral index
are from  the convergent accretion in the post-shock region
([68, 74, 238]) close to the black hole. The slope  due to the convergent
flow asymptotically approches $1.5$ for high accretion rates
and therefore could represent signature of rapid motion close to the
black hole. (Such a component is unexpected for a neutron
star accretion as discussed in \S 5.) The general variation of 
the soft ($L_S$), and hard luminosities in X-rays
and $\gamma$-rays ($L_X$ and $L_\gamma$), as well as the 
spectral index $\alpha$, as predicted by this model, (and as
observed as well) is presented in Table 1. The up-arrow shows the
correlation while down-arrow shows the anti-correlation. 
The smaller arrow shows a weaker dependence.

\noindent {\large TABLE 1}

\centerline{Variations of the Observed Quantities with Accretion Rates} 

\begin{center}
\begin{tabular}{|l|l|l|l|}
\hline
    & & & \\
\sl  & \sl Input & \sl ${\dot m}_d$ & \sl ${\dot m}_h$ \\
    & {\large $\rightarrow$} & & \\
\hline
Output & Row & & \\
{\large $\downarrow$} &  \&  & \sl (1) &  (2) \\
  & Column & & \\
\hline
    & & & \\
 $L_{S}$ & (1)  & {\Large $\uparrow$} &  {\small $\uparrow$}  \\
    & & & \\
\hline
    & & & \\
$L_X, \alpha$ & (2) & {\Large $\uparrow$}{\Large $\uparrow$}$^a$ 
& {\Large $\uparrow$}{\Large $\downarrow$}$^{b,c}$ \\
    & & & \\
\hline 
    & & & \\
$L_\gamma, \alpha$ & (3) & {\Large $\downarrow$}{\Large 
$\uparrow$}$^a$ & {\Large $\uparrow$}{\Large $\downarrow$}$^b$ \\
    & & & \\
\hline
\end{tabular}
\end{center}
\indent  $^a$ dependence is weaker for ${\dot m}_d \lsim 0.1$\\
\indent  $^b$ dependence is weaker for ${\dot m}_h \gsim 1$\\
\indent  $^c$ $\alpha_X \sim 1-1.5$, $L_X/L_S \leq 10^{-3}$ when shock 
is absent\\

The general agreement of the observed behaviours with the predictions 
from this more general model indicates we now understand the black hole
accretion more completely.

\noindent {\Large 7.3 Variability of AGN spectrum}

An interesting way to verify the possible presence of a black hole 
is to identify rapid variability in emitting gases from nearby regions. 
An extensive study of variability is now under way 
and there are excellent reviews on this topic [239]. 
Particularly interesting are the Blazars 
(the BL Lacertae objects and Optically Violent Variables),
which show high polarization and variability at almost all wavelengths, 
on time scales varying from roughly months (quiescent emission 
component) at wavelengths longer than $1$ cm
to weeks at sub-millimeter wavelengths to roughly days (flaring component)
in UV and smaller wavelengths. X-rays can also vary from days 
to months. Clear intra-day variabilities have also been reported [239-247, 298].
In objects such as OJ287, variabilities in time scale
of minutes may have been seen [299]. 

So far, no solid physical understanding of these variabilities has
emerged yet. Several phenomenological models have been proposed in terms of
multiple flares, hot-spots and unstable spiral shocks in disks. 
In modeling the optical and UV
data it has been shown that reasonable time scales for fluctuations can
be achieved with dozens of spots on both the standard `$\alpha$ disks' [11]
and ``$\beta$ accretion disks'' [136, 300] 

using sensible parameters. Very similar models (using $\alpha$ disks)
have been used to describe X-ray flickering as well [301]. 

Since the ``big blue bump" in the spectrum may be due to accretion disks,
it is reasonable to expect that fluctuations in the optical
and in the UV bands may be associated with accretion disks also,
though in BL Lac spectra no ``big blue bump"  is seen. 
Wandel \& Urry [302] 
have claimed that in the best studied case  of PKS 2155-304, a significant 
quasi-thermal, and thus, probably, accretion disk, component is present;
however, optical polarimetry of this source implies that the disk
contribution, if present at all, is quite small [303]. 
Another contribution to variability could come from the time dependent
behaviour of the formation and evolution of spiral density waves in outer 
disks. Such waves may be caused by some perturber massive compared to 
the disk, but probably less massive compared to the central black hole [237].

At least some of the variability features in blazars could be understood
assuming magnetic activities in the magnetized jets [304-305],
or in magnetic coronal activities in accretion disks close to the black hole
[141]. 
The general conclusion is that the emission must 
originate from a single compact region [306] 
of size $r\sim 10^{15}$ to $10^{17}$cm and the 
magnetic field, which is largely turbulent [307] 
is $B\sim 0.1 - 1$ G [308]. 
A large number of these objects, such as, OJ287, 3C 345, BL Lac, and 3C 454.3
etc. show clear evidence of a net magnetic field aligned orthogonal to the 
direction of the jet axis. All these points to the evidence that a strong
magnetic activity, very similar to what happens on the solar surface,
must be taking place at the funnel (or chimney) wall as well. 
Blandford \& K\"onigl [304] 
suggest that if a succession of mild shocks continuously re-accelerate the 
electrons, the flow should be roughly isothermal, and the field
perpendicular to the flow axis at the shock should vary as $B \propto r^{-1}$.
This condition produces a very high field of $\sim 1$G at the flaring region.
In a highly variable source, such as MK421 ($t_{var} \sim 10000$s) [309],
field varies as $B=B_0 r^{-1.4}$ with $B\sim 0.2$G. In PKS2155-304, field 
varies as $B=B_0 r^{-1.3}$ with $B_0 \sim 100$G close to the core to $70\ 
\mu G$ at $1$pc. Because of the strong evidence of the presence of magnetic 
fields, magnetic flares in the chimney can be considered to be major 
contributors to the variability. The energy deposited
at each flare is about $L^3B_f^2/4\pi$. Here, $L$ is the
length scale of a typical flare and $B_f$ is the average field
on the chimney surface. The energy is released in time $L/V_A$, where,
$V_A$ is the Alfv{\'e}n velocity $(B_f^2/4\pi\rho_f)^{1/2}$, $\rho_f$
being the density of matter inside the chimney. With reasonable estimates
of $L\sim r_g$, $B\sim 100$G, and $\rho \sim 10^{-14}$gm cm$^{-3}$,
one obtains the energy release, time-scale and the rate of release to be
given by $2 \times 10^{37}$erg, $0.3$d and $2 \times 10^{34}$ erg s$^{-1}$,
respectively for a black hole of $10^8 M_\odot$. Depending upon the size 
of the flares their occurrence rate will change. Thus, a large number of
frequent micro- and mini-flares are expected inside the chimney [141].

\noindent {\Large 7.4 Correlated Variabilities}

In some of the well-studied objects, such as NGC 5548, NGC 4151, and
Fairall 9, continuum and lines vary virtually simultaneously at optical and
ultraviolet frequencies [207, 310].
Molendi, Maraschi \& Stella [209] 
have pointed out that the propagation speed of the perturbation
have to be faster than $10$ per cent of the speed of light
to explain variabilities with so short time lags.
In a standard Keplerian accretion disk model
the typical time scales on which thermal and viscous instabilities
operate are very long (months or more) and thus the short time lag requirement
cannot be met. Molendi, Maraschi \& Stella [209] 
have proposed that possibly this could be due to re-processing of radiation
from a hot bulge component at the center of the disk. Alternatively,
it is proposed that to explain such a behaviour
flow has to be only partially disk-like [208].

In the presence of a strong shock wave in an accretion disk, 
the post-shock region becomes very hot, and the disk is puffed up
to become geometrically thick (Fig. 6.4, Fig. 7.3). 
Hard radiation produced at the shock would be reprocessed 
with $10-25$ percent of interception by the pre-shock flow of the disk. 
Since the cool and hot components of the disk reside side by side near
the shock wave, such a correlated variability may be naturally explained
by this generalized disk model (\S 7.2).

In the case of BL Lac 0716+714, it has been observed that variabilities
in radio and optical are well correlated [242, 311]
and some kind of quasi periodicities are
observed in time scales of $0.8d$ and $1.3$d in the first week and 
of $\sim 6$d in later weeks of observations. 
There is at present no reliable model 
of this correlation. It is possible that the strength of magnetic fields 
on the disk surface is well correlated with the pressure inside the disk
where the flux tube is originated [141]. 
If, for example, the optical variabilities are caused by the formation 
and disappearance of the hot-spots (non-axisymmetric density 
perturbations) inside the disk then the radio flares caused by magnetic fields 
anchored with these density waves should also follow a similar variability.
Indeed, if the anchored corona is also responsible for production
of $\gamma$-rays, then future observations of correlation with
$\gamma$-rays cannot be ruled out.

In the star-burst model  of quasars [266], most of the optical and UV 
luminosity is supposed to be due to young stars and the observed variabilities
correspond to the presence of a large number of compact supernovae 
remnants along the line of sight. However, this model cannot
explain uncorrelated variabilities observed in the four images of the lens
system Q2237+0305 [312] (which has no intrinsic variability)
nor can it explain short-time scale correlated variabilities 
$t_{var} \lsim 1$ week. This model also may not be able to produce
observed jets in quasars which has a very accurate directionality.

\noindent{\Large 7.5 Variability of Disk Line Emission}

Similar to the line emissions seen in Cataclysmic variables, one would 
expect that if the accretion disk around a black hole also emits 
lines, they should show double horned patterns [313, 314].
If there is a non-axisymmetric feature on the disk, the double 
horned pattern need not be symmetric [212, 315].
With the movement of the feature, the emitted pattern should also
move so that one obtains a temporal variation. Similarly, if the disk is well
resolved so that one could observe parts of the disk at a time, then 
one should also expect spatial variation of the line emission. These 
informations certainly would provide more detailed properties of the disk
as well as the properties of the central black hole.

\noindent {\large 7.5.1 Temporal Variation}

The first good candidate was the permitted $H_\beta$ emission 
from the broad line radio galaxy (BLRG) 3C 390.3 [316-317].
Soon thereafter, Chen, Halpern \& Fillipenko [318] 
and Chen \& Halpern [319] 
carefully analyzed H$\alpha$ emission line data for the BLRG ARP
102B and showed that the double-peaked emission could fit very
well using relativistic Keplerian disk models. The H$\alpha$ broad emission
line in yet another BLRG, 3C 322, could also be
explained successfully with a very similar model [320]. 
In these stationary relativistic disk models, Doppler boosting implies
that the short-wavelength (blue) side of the profile is brighter than
the long-wavelength (red) side. Recently, a comprehensive survey of
88 BLRGs is completed [321] in which nine additional objects
were found to emit line emissions which fit relativistic accretion disk model
emission profiles.
Figures 7.6(a-b) show fits of Chen and Halpern [319] 
of the broad line emissions of ARP 102B. In (a), the emissivity is assumed
to be power law throughout the disk:
$$
\epsilon (x) = \epsilon_0 (x/x_1)^q \ \ \ \  x < x_1  
$$
$$
\ \ \ \ \ \ \ \ \ = \epsilon_0 (x_1/x)^q e^{-(x-x_1)^2/x_2^2} \ \ \ \ 
x \geq x_1\\
$$
and in (b), local line broadening with a single power law,
$$
\epsilon (x) = \epsilon_0 x^{-q}
$$
is used. The fit of the broad lines is generally very good and indicates 
that possibly the broad lines are contributed by the disk itself.
In (a), the power law parameters are  $x_1 = 400GM/c^2$, $x_2 = 
900GM/c^2$, the emissivity exponent is  $q=3$, and the inclination is $i=32^o$. 
In (b), the inner edge is at $x_1=350 GM/c^2$, outer edge is at
$x_2=1000GM/c^2$, the emissivity exponent is $q=3.0$, the inclination angle is
$i=32^o$ and the broadening velocity is $\sigma = 850$ km s$^{-1}$.
Subsequently, it was pointed out that in some phases,
the red component from ARP 102B was found to be stronger than the blue [210].
It was also noted that the temporal variations in the blue and red
peaks were not well correlated; thus 
the possibility of an axi-symmetric disk could be ruled out [210].
Similar model of relativistic axi-symmetric standard Keplerian disk
is used to fit 3C390.3, for example. Figure 7.7 shows this fit
when the blue component is stronger then the red component [321].
The lower panel gives the residual spectrum obtained after subtraction
of the theoretical fits from the observed profile.
The inner and the outer edges of the emitting regions are $380GM/c^2$
and $1300GM/c^2$ respectively. The Gaussian profile broadening width
is assumed to be $\sigma=1300 $ km s$^{-1}$, the inclination angle is
assumed to be $i=26^o$ and the emissivity exponent is assumed to be $q=3.0$.
However, simple disk model for the broad emission lines from 3C390.3 was also 
shown to be inadequate because $H_\beta$ emission
lines are found to be quite variable. Veilleux \& Zheng [322] 
showed that the B/R ratio does not remain greater than unity for much of 
the extensive period between 1974--1988 and later a more complex,
bipolar origin of the temporal variability was postulated [323].
Even when $B/R >1$, a Keplerian disk is often unable to raise the 
blue component to the level that is required by observation (Fig. 7.7).
One possibility, similar to that used to
explain shifting B/R ratios in CVs [324-325] 
would be the presence of a large hot spot on a relativistic disk. 
It has been recently observed [326] that the
double-peaked Balmer broad emission lines for the LINER nucleus of NGC 
1097, the red peak exceeds the blue, which is inconsistent with the 
axi-symmetric relativistic disk model [318-319]. 

This type of temporal variability could be
naturally explained in disks which have non-axisymmetric features
which either advance (or trail) with the flow [212]. 
Using non-axisymmetric spiral shock models (\S 5.7), it is found that
one can fit emissions from 3C390.3 and ARP 102B at two different phases. 
The temporal variation seems to be
due to motion of the spiral shocks with respect to the line of sight.
Figures 7.8(a-b) show theoretical fits of the $H_\alpha$ data of 3C390.3 
[212] at two different
times. The solid curves are observed data and the dashed curves are
the model fits. In Fig. 7.8(a), the 1976 data is compared and in Fig. 7.8(b),
the 1980 data is compared. The inner and outer edges are chosen to be
$x_1 = 200 GM/c^2$, $x_2 = 900 GM/c^2$, the emissivity exponent $q=2.0$,
inclination angle $i=38^o$, and the broadening velocity $v_{th}=0.0015c$.

\noindent {\large 7.5.2 Spatial Variation}

Spatial resolution of the disk is now possible with the advent of
small aperture space telescopes. Recent imaging results of M87 [4] and the 
spectroscopic results of the disk [5] indicate that there are spiral features
in the disk. The spectroscopy was made pointing the telescope
at seven different sites. Figure 7.9 gives the spatial variation of the
line emission profiles from the spectrographic data (obtained by
the Faint Object Camera) at these positions. 
The image is plotted assuming the systemic velocity of 
$\sim 1300$Km of M87 from the sun. The prominent line features
include [O III]$\lambda\lambda 4959$\AA\ , $5007$\AA\ , [S II] $\lambda
\lambda 6716$\AA\ , $6731$\AA\ , [O I] $\lambda 6300$\AA\ , [$H_\beta$] $
\lambda 4861$\AA\  and the blending of [$H_\alpha$] $ \lambda 6563$\AA\
with [N II] $\lambda\lambda 6548$\AA\ , $6583$\AA\ . The velocity measurements
at position 5 and 6, which are taken
at opposite sides of the nucleus show that the velocity of the emitting
flows to be $\sim \pm 550$ km s$^{-1}$ respectively, indicating 
a significant rotational velocity around the nucleus. Assuming the velocities
observed to be Keplerian, the determined mass is $(2.4 \pm 0.7) \times
10^9 M_\odot$
and the inclination of the disk plane with the line of sight to be
$(42 \pm 5)^{o}$. These results could be further improved by
assuming that the spiral features are real and with a two armed
spiral shock solution (\S 5.7) the fits of lines are found
to be reproduced only if the mass of the central black hole
is chosen to be $(4.0 \pm 0.2) \times 10^9 M_\odot$ and the disk 
inclination angle $(42\pm 2)^{o}$. The disks with spiral shock waves are 
found to be sub-Keplerian and therefore a higher mass is required to
obtain the same frequency shift. Figure 7.10 shows the fit (solid) of the 
[O III]$\lambda 5007$\AA\ , lines (short dashed) emitted from different
positions. The same theoretical fit, upon suitable scaling, matches with 
observed [N II] $\lambda  6584$\AA\  profile (dot-dashed)
as well [6]. The general agreement indicates that the dissipative
shocks may be real
and an absence of a strong ionizing source in the system indicates
that the emission could be due to shock ionization. The agreement of the
spatial variation of the line profiles in M87 with theoretical
models, provide the strongest evidence so far of a
super-massive black hole at a galactic center. Masses could be
seriously underestimated due to the inefficient sub-Keplerian
disks.

Another important recent observation which provides a convincing proof
of the galactic black hole model is that due to Miyoshi et al [327].
Using Very Large Baseline Array (VLBA), they observe very strong 
water maser lines from the molecular torus which surrounds
NGC4258 and find evidence 
of mostly Keplerian velocity of about $1000$ km s${-1}$
at a distance of $0.13$pc from the galactic center. The estimated
mass of the central black hole is about $3.6 \times 10^7 M_{\odot}$.
In this system matter is found to have very little radial motion
($<10$ km s$^{-1}$) indicating the flow is primarily Keplerian
in the torus. A very complex structured spiral feature (identified as a jet)
has been observed in this system which also shows evidence of shock
ionization [328].

\noindent{\Large 7.6 Quasi-Periodic Oscillations}

Many low mass X-ray binaries and
X-ray pulsars are known to show quasi-periodic oscillations (QPOs). These
QPO behaviour is thought to be correlated with the $Z$-shaped spectral
states, namely, the horizontal, flaring and normal branches [329].
More recently, QPOs are observed in galactic black hole candidates
such as LMC X-1, [330], GX339-4, GS1124-68 [331-333] as well.
Whereas the frequencies of oscillations of the horizontal,
normal and flaring branches are $20-55$, $5-7$ and $7-20$Hz respectively,
the black hole candidates show evidences of generally much lower frequencies.
LMC X-1 has the centroid frequencies 
around $0.08$Hz [331], GX339-4 has around $6-8$Hz  and GS1124-68 has around
$3-8$Hz [332]. The modulated amplitude could be as high as $5-10$
percent or more.

Several models have been proposed to understand the QPOs. Beat frequency
model is more promising for the horizontal branch and X-ray pulsar
QPOs [334]. 
Particularly puzzling are the QPOs in black hole candidates, where 
there is no `hard' surface which would rotate at a different speed
than the inner edge of the disk. Thus, presently, there is no satisfactory
explanation for these QPOs. Some proposed models include the followings:\\
(a) Viscous pulsational instability model [335]:
In this case, the oscillation is
thought to be excited by axially symmetric acoustic modes at the transonic
region of the disk close to the black hole. The frequency
of oscillation is found to be around $10^3 (M/M_\odot)^{-1}$Hz which is
too high for QPOs observed. Furthermore, the energy in the acoustic
wave could be very low to produce the observed modulation of amplitude.
(b) Another mechanism is `one armed corrugation wave' generation at the
inner edge of the disk where the disk oscillates vertically as a whole [336].
The frequency of oscillation is around $1-0.02$ Hz for a critical accretion rate,
but the problem with the observed modulation persists.
(c) Yet another mechanism involves the trapped oscillation [337] which are
generated closer to the inner edge of the accretion disk. These oscillations
appear as there is a maxima in the epicyclic frequency. These oscillations
show 1-3 percent modulation in the absence of accretion. The computation
which includes accretion are to be performed in order to find the 
relevance of these modes.
(d) A fourth mechanism is due the oscillation of the 
shock waves [227] close to the black holes. Centrifugally supported
shock waves act as hard surfaces and the post-shock flow produces hard
radiations. When the cooling time of this region roughly
agrees with the in-fall time, the shock oscillates, modulating the 
hard-radiation by 5-15 percent. 

\noindent{\Large 7.7 Our Galactic Center}

Since our galactic center is only about $8.5$Kpc away, 
it would be the ideal candidate to look for the evidences
of a black hole [338].
However, observation of the galactic center is very difficult
because the solar system is close to the galactic plane
and one has to view the center through the interstellar dust and gas.
The best way to observe the center is to use wave-bands in which the
attenuation is minimum, such as in radio, infra-red, hard X-rays
and $\gamma$-rays.  

A very good and up-to-date review on the latest observational aspects 
is given in Genzel, Hollenbach \& Townes [7]. 
We shall restrict ourselves only to the basic conclusions regarding the
theoretical interpretation on the nature of the central object.
Figure 7.11 shows schematically  different identified structures as well
as distribution of mass computed from various arguments within $100$pc of the
galactic center. The resolution of the figure is $\sim 0.01$pc, roughly the 
size of the unresolved Sgr A$^*$. The major components 
are the stars and the stellar clusters, giant molecular
clouds, Sgr A east arm, the circum-nuclear ring or disk, and the mini-spiral.
Also shown are the locations of the Sgr A$^*$, the infra-red sources
IRS-7 and IRS-16. Mass within a radius of $0.1$pc, is found to be a few 
times $10^6M_\odot$ which is usually attributed to a black hole at the center.

There is indeed a major controversy regarding the nature of the massive 
entity (if it is a single one) in our galactic center. 
There is not yet a clear consensus even in identifying the galactic
center. Sanders [339] 
argues that among many stars observed in the
central parsec region, there are several very massive and luminous
blue stars with strong helium lines and fast winds. Particularly,
the brightest complex in the region, IRS-16 contains about $20$
such blue stars. Thus, the central activity
may be dominated by the ionizing radiation from IRS-16, rather than
anything associated with Sgr A$^*$. Even if it is assumed that Sgr A$^*$
is our galactic center, there are strong arguments for and against
it being a massive black hole. We wish
to close this Section by presenting a few such arguments. 

A major problem involves in understanding the nature and the origin of the
circumstellar wind which is inferred from the broad He-I emission lines from
the direction of IRS16. The broad lines indicate outflows 
from the galactic center with velocities of the order 
$v_{gw}=500-700 $ km s$^{-1}$ and a mass loss rate of ${\dot M}_{GW}
\sim 3-4 \times 10^{-3} M_{\odot}$ yr$^{-1}$ [340].
Yusuf-Zadeh \& Melia [341] 
argue that the tail associated with the giant star IRS 7 is due to the 
interaction between the galactic center wind and the wind from IRS 7. The 
origin of the galactic wind is IRS-16. There is a general opinion
that the wind originating from IRS-16 should be accreted onto Sgr A$^*$,
as it is only $0.04$-$0.06$pc away from IRS-16.
The argument goes as follows [342]:

The sound speed $a\sim (k T_{gw}/m_p)^{1/2} \sim 18$ km s$^{-1}$ $ (\frac
{T_{gw}}{40,000K})^{1/2}$ within the galactic center being smaller than the
outflow velocity of few hundred km s$^{-1}$, the accretion is initiated
by a bow shock and heats the gas to $T_0=(3 m_p /16 k) v_{gw}^2
\sim 7 \times 10^6 (v_{gw}/600)^2 $K. The plasma accretes almost
radially after passing through a shock and all plasma within 
the accretion radius $r_a=2GM/v_{gw}^2$, falls on the black hole. 
The accretion rate is about 
$$
{\dot M}  \sim 10^{22} {\rm gm}\ {\rm s}^{-1} {\cal W } (M/10^6 M_\odot)^2
\eqno{(7.1)}
$$
with 
$$
{\cal W} =  (\frac {v_{gw}}{600 \ {\rm km\ s}^{-1}})^{-4} 
(\frac{\dot M_{gw}}{ 3.5 \times 10^{-3} {\rm M_\odot yr^{-1}}}) (\frac {D}
{0.06 \ {\rm pc}})^{-2}
$$
Here $D$ is the deprojected distance $0.06$pc of the distance between the
Sgr A$^*$ and the galactic wind origin.

One can calculate the radiation emitted by the accreted matter assuming 
magnetic dissipation (\S 2). The heating rate is given by,
$$
\Gamma = \frac{{\dot M} c^8}{4^4 \pi G^3 M^3 }(\frac{r_g}{r})^4
\sim 3.3 \times 10^6 \beta_d W (\frac{M}{10^6 M_\odot})^{-1} (\frac{r_g}{r})^4
\ \ {\rm erg \ cm}^{-3} s^{-1} .
\eqno{(7.2)}
$$
The cooling of the gas is dominated by the synchrotron emission from the 
thermal electrons and bremsstrahlung emission due to e-e and e-p collisions 
(\S 2). Assuming that the electron and proton temperatures are equal, 
and the total gravitational potential energy goes into heating,
$$
T_e \sim 10^{13} \frac{r_g}{r} \ K ,
\eqno{(7.3)}
$$
till the flow reaches a distance of $r=r_i$ where heating and cooling are 
equal-- $\Lambda \sim \Gamma$, so that for $r <r_i$ the flow is nearly 
isothermal. Here, $r_g$ is the Schwarzschild radius of the black hole.

Figure 7.12 shows the fit (solid) of the observed intensity of radiation using
the simple model discussed above [342]. The mass of the central 
black hole is
chosen to be $0.9 \times 10^6 M_\odot$ and its distance is chosen to 
be $8.5$kpc. The flow is assumed to begin its spherical accretion at the
bow shock, which is also identified as the accretion radius.
The post (bow) shock gas temperature at the
accretion radius is $2\times 10^5 r_g$ is $T_e \sim T_p \sim 7\times 10^6$K. 
The flow becomes isothermal for $r<10^4 r_g$. 
Synchrotron emission dominates the spectrum
for $\nu_\infty \leq 10^{13}$Hz and luminosity at higher frequencies are
entirely due to thermal bremsstrahlung. The fit with
observational data [343] seems to be reasonable. This model was subsequently
improved by including  an exact treatment of the cyclotron/synchrotron 
emissivity that is valid for all temperatures, the actual determination
of the temperature distribution in the inflow, and some effect on 
the spectrum should the accreting plasma have a residual angular momentum, 
possibly forming a disk at small radii [344].  The most likely mass in this
model is found to be $2\pm 1 \times 10^6M_\odot$ which is a factor of two
higher than the earlier result.

Some authors suggest [345] that the luminosity of $7 \times 10^5 L_\odot$
of Sgr A$^*$ could be explained even by a standard disk model
around a  Kerr black hole ($a>0.9$) of mass $\sim 2 \times 10^6 M_\odot$
provided the accretion rate is about $10^{-8.5}M_\odot$ yr$^{-1}$ and the
disk is seen edge on. They rule out the possibility of having a very low
mass ($<10^3 M_\odot$) black hole.

A more critical approach to constrain 
the parameters ($M, \ {\dot M}$, and $\epsilon$ -- the efficiency)
is taken by Ozernoy [346] 
who considers diverse observations such as dynamics of spiral arm in the 
disk, stellar collisions close to the core, observation of high energy
photons and particles, radiation spectrum of accreting winds etc. to 
come to the conclusion that the mass of the black hole may be much less.

From Fig. 7.11, it is clear that 
the central couple of parsec region contains a circum-nuclear disk
of molecular gas, neutral hydrogen and dust. In it lies a
clumped, a mini-spiral shaped Sgr A West, presumably tidally
captured and sheared molecular cloud. Earlier attempts
to associate this  spiral arm with the underlying disk
reveals a mass of the central black hole to be $2 \times 10^6 M_\odot$.
However, in a recent work, Friedman et al. [347]
point out that the arm is probably made up of superposition of various modes
of density waves.
Assuming the observed pattern is a part of two or three armed spiral, and the
fact that the spiral could be traced as far in as $r \sim 0.1pc$, 
allows them to constrain the central point mass not to exceed $10^5 M_\odot$.

Another approach taken to tighten the constraint is to study the dispersion
of stellar velocities. A small core radius of size $\sim 0.15$ pc [348]
implies a very deep gravitational potential at the core in which 
stars with dispersion velocity $\sigma_0 \sim 100$ km s$^{-1}$ (which is much
smaller than the escape velocity of $600$ km s$^{-1}$) and number density 
$\sim 10^8$ pc$^{-3}$ constantly collide to produce massive stars. The 
luminosity of accretion of the ejected gas arising out of the collision
of the stars would seriously constrain the central mass to
$\sim 10^5 M_\odot$ [346]. 
Similar problem arises in explaining weak high energy
source at or around Sgr A$^*$ as observed by ART-P and EGRET. 
A comparison of the observations with current models of shock acceleration of
particles implies that the mass of the black hole may be as low as a 
few $\times 10^3 M_\odot$ [346]. Radio observations of Sgr A$^*$ reveals the 
Lorentz factor of the emitting electrons to be
of the order $10^2 $ to $10^3$, which is possible if the central
mass of the black hole is as low as about $5 \times 10^2 M_\odot$ [346].

Interpretations vary regarding the origin of
the X-ray flux. It may be due to bremsstrahlung [342] as mentioned
earlier, or it may be due to synchrotron-emitting relativistic pairs.
These pairs may result from the decay of pions originating in the vicinity 
of the shock inside the IRS-16 wind while accreting on the galactic center.
And the mass obtained by this constraint is again much less [349].

Another consideration is that the angular momentum 
$J_{gw} = r_a v_{gw} (2GMr_a)^{1/2}= 10^{25} M_6$ cm$^2$ s$^{-1}$ of the
in-falling gas at a distance $r=r_a =2GM/v_{gw}^2 \sim 7.5 \times
10^{16}M_6$ cm is much higher compared to the
limiting angular momentum of a black hole $r_g c \sim 10^{22} M_6$, 
a disk must form whose radiation efficiency is much higher
compared to the spherical Bondi accretion as assumed earlier [342].
Therefore, to explain the same efficiency, the mass of the black hole 
could be much smaller [346]. It is not clear however, that the
angular momentum is indeed so high.

Recently, it has been argued [350] 
that the black hole at the galactic center may produce enough shear in the
ionized flow that a weak seed field is amplified to equipartition value
of about $10^5$G very close to the black hole which is then taken 
to outer parts of the galaxy through winds. The time scale taken
to amplify to the equipartition value is around a thousand years, 
and therefore within the age of the galaxy, it is possible
to supply the flux needed to fill the galaxy if, to a lesser extent,
the local dynamo effect is allowed. One of the predictions of this model
would be to find signatures of strong fields close to the center. 
In a parsec scale, the field should be at least four orders of magnitude
higher compared to the field seen in the solar neighbourhood. Indeed,
in northern arm ($\sim  1.5$pc ) strong fields of the order of $10$mG
is observed [351]. In the circumnuclear disk ($\sim 5$pc),
the field of $2$mG is seen. The origin of these strong, structured
field could be understood within the framework of the shear-amplification
model [350]. 

So far, no model self-consistently studies the disk structure and radiation
emitted from it around our galactic center. 
The detailed study of the observational results indicates that the 
Sgr A$^*$, considered to be coincident with the center of the galaxy 
does not show any special feature which distinguishes it from 
nearby stars. However, the resolution is possibly close to $10^{16}$cm.
It is possible that, at this resolution, one is watching a very weakly active
and starved ionized disk around the central black hole. With the presence of 
about two dozens of stars, which are possibly  massive enough to 
launch density waves in the ionized disk, one would have expected some 
shock waves in the disk, which would be hot enough to 
excite line emissions. Detection of these lines may distinguish the 
disk around Sgr A$^*$ from the stars in nearby region. Presence of quite 
a few mass loosing stars close to the Sgr A$^*$ indicates that matter from
the winds may not have large angular momentum, since they could cancel 
(similar to Hoyle-Lyttleton accretion [45]) most of the angular momentum 
through collision. The apparent asymmetry of the bright sources (which may 
not be statistically significant, [7]) which supply matter 
to Sgr A$^*$ may cause the net angular momentum
not to nullify completely. But still, one would expect accretion
to take place with angular momentum orders of magnitude lower compared
to the Keplerian value. In this sub-Keplerian regime,
as discussed in \S 5, the transonic flow need not lose any energy at all.
The solution with advection is perfectly stable with a very little
energy emitted away. Hence the flow could be largely non-dissipative 
and not an efficient emitter (\S 5 -\S 6). This argument may be
valid for other galactic centers as well.

A second argument that the nucleus is massive, yet inactive
could be given from our knowledge of the behaviour of transonic
flows as a function of the accretion rate. From eq. (5.8),
it is clear that for a given mass flux,
${\dot M}$, ${\dot {\cal  M}}$ is inversely proportional to the
square of the mass of the black hole. Let the mass of the black 
hole be $M=m M_\odot$. Assuming an optimistic accretion rate of 
$10^{-5}M_\odot$yr$^{-1}$, and an intermediate radiation content,
$\beta \sim 0.1$, one has dimensionless ${\dot{\cal M}}\sim 10^{-16} /m^2$.
For any reasonable mass of the black hole, $m$, this is too
small to form an axisymmetric shock wave in the disk (Fig. 5.1).
Only if the flow very close to the black hole is
radiation dominated, say, if $\beta\sim 10^{-6}$ or so, 
${\dot{\cal M}} \sim 10^4/m^2$. This is a very important regime.
Because, if the mass of the black hole is around $10^6M_\odot$,
$\cm \sim 10^{-8}$, and the cooler flow falls onto the black hole
after passing through the outer sonic point and no
shock is expected. If the mass is around $10^5M_\odot$, 
${\dot{\cal M}}\sim 10^{-6}$ is in the region where shocks might form
for some range of specific angular momentum of the disk.
For even smaller black hole mass $m$, shock should not form but the
flow is hotter and passes through the inner sonic point.
Thus any indication of a strong X-ray radiation from regions 
of $10-100R_g \sim 10^{11-12}cm \frac{M}{10^5M_\odot}$ would
be from shock emissions.

\newpage
\noindent {\large\bf 8  Concluding Remarks and Acknowledgments}

In the past few Sections, several models of accretion flows such as the 
spherical Bondi flow and its variations, standard Keplerian thin disk
and its variations, thick accretion disks and transonic accretion disks
with or without shock waves have been discussed. There are a
large number of other disk models in the literature which are mostly 
variations of these models. We do not have space to discuss them here. 
Presently, we summarize this review  and add a few sentences on the future
trend in modeling accretion processes.

In the 1970s, the standard accretion disk models were constructed [11, 12]
largely to explain observations from the binary systems. In these systems,
angular momentum supplied at the outer edge is necessarily Keplerian. 
Viscosity drives the inflow by removing angular momentum outwards
and keeping the entire disk Keplerian in the process. Since then these 
disk models have also been used to explain the big blue bump observed
in the continuum of active galaxies and quasars and with some
success [274]. Occasionally, one invokes additional components along with 
the standard thin disks, such as corona, plasma cloud, absorbers, etc.
in order to explain the continuum as well as variable components of X-rays and 
$\gamma$-rays. As we mentioned in \S 7, there are now
several observations of almost zero time-lag correlated
variabilities between X-rays and optical data [207] which 
cannot be explained by simple Keplerian disk models. The production and
properties of the hard radiation and the occasional
change of states by the galactic black hole candidates cannot be explained 
using Keplerian disks [285-296]. We also mentioned that the temporal variation
of line profiles from objects, such as, ARP102B and 3C390.3 is impossible 
to explain by using axisymmetric disk models [212]. Recent HST observation 
of M87 shows clear evidence of non-axisymmetry in the ionized disk
around the black hole [4-6]. These examples may support
a hypothesis that accretion disks may have significant
sub-Keplerian components in them as well.

That the disks need not be of `standard' type was sensed by theoreticians
even in late '70s and throughout the '80s. We discussed thick accretion
disks (\S 4), transonic accretion disks and slim accretion disks (\S 3).
In thick disks, angular momentum is assumed to be almost constant
but the radial motion is ignored. In transonic disks [145], the flow is
thin but the radial motion is included. In slim disks [146], 
matter is allowed to
pass through the inner sonic point, thus improving on the standard
disk model. However, in many of these models, unsuccessful attempts were
made to match the flow with {\it Keplerian disks} at some distance and the disk 
structure depended strongly upon the matching radius and other parameters 
invoked. It is also easy to argue that if
matter had to accrete from a Keplerian disk from distances of parsecs
by using ordinary viscous process, it would take a time comparable to
Hubble time or more.

This intrinsic problem with these theoretical models, as well as problems
in explaining a large number observations with Keplerian disks, 
may actually disappear if one assumes that not only Keplerian (Roche lobe)
but a significant amount of sub-Keplerian component (either from
the wind or that originating from Keplerian disk itself) may co-accrete.
If the matter starts with highly sub-Keplerian angular momentum,
as probably is the case in active galaxies,
its subsequent behaviour depends strongly upon the accretion rate and the 
viscosity in the flow [\S\S 5-6]. If the accretion rate is small enough, 
the flow passes through the outer sonic point (just as a Bondi flow) and 
remains supersonic before falling onto a black hole. If the accretion rate 
is high, matter would pass through the inner sonic 
point. This description is valid if the entropy remains
almost constant, i.e., the viscosity is small enough.
Even when the flow starts with small entropy, some entropy could be
generated at a shock or in the flow by viscosity which is advected
along with the flow and allows the flow to pass through the inner sonic point 
as well. If the entropy is higher (for a given accretion rate) it is likely
that strong winds may be formed from flows with positive energy [259].
Due to strong advection, the flow close to the black hole
radiates with a very low efficiency as in a Bondi flow (\S 2, \S 5). 
This means that central black hole masses could be considerably
(by a factor of $10$ or more) than what is obtained from Keplerian
flow considerations.

Contrary to a Newtonian star, a black hole has no hard 
surface. If the flow is unable to lose angular momentum efficiently, the
centrifugal barrier causes the flow to have a shock close to the black hole 
(\S 5). The post-shock flow in a black hole accretion has similar properties
as that of a boundary layer. This is clearly the case when the viscosity is
small enough. The post-shock flow also has all the features of a
thick accretion disk only more self-consistent since the
radial motion is also included [218, 259]. 
The pre-shock flow is mainly advected
towards the black hole, just as a Bondi flow, but the post-shock flow
has very little radial velocity. Further in,
the flow picks up radial motion and supersonically enters into the
black hole. If the viscosity is high ($\alpha>\alpha_c$, where, $\alpha_c$
could be any number between $0$ and $1$ depending on flow and cooling
parameters), the {\it stable} shock
disappears and the disk passes through only inner
sonic point eventually becoming Keplerian at a far distance
(Fig. 5.7, Fig. 6.10a-b) except close to 
the inner boundaries. That a Keplerian disk can form out of
sub-Keplerian initial matter is no surprise. 
Matter accreted onto a black hole can have less specific angular momentum
than what the incoming matter at the outer edge has and therefore
the excess could be {\it redistributed } by the viscosity to 
produce a Keplerian or quasi-Keplerian disk. This also solves the so-called
angular momentum problem in active galaxies and possibly requires
jet models whose characteristics are not strongly tied with the Keplerian 
distribution in the disk. Similarly, depending on heating, cooling and viscous
effects, only a sub-Keplerian disk could form, representing quiescent
states of novae such as A0620-00 [352]

Thus, one may then have a thick, thin, slim or transonic (with or 
without shocks) disk from an initially Keplerian or
sub-Keplerian inflow at large distance depending upon accretion 
rate and viscosity. Based on this experience, in 
order to explain the observed features across the electromagnetic spectrum, 
a generic model could be constructed for the accretion disk (\S 7.2)
which has all the `desirable' properties of an accretion disk around a
black hole. A disk of this kind would form if there is a significant 
reduction of viscosity with height in the vertical direction inside the disk. 
Higher viscosity on the 
equatorial plane produces an optically thick standard Keplerian disk, which is
vertically flanked by warm, optically thin halo of low angular momentum gas. 
Alternatively, if at the outer boundary, an admixture of Keplerian 
and non-Keplerian matter is supplied, the situation would be similar. 
The virtually inviscid halo forms a standing shock close to the black hole 
($\gsim 10\ R_g$). 
The post-shock flow ($T_p \sim 10^{11}\ K$) heats up soft-photons  
coming from the disk to produce observed hard X-rays and $\gamma$-rays. 
Such a model has the potential to explain 
most of the steady state as well as time dependent behaviour of the continuum
and line emissions [297]. Again, the shocked halo could be replaced by a hot 
magnetic corona, provided one can build up a model (with 
entropy structure similar to that of the sun) where 
self-consistently coronae could be anchored [141].
Shock or no shock, we stress that the sub-Keplerian component
must be present close to a black hole. This is a general
feature of any viscous transonic flows [68, 76, 224, 229]. The
observations seems to indicate its presence as well [Fig. 7.5, 68, 297].

Is it possible to provide THE
signature of a black hole? In other words, is there 
a way to distinguish between a black hole candidate from a neutron star
purely from the observational point of view? We have briefly
discussed this issue in terms of presence and absence of the inner
sonic point [\S 5, 77, 218]. In the case of
black hole accretion the flow must be supersonic, and the bulk relativistic
motion can Comptonize cooler soft-photons in an optically thick flow 
(soft state) to produce a power-law hard spectra [68, 297, 238, \S 5].
In the case of a Newtonian star, the inner sonic point
does not exist, and 
the flow on the neutron star surface is subsonic. Thus such a hard component 
in the soft state should not be produced. Some observations already
suggest that this weak hard component is indeed observed [353]. 
Perhaps there are black holes after all!

The present review article by no means claims to be a complete one. 
Against the wish of the author, many works had to be
left out, partly due to the spatial limitation of the
article size and the temporal limitation of the author
and the rest is due to authors own limitation of knowledge
in many of these topics. During the past years while preparing the manuscript
the author is greatly benefited from discussions with several of his colleagues.
He especially thanks M. Camenzind, D. Kazanas, J. Katz, R. Khanna, M. Livio,
M. Malkan, L. Maraschi, F. Mayer, D. Molteni, J. Ostriker, L. Ozernoy,
B. Paczy\'nski, M. Rees, R. Rosner, H. Sponholz, L. Titarchuk, K. Thorne, 
S. Vainshtein, E. Vishniac and P. Wiita for helpful discussions.
He is very grateful to his wife Sonali Chakrabarti for patiently reading the
manuscript. Some of the material for this review has been compiled while 
the author was visiting the University of Chicago, Max Planck Institute of 
Astrophysics (Garching), Landessternwarte (Heidelberg) and
the Goddard Space Flight Center. The author
thanks several of his colleagues for giving him permission to reproduce
their published or unpublished figures. They are: M.J. Rees (Fig. 1.1),
J. Kormendy (Fig. 1.2), H. Sponholz (Fig. 1.3), L. Maraschi (Figs. 2.4-2.6,
Fig. 4.1), J. Ostriker (Fig. 2.11), D. Kazanas (Figs. 2.13-2.15), S. Shapiro
(Fig. 3.2), J.P. Lasota (Fig. 3.6), R. Khanna (Figs. 3.8-3.9), R. Blandford
(Fig. 3.12), P. Madau (Fig. 4.7-4.10), M. Livio (Fig. 6.1), 
J. Hawley (Fig. 6.2), I. Hachisu (Fig. 6.7),
J. Katz (Figs. 6.12-6.13), M. Malkan (Fig. 7.1), L. Titarchuk (Fig. 7.2),
J. Halpern (Fig. 7.6), M. Eracleous (Fig. 7.7), A. Kochhar (Fig.
7.9) and F. Melia (Fig. 7.12). Also permissions from respective publishers
are gratefully acknowledged. He thanks the members of the
photography section of the Tata Institute of Fundamental Research
for producing several of the glossy prints very promptly. The author thanks
NASA Space Physics Theory Programme for the support of his visit to 
the University of Chicago and Indian National Science Academy and Deutsche
Forschungsgemeinschaft for the support of his visit to Germany.

\newpage
\noindent {\large\bf Note Added In Proof}

\noindent (A) Effect of non-Keplerian accretion disk on gravitational
wave emission from a binary black hole system:

Study of gravitational wave emission from compact binary
systems has received a significant boost in recent years
because of the realization that the detection  of gravitational
waves would directly identify the compact and
strongly gravitating bodies, such as neutron stars and black holes.
Laser Interferometric Gravitational Wave Observatory (LIGO) and 
Laser Interferometer Space Antenna (LISA) project instruments are
being constructed  to achieve these goals [354, 355].
In binary systems composed of only neutron stars and stellar black holes,
quadrupole radiations from binary systems are found to be adequate [356].
However, around a massive black hole ($M_1$), 
one needs to consider the exchange of
angular momentum of the infalling compact object 
(in a Keplerian orbit) of mass $M_2$ with the
accretion disk, since the disk usually deviates from Keplerian (Fig. 5.7).
The orbital angular momentum loss rate 
due to gravitational wave would be, $ R_{gw}=\frac{dL}{dt}|_{gw}$
In presence of an accretion disk co-planer with the 
orbiting companion, matter from the disk (with local
specific angular momentum $l(r)$, Fig. 5.7) will be accreted onto the companion
at a rate close to its Bondi accretion rate  (eq. 2.5). 
The rate at which angular momentum of the companion will be changed 
due to Bondi accretion from disk will be, $ R_{disk}=\frac{dL}{dt}|_{disk}=
{\dot M_2} (l_{Kep} (x) -l_{disk} (x) ) $
Here, $l_{Kep}$ and $l_{disk}$ are the local Keplerian and disk
angular momenta respectively. 
Consider a special case where, $M_2 <<M_1$ and $l_{disk}<<l_{Kep}$. 
In this case, the ratio $R$ of these two rates is,
$$
R=\frac{R_{disk}}{R_{gw}}=1.518\times 10^{-7} \frac{\rho_{-10}}{{T_{10}}^{3/2}}
{x^4}{M_8}^2
$$
Here, $x$ is the companion orbit radius
in units of the Schwarzschild radius of the primary,
$M_8$ is in units of $10^8 M_\odot$, $\rho_{10}$ is the density in units
of $10^{-10}$ gm cm$^{-3}$ and $T_{10}$ is the temperature of the
disk in units of $10^{10}$K. It is clear that, for instance, at $x=10$,
and $M_8=10$, the ratio $R\sim 0.015$ suggesting the effect of the
disk could provide a significant correction term to the general relativistic
loss of angular momentum. Thus, the gravitational wave emitted
from such a system will also be affected.
In the above example, both the disk and the
gravitational wave work in the same direction in reducing angular
momentum of the secondary. Alternatively,  when $l_{disk} > l_{Kep}$
they act in opposite direction and may slow down the loss of angular 
momentum due to gravity waves [62]. In either case, the ratio $R$ is independent
of the mass of the companion black hole, as long as $M_2 <<M_1$.

(B) Spectral signatures of bulk motion Comptonization close
to a black hole:

As accreting gas comes closer to a black hole, it is accelerated and
soft photons intercepted from Keplerian accretion disk (or, generated
within by synchrotron radiation) are up-scattered by this flow 
[74, 238]. Signature of 
this up-scattered hard component may have been observed in soft-states
of black hole candidates (Figs. 7.4-7.5; [68]). 

Consider a low energy photon diffusing outwards through an accelerated
inflowing electron and getting scattered. The photons within trapping
radius $r_{trap} \sim {\dot M}$ (Section 2) gain energy on an average
and produce a hard tail. Physically, at a given scattering, the 
relative change in the photon energy due to bulk motion is 
$\Delta E/E \sim d(v/c)/d\tau_T \sim 1/{\dot m} $ but the energy 
loss due to recoil effect (Compton scattering) is $\Delta E/E \sim 
-E/m_e c^2$. For very soft photons, $E/m_ec^2 <<1$ the energy gain 
is important and the resulting spectrum is a power law of spectral slope
$\alpha \sim 1.5$. When these 
two effects are comparable, a sharp cut-off is produced at 
$E\sim m_e c^2/{\dot m}$. In the case of neutron star accretion,
the velocity $v$ becomes zero on the star surface, which reduces
the spectral slope to $\alpha\sim 0$.

\newpage
\bibliographystyle{plain}
{}

\newpage
\centerline{\large Figure Captions}

\noindent Fig. 1.1: 
Possible evolutionary routes which produce massive black holes
at galactic centers (reproduced with permission from the Annual
Review of Astronomy and Astrophysics [2]).

\noindent Fig. 1.2: Velocity dispersion $\sigma$ and stellar rotation velocity
$V$ along the major axis and the nuclear bulge component of M31.
Rapid increase of dispersion velocity at $r\sim 0$ indicates the
presence of a massive compact object of at least ten million solar mass.
The Figures are reproduced from Kormendy [8].

\noindent Fig. 1.3: Gradual deformation of a star of mass $1M_\odot$
passing close to a black hole of mass $10^4M_\odot$. Here, $\beta$ is the
ratio of the Roche radius to the periastron distance $r_p$. The
star is totally disrupted when $\beta$ is close to or greater than $1$ [37].

\noindent Fig. 1.4: 
Effective potential felt by an orbiting particle with various specific
angular momenta around a Schwarzschild black hole (solid and short-dashed 
curves) and a Newtonian Star (long-dashed curve). Solid curves are drawn
with marginally stable (lower) and marginally bound (upper) angular
momenta respectively.

\noindent Fig 1.5: Effective potential felt by an orbiting particle with
various specific angular momenta around a Kerr black hole ($a=0.95$). 
Solid curves are drawn with marginally stable (lower) and marginally bound 
(upper) angular momenta respectively.

\noindent Fig. 2.1(a-b): Variation of Mach number with distance
in a Bondi flow around a Newtonian star. (a) $n\!=\!3$, (b) $n\!=\!3/2$. 
Contours are of constant $\cm$. $\ce$ is chosen to be $0.01$ throughout.
In (a), only curves $ABC$ and $A'BC'$ are physically significant
transonic solution.
The other solutions such as $DD'$ or $EE'$ are not transonic.
$FF'$ and $GG'$ are unphysical. In (b), the critical curve 
passes through $r\!=\!0$.

\noindent Fig. 2.2: Various types of critical points are schematically drawn.
A critical point is `center' type or `O'-type if the trajectories in the phase
space are closed around that point. If the trajectories intersect with slopes
of opposite signs, the critical point is `saddle' type, or `X'-type; if
slopes are of same sign it is of `nodal' type. If trajectories spiral 
around the point indefinitely, it is a `spiral' type point.

\noindent Fig. 2.3: Nature of critical points in the parameter space 
spanned by $b$ and $c$ for a harmonic oscillator problem. 
Parabolic curve $D\!=\!b^2-4ac\!=\!0$ separates various types
of critical points (when in $x-{\dot x}$ plane). $R$ 
and $I$ represent real and imaginary axes. Dots drawn in the $R-I$ plane
represent locations of eigenvalue $\lambda$ on that plane. 
For $D\! <\! 0$, points 
are center type or `O'-type if $b\!=\!0$ and spiral type if $b\!\ne\! 0$.
For $D \!=\! 0$, points are inflected nodal type where the slopes of both 
the trajectories are the same. For $D\!>\!0$, points are `nodal' type 
if $c\!>\!0$ (both slopes are of same sign), straight line $c\!=\!0$ 
(one slope is zero), saddle type or `X'-type if $c\! <\! 0$ (slopes are of 
opposite signs). Gravity and viscosity in an accretion flow 
play roles similar to negative stiffness and damping. 

\noindent Fig. 2.4(a-c): Temperature variation of the electrons and 
protons as functions of the logarithmic radial distance. 
The black hole mass is chosen to be $10M_\odot$ and
(a) ${\dot M}=10^{16}$ gm s$^{-1}$, (b-c) ${\dot M}=10^{18}$ gm s$^{-1}$
In (c), the electron proton coupling term is artificially 
increased by a factor $20$ to show the effect of strong cooling.
Figures are reproduced from Colpi, Maraschi \& Treves [95].

\noindent Fig. 2.5(a-d): Spectrum of radiation emitted from electrons 
in a spherical accretion flow at a rate of (a)  $10^{16}$ gm s$^{-1}$
(b) $10^{18}$ gm s$^{-1}$ on a black hole of mass (a-b) $10\ M_\odot$ 
and at a rate of (c)  $10^{23}$ gm s$^{-1}$
(d) $10^{25}$ gm s$^{-1}$ on a black hole of mass (c-d) $10^8\ M_\odot$.
Also shown in the inset is the $\gamma$-ray spectrum due to $\pi^0$ decay.
Figures are reproduced from Colpi, Maraschi \& Treves [95].

\noindent Fig. 2.6: Variation of $L_\gamma$ with the 
accretion rate in a spherical accretion onto a $10\!M_\odot$ black hole. 
The maximum occurs at around $\tau_T \approx 0.3$.
Figures are reproduced from Colpi, Maraschi \& Treves [95].

\noindent Fig. 2.7: A typical accreting solution in optically thick flows
near the critical point is shown in dashed line [77, 89].
The solution of ${\hat L}_s$ (solid curves)
blows off at $u\!=\!a_s$ as $k\!=\!0$ at that point.
A complete solution which satisfies the
condition that it should pass through the point $u\!=\!a_s$ where $C\!=\!0$,
asymptotically merges with the ${\hat L}_s$ curves.

\noindent Fig. 2.8: A complete solution of general relativistic optically thick
Bondi Flow [77, 89]. Cross marks are the
critical points and the open circle marks are the sonic points. 
The parameters chosen in CGS units are $M\!=\! 3 M_\odot$, ${\dot M}\!=\!2.954
\times 10^{-5} M_\odot yr^{-1}$, ${\hat L}_\infty \!=\! 9.795 \times 10^4
L_\odot$, ${\hat L}_{s,\infty}\!=\!9.795 \times 10^4
L_\odot$, $\rho_\infty \!=\! 1.878 \times 10^{-9} g \ cm^{-3}$, $T_\infty\!=\!
4.313 \times 10^5 K$, $a_{s,\infty} \!=\! 8.303 \times 10^{-3} \times c$,
$a_{c,\infty}\!=\! 2.814 \times 10^{-4} \times c$, $[p_{gas}/p_{rad}]_\infty
\!=\!1.532 \times 10^{-3}$.

\noindent Fig. 2.9: A comparison of results of the `self-consistent'
spherical accretion models in the literature.
The dimensionless accretion rate ${\dot m}$ is plotted against 
the dimensionless luminosity. See text for the explanation of the symbols.

\noindent Fig. 2.10: Region of the parameter space showing
stationary spherical accretion solutions which include shock waves [119]
(open circles). Also presented are 
the self-consistent shock solutions of Babul, Ostriker and M\'esz\'aros [121] 
(filled circles). The high and low (or no) pair density regions are as well.

\noindent Fig. 2.11(a-d): Variations of (a) velocity, (b) temperature,
(c) density and (d) Mach numbers in a specific case, clearly showing the 
formation of a standing shock at a few $\times 10^6 r_g$. 
Black hole of $1 M_\odot$, luminosity (in units
of Eddington luminosity) of $0.01$ and efficiency $\eta = 0.003$ are
chosen. Dashed curve in (a) shows free-fall velocity and the
dashed curve in (b) shows the virial temperature for comparison. Figures
are reproduced from Chang \& Ostriker [119]. 

\noindent Fig. 2.12: A cartoon diagram of a plane parallel shock wave.
$u_-$ and $u_+$ are the mean velocities of the scattering centers
in the pre-shock and post-shock regions.

\noindent Fig. 2.13: Variation of the compression ratio $r=\beta_-/ \beta_+$ 
as a function of the Mach number (solid) when losses are included. Also shown
are the ratio $r$ for adiabatic flows with $\gamma=4/3$ (dotted) and 
$\gamma=5/3$ (dashed). The Figure is reproduced from Kazanas \& Ellison [122].

\noindent Fig. 2.14: Variation of the efficiency of conversion of kinetic 
energy flux to the flux of relativistic particles ($Q$), 
and the ratio of downstream relativistic particle pressure to ram pressure
($\eta$) as function of the Mach number $M_1$. 
The Figure is reproduced from Kazanas \& Ellison [122]. 

\noindent Fig. 2.15:
Comparison of luminosities as predicted by the shock acceleration
model (solid lines) as a function of the shock location $X_1$
and the model (open circles for quasars and crosses for Seyferts)
of Wandel \& Yahil [132] as functions of the central black hole mass.
The Figure is reproduced from Kazanas \& Ellison [122].

\noindent Fig. 3.1: Equipotential surfaces
of a compact binary system with mass ratio $M_1/M_2=1/3$. 
Distances are in units of $GM_1/c^2$. Five points of four distinct types 
marked as $L_1$, $L_2$, $L_3$ and $L_4$ are the so-called Lagrange points where 
$\Phi_{eff}$ is locally or globally an extremum. 
Roche lobe overflow occurs when matter from $M_2$
fills its lobe (right section of the figure-of-eight formed by the
innermost contour) and passes through $L_1$ to the star $M_1$ on the left.

\noindent Fig. 3.2: Two annular sections of a thin accretion disk are drawn at
radii $r$ and $r+\Delta r$ to illustrate how matter is accreted
from $r+\Delta r$ to $r$ after angular momentum is transported from $r$
to $r+\Delta r$ through the action of the viscous stress $f_\phi$. The Figure
is reproduced from Shapiro \& Teukolsky [70].

\noindent Fig. 3.3: Integral radiation spectrum of the standard disk for
different accretion rates ${\dot M}$ and viscosity parameters $\alpha$.
Upper curves are for the critical rate ${\dot M}_{crit}=3 \times 10^{-8} 
\frac{0.06}{\eta} M_\odot$ yr$^{-1}$ with $\eta=0.06$ and the lower
curves are for an accretion rate a hundred times smaller.
`BB' refers to the emission of black body radiation at local disk temperature.

\noindent Fig. 3.4: Region of a disk having a negative slope  ($d(\nu \Sigma)
/d\Sigma < 0$) would be viscous unstable and disk accretion takes a limit cycle
behaviour $ABCD$. In $AB$ and $CD$ the slopes are positive and the
disk is stable. 

\noindent Fig. 3.5: Schematic diagram showing the origin of Balbus-Hawley
instability inside a differentially rotating disk in presence of
vertical and toroidal flux tubes.

\noindent Fig. 3.6: 
A few self-consistent angular momentum distributions inside disks which 
are supposed to be Keplerian at large distances. 
Accretion rate ${\dot m}=\frac{\dot M}{\eta {\dot M}_{Edd}}=50$ is
chosen. The distribution depends upon the radial distance where 
the disk is matched with a Keplerian disk. The Figure is taken from
Abramowicz et al. [146].

\noindent Fig. 3.7: Meridional cross section of a self-similar disk
(long dashed), the Alfv{\'e}n surface (short dashed) and the field lines
and associated outflows (solid) in the $\psi,\ \xi$ plane for the 
parameters: $n=0.5$, $C_1=4 \times 10^{13}$, $r_0=3 \times 10^{13}$, 
$\rho_{i0}=10^{-16}$gm cm$^{-3}$, $B_0=0.019$ G, $\alpha=9.0$, $\beta=9.0$, 
and $\epsilon=0.1$.  The field lines are advected with the flow inside the
disk, pushed away by centrifugal force of the outgoing matter just outside
the disk and subsequently collimated due to `hoop' stress [150].

\noindent Fig. 3.8(a-b):
Calculated poloidal fields $B_p$ (solid lines) and currents $j_P$ (dashed lines)
with (a) even and (b) odd symmetries inside a disk. These are examples
of solutions which do not require dynamo if sufficient fields are fed
from outer edge. They can probably match jet solutions outside the disk
as is schematically shown here. The Figure is reproduced from Khanna \&
Camenzind [157].

\noindent Fig. 3.9:
Poloidal flux $\Psi$ at $t=4.6$ times the diffusion time scale as obtained
from self-excited dynamo effects around a Kerr black hole ($a=0.98$). 
The box represents the disk. These solutions may match with 
jet solutions outside the disk. 
The Figure is reproduced from Khanna \& Camenzind [158]

\noindent Fig. 3.10: Contours of constant ${\cal E}$
and $L$ in the $(r_c, u_c)$ plane showing the existence of five critical
points in a magnetized flow around a Schwarzschild black hole. Solid 
contours are drawn for $L\!=\! 1.2, \  1.8,\ 2.2345  ,\ 3.0 ,$ and the
dashed contours are drawn for ${\cal E}\!=\!1.2  ,\ 1.8 , \  2.355,
\ 2.8$ respectively [77, 161].

\noindent 3.11(a-f): Contours of constant ${\dot{\cal M}}$
depicting all the solution topologies of the  magnetized disks and winds
[77, 161]. (a) $L\!=\!10^{-4}$, ${\cal E}\!=\!2.0$,
(b) $L\!=\!1.0$, ${\cal E}\!=\!-0.5$,
(c) $L\!=\!1.0$, ${\cal E}\!=\!1.8871$,
(d) $L\!=\!1.7$, ${\cal E}\!=\!-2.0$, 
(e) $L\!=\!1.7$, ${\cal E}\!=\!0.0$, 
(f) $L\!=\!1.7$, ${\cal E}\!=\!4.0$. 
Important point to note here is 
that the fast magnetosonic point is located at a finite distance (c)
which causes outflows to be superfast supermagnetosonic at a finite
distance. This property is  important for magnetized jet solutions.

\noindent Fig. 3.12: Schematic diagram showing the Blandford-Znajek
process of extraction of energy from the black hole. See text
for details. The Figure is reproduced from Blandford \& Znajek [168].

\noindent Fig. 4.1(a-b): Variation of angular velocity $\omega$
and radial velocity $v=v_r$ in units of the Keplerian angular velocity
($\omega_K=(GM/r_0^3)^{1/2}$) and the free-fall velocity $v_K=(GM/r_0)^{1/2}$ 
at the inner edge of the disk $r_0$.
(a) ${\dot M}_c = L_{Edd}/\eta$ with $\eta=GM/2r_0$, the efficiency factor. 
(b) ${\dot M}_c$ is $25$ times higher. The dashed curves are Keplerian
for comparison. The Figures are reproduced from Maraschi et al. [171].

\noindent Fig. 4.2: Schematic representation of the typical angular
momentum distribution inside a thick accretion disk around a black hole
(dashed). Keplerian distributions around a Schwarzschild black hole (solid)
and around a Newtonian star (dash dotted) are shown for comparison. Inner edge
of the disk is located at $r_i$ and the center of the disk is located at $r_c$.

\noindent Fig. 4.3:
An equipotential surface of a barotropic thick accretion disk is schematically
shown along with the chimney (or funnel) along the axis. The directions of the
gravitational force, centrifugal force, the net force along the effective
gravity, and the force due to pressure gradient are shown with arrows.
An equipotential surface is formed by the tangent vectors normal to the
directions of the pressure balance.

\noindent Fig. 4.4(a-b): Equipotential surfaces in Schwarzschild
geometry (a) and in Newtonian geometry (b) are shown. $I$ 
and $C$ denote the inner edge (cusp) and the center of the disk respectively.
In Newtonian geometry, the equipotentials are always closed and no cusp 
is formed.

\noindent  Fig. 4.5: Central temperature and density of a thick accretion 
disk [185] as functions of the ratio of the gas pressure to the total pressure
$\beta$. In abscissa, mass of the central black hole is shown. In the region
marked $\tau <1$, the optical depth (due to Thomson scattering) is less than
unity and the radiation pressure supported disk is
impossible. In the region marked `self-gravity',
the mass of the disk is higher compared to the mass of the central black hole
and various instabilities may develop in the disk.

\noindent Fig. 4.6: Keplerian distribution of an accretion disk
when the mass of the disk is three percent (short dashed), five percent
(medium dashed) and ten percent (long dashed) as heavy as the central 
compact object [188]. Here, the entire mass is assumed to be concentrated
close to the disk center located at $r_c \sim 9.0$.

\noindent Fig. 4.7: 
Temperature distributions in a thin disk (dotted curve) and in thick
disk (dashed curve) are compared for the same central 
mass and accretion rate. This Figure is reproduced from Madau [193]. 

\noindent Fig. 4.8:
Disk spectra at inclination angles (a) $90^o$, (b) $50^o$,
(c) $25^o$ and (d) $0^o$ are compared showing the obscuring effects
of the funnel. This Figure is reproduced from Madau [193]. 

\noindent Fig. 4.9: Specific luminosities at various observed frequencies
are compared:
$log \, (\nu)$ is equal to (a) $15.5$, (b) $16.0$, (c) $16.5$ and (d) $17$
respectively. In optical and ultraviolet, the specific luminosity is almost
independent of the angle, but at higher frequencies the dependence is
very strong due to the funnel. This Figure is reproduced from Madau [193]. 

\noindent Fig. 4.10: Comparison of the thin disk spectrum (dashed curve) 
with the spectra of a thick disk: the spectrum without the reflection
effects (solid), with reflection effects (dotted), with a sum-of-black-body 
assumption (dash-dotted). Compared to a thin disk, the thick disk (with 
reflection effects) produces orders of magnitude more emission at higher 
frequencies. This Figure is reproduced from Madau [193]. 

\noindent Fig. 5.1:
Variation of the energy $\ce$ with the accretion rate $\cm$ 
(a measure of entropy for a given mass flux) when
the angular momentum is kept fixed and is equal to $1.675$.
Flow with parameters from the segment $AB$ passes through the innermost
`X'-type critical point and flow with parameters from the segment
$CD$ passes through the outermost `X'-type critical point.
Typical Rankine-Hugoniot shock transitions in accretion ($s_1 s_2$) and in
winds ($w_1w_2$) are shown [77, 218].
The ``swallow-tail" singularity is produced due to the projection effects.

\noindent Fig. 5.2:
Example of the solution of an accretion flow. 
At the sonic point, the Mach number is $0.93$.
Contours are of constant accretion rate ${\dot{\cal M}}$.
Shock locations are: $x_{s_1}\!=\!2.437, \ x_{s_2}\!=\!3.419, \
x_{s_3}\!=\!18.862, \ x_{s_4}\!=\!57.592$. The inner and the outer critical
points, denoted by $I$ and $O$, are located at
$x_i\!=\!2.800$ and $x_o\!=\!34.580$ respectively. The solution which includes
a shock is shown by arrows [77, 218]. The shock transition at $x_{s2}$ is
unstable. $X_{s1}$ is possible of neutron stars. $x_{s4}$ is unphysical. 

\noindent Fig. 5.3(a-d): Phase space trajectories and the shock locations 
in a viscous isothermal disk. (a) $\alpha\!=\!0.01$, and 
(b) $\alpha\!=\!0.02$. (c-d): Plot of $M$ and $\frac{1}{M}$ as functions
of the logarithmic distance to show the shock locations in a dissipative
flow. Parameters are the same as in (a) and (b) respectively. In (c),
there are two shock locations, the one close to the black hole
is unstable in accretion and the one away from the black hole is stable. 
In (d), when the viscosity is higher, only the unstable one remains [77, 224].
Also shown in (d), two shock-free branches of the solution out of which
the flow chooses the one passing through the inner sonic point.

\noindent Fig. 5.4(a-b): Classification of various disks around a
black hole into a single framework [218]. Regions I,W,A,O
respectively describe the parameter space for which (I)
the flow passes through 
only the inner sonic point, (W) may have shocks in winds but no-shock
in accretion, (A) may have shocks in accretion but no-shock in winds, and 
(O) only through outer sonic points. In (b), A and W are further 
classified. Region between long dashed curve and central curve BF
will have shocks and the post-shock region will lie between the dotted
curve and the central solid curve BF [218].

\noindent Fig. 5.5 Energy vs. entropy density is plotted for $\lambda_{in}
=1.65$, viscosity parameter $\alpha=0,\  0.4,\  0.8$ and 
cooling parameter $f=(Q_+-Q_-)/Q_+=0,\ 1$
(as marked) and for polytropic index $\gamma=4/3$ (left) and $5/3$ (right,
moved towards left by a factor of $10$) 
respectively at all possible sonic points.
The branches of type $AMB$ pass through inner sonic point $x_{in}$ and the
branches of type $CMD$ pass through outer sonic point $x_{out}$ and a
solution with a shock must connect two such points with the condition that
$E(x_{in})\leq E(x_{out})$ along with $s(x_{in})\geq s(x_{out})$ [226].

\noindent Fig. 5.6: Example of a shock solution in presence of viscosity,
heating and cooling. Here $Q_-/Q_+=0.5,\ \alpha=0.05$ and angular
momentum at the inner edge $1.6$ is chosen. The vertical dashed curve
at $X_{s3}=13.9$ represents the stable shock. $\alpha_{c}=0.1$ and $\gamma=4/3$
in this case.

\noindent Fig. 5.7: Ratio of the disk angular momentum to Keplerian
angular momentum in three different cases. In case marked `A', the flow becomes
sub-Keplerian before becoming super-Keplerian close to the black hole.
In case marked `B', the disk becomes sub-Keplerian and remained so
before entering the black hole. In case marked `C', the disk
becomes Keplerian, then passed through a standing shock  at $r=13.9r_g$
before entering the black hole [68].
 
\noindent Fig. 5.8: Examples of shock solution (logarithmic Mach number against
radial distance) in presence of power
law cooling effects: $\Lambda \sim \rho^2 T^\alpha_b$. The solid curve is
for $\alpha_b=0.5$ and the long-dashed curve is for $\alpha_b=0.6$.
The short-dashed curve is without cooling effects. Higher cooling produces 
weaker shocks closer to the black hole.

\noindent Fig. 5.9: Examples of MHD shock formation in accretion.
Solutions are shown in arrows and shock transitions are shown 
in vertical dashed lines. The parameters are:
$L\!=\!1.45$, ${\cal E}\!=\!0.2$, ${\dot{\cal M}}_-\!=\!0.4857$,
${\dot{\cal M}}_+\!=\! 0.5074$, $r_s\!=\!1.0995$, $r_s$ being the location 
of the shock. The dashed curves are the contours of constant
slow magnetosonic wave velocities [77, 161]. 

\noindent Fig. 5.10: Relative positions of two spiral shocks (thick 
solid curves) at $\psi_{s1}$ and $\psi_{s2}$ and two sonic surfaces 
(thin dashed curves) at $\psi_{c1}$ and $\psi_{c2}$ are schematically shown. 
The shocks as well as the sonic surfaces are separated by
$\delta \psi\!=\! \pi$ ($n_s\!=\!2$), and $0 \leq \epsilon \leq 1$ 
measures the position of the shock relative to the sonic surface downstream
[77, 248].

\noindent Fig. 5.11(a-b): Example of a two-armed spiral shock
solution. (a) Variation of Mach number with the spiral angle
at a given radial distance. Dashed vertical lines (at $\psi_{s1}$ and 
$\psi_{s2}$) are the shock transitions.
Crosses represent locations of the sonic surfaces 
(at $\psi_{c1}$ and $\psi_{c2}$). (b) Variation of the velocity
coefficients $q_j $($\times 10$) with the spiral angle. The parameters are:
$B\!=\!-1.0, \ q_{2c}\!=\!0.02, \ n\!=\!6.1840, \ \epsilon\!=\!0.7274, 
\ M_-\!=\!1.0394, \ M_+\!=\!0.8909$ [77, 248]. The flow is seen to be
highly sub-Keplerian (for Keplerian, $q_2=1$).

\noindent Fig. 6.1: Density contours  at different times
during the accretion of nearly isothermal winds ($\gamma=1.005$).
The shock cone is seen to exhibit `flip-flop' property before settling down
in the symmetry axis. The Figure is reproduced from Matsuda et al. [254].

\noindent Fig. 6.2: An example of early simulation 
of an inviscid, constant angular momentum ($l=3.77GM/c$) 
flow showing the formation of a thick accretion disk. The contours of constant 
density and the velocity vectors are shown. The figure is drawn at 
$t=300 GM/c^3$  and is taken from Hawley, Smarr and Wilson [216].  

\noindent Fig. 6.3: An example of the simulation of a thin accretion disk
which includes a standing shock wave [220].
Mach number of the flow is plotted against the radial distance.
Results at different times are shown in solid curves. Analytical
solution is shown as a dashed curve with vertical shock transition.
The flow forms the shock at the predicted location.

\noindent Fig. 6.4: Pseudo-particles in a smoothed particle hydrodynamics
simulation of a two-dimensional sub-Keplerian angular momentum flow. 
A standing shock is formed at $X\sim 16$. 
Note the presence of the oblique shocks also. The specific energy and 
angular momentum used were $0.006 c^2$ and $3.3 GM/c$ respectively.
There are $60,000$ particles in this simulation [259].

\noindent Fig. 6.5: Contours of constant Mach number in the two dimensional
flow. Note that the flow
which is subsonic after the shock, becomes supersonic very close to the 
black hole. Also, the wind originated subsonically on the disk surface
becomes supersonic farther away [259].

\noindent Fig. 6.6: Contours of constant temperature (in geometric 
units). At the shock location, the temperature of the flow becomes high and the
velocity becomes very low thus satisfying all the conditions of a
thick accretion disk [75]. The contours in the immediate vicinity of the 
post-shock region resemble that of a thick accretion disk (Fig. 4.4).
The pre-shock flow is primarily advected like Bondi flow
whereas the post-shock flow is rotation dominated [259].

\noindent Fig. 6.7(a-b): Examples of the spiral 
shock waves in early numerical simulations. Density contours are plotted. 
The ratio ($Q$) of the compact object to the companion is chosen to be unity.
In (a), $\gamma=1.2$ and a two armed spiral is formed, whereas in (b),
$\gamma=5/3$, a three-armed spiral is formed.
Number of arms depends on boundary conditions of the problem.
Matter is injected at the inner Lagrange point $L_1$. These figures are
taken from Sawada, Matsuda \& Hachisu [142].

\noindent Fig. 6.8(a-c): Evolution of the density waves for $Q=1$ and
$\gamma=1.2$ case at times (a) $T=2.3$, (b) $3.2$, and (c) $3.7$ respectively.
This case is similar to that shown in Fig. 6.7(a) above 
but the injected gas is somewhat cooler and matter is injected all over 
the outer boundary (at $0.65$). 
The number of arms changes from $2$ (a) to $3$ (b) to $2$ (c). 
In the same time unit, Keplerian periods at a radial distance $\sim 0.2$
(where the shocks are fully developed) and at the outer boundary
are $0.56$ and $3.3$ respectively.

\noindent Fig. 6.9(a-b): Comparison of (a) Mach number and 
(b) angular momentum variations in viscous (solid) and inviscid (dashed)
isothermal thin accretion disks. The viscosity parameter 
$\alpha_s=0.01$ is chosen everywhere in the simulation.
Note that the shock in the viscous disk forms farther out and is weaker
and wider. The solid curve in (b) marked `Keplerian'
is the Keplerian distribution
plotted for comparison.  Due to the inefficiency of transfer of
angular momentum in the pre-shock flow, a mixed-shock forms with
higher angular momentum in the post-shock flow [226].

\noindent Fig. 6.10(a-b): Evolution of (a) Mach number and (b) angular momentum
when viscosity parameter is high ($\alpha_s=0.1$). Higher efficiency of 
transport of angular momentum in the post-shock region causes its accumulation
in the post-shock region which  drives the shock outwards. This makes the
entire flow outside the inner sonic point to be subsonic, 
and the angular momentum distribution becomes almost Keplerian. Dashed 
curve marked $\alpha=0$ represents the solution for inviscid flows [225].
In (b), Keplerian distribution is shown for comparison.

\noindent Fig 6.11: Formation of an isothermal Keplerian disk.
In this simulation, particles are injected at the outer edge of the disk
at $r_{o}=100$ with angular momentum $7.00$ which is close to 
Keplerian value at $r_o$. The (constant) sound speed
is chosen to be $a_0=0.005$ and the injection velocity was $v_0=0.003$.
Shakura-Sunyaev viscosity parameter $\alpha_s=0.25$ was used in this
simulation [226]. The dashed Keplerian distribution is shown for comparison.

\noindent Fig. 6.12 Evolution of a sub-critical, initially radiation
pressure supported, Keplerian disk. Density contours
of the simulation at different times are shown in different panels
in which the time $t$ (since the commencement of the simulation) is
also noted. Length and time units are $GM/c^2$ and $GM/c^3$ respectively.
In the first panel, the disk is radiation pressure supported, but 
subsequently it becomes cooler, dense and gas pressure supported.
The Figure is reproduced from Eggum, Coroniti and Katz [263].

\noindent Fig. 6.13: Evolution of a super-critical viscous flow which 
produces a geometrically thick disk. ${\dot M}= 4 {\dot M}_{Edd}$ was used.
Density contours and unit velocity vectors are shown at $t\sim 6000$ unit
after the commencement of the simulation. Contours are labeled with 
densities in CGS units. Four regions: (A) a convective core,
(B) an accretion zone, (C) a photocone and (D) a jet 
are distinguished and marked. The Figure is reproduced from Eggum,
Coroniti and Katz [264].

\noindent Fig. 6.14(a-b): Example of `breathing' of accretion disks
in presence of bremsstrahlung cooling 
and accretion shock. The post-shock hot corona
cools in a time scale comparable to the in-fall time and the shock 
oscillates [227]. The figures show two different epochs of this
oscillation. Upper half shows velocity vectors and Mach number contours
and the lower half shows the velocity contours.

\noindent Fig. 7.1: Examples of fitting optical/UV region 
of nine typical quasars using emissions from a thin standard Keplerian disk, 
and contributions from the dust (modeled as a power law in infra-red region),
from starlight, from re-combination lines and Fe II lines. This Figure
is reproduced from Malkan [277]. 

\noindent Fig. 7.2(a-b): Fits of hard X-rays and $\gamma$-ray data
with the Comptonization models: (a) EXOSAT, GRANAT and OSSE observation 
of Cyg X-1 [290] and (b) OSSE and Ginga data of NGC4151 
[291-292]. In (a), histogram
is the result of [288], dashed curve is the result of [289], solid curve
is the analytical solution [287], the dash-dot-dot-dot curve is from
[285]. In (b), the histogram is from [287]. The Figures are reproduced
from Titarchuk [287]. The thermal model provides sufficiently good fit
in both of these cases.

\noindent 7.3: A possible model of the accretion flow which
is produced by either (a) a Keplerian/sub-Keplerian flow with
vertical viscosity variation, or (b) an admixture of 
Keplerian and sub-Keplerian flow at the outer boundary. Soft emission
comes from the Keplerian equatorial disk, whereas the X-rays to $\gamma$-rays
are from the shocked component of the halo which cools by re-processing
soft photons from the standard disk. Existence of various components
will depend upon the accretion rate, viscosity and angular momentum [296, 297].
 
\noindent Fig. 7.4:
Variation of the spectral properties as the accretion rate
of the disk is changed: ${\dot m}_d =0.001$ (solid), ${\dot m}_d=0.01$ (long
dashed), ${\dot m}_d=0.1$, (short dashed) and ${\dot m}_d=1.0$
(dotted). The accretion rate in the sub-Keplerian  halo component is
kept fixed at ${\dot m}_h =1.0$. The hard component softens 
and the soft component brightens with ${\dot m}_d$.
This figure is drawn for $M=5M_\odot$ but should remain similar
for any black hole candidate [68, 297]. The dash-dotted curve over-laid
on the dotted curve represents the weaker hard component due to
the convergent inflow near the black hole horizon (Newtonian computation).

\noindent Fig. 7.5: Variation of the energy index $\alpha$
(of the hard component between $\sim 2-50$keV) as the accretion
rate of the disk ${\dot m}_d$ is changed. Halo rates are
marked. Plots on the left-side are 
due to post-shock hot flow, and those on the right side are
due post-shock flow cooled by high disk accretion rate. Dashed
curves are drawn where the effects of both the components
may be present [68, 297].

\noindent Fig. 7.6(a-b): Fits of broad line emission features from ARP 102B
using a relativistic Keplerian disk model. In (a), the continuous
emissivity is used and in (b) local broadening is used. Figures
are reproduced from Chen \& Helpern [319].

\noindent Fig. 7.7: Fit of 3C390.3 is presented using relativistic
Keplerian disk model. This Figure is taken from Eracleous \& Helpern
[321]. Lower panel shows the residual spectrum.

\noindent Fig. 7.8(a-b): Fits (dashed) of the $H_\alpha$ data (solid)
of (a) 1976 and of (b) 1980. The parameters are:
$x_1 = 200 GM/c^2$, $x_2 = 900 GM/c^2$, $q=2.0$, $i=38^o$, 
$v_{th}=0.0015c$ [212].

\noindent Fig. 7.9: Spatial variation of the
line emission profiles from the ionized disk of M87
taken by the Faint Object Camera (FOS) of Hubble space telescope. This Figure
is reproduced from Harms et al. [5].
The vertical dashed lines are drawn at positions [O III]$\lambda\lambda 
5007$\AA\ , [N II] $\lambda\lambda  6584$\AA\  to show that
red and blue shifts of lines at different positions.

\noindent Fig. 7.10: Fits (solid) of the [O III]$\lambda\lambda 5007$\AA\ , 
lines (short dashed) emitted from different positions. The
same theoretical fit, upon suitable scaling, matches with 
observed [N II] $\lambda\lambda  6584$\AA\  profile (dot-dashed)
as well. The good fit indicates that the flow may be strongly 
non-Keplerian [6].

\noindent Fig. 7.11: Various identified structures
and mass distribution within $100$pc of our galactic center. 

\noindent Fig. 7.12: Fit of the observed luminosity [343]
using magnetized spherical accretion model [342]. 
Emission dominates the spectrum for $\nu_\infty \leq 10^{13}$Hz and 
luminosity at higher frequencies are entirely due to thermal bremsstrahlung. 
Mass of the central black hole is chosen to be $0.9 \times 10^6 M_\odot$. 
The Figure  is reproduced from Melia [342].

\end{document}